\documentclass[a4paper,11pt]{article}
\pdfoutput=1
\usepackage{jheppub}

\usepackage[utf8]{inputenc}
\usepackage[T1]{fontenc}
\usepackage{lmodern}
\usepackage{textcomp}
\usepackage{microtype}

\usepackage[english]{babel}
\usepackage[autostyle, english = american]{csquotes}
\MakeOuterQuote{"}

\usepackage{amsmath}
\usepackage{mathtools}
\usepackage{upgreek}
\usepackage{units}
\usepackage{slashed}
\usepackage{physics}

\usepackage{authblk}
\usepackage[utf8]{inputenc}

\usepackage{subcaption}
\usepackage{booktabs}
\usepackage{multirow}
\usepackage{graphicx}
\usepackage{adjustbox}
\usepackage{xcolor}

\usepackage{rotating}
\usepackage{lineno}

\graphicspath{{.}}

\definecolor{darkblue}{rgb}{0.,0.,0.4}
\definecolor{darkred}{rgb}{0.5,0.,0.}
\RequirePackage{doi}

\begin{document}


\begin{titlepage}

\pagenumbering{roman}

\vspace*{-2.5cm}
\centerline{\large EUROPEAN ORGANIZATION FOR NUCLEAR RESEARCH (CERN)}
\vspace*{0.2cm}
\hspace*{-15mm}
\begin{tabular*}{16.5cm}{lc@{\extracolsep{\fill}}r}
  \vspace*{-10mm}
    \includegraphics[width=0.10\linewidth]{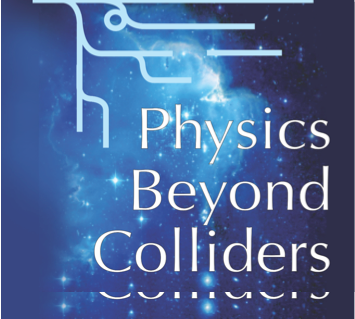} & & \\
 & & CERN-PBC-REPORT-2018-007 \\  
 & & \\
\hline
\end{tabular*}

\vspace*{3.0cm}

{\bf\boldmath\Large
  \begin{center}
    Physics Beyond Colliders at CERN \\
    Beyond the Standard Model Working Group Report 
\end{center}
}

\vspace*{1cm}
\begin{center}

\vskip 2mm
J.~Beacham$^{1}$, C.~Burrage$^{2,*}$,
D.~Curtin$^{3}$, A. De~Roeck$^{4}$, J.~Evans$^{5}$, J.~L.~Feng$^{6}$, C.~Gatto$^{7}$, S.~Gninenko$^{8}$, A.~Hartin$^{9}$,
I.~Irastorza$^{10}$, J. Jaeckel$^{11}$, K. Jungmann$^{12,*}$,  K.~Kirch$^{13,*}$, F.~Kling$^{6}$,
S.~Knapen$^{14}$, M.~Lamont$^4$, G.~Lanfranchi$^{4,15,*,**}$,
C.~Lazzeroni$^{16}$, A.~Lindner$^{17}$, F.~Martinez-Vidal$^{18}$, M.~Moulson$^{15}$, N.~Neri$^{19}$,
M.~Papucci$^{4,20}$, I.~Pedraza$^{21}$, K.~Petridis$^{22}$, M.~Pospelov$^{23,*}$,
A.~Rozanov$^{24,*}$, G.~Ruoso$^{25,*}$, P.~Schuster$^{26}$, Y.~Semertzidis$^{27}$, T.~Spadaro$^{15}$, C.~Vall\'ee$^{24}$,
and G.~Wilkinson$^{28}$.

\end{center}


\vspace*{2cm}
\begin{abstract}
  \noindent
  {\bf Abstract:} The Physics Beyond Colliders initiative is an exploratory study aimed at exploiting the full scientific potential
  of the CERN's accelerator complex and scientific infrastructures through projects complementary to the LHC and other
  possible future colliders. These projects will target fundamental physics questions in modern particle physics.
  This document presents the status of the proposals presented in the framework of the Beyond Standard Model
  physics working group,  and explore their physics reach and the impact that CERN could have in the next 10-20 years
  on the international landscape.
\end{abstract}

\vspace*{3.0cm}
\noindent
    { \small $^{*}$ PBC-BSM Coordinators and Editors of this Report}\\
    {\small $^{**}$ Corresponding Author: Gaia.Lanfranchi@lnf.infn.it}

\clearpage
\bigskip
    {\it \footnotesize
      \noindent
      $^{1}$ Ohio State University, Columbus OH, United States of America\\
      $^{2}$ University of Nottingham, Nottingham, United Kingdom \\
      $^{3}$ Department of Physics, University of Toronto, Toronto, Ontario M5S 1A7 Canada \\
      $^{4}$ European Organization for Nuclear Research (CERN), Geneva, Switzerland \\
      $^{5}$ Department of Physics, University of Cincinnati, Cincinnati, Ohio 45221, USA \\
      $^{6}$ Department of Physics and Astronomy, University of California, Irvine, CA 92697-4575, USA\\      
      $^{7}$ Sezione di Napoli, INFN, Napoli (Italy) and Northern Illinois University (US)\\
      $^{8}$ Institute for Nuclear Research of the Russian Academy of Sciences, 117312 Moscow, Russia \\
      $^{9}$ University College London, Gower Street, London WC1E 6BT, UK\\
      $^{10}$ Grupo de F\'{\i}sica Nuclear y Altas Energ\'{\i}as, Universidad de Zaragoza, Zaragoza, Spain \\      
      $^{11}$ Theory Institute of University of Heidelberg, Heidelberg, Germany \\      
      $^{12}$ VSI (Van Swinderen Institute), University of Groningen, Groningen, the Netherlands\\      
      $^{13}$ ETH Zurich and Paul Scherrer Institute, Villigen, Switzerland \\
      $^{14}$ Institute for Advanced Study, Princeton, NJ, USA \\      
      $^{15}$ Laboratori Nazionali di Frascati, INFN, Frascati (Rome), Italy\\
      $^{16}$ University of Birmingham, Birmingham, United Kingdom\\
      $^{17}$ DESY, Hamburg, Germany\\
      $^{18}$ IFIC/University of Valencia-CSIC, Valencia, Spain \\
      $^{19}$ NFN Sezione di Milano and Universit\`a di Milano, Milano, Italy\\
      $^{20}$ Lawrence Berkeley National Laboratory and UC Berkeley, Berkeley, CA \\
      $^{21}$ Benemerita Universidad Autonoma de Puebla, Mexico\\      
      $^{22}$ University of Bristol, Bristol, United Kingdom \\      
      $^{23}$ Perimeter Institute, Waterloo and University of Victoria, Victoria, Canada \\      
      $^{24}$ CPPM, CNRS-IN2P3 and Aix-Marseille University, Marseille, France \\
      $^{25}$ Laboratori Nazionali di Legnaro, INFN, Legnaro, Italy \\
      $^{26}$ SLAC National Accelerator Laboratory, Menlo Park, CA 94025, USA \\
      $^{27}$ KAIST/IBS, Daejeon, Korea \\
      $^{28}$ University of Oxford, Oxford, UK \\
}

\end{titlepage}

\pagestyle{empty}  

\tableofcontents

\cleardoublepage
\pagestyle{plain} 
\setcounter{page}{1}
\pagenumbering{arabic}

\clearpage
\noindent{\large \bf Executive Summary}

\vskip 2mm
The main goal of this document follows very closely the mandate of the Physics Beyond Colliders (PBC) study group,
and is {\it  ``an exploratory study aimed at exploiting the full scientific potential of CERN's
  accelerator complex and its scientific infrastructure through projects complementary to the LHC, HL-LHC and
  other possible future colliders. These projects would target fundamental physics questions that are similar
  in spirit to those addressed by high-energy colliders, but that require different types of beams and
  experiments\footnote{See https://pbc.web.cern.ch }}''.

\vskip 2mm
Fundamental questions in modern particle physics as the origin of the neutrino masses and oscillations,
the nature of Dark Matter and the explanation of the mechanism that drives the baryogenesis
are still open today and do require an answer.

\vskip 2mm
So far an unambiguous signal of New Physics (NP) from direct searches at the Large Hadron Collider (LHC),
indirect searches in flavour physics
and direct detection Dark Matter experiments is absent. Moreover, theory provides no clear guidance on the NP scale.
This imposes today, more than ever, a broadening of the experimental effort in the quest for NP.
We need to explore different ranges of interaction strengths and masses with respect to what is already
covered by existing or planned initiatives.

\vskip 2mm
Low-mass 
and very-weakly coupled particles represent an attractive possibility, theoretically and phenomenologically well motivated,
but currently poorly explored: a systematic investigation should be pursued in the next decades both at accelerator-based experiments
and with proposals aiming at detecting axions and axion-like particles with terrestrial experiments.

Very high energy scales ($\sim 100$~TeV)
will never be directly reachable with colliders that exist now or in any foreseeable future
and can be explored only using extremely rare or forbidden decays
as probe to the NP in the multi-TeV range. Electric dipole moments are simultaneously probes of NP in the
extremely low ($< 10^{-15}$~eV) and in the very large ($> 100$~TeV) mass scale range.

\vskip 2mm
The CERN laboratory could offer an unprecendented variety of high-intensity, high-energy beams
and scientific infrastructures that could be exploited to this endevour. This effort would nicely complement and further broaden the
already rich physics programme ongoing at the LHC and HL-LHC.

\vskip 2mm
This document presents the status of the proposals presented in the framework of the PBC Beyond the Standard Model (BSM)
physics working group, and explore their physics reach and the consequent
impact that CERN could have in the next 10-20 years on the international landscape.

\clearpage
\section{Introduction}

The Physics Beyond Colliders BSM study group has considered about 15 different proposals aiming at exploiting the CERN
accelerator complex and scientific infrastructure. These proposals will be sensitive to New Physics in a range of
masses and couplings unaccessible to other existing or planned initiatives in the world, as the experiments at
the Large Hadron Collider (LHC) or at a Future Circular Collider (FCC), Dark Matter (DM) direct detection experiments
and flavor physics initiatives.

\vskip 2mm
This document focusses on the searches for Physics Beyond the Standard Model (BSM) also known as NP.
It introduces the physics motivations and the complementarity of the proposals presented within the PBC-BSM
activity with respect to the LHC and other initiatives in the world in the quest for NP.
NP is required to answer open questions in modern particle physics, as the origin of the neutrinos masses and oscillations,
the baryogenesis and the nature of Dark Matter.
A viable possibility is so called {\it hidden sector physics}, that comprises new particles with masses below the electro-weak (EW)
scale that couple very weakly to the Standard Model (SM) world via {\it portals}.
Another viable possibility is that NP is well above the EW scale (and therefore well beyond the direct
reach of the LHC and any other future high-energy collider), and can be only probed via indirect effects
in extremely rare or forbidden processes in the SM or by testing the presence of electric dipole moments (EDMs)
either in elementary particles (such as proton and deuteron) or in more complex systems.

\vskip 2mm
Three main categories of experiments have been identified, following the NP mass range they are sensitive to:
\begin{itemize}
\item[-] {\it experiments sensitive to NP with mass in the sub-eV range and very weakly coupled to SM particles:}
these are mostly experiments searching for axions and axion-like particles using a large variety of experimental techniques;
\item[-] {\it experiments sensitive to NP with mass in the MeV-GeV range and very weakly coupled to SM particles:}
these are accelerator-based experiments that could exploit the large variety of high-intensity high-energy beams currently available or proposed at CERN;
\item[-] {\it experiments sensitive to NP with mass in the multi-TeV range and strongly coupled to SM particles:}
these are experiments searching for extremely rare or forbidden processes, that could be produced via high-intensity beams.
\end{itemize}

The document is organized as follows.
Section~\ref{sec:phys_mot} presents a brief summary of the main physics motivations. In particular in Section~\ref{sec:portals}
it discusses in detail portals to a hidden sector along with a set of benchmark cases that have been identified
as theoretically and phenomenologically motivated target areas to explore the physics reach of the PBC proposals
and put them in the world wide landscape. The proposals presented in the framework of the PBC-BSM study group
are briefly described in Section~\ref{sec:exps} and classified in terms of their sensitivity to a given mass range and to a
set of benchmark cases. A more detailed description is then given in Sections~\ref{ssec:exps_sub_eV}-\ref{ssec:exps_multi_TeV}
ordered along the identified main mass ranges.
The physics reach of the PBC-BSM proposals is shown in Sections~\ref{sec:phys-reach}-\ref{sec:phys-reach-multi-TeV}
along with the current status of these searches at ongoing and/or planned initiatives in the world that are
or will be important players on the same timescale of the PBC proposals.
Brief conclusions are drawn in Section~\ref{sec:conclusion}.

\clearpage
  \section{Physics Motivations}
  \label{sec:phys_mot}
  \noindent
With the discovery at the LHC of the Higgs boson~\cite{Aad:2012tfa,Chatrchyan:2012xdj},
the last missing piece for the experimental validation
of the SM is now in place. An additional LHC result of great importance is that a large
new territory has been explored and no unambiguous signal of NP has been found so far. These
results, together with several constraints from flavor phenomenology and the absence of any charged lepton
flavour violation process, indicate that there might be no NP with a direct and sizeable coupling to SM particles
up to energies $\sim 10^5$ TeV unless specific flavour structures/symmetries are postulated.

\vskip 2mm
The possibility that the SM holds well beyond the electroweak (EW) scale must now seriously
considered. The SM theory is renormalizable and predictive and the measured masses of the Higgs boson and
the top quark fall into a narrow region of parameters where consistency of the SM does not require new
particles up to a very high energy scale, possibly all the way up to the
Planck scale~\cite{Degrassi:2012ry,Buttazzo:2013uya,Bezrukov:2012sa}.
However some yet unknown particles or interactions are required to explain a number of observed phenomena in particle
physics, astrophysics, and cosmology as the neutrino masses and oscillations, the baryon asymmetry of the universe, the
Dark Matter (DM) and the cosmological inflation.

\begin{itemize}
\item[-] {\it Neutrino oscillations} \\
Propagating neutrinos have been seen to oscillate between different flavours.
This implies the existence of a neutrino mass matrix which differentiates the flavour
eigenstates from the mass eigenstates.
This is absent in the SM. It is, additionally, challenging to explain
why the observed neutrino masses are so much smaller than the masses of other leptons.
One common mechanism to generate such a mass matrix is the, so called,
{\it seesaw} mechanism, which introduces one or more heavy sterile neutrinos. This heavy mass scale,
combined with the SM scales, allows for the generation of very light mass eigenstates for the
electroweak neutrinos.
Estimates for the mass of these additional neutrinos range from $10^{-9} - 10^{15}~\mbox{ GeV}$.

\item[-] {\it Abundance of matter, absence of anti-matter} \\
All of the structure that we see in the universe is made of matter, and there is very little indication of the
presence of significant amounts of anti-matter.

The dominance of matter over not anti-matter can be explained by  processes in the early universe
violating $B$-number conservation, as well as $C$ and $CP$ symmetries, and occuring out of equilibrium.
 These Sakharov conditions~\cite{Sakharov:1967dj}
 are necessary to generate the baryon asymmetry when assuming symmetric initial conditions and $CPT$ conservation.
 Neither the $CP$-violation nor the out-of-equilibrium condition can be accomodated without extending the SM in some way.
 In particular our new understanding of
 the Higgs mechanism means that we now know that the electroweak phase transition is
 not a strong first order transition, and so cannot be the explanation for the asymmetry
 between matter and antimatter that we see in the present universe
 \footnote{An alternative model assumes $CPT$ and $B$-number violation. It could create a matter anti-matter
 asymmetry in thermal equilibrium~\cite{Coleman:1998ti, Jackiw:1999yp}.
 An active field with a multitude of experimental searches for $CPT$ violating processes exists worldwide,
 among which leading activities are located at the CERN AD facility.
 They have yielded many tight bounds already on Lorentz and $CPT$ violation~\cite{Kostelecky:2008ts}.}.

\vskip 8mm
\item[-] {\it Dark Matter} \\
Evidence that the particles of the SM are not abundant enough to account for all of the matter in the
universe comes from a multitude of galactic and cosmological observations.
Two key observations are galactic dynamics and the Cosmic Microwave Background (CMB).
The stability of spiral galaxies, and their observed rotation curves require an
additional (cold) matter component to be clustered on galactic scale.  This additional component contains a significant
fraction of the total mass of the galaxy and has a greater spatial extent than the visible galactic matter.
Observations of the CMB tell us about the average properties of the universe that these microwave photons have
passed through since the epoch of decoupling. Again this tells us that, on average, SM matter can only
account for approximately 5\% of the universe that we see, and that there is an additional $25\%$ of our current
universe which appears to be  cold and dark non-relativistic matter. 

There are many proposed models of DM which would be compatible with these observations,
ranging from ultra-light scalars
with masses $10^{-31} \mbox{GeV}$ to a distribution of black holes with masses up
to 10~$M_{\rm sun}$, being $M_{\rm sun}$ the mass of the sun.

\item[-] {\it Cosmological inflation and dark energy} \\
Additionally, observations of the CMB indicate that our universe began with a period of exponential inflation, and is currently undergoing a second period of accelerated expansion.  No explanation for either of these periods of the universe's evolution exists
within the SM. 

\end{itemize}

\noindent
In addition to the evidence described above there are a number of other hints that physics beyond the SM is required.
These are typically unusually large fine tunings of parameters which are challenging to explain within the SM framework.
These should not be taken to have the same status, regarding motivating NP, as the observational evidence
described above, but rather as possible sign posts to parts of the model which are not yet fully understood. 

\begin{itemize}
\item[-] {\it Higgs mass fine tuning} \\
  The Higgs boson is the only scalar field present in the SM.
  In contrast to the other particles we observe, it is not understood how to protect the mass of the scalar Higgs field
  from quantum corrections driving it to a much higher scale without a high degree of fine tuning. 
  Possible solutions to this problem include low-scale supersymmetry, the existence of extra spatial dimensions,
  and dynamical relaxation mechanisms.

\item[-] {\it Strong CP problem} \\
  There is no reason to expect that the strong sector of the SM would respect $CP$ symmetry.
  Without a large degree of fine tuning, this level of $CP$ violation would produce an electric dipole moment
  for the neutron at an observable level. Unlike the other fine tuning problems we discuss here, it is not even possible
  to make an anthropic argument for why the degree of $CP$ violation in the strong sector should be unobservably small. 

  The most common explanation for this degree of fine tuning, is the introduction of a pseudo scalar field,
  the axion, which dynamically relaxes the degree of $CP$ violation to small values.
  With an appropriately chosen mass the axion may also make up all or part of the DM in our universe.


\item[-] {\it Cosmological constant and dark energy} \\
  As mentioned above, the CMB combined with other cosmological observations, in particular of Type 1a supernovae,
  indicates that approximately $70\%$ of the energy density in our current universe is due to a cosmological constant,
  or something that behaves very similarly. A cosmological constant term in the Einstein equations is naturally generated by
  quantum fluctuations of the vacuum, but unfortunately this is many orders of magnitude too large to be compatible with
  cosmological observations.  Explaining why such a large cosmological constant is not seen typically requires a
  significant amount of fine tuning. 

\end{itemize}

There is a vast landscape of theoretical models to address some, or all, of the above mentioned motivations for NP.
This often involves introducing new particles which can be bosons or fermions, heavy or light, depending on
the theory and the problems it  addresses. There are theories that aim to make the most minimal modification
possible to the SM, whilst still addressing all of the motivations for new physics that we have
described here, as well as model independent approaches, which try to parametrize all of the possible ways
certain types of new physics could extend the SM.
Here we will outline the most popular classes of current theoretical ideas for BSM physics.

\begin{itemize}
\item[-] {\em New physics at the TeV scale and beyond }\\
If there is an intermediate scale between the EW and the Planck scale,
it is necessary to introduce a mechanism to protect the Higgs mass from receiving large quantum corrections.
The most commonly studied possibility, by far, is the introduction of supersymmetry.
No compelling hints for supersymmetry have yet been seen at the LHC, which suggests that,
if this symmetry is realized in nature, it may only be restored at an energy scale much higher
than can currently be reached with collider experiments. 
We will see that precision measurements, such as Kaon physics, $B$ physics, and EDM measurements,
can indirectly search for NP
at a much higher scale than can be directly probed with the LHC or any future high-energy collider. 

\item[-] {\em Right handed neutrinos} \\
The introduction of right handed is motivated by explanations of neutrino masses, in particular their smallness
via the see-saw mechanism.
However, it can also be a useful ingredient for generating baryon asymmetry via leptogenesis.
If the new neutrino masses are at the GeV scale, they could also generate this asymmetry directly through baryogenesis.
The introduction of such right handed neutrinos can generate $CP$ violation, but as yet the scale at which this happens is not
constrained, if it lies near the electroweak scale it could lead to observable EDMs.
The masses of the neutrinos can lie anywhere from the GUT scale down to $\sim$100 MeV.

A viable example including right-handed neutrinos 
is the {\it Neutrino Minimal Standard Model} ($\nu$MSM)~\cite{Asaka:2005an, Asaka:2005pn}
which accounts for neutrino masses and oscillations, for the evidence of DM and for the baryon asymmetry of the Universe.
This model adds to the SM  only three right-handed singlet sterile neutrinos or Heavy Neutral Leptons (HNLs), one with a mass in the
keV range that acts as DM candidate  and the other two with a mass in the GeV range and Yukawa
couplings in the range $10^{-11}-10^{-6}$.

\item[-] {\em WIMP dark matter models} \\
The idea that the DM is a thermal relic
from the hot early universe motivates non-gravitational interactions between dark and ordinary matter. The
canonical example involves a heavy particle with mass between 100-1000 GeV interacting through the weak
force (WIMPs), but so far no WIMP has been observed. However a thermal origin is equally compelling even
if DM is not a WIMP: DM with any mass from a MeV to tens of TeV can achieve the correct relic
abundance by annihilating directly into SM matter. Thermal DM in the MeV-GeV range with SM
interactions is overproduced in the early Universe so viable scenarios require additional SM neutral mediators
to deplete the overabundance~\cite{Lee:1977ua,Boehm:2002yz,Boehm:2003hm, Pospelov:2007mp,Feng:2008ya, Feng:2008mu,
Pospelov:2008zw, ArkaniHamed:2008qn, Pospelov:2008jd}.
The sub-GeV range for the dark matter mediators can  additionally provide a solution to some outstanding cosmological
puzzles including an explanation of why the mass distribution at the center of a galaxy is smoother than expected.

\item[-]{\em Axion dark matter models} \\
Axions are another well motivated DM candidate, that may simultaneously solve the $CP$
problems of QCD. Axion DM particles are sufficiently light that they must be produced
non-thermally through a gravitational, or misalignment production mechanism.
Alternatively axions may be heavy and thermally produced in the early universe.
The minimal axion model relates the mass and coupling constant of the axion.
If this condition is relaxed the theory can be generalized to one of axion-like-particles (ALPs) and
such a generalization may also be motivated from string theory.
The search for axions and ALPs in the sub-eV mass range comprises a plethora of different experimental techniques
and experiments as {\it haloscopes}, {\it solar helioscopes} and {\it pure laboratory experiments} among which, for example,
regeneration or {\it light-shining-through a wall} (LSW) experiments.
ALPs with masses  in the MeV-GeV range can be produced, and possibly detected, at accelerator-based experiments.

\end{itemize}

\vskip 2mm
So far the experimental efforts have been concentrated on the discovery of new particles with masses at or
above the EW scale with sizeable couplings with SM particles. Another viable possibility, largely
unexplored, is that particles responsible of the still unexplained phenomena beyond the SM are below
the EW scale and have not been detected because they interact very feebly with the SM particles.
Such particles are thought to be linked to the so called {\it hidden sector}.
Given the exceptionally low-couplings, a high intensity source is necessary to produce them at a detectable rate:
this can be astrophysical sources, or powerful lasers, or high-intensity accelerator beams.
The search for NP in the low-mass and very low coupling regime at accelerator beams is what is currently called
the {\it intensity frontier}.

\vskip 2mm
{\it Hidden Sector particles} and {\it mediators} to the SM can be light and long-lived. They interact very weakly with
SM fields that do not carry electromagnetic charge, like the Higgs and the $Z^0$ bosons,
the photon and the neutrinos. They are singlet states under the SM gauge interactions and
the couplings between the SM and hidden-sector particles arise via mixing of the hidden-sector field
with a SM ``portal'' operator. In the following Section we will present the generic framework for
{\it hidden sector portals} along with a set of specific benchmark cases that will be used in this document
to compare the physics reach of a large fraction of proposals presented within this study.

\subsection{Hidden Sector portals}
\label{sec:portals}

The main framework for the BSM models, the so-called {\it portal} framework, is given by the following generic setup 
(see {\em e.g.} Refs. \cite{Patt:2006fw,Batell:2009di,Alekhin:2015byh}). Let $O_{\rm SM}$ 
be an operator composed from the SM fields, and $O_{\rm DS}$ is a corresponding counterpart
composed from the dark sector fields. 
Then the portal framework combines them into an interaction Lagrangian, 
\begin{equation}
{\cal L}_{\rm portal} = \sum  O_{\rm SM}\times O_{\rm DS}.
\end{equation}
The sum goes over a variety of possible operators and of different composition and dimension. 
The lowest dimensional renormalisable portals in the SM can be classified
into the following  types:

\begin{center}
\begin{tabular}{rl}
Portal & Coupling \\
\hline
\smallskip
Dark Photon, $A_{\mu}$ &  $-\tfrac{\epsilon}{2 \cos\theta_W} F'_{\mu\nu} B^{\mu\nu}$ \\ \smallskip
Dark Higgs, $S$  & $(\mu S + \lambda S^2) H^{\dagger} H$  \\ \smallskip
Axion, $a$ &  $\tfrac{a}{f_a} F_{\mu\nu}  \tilde{F}^{\mu\nu}$, $\tfrac{a}{f_a} G_{i, \mu\nu}  \tilde{G}^{\mu\nu}_{i}$,
${\partial_{\mu} a \over f_a }\overline{\psi} \gamma^{\mu} \gamma^5 \psi$\\ \smallskip
Sterile Neutrino, $N$ & $y_N L H N$ \\
\end{tabular}
\end{center}

Here, $F'_{\mu \nu}$ is the field strength for the dark photon, which couples to the hypercharge field,
$B^{\mu\nu}$; $S$ is a new scalar singlet that couples to the Higgs doublet,
$H$, with dimensionless and dimensional couplings, $\lambda$ and $\mu$;
$a$ is a pseudoscalar axion that couples to a dimension-4 diphoton, di-fermion or digluon operator;
and $N$ is a new neutral fermion that couples to one of the left-handed doublets of the SM and the Higgs
field with a Yukawa coupling $y_N$.

\vskip 2mm
According to the general logic of quantum field theory, the lowest canonical dimension operators are the most important.
All of the portal operators respect all of the SM gauge symmetries.
Even the global symmetries are kept in tact with the only exception being the (accidental) lepton number conservation if the
HNL is Majorana. The kinetic mixing and $S^2 H^\dagger H$ operators are generically generated at loop level
unless targeted symmetries are introduced to forbid them\footnote{E.g. a $Z_2$ symmetry for the hidden photon field.}.

The PBC-BSM working group has identified the main  {\it benchmark physics cases}, presented the corresponding Lagrangians,
and defined the parameter space to be examined in connection with experimental sensitivities.
In the subsequent Sections, we formulate the benchmark models in some detail.

\subsubsection{Vector portal models}
\label{ssec:portals_vector}

A large class of BSM models   includes interactions with light new vector particles.
Such particles could result from extra gauge symmetries
of BSM physics. New vector states can mediate interactions both with the SM fields, and extra fields in the dark
sector that {\em e.g.} may represent  the DM states. 

The most minimal vector portal interaction can be written as 
\begin{equation}
\label{vector}
{\cal L}_{\rm vector} = {\cal L}_{\rm SM} + {\cal L}_{\rm DS} -\frac{\epsilon}{2\cos\theta_W} F'_{\mu\nu} B_{\mu\nu},
\end{equation}
where ${\cal L}_{\rm SM}$ is the SM Lagrangian, $B_{\mu\nu}$ and $F'_{\mu\nu}$ are the field stengths of hypercharge and
new $U(1)'$ gauge groups, $\epsilon$ is the so-called kinetic mixing parameter \cite{Holdom:1985ag}, and 
${\cal L}_{\rm DS}$ stands for the dark sector Lagrangian that may include new matter fields $\chi$ charged under $U'(1)$,
\begin{equation}
{\cal L}_{\rm DS} = -\frac{1}{4} (F'_{\mu\nu})^2 + \frac{1}{2}m_{A'}^2(A'_\mu)^2+   |(\partial_\mu + ig_DA'_\mu )\chi|^2 +...
\end{equation}
If $\chi$ is stable or long-lived it may constitute a fraction of entirety of dark matter. At low energy this theory contains 
a new massive vector particle, a dark photon state, coupled to the electromagnetic current with $\epsilon$-proportional strength,
$A'_{\mu} \times \epsilon J^\mu_{EM}$. 

We define the following important benchmark cases (BC1-BC3) for the vector portal models.

\begin{itemize}

\item {\em BC1, Minimal dark photon model}: in this case the SM is augmented by a single new state $A'$. DM is assumed to be either 
heavy or contained in a different sector. In that case, once produced, the dark photon decays back to the SM states. The parameter 
space of this model is then $\{ m_{A'}, \epsilon \}$.

\item{\em BC2, Light dark matter coupled to dark photon}: this is the model where minimally coupled viable WIMP dark matter 
  model can be constructed \cite{Boehm:2003hm,Pospelov:2007mp}.
  The preferred values of dark coupling $\alpha_{D} = g_D^2/(4\pi)$ are such that the decay of $A'$ occurs predominantly 
  into $\chi\chi^*$ states. These states can further rescatter on electrons and nuclei due to $\epsilon$-proportional
  interaction between  SM and DS states mediated by the mixed $AA'$ propagator \cite{Batell:2009di,Izaguirre:2014dua}.
  
  The parameter space for this model is  $\{ m_{A'}, \epsilon, m_\chi, \alpha_D \}$ with further model-dependence associated
  with properties of $\chi$ (boson or fermion). The suggested choices for the PBC evaluation are
  1. $\epsilon$ vs $m_{A'}$ with $\alpha_D \gg \epsilon^2 \alpha$ and $2m_\chi <m_{A'}$, 2. $y$ vs. $m_\chi$ plot
  where the {\it yield} variable $y$,  $y = \alpha_D \epsilon^2 (m_\chi/m_{A'})^4$, is argued  \cite{Izaguirre:2015yja}
  to contain a combination of parameters relevant for the freeze-out and 
  DM-SM particles scattering cross section. One possible choice is $\alpha_D = 0.1$ and $m_{A'}/m_\chi = 3$. 

\item{\em BC3, Millicharged particles}: this is the limit of $m_{A'}\to 0$, in which case $\chi$ of $\bar\chi$ have
  an effective electric charge of 
$|Q_\chi| = |\epsilon g_D e|$ \cite{Holdom:1985ag,Jaeckel:2010ni}. The suggested choice of parameter space is 
$\{m_\chi, Q_\chi/e\}$, and $\chi$ can be taken to be a fermion. 

\end{itemize} 

The kinetic mixing coupling of $A'$ to matter is the simplest and most generic, but not the only possible vector portal. 
Other cases considered in the literature include gauged $B-L$ and $L_\mu-L_\tau$ models, and somewhat less motivated
leptophylic and leptophobic cases, when $A'$ is assumed to be coupled to either total lepton current, or total baryon current 
with a small coupling $g'$.

Such other exotic vector mesons however, generically mix with the SM photon at one loop, which is often enhanced by the number of flavors
and/or colors of the quarks/leptons running in the loop. This means that the kinetically mixed dark photon benchmarks
outlined above also cover these scenarios, to some extent.

\subsubsection{Scalar portal models}
\label{ssec:portals_scalar}
The 2012 discovery of the BEH mechanism, and the Higgs boson $h$, prompts to investigate the so called scalar
or Higgs portal, that couples 
the dark sector to the Higgs boson via the bilinear $H^\dagger H$ operator of the SM. 
The minimal scalar portal model operates with one extra singlet field $S$ and two types of couplings, $\mu$ and $\lambda$ 
\cite{OConnell:2006rsp},
\begin{equation}
\label{scalar}
{\cal L}_{\rm scalar} = {\cal L}_{\rm SM} + {\cal L}_{\rm DS} - (\mu S+ \lambda S^2)H^\dagger H.
\end{equation}
The dark sector Lagrangian may include the interaction with dark matter $\chi$, ${\cal L}_{\rm DS}= S\bar\chi \chi+...$. 
Most viable dark matter models in the sub-EW scale range imply $2 \cdot m_\chi > m_S$ \cite{Krnjaic:2015mbs}.

At low energy, the Higgs field can be substituted for $H = (v + h)/\sqrt{2}$, where $v = 246$\,GeV
is the the EW vacuum expectation value, and  $h$ is the field corresponding to the physical 125\,GeV Higgs boson.
The non-zero $\mu$ leads to the mixing of $h$ and $S$ states. In the limit of small mixing it can be written as 
\begin{equation}
\theta = \frac{\mu v}{m_h^2- m_S^2}.
\end{equation}
Therefore the linear coupling of $S$ to SM particles can be written as $\theta_S\times\sum_{\rm SM}  O_h$,
where $O_h$ is a SM operator  to which Higgs boson is coupled and the
the sum goes over all type of SM operators coupled to the Higgs field. 

The coupling constant $\lambda$ leads to the coupling of $h$ to a pair of $S$ particles, $\lambda S^2$.
It can lead to pair-production of $S$ but cannot induce its decay.
An important property of the scalar portal is that at loop level 
it can induce flavour-changing transitions, and in particular lead 
to decays $K\to \pi S $, $B \to K^{(*)}S$ etc \cite{OConnell:2006rsp,Batell:2009jf,Bezrukov:2009yw} and similarly for the $hS^2$
coupling \cite{Bird:2004ts}. 
We define the following benchmark cases for the scalar portal models:

\begin{itemize}
\item {\em BC4, Higgs-mixed scalar: } in this model we assume $\lambda = 0$, and all production and decay are controlled by the same 
parameter $\theta$. Therefore, the parameter space for this model is $\{\theta, m_S\}$.

\item {\em BC5, Higgs-mixed scalar with large pair-production channel: }
  in this model the parameter space is $\{\lambda, \theta, m_S\}$, and $\lambda$ is assumed to dominate
  the production via {\em e.g.} $h\to SS$, $B \to K^{(*)}SS$, $B^0 \to SS$ etc.
  In the sensitivity plots shown in Section~\ref{ssec:scalar_portal} a value  of the branching fraction $BR({h \to SS})$
  close to $10^{-2}$ is assumed in order to be complementary to the LHC searches for the Higgs to invisible channels.
\end{itemize}

We also note that while the 125 GeV Higgs-like resonance has properties of the SM Higgs boson within errors, 
the structure of the Higgs sector can be more complicated and include {\em e.g.} several scalar doublets. In the two-Higgs 
doublet model the number of possible couplings grows by a factor of three, as $S$ can couple to 3 combinations of Higgs field 
bilinears, $H_1^\dagger H_1$, $H_2^\dagger H_2$ and $H_1 H_2$. Therefore, the experiments could investigate their sensitivity to 
a more complicated set of the Higgs portal couplings that are anyhow beyond the present document. 

\subsubsection{Neutrino portal models}
\label{ssec:portals_neutrino}

The neutrino portal extension of the SM is very motivated by the fact that it can be tightly related with the neutrino mass generation 
mechanism. The neutrino portal operates with one or several dark fermions $N$, that can be also called {\it heavy neutral leptons} or HNLs. 
The general form of the neutrino portal can be written as 
\begin{equation}
\label{neutrino}
{\cal L}_{\rm vector} = {\cal L}_{\rm SM} + {\cal L}_{\rm DS} + \sum F_{\alpha I}(\bar L_\alpha H) N_I
\end{equation}
where the summation goes over the flavour of lepton doublets $L_{\alpha}$, and the number of available HNLs, $N_I$.
The $F_{\alpha I}$ are the corresponding Yukawa couplings.
The dark sector Lagrangian should include the mass terms for HNLs, that can be both 
Majorana or Dirac type. For a more extended review, see Ref. \cite{Gorbunov:2007ak,Alekhin:2015byh}.
Setting the Higgs field to its v.e.v., and diagonalizing mass terms for neutral fermions,
one arrives at $\nu_i-N_J$ mixing, that is usually parametrized by a matrix called $U$.
Therefore, in order to obtain interactions of HNLs, inside the SM interaction terms,
one can replace $\nu_\alpha \to \sum_I U_{\alpha I} N_I$.
In the minimal HNL models, both the production and decay of an HNL are controlled by the elements of matrix $U$. 

PBC has defined the following benchmark cases: 

\begin{itemize}

\item{\em BC6, Single HNL, electron dominance:}
  Assuming one Majorana HNL state $N$, and the predominant mixing with electron neutrinos, 
all production and decay can be determined as function of parameter space $(m_N, |U_e|^2 )$. 
\item{\em BC7, Single HNL, muon dominance:}  Assuming one Majorana HNL state $N$, and the predominant mixing with muon neutrinos, 
all production and decay can be determined as function of parameter space $(m_N, |U_\mu|^2 )$. 

\item{\em BC8, Single HNL, tau dominance:}  One Majorana HNL state with predominant mixing to tau neutrinos. Parameter space is 
$(m_N, |U_\tau|^2 )$.
 
\end{itemize} 

These are representative cases which do not exhaust all possibilities.
Multiple HNL states, and presence of comparable couplings to different 
flavours can be even more motivated than the above choices. The current choice of benchmark cases is motivated by simplicity. 

\subsubsection{Axion portal models}
\label{ssec:portals_axion}
QCD axions are an important idea in particle physics \cite{Peccei:1977hh,Weinberg:1977ma,Wilczek:1977pj} 
that allows for a natural solution to the strong CP problem, or apparent lack of 
CP violation in strong interactions. Current QCD axion models are restricted to the sub-eV range of axions.
However, a generalization of the  minimal model to {\em axion-like particles} (ALPs) can be made \cite{Jaeckel:2010ni}. 
Taking a single pseudoscalar field $a$ one can write a set of its couplings 
to photons, quarks, leptons and other fields of the SM. In principle, the set of possible couplings
is very large and in this study we consider only the flavour-diagonal subset, 
\begin{equation}
\label{axion}
{\cal L}_{\rm axion} = {\cal L}_{\rm SM} + {\cal L}_{\rm DS} + \frac{a}{4f_\gamma} F_{\mu\nu} \tilde F_{\mu\nu} + \frac{a}{4f_G} {\rm Tr}G_{\mu\nu} \tilde G_{\mu\nu} +
\frac{\partial_\mu a} {f_l} \sum_\alpha  \bar l_\alpha \gamma_\mu\gamma_5 l_\alpha +
\frac{\partial_\mu a} {f_q} \sum_\beta  \bar q_\beta \gamma_\mu\gamma_5 q_\beta 
\end{equation}
The DS Lagrangian may contain new states that provide UV completion to this model (for the case of the QCD axion they are called 
the PQ sector). All of these interactions do not lead to large additive renormalization of $m_a$, making this model
technically natural. 
Note, however, that the coupling to gluons does lead to the non-perturbative contribution to $m_a$. 

The PBC proposals have considered the following benchmark cases:

\begin{itemize}
\item  {\em BC9, photon dominance:} assuming a single ALP state $a$, and the predominant coupling to photons, 
all phenomenology (production, decay, oscillation in the magnetic field) can be determined as functions on  $\{m_a, g_{a\gamma\gamma} \}$
parameter space, where $g_{a\gamma\gamma}=f^{-1}_\gamma$ notation is used.

\item {\em BC10, fermion dominance:} assuming a single ALP state $a$, and the predominant coupling to fermions, 
all phenomenology (production and decay) can be determined as functions on  $\{m_a, f^{-1}_l, f^{-1}_q \}$.
Furthermore, for the sake of simplicity, we take $f_q=f_l$.  
 
 \item  {\em BC11, gluon dominance: } this case assumes an ALP coupled to gluons. The parameter space is $\{m_a, f^{-1}_G \}$. 
 Notice that in this case the limit of $m_a < m_{a,QCD}|_{f_a=f_G}$ is unnatural as it requires fine tuning and therefore is less motivated.
 
\end{itemize} 

The ALP portals, $BC9-BC11$, are {\em effective} interactions, and would typically require UV completion at or below $f_i$ scales. This is fundamentally different from vector, scalar and neutrino portals that do not require external UV completion.  
Moreover, the renormalization group (RG) evolution is capable of inducing new couplings.
All the sensitivity plots shown in Section~\ref{sec:phys-reach} assume a cut-off scale of $\Lambda =1$\, TeV.
Details about approximations and assumptions assumed in computing sensitivities for
the BC10 and BC11 cases are reported in Appendices~\ref{sec:A} and ~\ref{sec:B}.

\clearpage
\section{Experiments proposed in the PBC context}
\label{sec:exps}

The PBC-BSM working group has considered
about 15 different initiatives which aim at exploiting
the CERN accelerator complex and scientific infrastructure  with a new, broad and compelling physics programme
that complement the quest of NP at the TeV scale performed at the LHC or other initiatives in the world.
The proposals have been classified in terms of their sensitivity to NP in a given mass range, as reported below.

\begin{enumerate}
\item {\it Sub-eV mass range} \\
  Axions and  ALPs with gluon- and photon-coupling can have masses ranging from $10^{-22}$~eV to $10^{9}$ eV.
  Axions and ALPs with gluon-coupling in the sub-eV mass range can generate a non-zero  oscillating electric dipole moment (oEDM)
  in protons. 
  The PBC proposal related to the study of
  oEDMs in protons is  {\it CP-EDM}.

  The search for axions and ALPs with photon-coupling and mass in the sub-eV range comprises a
  plethora of different experimental techniques
  and experiments as {\it haloscopes}, {\it solar helioscopes} and {\it pure laboratory experiments} among which, for example,
  regeneration or {\it light-shining-through a wall} (LSW) experiments.
  Two experiments have been proposed in the framework of the PBC-BSM study:
  the {\it International AXion Observatory (IAXO)} aims at searching axions/ALPs coming from the sun by using the axion-photon coupling,
  and the {\it JURA} experiment, considered as an upgrade of the ALPS II experiment, currently under
  construction at DESY, and exploiting the LSW technique.
   
\item {\it MeV-GeV mass range}\\
  Heavy neutral leptons, ALPs, Light Dark Matter (LDM) and corresponding light mediators (Dark Photons, Dark Scalars, etc.)
  could have masses in the MeV-GeV range and can be searched for using the interactions of proton, electron and muon
  beams available (or proposed) at the PS and SPS accelerator complex
  and at the LHC interaction points.
  Ten proposals discussed in the PBC-BSM working group are aiming to search for hidden sector physics in the MeV-GeV range
  and are classified in terms of the accelerator complex they want to exploit:
  \begin{itemize}
  \item[-] {\it PS extracted beam lines:} {\it REDTOP}.
  \item[-] {\it SPS extracted beam lines:} {\it  NA62$^{++}$ or NA62 in dump mode} at the K12 line currently
    used by the NA62 experiment;
    {\it NA64$^{++}(e)$} and {\it NA64$^{++}(\mu)$}
    proposed at the existing H4 and M2 lines of the CERN SPS;
    {\it  LDMX} at a proposed slow-extracted primary electrons line at the SPS;
    {\it SHiP} at the proposed {\it Beam Dump Facility (BDF)} at the SPS, and {\it AWAKE} at the IP4 site of the SPS.
  \item[-] {\it LHC interaction points:}  {\it MATHUSLA}, {\it FASER},
    {\it MilliQan},  and {\it CODEX-b} at the ATLAS/CMS, ATLAS, CMS and LHCb interaction points of the LHC, respectively.
    These experiments probe New Physics from below the MeV to the TeV scale, but their physics case is beyond the scope of this document.
    We focus on comparing their reach to NP in the MeV-GeV range to the other proposals at the PS and SPS lines.

\end{itemize}
      
\item {\it $>>$TeV mass range}\\
  The search for new particles at a very high mass scale is traditionally performed by studying
  clean and very rare flavor processes, as for
  example $K^+ \to \pi^+ \nu \overline{\nu}$ and $K_{\rm L} \to \pi^0 \nu \overline{\nu}$ rare decays
  or lepton-flavor-violating (LFV)
  processes as $\tau \to 3 \mu$. The {\it KLEVER} project aims at measuring the
  branching fraction of the very rare and clean decay $K_L \to \pi^0 \nu \overline{\nu}$ using an
  upgraded P42/K12 line at the SPS;
  {\it TauFV} is a fixed-target experiment proposed at the BDF
  to search for the LFV decay $\tau\to 3\mu$ and other lepton-flavour-violating (LFV) $\tau$ decays
  produced in the interactions
  of a primary high-energy proton beam with an active target.
  Proposals searching for permanent EDMs in protons, deuterons or charmed hadrons,
  can be also probe NP at the ${\mathcal{O}}$(100)~TeV scale,
  if the EDMs is originated by new sources of CP violation. 
  PBC proposals aiming at studying permanent EDM in proton and deuteron, and EDMs/MDMs in charmed and strange hadrons are
  {\it CPEDM} and {\it LHC-FT}, respectively.
\end{enumerate}

Table~\ref{tab:experiments_per_mass_range} summarizes the projects presented in the PBC-BSM study group framework
divided on the basis of their sensitivity of NP at a given mass scale, along with their main physics cases and the characteristics
of the required beam, whenever is applicable.

\begin{figure}[h]
\centerline{\includegraphics[width=0.7\linewidth]{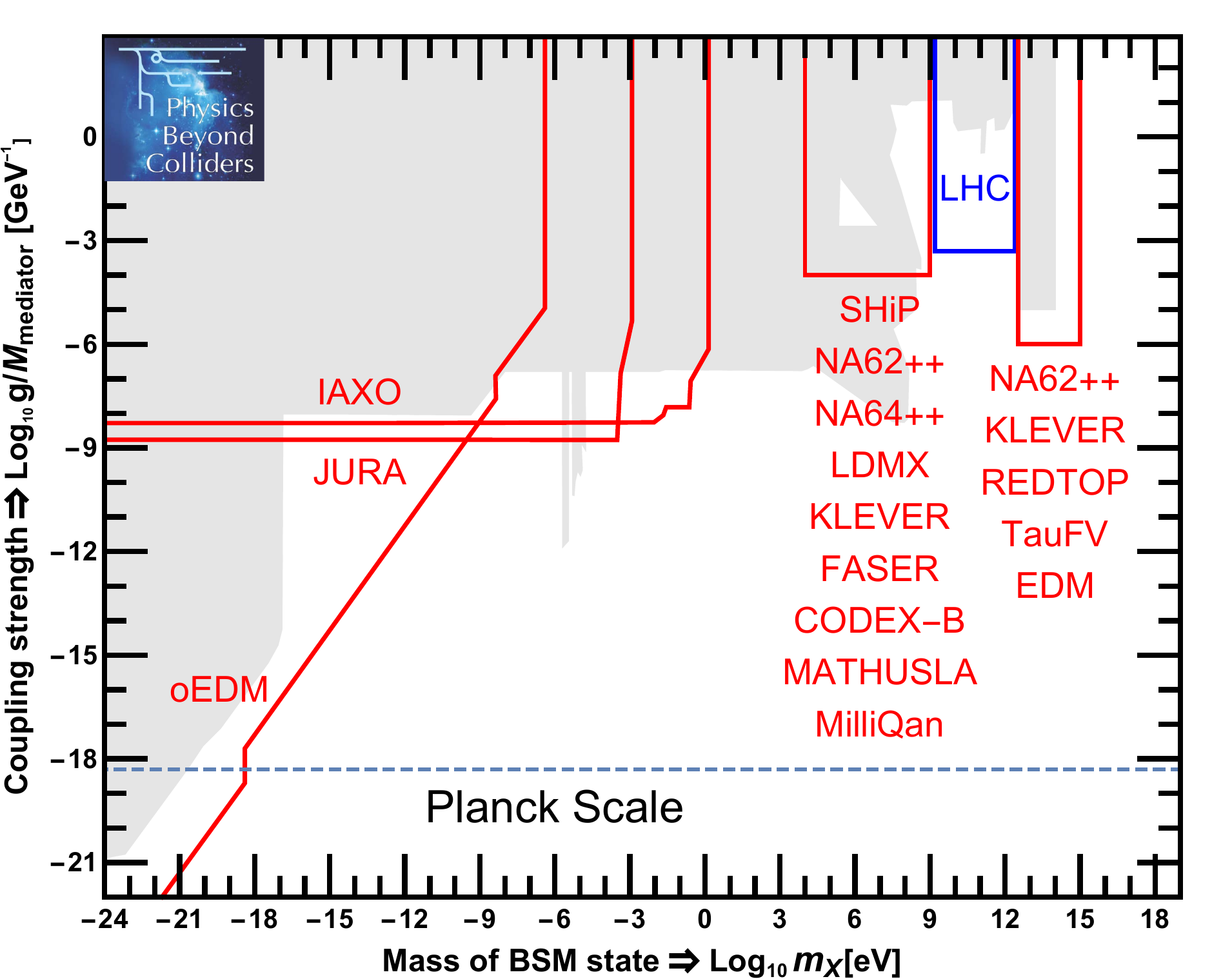}}
\caption{
  Schematic overview of the BSM landscape, based on a selection of specific models,
  with a rough outline of the areas targeted by the PBC experiments.
  The $x-$axis corresponds to the mass  $m_X$ of the lightest BSM state, and the $y-$axis to the scale of the effective
  new interaction $f = M_{\rm Mediator}/g$, where $M_{\rm Mediator}$ is the mass of a heavy mediator and $g$ its (dimensionless)
  coupling constant to the Standard Model. The grey shaded area outlines the currently excluded regions for a class of models
  corresponding to the benchmarks BC9 and BC11 (see Refs~\cite{Abel:2017rtm,Jaeckel:2010ni,Irastorza:2018dyq}).}
\label{fig:all_couplings}
\end{figure}

The physics reach of the PBC BSM projects is schematically shown in Figure~\ref{fig:all_couplings} in a generic plane of coupling versus mass,
along with the parameter space currently explored at the LHC:
the PBC-BSM projects will be able to explore a large variety of ranges of NP couplings and masses using very different experimental
techniques and are fully complementary to the exploration currently performed at the high energy frontier and at Dark Matter
direct detection experiments.

\clearpage

\begin{sidewaystable}
  \centering
  \caption{Projects considered in the PBC-BSM working group categorized in terms of their sensitivity to a set of benchmark models
    in a given mass range. The characteristics of the required beam lines, whenever applicable, are also displaied.}
  \label{tab:experiments_per_mass_range}

      \begin{tabular}{ccccc} 
        {\rm Proposal} &  Main Physics Cases & Beam Line &  Beam Type & Beam Yield    \\
        \hline \hline
               {\bf sub-eV mass range: } &  &  & \\ \hline
               {\small IAXO}   & {\small axions/ALPs (photon coupling)}   & -- &  {\small axions from sun} & -- \\
               {\small JURA} & {\small axions/ALPs (photon coupling)} & {\small laboratory}  & eV photons & -- \\ 
               {\small CPEDM}    & {\small $p,d$ oEDMs} &  {\small EDM ring}   & $p, d$ & -- \\
               & {\small axions/ALPs (gluon coupling)} &    & $p, d$ & -- \\
               {\small LHC-FT }  & {\small charmed hadrons oEDMs }  & {\small LHCb IP} & {\small 7 TeV $p$} &   --\\               
               {\bf MeV-GeV mass range: } &  &  & \\ \hline
               {\small SHiP}    & {\small ALPs, Dark Photons, Dark Scalars} & BDF, SPS & {\small 400 GeV $p$} & {\small $2\cdot 10^{20}$/5 years} \\
               & {\small LDM, HNLs, lepto-phobic DM, ..} &  &  &  \\        
               {\small NA62$^{++}$} & {\small ALPs, Dark Photons,}   & K12, SPS &{\small  400 GeV $p$}  & {\small up to $3\cdot 10^{18}$/year} \\
               & {\small Dark Scalars, HNLs}   &  &   & \\   
               {\small NA64$^{++}$} & {\small ALPs, Dark Photons, }   & H4, SPS  & {\small 100 GeV $e^-$} & {\small $5\cdot 10^{12}$ eot/year} \\
               & {\small Dark Scalars, LDM }   &   &  &  \\    
               & + {\small $L_{\mu} - L_{\tau}$}                     & M2, SPS  & {\small 160 GeV $\mu$  }   & {\small $10^{12}-10^{13}$ mot/year} \\
               & + CP, CPT, leptophobic DM                         & H2-H8, T9 & {\small $\sim$ 40 GeV $\pi,K,p$} & {\small $5\cdot 10^{12}$/year} \\
               {\small LDMX }   & {\small Dark Photon, LDM, ALPs,...  }  & eSPS & {\small 8 (SLAC) -16 (eSPS) GeV $e^-$} & {\small $10^{16}-10^{18}$ eot/year} \\
               {\small AWAKE/NA64} & {\small Dark Photon}          & AWAKE beam  & {\small 30-50 GeV $e^-$} & {\small $10^{16}$ eot/year} \\
               {\small RedTop  }       & {\small Dark Photon, Dark scalar, ALPs}    & CERN PS    &  {\small 1.8 or 3.5 GeV}   & {\small $10^{17}$ pot}\\
               {\small MATHUSLA200 }      & {\small weak-scale LLPs, Dark Scalar, }    & {\small ATLAS or CMS IP}  & {\small 14 TeV $p$ } & {\small 3000 fb$^{-1}$} \\
                                       & {\small Dark Photon, ALPs, HNLs}    &    &    &  \\
               {\small FASER   }       & {\small Dark Photon, Dark Scalar, ALPs,}    & ATLAS  IP       & {\small 14 TeV $p$}  & {\small 3000 fb$^{-1}$} \\
                                       & {\small HNLs, $B-L$ gauge bosons} &   &   &   \\
               {\small MilliQan  }     & {\small milli charge }   &  CMS IP  & {\small 14 TeV $p$} & {\small 300-3000 fb$^{-1}$ } \\
               {\small CODEX-b }       & {\small Dark Scalar, HNLs, ALPs, } &   {\small LHCb IP} & {\small 14 TeV $p$} & 300 fb$^{-1}$ \\ 
                                       & {\small LDM, Higgs decays }        &          &        & \\ \hline
               {\bf $>>$ TeV mass range:} & & & \\ \hline
               {\small KLEVER}         &  {\small $K_{\rm L} \to \pi^0 \nu \overline{\nu}$} & P42/K12 & {\small 400 GeV $p$} & {\small $5\cdot 10^{19}$ pot /5 years} \\
               {\small TauFV }         & {\small LFV $\tau$ decays} &  BDF  & {\small 400 GeV $p$} &  ${\mathcal{O}}(2\%)$ of the BDF proton yield \\
               {\small CPEDM}    & {\small $p,d$ EDMs} &  {\small EDM ring}   & $p, d$ & -- \\
               & {\small axions/ALPs (gluon coupling)} &    & $p, d$ & -- \\    
               {\small LHC-FT }  & {\small charmed hadrons MDMs, EDMs }  & {\small LHCb IP} & {\small 7 TeV $p$} &   --\\               
               \hline \hline
      \end{tabular} 
    \end{sidewaystable}

\clearpage

\section{Proposals sensitive to New Physics in the sub-eV mass range}
\label{ssec:exps_sub_eV}

Axions and ALPs have been searched for in dedicated experiments since their proposal, however to date no detection
has been reported and only a fraction of the available parameter space has been probed. Indeed, nowadays there are
experiments or proposals that studies masses  starting from the lightest possible value of $10^{-22}$ eV up to several GeV.
The apparata employed in such a search are highly complementary in the mass reach and use detection techniques that are
not common, taking advantage for example of solid state physics, optical and microwave spectroscopy,
resonant microwave cavities, precision force measuring system, highly sensitive optical polarimetry.
A relevant point which characterizes the detector is the choice of the axion source: in fact, due to the extremely
weak coupling with ordinary matter, axion production in a laboratory will be suffering from extremely small fluxes
compared with possible natural sources like the Sun or the Big Bang.

Different experiments can probe different couplings,
but the majority of the running or proposed experiment are actually exploiting the coupling of the axion to two photons
through the Primakoff effect. The following categories can then be identified:

\begin{description}
\item{\it - Dark matter haloscopes}\\
  Taking advantage of the large occupation number for the axion in the local dark matter halo,
  an axion haloscope searches for the reconversion of dark matter axions into visible photons inside a
  magnetic field region. A typical detector is a resonant microwave cavity placed inside a strong magnetic
  field~\cite{PhysRevLett.51.1415}.
  The signal would be a power excess in the cavity output when the cavity resonant frequency matches the axion mass.
  Current research is  limited in range to a few $\mu$eV masses, but several new proposals are on the way.

\item{\it - Solar helioscopes}\\
  Axion and ALPS can be efficiently produced in the solar interior with different reactions: Primakoff conversion of
  plasma photons in the electrostatic field of a charged particles, thus exploiting the axion to photon coupling;
  Axio-recombination, Bremsstrahlung and Compton are other possible channels based on the axion electron coupling.
  Solar axions escape from the sun and can be detected in earth laboratory by their reconversion into photons (X-rays)
  in a strong electromagnetic field. 

\item{\it - Pure laboratory experiment}\\
  Laboratory searches for axions can be essentially divided into three categories: polarization experiments~\cite{MAIANI1986359},
  regeneration experiments (light-shi\-ning-thro\-ugh wall - LSW)~\cite{PhysRevLett.59.759}
  and long range forces experiments~\cite{PhysRevD.30.130}.
  The key advantage for this apparata is the model independency of the detection scheme, however fluxes
  are so low that only ALPs coupling can be probed. Among others, the LSW type apparatus feature an axion source,
  for example a powerful laser traversing a dipolar magnetic field, and an axion reconverter placed after a barrier,
  again based on a static e.m. field. Reconverted photons can the be detected with ultra low background detectors.

\end{description}

Table~\ref{tab:axion_methods} compares the physics reach, the model dependency, the mass range of a possible axion or ALP particle,
the intensity of the expected flux and the wavelength of the detected photons for three categories of  experiments
sensitive to axions/ALPs with photon-coupling.

\begin{table}[htb]
\caption{Comparison between the main techniques employed in the search for axion like particles in the sub eV range.}
\begin{center}
\begin{tabular}{|c||c|c|c|}
\hline \hline
 Category & Haloscopes & Helioscopes & Lab experiments\\
  \hline \hline
  Physics reach & ALPs \& QCD axion & ALPs \& QCD axion & ALPs \\
    \hline
  Model dependency & Strong & Weak & Absent \\
  \hline
  Ranges & Resonance detector & Wide band & Wide band \\
  \hline
  Flux & Very high & high & low \\
  \hline
  Typical photon & Microwave & X-rays & Optical \\
  \hline \hline
  
\end{tabular}
\end{center}
\label{tab:axion_methods}
\end{table}%

\subsection{Solar axions helioscopes: IAXO }

\noindent
{\it Brief presentation, unique features}

The International Axion Observatory (IAXO) is a new generation axion helioscope~\cite{Armengaud:2014gea},
aiming at the detection of solar
axions with sensitivities to the axion-photon coupling $g_{a \gamma}$ down to a few 10$^{-12}$ GeV$^{-1}$, a factor of 20
better than the current best limit from CAST (a factor of more than 10$^4$ in signal-to-noise ratio).
Its physics reach is highly complementary to all other initiatives in the field, with unparalleled sensitivity to
highly motivated parts of the axion parameter space that no other experimental technique can probe.
The proposed baseline configuration of IAXO includes a large-scale superconducting multi-bore magnet,
specifically built for axion physics, together with the extensive use of X-ray focusing based on cost-effective
slumped glass optics and ultra-low background X-ray detectors.  
The unique physics potential of IAXO can be summarized by the following statements:
\begin{itemize}
\item[-]	IAXO follows the only proposed technique able to probe a large fraction of QCD axion models in the meV to eV
  mass band. This region is the only one in which astrophysical, cosmological (DM) and theoretical (strong CP problem)
  motivations overlap. 
\item[-]	IAXO will fully probe the ALP region invoked to solve the transparency anomaly, and will largely probe the axion
  region invoked to solve observed stellar cooling anomalies. 
\item[-]	IAXO will partially explore viable QCD axion DM models, and largely explore a subset of predictive ALP models
  (dubbed {\it ALP miracle}) recently studied to simultaneously solve both DM and inflation.
\item[-] The above sensitivity goals do not depend on the hypothesis of axion being the DM, i.e. in case of non-detection,
  IAXO will robustly exclude the corresponding range of parameters for the axion/ALP.
\item[-] IAXO relies on detection concepts that have been tested in the CAST experiment at CERN. Risks associated
  with the scaling up of the different subsystems will be mitigated by the realization of small scale prototype BabyIAXO. 
\item[-] IAXO will also constitute a generic infrastructure for axion/ALP physics with potential for additional
  search strategies (e.g. the option of implementing RF cavities to search for DM axions).
\end{itemize}

\noindent
{\it Key requirements}

\noindent
The main element of IAXO is a new dedicated large-scale magnet, designed to maximize the helioscope figure of merit.
The IAXO magnet will be a superconducting magnet following a large multi-bore toroidal configuration,
to efficiently produce an intense magnetic field over a large volume. The design is inspired by the ATLAS barrel
and end-cap toroids, the largest superconducting toroids ever built and presently in operation at CERN. Indeed the
experience of CERN in the design, construction and operation of large superconducting magnets is crucial for the project.
IAXO will also make extensive use of novel detection concepts pioneered at a small scale in CAST, like X-ray focusing
and low background detectors. The former relies on the fact that, at grazing incident angles, it is possible
to realize X-ray mirrors with high reflectivity. IAXO envisions newly-built optics similar to those used
onboard NASA's NuSTAR satellite mission, but optimized for the energies of the 
solar axion spectrum. Each of the eight $\sim$60 cm diameter magnet bores will be equipped with such optics.
For BabyIAXO, using existing optics from the ESA's XMM
mission is being considered. At the focal plane of each of the optics, IAXO will have low-background X-ray detectors.
Several technologies are under consideration, but the most developed one are small gaseous chambers read by pixelised
microbulk Micromegas planes. They involve low-background techniques typically developed in underground laboratories,
like the use of radiopure detector components, appropriate shielding, and the use of offline discrimination algorithms.
Alternative or additional X-ray detection technologies are also considered for IAXO, like GridPix detectors,
Magnetic Metallic Calorimeters, Transition Edge Sensors, or Silicon Drift Detectors. All of them show promising prospects
to outperform the baseline Micromegas detectors in aspects like energy threshold or resolution, which are of interest,
for example, to search for solar axions via the axion-electron coupling, a process featuring both lower energies
that the standard Primakoff ones, and monochromatic peaks in the spectrum.

\vskip 3mm
\noindent
{\it Open questions, feasibility studies}

\noindent
As a first step the collaboration pursues the construction of BabyIAXO, an intermediate scale experimental infrastructure.
BabyIAXO will test magnet, optics and detectors at a technically representative scale for the full IAXO, and,
at the same time, it will be operated and will take data as a fully-fledged helioscope experiment,
with sensitivity beyond CAST and potential for discovery.

\vskip 3mm
\noindent
{\it Status, plans and collaboration}

\noindent
After a few years of preparatory phase, project socialization and interaction with funding bodies,
the IAXO collaboration was eventually formalized in July 2017. A collaboration agreement document (bylaws)
was signed by 17 institutions from Croatia, France, Germany, Italy, Russia, Spain, South Africa, USA, as well as CERN.
They include about ~75 physicists at the moment.
It is likely that this list will increase with new members in the near future.
A collaboration management is already defined and actively implementing steps towards the BabyIAXO design and construction.
The experiment will likely be sited at DESY, and it is expected to be built in 2-3 years, entering into data taking
in 3-4 years.

\vskip 2mm 
The collaboration already nicely encompasses all the know-how to cover BabyIAXO needs,
and therefore a distribution of responsibilities in the construction of the experiment exists already.
The magnet of (Baby)IAXO is of a size and field strength comparable to that of large detector magnets typically
built in high energy physics. For this IAXO relies on the unique expertise of CERN in large superconducting magnets.
The CERN magnet detector group has already led all magnet design work so far in the IAXO CDR.
The technical design of the BabyIAXO magnet, for which CERN has allocated one Applied Fellow, has started in January 2018.
Further CERN participation is expected in terms of, at the least, allocation of expert personnel
to oversee the construction of the magnet, as well as the use of existing CERN infrastructure.
Other groups with magnet expertise in the collaboration are CEA-Irfu and INR.
The groups of LLNL, MIT and INAF are experts in the development and construction of X-ray optics,
in particular in the technology chosen for the IAXO optics. Detector expertise exists in many of the collaboration
groups, encompassing the technologies mentioned above. Experience in general engineering, large infrastructure
operation and management is present in several groups and in particular in centers like CERN or DESY.
Many of the groups have experience in axion phenomenology and the connection with experiment, and more
specifically experience with running the CAST experiment. Following these guidelines the collaboration board
is in the process of defining a collaboration agreement (MoU) to organize the distribution of efforts and
commitments among the collaborating institutes. 

IAXO has also submitted a separate document to be considered in the update of the ESPP.

\subsection{Laboratory experiments: JURA}
 \noindent
{\it Brief presentation, unique features}

\noindent
The pioneer LSW experiment was conducted in Brookhaven by the BFRT collaboration~\cite{Cameron:1993mr},
and the two most recent results are those of the experiments ALPS~\cite{Ehret:2010mh}
and OSQAR~\cite{PhysRevD.92.092002}.
ALPS is DESY based and used a decommissioned HERA magnet.
ALPS is currently performing a major improvement to phase II, where a set of 10 + 10 HERA magnets will be coupled to two 100~m long Fabry Perot cavities.
ALPS~II~\cite{1748-0221-8-09-T09001}
will in fact take advantage of a resonant regeneration apparatus,
thus expecting a major improvement of the current limit on LSW experiment given by OSQAR.
ALPS~II will represent the current state of the art LSW experiment, and for this reason its activities are monitored
with interest by the PBC since they will give key elements to judge the proposal JURA (Joint Undertaking on the Research for Axion-like particles).

\vskip 2mm
ALPS~II aims to improve the sensitivity on ALP-photon couplings by three orders of magnitude compared to
existing exclusion limits from laboratory experiments in the sub-meV mass region.
ALPS~II will inject a 30~W laser field into the 100~m long production cavity (PC) which is immersed in a 5.3~T
magnetic field. The circulating power inside the PC is expected to reach 150 kW.
The 100~m long regeneration cavity (RC) on the other side of the wall will have a finesse of 120,000.
The RC is also placed inside a similar 5.3 T magnetic field. The employed two different photon detection concepts are expected
to be able to measure fields with a photon rate as low as $\sim10^{-4}$ photons per second.
A next generation experiment for a LSW techniques will mainly rely on improved magnetic field structure,
since from the optical part only limited improvements seems to be feasible. The project {\it JURA} basically
combines the optics and detector development at ALPS II with dipole magnets for future accelerators
under development at CERN. 

The sensitivity of ALPS II in the search for axion-like particles is mainly limited by the magnetic field strength
and the aperture (which limits the length of the cavities) of the HERA dipole magnets.
JURA assumes the usage of magnets under development for an energy upgrade of LHC or a future FCC. 

\vskip 3mm
\noindent
    {\it Key requirements}
    
Several options of these future magnets are of interest to the JURA initiative. In one of them the inner high
temperature superconductor part would be omitted, so that magnets with a field of about 13 T and 100 mm aperture
would be available (the modified HERA dipoles provide 5.3 T and 50 mm). In Table \ref{Tab:JURA}
experimental parameters of ALPS II and this option of JURA are compared.
They follow from assuming the installation of optical cavities inside the magnet bore in a (nearly) confocal configuration. 

\begin{table}[htb]
\caption{Comparison of experimental parameters of ALPS II at DESY and the JURA proposal}
\begin{center}
\begin{tabular}{|c|c|c|c|c|}
\hline
Parameter & Sensitivity & ALPS~II & JURA & Rel. sensitivity \\ 
 & & & & JURA / ALPS~II \\
\hline
Magnet aperture & &  50 mm & 100 mm & \\
\hline
Magnetic field amplitude B & $g_{a \gamma} \propto B^{-1}$ & 5.3 T & 13 T & 2.5 \\
\hline
Magnetic field length L & $g_{a \gamma} \propto L^{-1}$ & 189 m & 960 m & 5.1 \\
\hline
Effective laser power P & $g_{a \gamma} \propto P^{-1/4}$ & 0.15 MW & 2.5 MW & 2.0 \\
\hline
Regeneration build up & & & & \\ (finesse F) 
& $g_{a \gamma} \propto F^{-1/4}$ & 40 k & 100 k & 1.3 \\
\hline
Detector noise rate R & $g_{a \gamma} \propto R^{1/8}$ & $10^{-4}$ Hz & $10^{-6}$ Hz & 1.8 \\
\hline
Total gain & & & & 56 \\
\hline
\end{tabular}
\end{center}
\label{Tab:JURA}
\end{table}

\noindent
{\it Open questions, feasibility studies}

\noindent
The project JURA is a long term development, for which the experiment ALPS II can be considered as a feasibility study,
especially for the resonant regeneration scheme.
There are in fact some open questions: for example, the possibility of running cavities of very high finesses
for distances of the order of several hundreds meters is still open.
The linewidth of such cavities is in fact of the order of a few Hz, about one order of magnitude better
than current state of the art. Another issue is the detector noise, however recent development using coherent
detection schemes seems to be very promising.  Of course, the development of
new magnets at CERN is not related to JURA, and thus this project will just rely on other projects' results. 

JURA in the abovementioned configuration would surpass IAXO by about a factor of 2 in the photon-ALP coupling. 
It would allow to determine the photon-coupling of a lightweight ALP discovered by IAXO unambiguously and in a
model-independent fashion or probe a large fraction of the IAXO parameter space model independently in case IAXO
does not see anything new. 

\vskip 3mm
\noindent
{\it Status, plans and collaboration}

\noindent
ALPS II is currently being constructed at DESY in the HERA tunnels. The tunnels and hall are currently being cleared
and  magnet installation will begin early 2019. The optics installation will begin at the end of 2019
and first data run is scheduled for 2020. About two years of operation is then expected. The time schedule for JURA
is foreseen to be for a 2024-2026 starting time by using a LHC dipole magnet in its first phase.
At the moment there is no real collaboration and JURA
might be considered an idea to for a possible experiment which should grow within the years to come.

\clearpage
\section{Proposals sensitive to New Physics in the MeV-GeV mass range}
\label{ssec:exps_MeV_GeV}

\vskip 2mm
Feebly-interacting  particles with masses in the MeV-GeV region can be produced in the decay of beauty, charm and strange hadrons
and by photons produced in the interactions of a proton, electron or muon beam with a dump or an active target.
Their couplings to SM particles are very suppressed leading to exceptionally low expected production rates, and therefore
high-intensity beams are required to improve over the current results.

\vskip 2mm
Accelerator experiments represent a unique tool to test models with light dark matter (LDM) 
in the MeV-GeV range, under the hypothesis that DM annihilates directly to SM particles
via new forces/new dark sector mediators.
The advantage of accelerator experiments
is that the DM is produced in a relativistic regime, and therefore its abundance
depends very weakly on the assumptions about its
specific nature, while the rates can be
predicted from thermal freeze-out.

\vskip 2mm
In addition, accelerator based experiments can probe the existence
of Heavy Neutral Leptons (HNLs) with masses
between 100 MeV and $\sim$10 GeV in a range of couplings phenomenologically motivated and
challenge the see-saw mechanism in the freeze-in regime.

\vskip 2mm
Hidden sector physics in the MeV-GeV mass range can be studied at fixed-target, dump and colliders experiments.
The focus of this document is on initiatives that want to exploit the CERN accelerator complex beyond the LHC,
as eg extracted beam lines at the PS and SPS injectors,
however proposals designed to be operated at or near the LHC interaction points have been included in the study to
provide a complete landscape scenario of the physics reach at CERN achievable in the next 10-20 years.
Several experimental approaches can be pursued to search for HNLs, ALPs, LDM and corresponding light mediators,
depending on the characteristics of the available beam line and the
proposed experiment. These can be classified as follows:

\begin{itemize}
  \item[-] {\it Detection of visible decays}:\\
  HNLs, ALPs and LDM mediators are very weakly coupled to the SM particles and can therefore decay to visible
  final states with a probability that depends on the model and scenario.
  The detection of visible final state is a technique mostly used in beam-dump experiments
  and in collider experiments (Belle, ATLAS, CMS and LHCb),
  where typical signatures are expected to show up as narrow resonances over an irreducible background. 
  This approach is of particular importance when the light mediator has a
  mass which is less than $2 m_{\chi}$, being $m_{\chi}$ the mass
  of the LDM, in which  case the mediator can decay only to visible final states.

  \item[-] {\it Direct detection of LDM scattering in the detector material}:\\
  LDM produced in reactions of electrons and/or protons with a dump
  can travel across the dump and be detected via the scattering with electrons and/or protons of a heavy material.
  This technique has the advantage of probing directly the DM production processes but requires
  a large proton/electron yield to compensate the small scattering probability.
  Moreover the signature is very similar to that produced by neutrino interactions. This is a limiting factor unless
  it is possible to use a bunched beam and time-of-flight techniques.
  
\item[-] {\it Missing momentum/energy techniques:}\\
Invisible particles (as LDM or HNLs, ALPs, and light mediators with very long lifetimes)
can be detected in fixed-target reactions as, for example,  $e^- Z \to e^- Z A'$  with $A' \to \chi \overline{\chi}$
by measuring the missing momentum or missing energy carried away from the escaping invisible particle(s).
Main challenge of this approach is the very high background rejection that must be achieved, that relies heavily on the
detector hermeticity and, in some cases, on the exact knowledge of the initial and final state kinematics.
This technique guarantees an intrinsic better sensitivity for the same luminosity than the technique based on the
detection of HNLs, ALPs and light mediator going to visible decays
or based on the  direct detection of LDM scattering in the detector, as it is independent of the
probability of decays or scattering. However it is much more model-dependent and more challenging as far as the background is concerned.
Moreover, if the mediator decays promptly or with a short lifetime to detected SM particles, these techniques have no sensitivity.

\item[-] {\it Missing mass technique}: \\
This technique is mostly used to detect invisible particles (as DM candidates) in reactions with a well-known initial state,
as for example $e^+ e^- \to \gamma A'$ with $A' \to \chi \overline{\chi}$. This technique requires detectors with very
good hermeticity that allow to detect all the other
particles in the final state. Characteristic signature of this reaction is the presence of a narrow resonance
emerging over a smooth background in the distribution of the missing mass.
Main limitation of this technique is the knowledge of the background arising
from processes in which particles in the final state escape
the apparatus without being detected.
\end{itemize}

\begin{figure}[htb]
\centerline{\includegraphics[width=\linewidth]{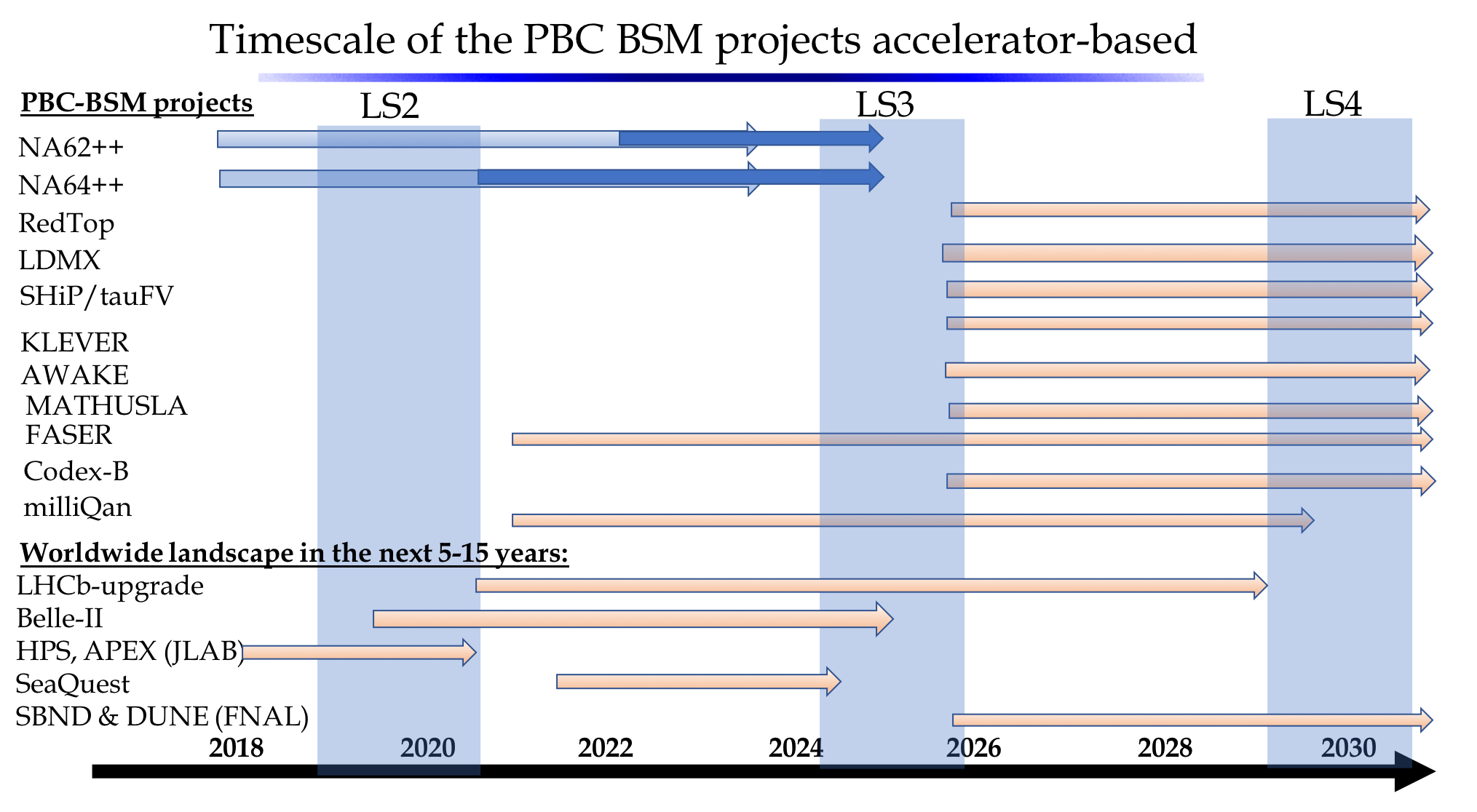}}
\caption{Tentative timescale for PBC projects exploring the MeV-GeV mass range compared to other similar
  initiatives in the world that could compete on the same physics cases.}
\label{fig:timescale}
\end{figure}

The timescale of the PBC-BSM projects that will explore the MeV-GeV mass range
is shown in Figure~\ref{fig:timescale} and compared with other similar initiatives in the world.
A concise description of each proposal along with beam request, key requirements for the detectors,
open questions and feasibility studies, is shown in the following Sections.

\clearpage
\subsection{Proposals at the PS beam lines}
\label{sssec:PS}

\subsubsection{REDTOP}
\label{ssssec:redtop}
\noindent
{\it Brief presentation, unique features} \\
REDTOP is a fixed target experiment searching for physics BSM primarily 
in ultra-rare decays of the $\eta$ and $\eta'$ mesons
produced in the interactions of the high-intensity, low-energy (few GeV) proton beam
with a target.
REDTOP was originally proposed at FNAL~\footnote{http://redtop.fnal.gov/wp-content/uploads/2016/02/REDTOP\_EOI\_v10.pdf.}
but recently expressed interest to be hosted at CERN.
The experiment requires to collect approximately $10^{13}$ $\eta$ ad $10^{11}$ $\eta'$ mesons
produced in the interactions  of 10$^{17}$ protons with energy
in the range 1.7-1.9 GeV (for $\eta$ production) and about the same number of protons
with an energy of about 3.5 GeV (for $\eta'$ production).
A fast detector, blind to most hadrons and baryons produced from the inelastic scattering
of the beam, surronds the target systems and covers about 98\% of
the solid angle.

The $\eta$ and $\eta'$ mesons are quite unique in nature. The additive
quantum numbers for these particles are all zero, the same as for
the vacuum and the Higgs, with the exception of their negative parity,
leading to the suppression of SM decays. An attractive feature of
the $\eta$ and $\eta'$ mesons is that they are flavor-neutral, so its
SM \emph{C}- and \emph{CP}-violating interactions are known to be
very small. 

Thus, rare $\eta$ /$\eta'$ decays are a promising place
to look for BSM effects.
They complement analogous searches performed
with $K$ and $B$ mesons with the unique feature that their decays are flavor-conserving.
Such decays, therefore, can provide
distinct insights into the limits of conservation laws, and open unique
doors to new BSM models at branching fraction sensitivity levels typically
below 10$^{-9}$. Notably, however, current experimental upper limits
for $\eta$ decays are many orders of magnitude larger, so $\eta$
decays have not been competitive with rare decays of flavored mesons.

\vskip 2mm
Rare $\eta/\eta'$ decays can be also exploited to search for dark photons as, eg, in  the process
$\eta \to \gamma A', A' \to \mu^+ \mu^-$.
ALPs and Dark Photons could be radiated from the beam trough
multiple processes~\cite{Dobrich:2015jyk} (Primakoff effect, Drell-Yan, proton
bremsstrahlung, etc). Such models indicate that the production cross
section of the ALPS increases at low beam energy, making such searches
more advantageous~\cite{Dobrich:2015jyk}.

\vskip 0.5cm
\noindent
 {\it {Beam, beam time}}\\
In order to generate 10$^{13}$ $\eta$ mesons on the 10-foils target
systems of the experiment, approximately 10$^{17}$ protons with energy
in the range 1.7-1.9 GeV are required. The same number of protons
with an energy of about 3.5 GeV would generate appriximately 10$^{11}$
$\eta'$ mesons. These yields give enough sensitivity for exploring
physics BSM as they correspond to sample of mesons a factor about
10$^{4}$ larger than the existing world sample. A near-CW beam is
necessary in order to limit the pile-up of events and to suppress
the combinatorial background. Only about 0.5\% of the beam interacts
inelastically with the target systems. Consequently, the power dissipatated
in the latter corresponds to only 15 mW total (1.5 mW/target foil)
for a 1.8 GeV proton beam and 24 mW total (2.4 mW/target foil) for
a 3.5 GeV proton beam. The remaining (99.5\% ) of the beam is unaffected
and it could be deviated toward other experimental apparatuses downstream
of REDTOP.

\vskip 2mm
The collaboration aims to integrate
about $10^{17}$~pot at 1.8 GeV ($\eta-factory)$ and $10^{17}$~pot
at 3.5 GeV ($\eta'-factory)$ . These yields could be provided in one or multiple
years, depending on the availaility of such beam at CERN.

\vskip 0.5cm
\noindent
{\it {Key requirements for detector}}\\
The REDTOP detector is being designed to sustain a maximum inelastic
interaction rate of about 5$\times$10$^{8}$ evt/sec. These capabilities
exceed the event rate expected at CERN by about one order
of magnitude and could portend to running the detector at future,
high-intensity proton facility (for example, PIP-II at Fermilab).
In order to sustain such an event rate, the detector must be: a) very
fast; b) blind to the baryons. The latter are produced within the
target with a multiplicity of about 5/event and could easily pile-up
if detected. On the other hand, since BSM physics is being searched
for mostly in channels with charged leptons in the final state, the detector
must have good efficiency to electrons and muons and excellent PID
capabilities. The above requirements are fullfilled by adopting an
Optical-TPC~\cite{Oberla:2015oha} for the tracking systems and
a high-granularity, dual-readout ADRIANO calorimeter~\cite{Gatto:2015gna}.

A fiber tracker, with identical features as that under construction
for the LHCb experiment~\cite{Kirn:2014cza}, has been recently included in
the detector layout.

\begin{figure}[h]
  \centerline{\includegraphics[width=0.5\linewidth]{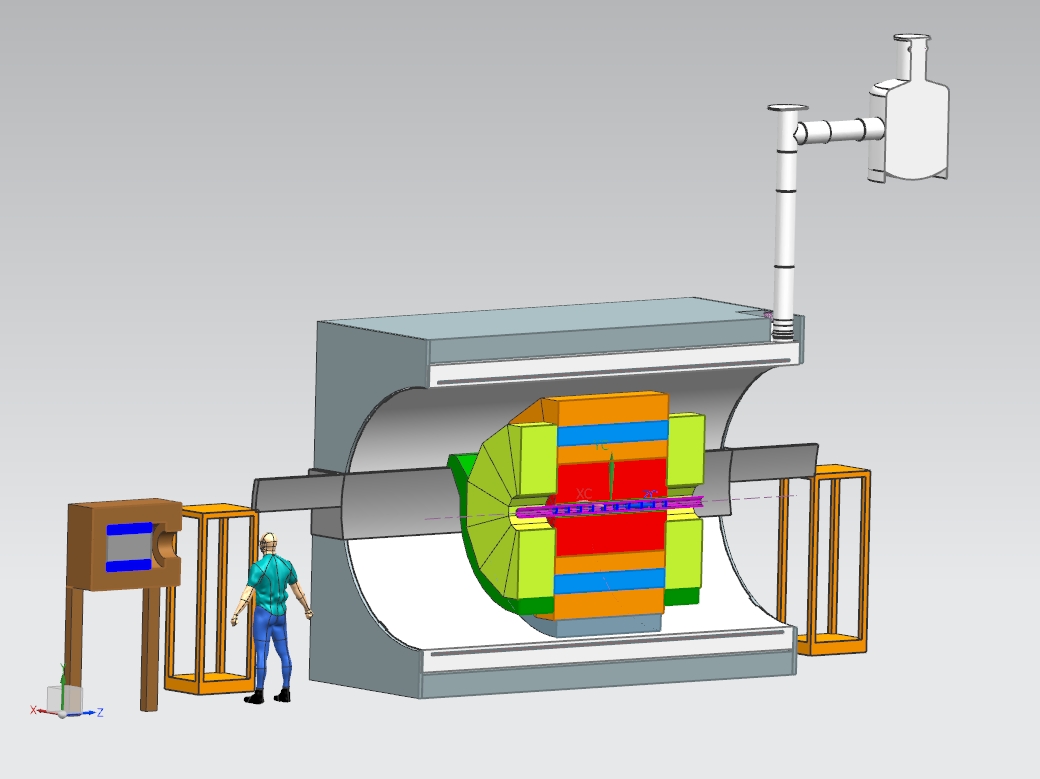}}
  \caption{Schematic layout of REDTOP detector.}
  \label{fig:redtop}
\end{figure}

The schematic layout of the detector is shown in Figure~\ref{fig:redtop}.

\vskip 0.5cm
\noindent
{\it {Open questions, feasibility studies}} \\
Few open questions exist, at this stage, for REDTOP. The largest unknown
is related to the available accelerator complex and the experimental
hall where the experiment could be operated.
Both LEAR and Booster were considered as options for REDTOP and rejected.
A possibility could be to use the 24 GeV, T8 proton beam line that currently serves CHARM and IRRAD facilities with a
maximum intensity of $6.5 \times 10^{11}$ ppp over 0.4 s. REDTOP would require lower kinetic energy
(2-3 GeV) and a much longer flat-top. No showstoppers have been identified but machine studies would be required
and, in any case, the impact on the rest of the CERN physics program would be significant.

\vskip 2mm
The second unknown
is related to the Detector R\&D still necessary to complete the design
of the apparatus. In fact, while a multi-year R\&D has been in place
for ADRIANO and for the fiber tracker, very little has been done for
the moment toward the development of an Optical-TPC prototype. The
latter is conditional to the availability of R\&D funding which, at
present, is still not in place. The Collaboration is considering,
meanwhile, to launch a simulation campaign to understand if alternative,
more conventional solutions could be found that are compatible with
the event rate expected at REDTOP.

\vskip 0.5cm
\noindent
{\it Timeline}\\
The Collaboration has estimated that about two years of detector
R\&D are necessary (dominated by the R\&D on the Optical-TPC)
and about 1 year for the construction and installation of the detector.
The solenoid and the lead-glass required for the Cerenkov component
of ADRIANO are readily available from INFN while the fibers for Tracker
and for the Scintillating component of ADRIANO are commercially available
with short lead times. The low cost, large area photo-detectors required
for the O-TPC are becoming commercially available at Incom   
and the production of about 100 units for REDTOP seems not to represent
a problem for the company.

\vskip 2mm
Under the assumption that
the funding for the Optical-TPC is available starting in 2020, REDTOP
would be ready to install in 2022 and run in 2023, one year before
LS3. The proposed schedule is very agressive but considered feasible by the Collaboration.
However, the PBC coordinators decided to have a conservative approach and consider REDTOP a proposal
for Run 4.
A full proposal will be presented to the SPSC immediately after
the conclusion of the ESPP process (mid-2020).
A coincise document will also be submitted by December 18th for the next update of the ESPP.

\vskip 0.5cm
\noindent
{\it Status of the collaboration}\\
REDTOP Collaboration counts, presently, 23 Institutions and 67 Collaborators,
as reported here: http://redtop.fnal.gov/collaboration/.

\subsection{Proposals at the SPS beam lines}
\label{sssec:SPS}

\subsubsection{NA64$^{++}$}
\label{ssssec:na64}
\noindent
{\it Brief presentation, unique features}\\
      The NA64 is a hermetic general purpose detector  to search for dark sector particles in missing energy events
      from  high-energy ($\sim$ 100 GeV) electrons, muons, and hadrons scattering off nuclei in an active dump.
      A high energy electron beam, for example, can be used to produce a vector mediator, e.g. dark photon $A'$,
      via the reaction
      $e^- Z \to e^- Z A'; \; A' \to \chi \overline{\chi}$ where $A'$ is produced via kinetic mixing with bremsstrahalung photons and then
      decay promptly and invisibly into light (sub-GeV) DM particles~\cite{Gninenko:2013rka,Andreas:2013lya} in a
      hermetic detector~\cite{Gninenko:2016kpg, Gninenko:2017yus}.
      The signature of possible $A'$ would appear as a single isolated
      electromagnetic  shower in the active dump with detectable energy  accompanied by missing energy
      in the rest of the detector.

\vskip 2mm
      The advantage of this technique compared to traditional beam dump experiments is that the sensitivity to $A'$
      scales as $\epsilon^2$
      instead of $\epsilon^4$, $\epsilon$ being the kinetic mixing strength, as the $A'$ is required
      to be produced but not detected in the far apparatus.
      Another advantage of the NA64 approach is the high energy of the incident beam, that boosts the 
      centre-of-mass system relative to the laboratory system: this results in an enhanced hermeticity of the detector
      which provides a nearly full solid angle coverage.
      
\vskip 2mm
      The missing energy technique can be used also with high energy muon and hadron beams.
      The reaction of muon scattering off nuclei $\mu +  Z \to \mu + Z + Z'_{\mu}$ is sensitive to
      dark sector particles predominantly
      coupled to muons \cite{Gninenko:2014pea, Gninenko:2018num}, and, as such,
      is fully complementary to the dark photon searches.
      This search is quite appealing and very timely in particular in connection
      to the $g_{\mu}-2$ anomaly~\cite{Gninenko:2018tlp}, and will be competing with other  proposal
      at Fermilab~\cite{Kahn:2018cqs} and
      elsewhere (see, e.g. Ref.~\cite{Chen:2018vkr} for a review).
      A $Z_{\mu}$ model gauging the $L_{\mu}- L_{\tau}$ lepton number could also explain the hints of
      LFU violations in
      $R_K$ and $R_{K^*}$ ratios observed by LHCb~\cite{Aaij:2014ora, Aaij:2017vbb}.
      The sensitivity to a $Z_{\mu}$ particle
      compatible with the observed $B$-anomalies and other constraints is currently under study by the
      Collaboration.
      
      High energy hadron beams can be used to search for dark sector particles in the decays
      $K_L, K_S, \pi^0, \eta, \eta' \to$ invisible, where the neutral mesons $M^0$ are produced via the charge-exchange reactions
      $\pi(K) p \to M^0 n + E_{\rm miss}$~\cite{Gninenko:2015mea, Gninenko:2014sxa, Barducci:2018rlx}.
      This type of search with neutral kaons is also quite complementary, see eg. Refs~\cite{Barducci:2018rlx,Abada:2016plb}
      to the current CERN  and the proposed PBC
      program in the kaon sector.

\vskip 0.5cm
\noindent
{\it Key requirements for detector, beam, beam time, timeline}\\
      NA64 is currenly taking data at the H4 beam line of the SPS~\cite{Banerjee:2016tad, Banerjee:2017hhz, Banerjee:2018vgk}.
      The beam line is a 100~GeV electron beam with a maximum intensity of $\sim 10^7~e^-$ per SPS spill.
      Beam intensity and transverse size have been optimized to guarantee an efficient detection of the synchrotron radiation
      during NA64 operation. The detection of synchrotron radiation is necessary to reach the electron beam required purity.
      
      NA64 has collected about $3\times 10^{11}$  eot before LS2, and aims at reaching $5 \times 10^{12}$ eot during Run 3. 

\vskip 2mm
      The NA64 detector is a spectrometer with a low material budget tracker, micro-MEGA,
      GEM and straw-tubes based, followed by an active target, which is a hodoscopic
      electromagnetic calorimeter (ECAL), Shashlik-type, for the measurement of the energy  of the recoil electrons.
      A high-efficient veto counter and a massive, hermetic hadronic calorimeter
      are positioned just after ECAL to efficiently detect muons or hadronic secondaries produced in the
      $e^-A$ interactions in the active target.

      \vskip 2mm
      The key feature of the NA64 apparatus is the detection of the synchrotron radiation from 100 GeV
      electrons in the magnetic field to significantly enhance electron identification and suppress
      background from the hadron contamination in the beam. The addition of a  compact tungsten calorimeter
      after the syncrotron radiation detector as a active target for the production of energetic $A'$ or $X$-boson
      explaining the $^* Be$ anomaly~\cite{Feng:2016jff,Feng:2016ysn},
      enables the search of $A' \to e^+ e^-$ visible decays.
      The first results obtained in 2016-2017 for the both $A' \to \chi \overline{\chi}$ and $A' \to e^+ e^-$ decay
      modes~\cite{Banerjee:2016tad, Banerjee:2017hhz, Banerjee:2018vgk} confirm the
      validity and sensitivity of the NA64 technique
      for searching for dark sector physics.   

      \vskip 2mm
      The NA64$^{++}$ experiment proposed in the PBC context aims at using high-energy electron, muon and hadron beams
      extracted at the SPS and currently available at the CERN North Area, starting in Run 3.

      \begin{itemize}
      \item[-]{\it NA64$^{++}(e)$:} \\
        NA64 plans to continue the data taking after LS2 with the main goal to integrate up to $5\times 10^{12}$ eot
        at the H4 line in about $(6-8)$ months. The preparation of an area able to host a quasi-permanent
        installation of NA64 began in 2018.

        An upgrade of the detector is also needed in order to cope with a high intensity beam: this includes the
        replacement of the electro-magnetic calorimeter electronics, the addition of a zero-degree hadron calorimeter, and the
        upgrade of the data acquisition system.

     \item[-] {\it NA64$^{++}(\mu)$: }  \\     
       A new detector served by the M2 beam line and located in the EHN2 experimental hall in the CERN North Area
       is proposed to be built after LS2  to investigate  dark sector predominantly coupled to the second and third generation
       and Lepton-Flavor-Violating (LFV) $\mu-\tau$ conversion
       with a high energy muon beam. 
       The M2 line, currently serving the COMPASS experiment, is able to provide muons with momentum of $\simeq (100-160)$ GeV/c,
       and intensity up to $\sim 10^8$~$\mu$/spill.
       
       \vskip 2mm
       The detector setup follows closely the one currently operating with $e^-$ beam: an active (muon) target followed by a large
       hadron calorimeter located in a magnetic field, which is used both to measure the outgoing muon momentum and to veto events
       with associated hadrons.
       The signal consists of a muon with outgoing momentum significantly lower than the incoming one and no energy deposition
       in the rest of the detector.

       \vskip 2mm

       A key issue is the purity of the incoming muon beam: a background  study performed in 2017 shows that the
       level of the hadron contamination in the muon beam can be reduced down to the negligible
       level $\leq 10^{-6}$ by using nine $Be$ absorbers.
       Another key issue is the precise measurement of the momentum of the outgoing muon and its identification with high purity.
       
       Some modification of the M2 upstream part are also foreseen,
       as described in the PBC Conventional Beams WG Report~\cite{Pbc:002}.
       Assuming a muon beam intensity of $\sim 3 \times 10^7$ $\mu$/spill,
       with  $\sim 4 \times 10^3$ spills per day, about 1.5 years are necessary
       to accumulate $\sim 5 \times 10^{13}$ mot.
       NA64 has submitted in October the addendum  for the SPSC\footnote{CERN-SPSC-2018-024/SPSC-P-348-ADD-3.} for
       the Phase 1 of NA64$^{++}(\mu)$, which requires $10^6$~muons/s at 100 GeV.       
       
     \item[-]NA64$^{++}(h)$:
     
     the NA64 studies with hadron beams are less advanced and will continue during the coming years.
     Integrated luminosities of $10^{13}$ pions-on-target,
     $10^{12}$ kaons-on-target,
     and $10^{12}$ protons-on-target could investigate dark sector models complementary to the dark photon one.
     These searches would require $(20-50)$ GeV hadron beams that could be provided by the H4 line without modification.
     The Collaboration aims to start data taking with hadron beams after LS3.
     
      \end{itemize}

\vskip 0.5cm
\noindent
{\it Open questions, planned feasibility studies}\\
      The main open question for NA64$^{++}(e)$ is the detector ability to cope with the higher beam intensity,
      which is already available, and hence increased pile-up: this has already been positively answered based on the
      preliminary analysis of the data sample collected during the 2018 run where an intensity close to $\sim 10^7 e$/spill
      has been reached. With such an intensity and 4000/spills/day, about four months will be required to collect $4 \times 10^{12}$ eot.
      Upgrades of the detector and data acquisition system
      are planned during LS2. As for NA64$^{++}(e)$, key issues for NA64$^{++}(\mu)$ are the beam purity and beam momentum
      measurement, and detector hermeticity. In both cases, the time sharing in the two (highly demanded) beam lines (H4 and M2)
      with other potential users (eg. COMPASS, MUonE, etc.) is an issue and will require a careful planning
      and prioritization of the operations.
      
\vskip 0.5cm
\noindent
{\it Status of the Collaboration} \\
The collaboration currently consists of $\simeq 50$ participants representing  14 Institutions
from Chile, Germany, Greece, Russia, Switzerland, and USA.
An updated list of authors and institutions can be found at: https://na64.web.cern.ch.
The NA64 Collaboration has also submitted a separate document for the next update of the ESPP.

\subsubsection{NA62$^{++}$}
\label{ssssec:na62}
\noindent
    {\it Brief presentation, unique features}\\
    NA62~\cite{Gonnella:2017hsz} is a fixed target experiment at the CERN SPS with the main goal of measuring the BR of the
    ultra-rare decay $K^+ \to \pi^+ \nu \overline{\nu}$ with 10\% precision.
    It is currently taking data at the K12 beam line at the CERN SPS.
    The NA62 long decay volume, hermetic coverage, low material budget, full PID capability and
    excellent tracking performance, make NA62 a suitable detector for the search for hidden particles.
    The possibility of dedicating part of Run 3 to this physics case is timely,
    since the projected sensitivity surpasses that of competitive experiments
    in the same time range. NA62 proposes to integrate  $\sim 10^{18}$ pot operating the detector in dump mode for few months
    during Run~3.

    \vskip 0.5cm
\noindent
    {\it Location, beam requirements, beam time, timeline}\\
    NA62 is currently operating at the K12 beam line in the North Area.
    At full intensity, a beam of $3 \times  10^{12}$ protons-per-pulse (ppp), 400 GeV momentum, in 
    3.5 s long effective spills from the SPS hit a beryllium
    target to produce a 75 GeV momentum-selected 750-MHz intense secondary beam of positive particles, 6\% of
    which are charged kaons.
    The beryllium target used by NA62 is followed by two 1.6~m long, water-cooled, beam-defining copper
    collimators (TAX) which can act also as a dump of $\sim 10.7$ nuclear interaction lengths each.
    In the standard NA62 operation, roughly 50\% of the beam protons punch through the beryllium target
    and are absorbed by the TAX collimators.

    \vskip 2mm
    At the NA62 nominal beam intensity, $10^{18}$ pot can be acquired in ${\mathcal{O}}(3)$ months of
    data taking. The dump-mode operation can be obtained by lifting the NA62 Beryllium target away from the
    beam line and by closing the first TAX collimator, placed $\sim$ 22 m downstream of the target.
    The muon halo emerging from the dump is partially swept away by the existing muon clearing system.
    The  switching from the standard beam mode to the beam-dump mode takes a few minutes and it is already 
    done regularly. About $3\times 10^{16}$ pot in dump mode have already been collected and are being analysed
    for background studies.

    \vskip 2mm
    The NA62 Collaboration is preparing a thorough plan for running after the end of LS2
    with a fraction of the beam time in dump mode during Run 3 (2021-2023). A possible sharing could be
    two years in beam mode to complete the measurement of the branching fraction (BR) of the $K^+ \to \pi^+ \nu \overline{\nu}$
    mode and ${\mathcal{O}}(1)$ year in beam dump mode.
    The proposal will take into account the results obtained on the measurement of the  $BR(K^+ \to \pi^+\nu \overline{\nu})$
    based on the analysis of data taken in current (2016-2018) run.
    
\begin{figure}[htb]
\begin{center}
  \includegraphics[width=0.9\linewidth]{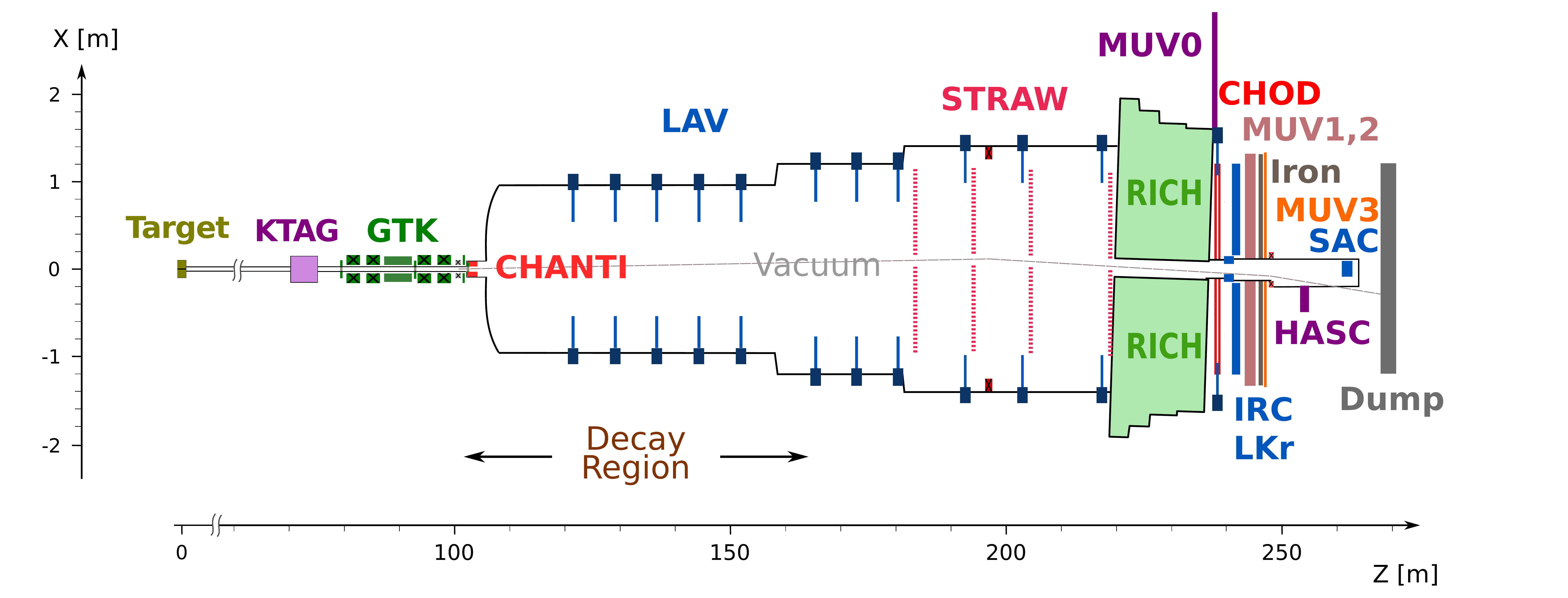}
  \caption{Layout of the NA62 experiment.}
\label{fig:na62}
\end{center}
\end{figure}

\vskip 0.5cm
\noindent
    {\it Detector description, key requirements for detector}\\
    A schematic layout of the NA62 detector is shown in Figure~\ref{fig:na62}.

    The secondary positively charged hadron beam of 75 GeV/c momentum reaches the 120~m long, 2~m diameter,
    in-vacuum decay volume, placed 100~m downstream of the target.
    A Cherenkov counter (KTAG) filled with $N_2$ along the beam line identifies and timestamps kaons,
    which are about 6\% of the hadron beam.
    Three silicon pixel stations (Gigatracker, GTK) measure the momentum and the time of all the particles in the
    beam at a rate of 750 MHz. A guard ring detector (CHANTI) tags hadronic interactions in the last GTK station
    at the entrance of the decay volume. Large angle electromagnetic calorimeters (LAV) made of lead glass blocks
    surround the decay vessel can be used to veto  particles up to 50~mrad.
    A magnetic spectrometer made of straw tubes in vacuum measures the momentum of the charged particles.

\vskip 2mm
    A 17~m long RICH counter filled with Neon separates $\pi$, $\mu$ and $e$ up to 40 GeV/c.
    The time of charged particles is measured both by the RICH and by scintillator hodoscopes
    (CHOD and NA48-CHOD) placed downstream to the RICH. The electromagnetic calorimeter filled with liquid krypton (LKr)
    covers the forward region and complements the RICH for the particle identification.
    A shashlik small-angle calorimeter (IRC) in front of LKr detects
    $\gamma$ directed on the inner edges of the LKr hole around the beam axis. The hadronic calorimeter
    made of two modules of iron-scintillator sandwiches (MUV1 and MUV2) provides further $\pi - \mu$
    separation based on hadronic energy. A fast scintillator array (MUV3) identifies muons with sub-nanosecond time resolution.
    A shashlik calorimeter (SAC) placed on the beam axis downstream of a dipole magnet bending off-axis
    the beam at the end of the NA62 detector, detects $\gamma$ down to zero angle.
    A multi-level trigger architecture is used when operated in beam mode.
    The hardware-based level-0 trigger uses timing information from CHOD, RICH and MUV3, and calorimetric variables
    from electromagnetic and hadronic calorimeters. Higher-level software-based trigger requirements
    are based on variables from KTAG, LAV and magnetic spectrometer.

\vskip 2mm
    Such a setup is perfectly suited to perform a comprehensive search for hidden particles in a large variety of visible
    final states.

\vskip 0.5cm
\noindent
    {\it Open questions, feasibility studies}\\
    The operation of NA62 in dump mode does not pose particular problems and no show-stopper have been identified.
    The analysis of ${\mathcal{O}}(10^{16})$ pot collected in dump mode shows that the background can be kept under control
    for hidden particles decaying to final states that are then fully reconstructed.
    The addition of an Upstream Veto at the front of the fiducial volume is currently being studied:
    this detector should be able to further reduce the combinatorial di-muon background coming from
    random combinations of halo muons  and to open the possibility
    of detecting also partially reconstructed final states.
    In normal operation mode half of the protons do not interact with the Be target and impinge upon the TAXes:
    these data are used for some specific background studies, namely for the di-muon background.

    \vskip 2mm
    Minor modifications to the beam line are possible, too, aimed at reducing the upstream-produced 
    background (mainly again halo muons).
    A full GEANT4-based simulation of the beam line has been implemented and is being used to study
    optimized settings of the existing magnetic elements of the line and possibly an optimized new layout
    for the beam-dump operation.
    Preliminary studies show that the component of the muon flux above 20 GeV can be reduced by
    two orders of magnitude with an appropriate setting of the magnetic elements of the beam line.
    The maximum intensity achievable is under study, as well, with some prospects of
    increase beyond the present nominal one. These aspects are under study within the PBC
    Conventional Beams working group.

\vskip 0.5cm
\noindent
    {\it Status of the collaboration}\\
    The NA62 collaboration is made of 213 authors from 31 Institutions.
    An updated list of authors and institutions can be found at:
    https://na62.web.cern.ch.

\subsubsection{LDMX @ eSPS}
\label{ssssec:ldmx}
\noindent
{\it Brief presentation, unique features}\\
 The Light Dark Matter eXperiment (LDMX) aims to probe Dark Matter (DM) parameter
 space far below expectations from the thermal freeze-out mechanism by exploiting the missing-energy
 missing-momentum technique in a fixed-target experiment with a primary electron beam of modest GeV-range energy,
 low current and high duty-cycle.
 LDMX is the only experiment exploiting this technique among those presented in the PBC framework,
 and it has a unique physics reach. Apart from its unparalleled
sensitivity to sub-GeV DM scenarios over a wide mass range, it will have sensitivity to a variety of
other BSM phenomena~\cite{Akesson:2018vlm}.

\vskip 2mm
A high-intensity primary electron beam can be provided via an X-band  70 m long linac  based on CLIC technologies
that could accelerate electrons to 3.5 GeV and fill the SPS in 1-2 sec. The beam could be further accelerated up to 16 GeV
by the SPS and then slowly extracted to a Meyrin site. The eSPS collaboration has recently submitted an expression of interest
to the SPSC~\cite{Akesson:2018yrp}.

 \vskip 2mm
 The design of the experiment is driven by two main goals: to {\it measure} the distinguishing properties of DM
production and to efficiently reject potential backgrounds, in particular photo-nuclear reactions of
bremsstrahlung photons. The signal signature has three main features: (i) a reconstructed recoiling
electron with energy substantially less than the beam energy but also (ii) detectable, with
measurable transverse momentum, and (iii) the absence of any other activity in the final state. A
constraint on the DM particle production rate can be transferred into robust bounds on the
interaction strength which in turn can be compared to direct freeze-out rates that would yield the
observed cosmic DM abundance.
 
 \vskip 2mm
 The missing-momentum approach has distinct advantages
 compared to other techniques such as missing mass (requires the reconstruction of all final state
 particles and allows only much lower luminosity), beam-dump experiments (have to pay the
 penalty of needing an additional interaction of the DM in the detector), or missing-energy only
 (suffers from higher backgrounds due to fewer kinematic handles and lack of discrimination
 between electrons and photons).

\vskip 0.5cm
\noindent
{\it Key requirements for detector, beam, beam time, timeline}\\
Reaching the full potential of the missing-momentum technique places demanding constraints on
the experiment and the beamline supporting it. A high repetition rate of electrons is required (as
much as $\sim 10^9$ electrons-on-target (eot) per second)
in order to reach the envisaged integrated luminosities of $10^{14}-10^{16}$ eot, while keeping an extremely low
electron density per bunch ($1-5 e^-$/bunch).

\begin{figure}[h]
\centerline{\includegraphics[width=0.6\linewidth]{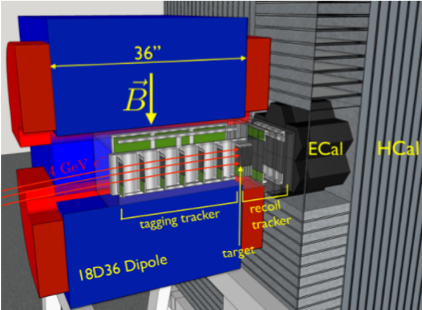}}
\caption{The LDMX experiment layout.}
\label{fig:LDMX}
\end{figure}

\vskip 2mm
This requires a fast detector that can individually resolve the energies and angles of incident
electrons, while simultaneously rejecting a variety of potential background processes that vary in
rate over many orders of magnitude.
The LDMX design makes use of a low-mass, silicon-based tracking system in a 1.5~T dipole
magnet to measure the momentum of the incoming electrons, and to cleanly reconstruct electron
recoils, thereby providing a measure of missing momentum. A high-speed, high-granularity $Si-W$
calorimeter with MIP sensitivity is used to reject potential high rate bremsstrahlung background at trigger level,
and to work in tandem with a scintillator-based hadron calorimeter to veto rare photo-nuclear reactions.
The design leverages new and existing calorimeter technology under development for the HL-LHC, as well
as existing tracking technology and experience from the HPS experiment~\cite{Adrian:2015hst}.
The experiment  is fairly small-scale for HEP standards. Thus it could be built, commissioned and run over the course of a
few years. A rendering of the proposed experimental design is shown in Figure~\ref{fig:LDMX}.

\vskip 0.2cm
The scenario for a CERN SPS beam outlined below envisages a beam energy between 3.5 and 16 GeV~\cite{Akesson:2018yrp}.
Further requirements for the beam are a low current and large beam-spot to ease the identification of individual electrons,
paired with a high duty factor for large integrated luminosity.

\vskip 1mm
All of this can be provided at CERN in three basic steps:
  a new LINAC providing electrons with 3.5 GeV, injecting into the SPS where the electrons are accelerated to up to 16 GeV,
  followed by a slow extraction of electrons to be delivered to the experiment.
  The bunch spacing can be any multiple of 5~ns up to 40~ns, the average number of electrons per bunch can range from $<$1 up
  to anything that can be tolerated by the experiment, and there is a high flexibility in the beam size,
  such that for example a beam spot of 2~cm$\times$ 30~cm is perfectly feasible.
  To achieve $10^{16}$ eot in one year would require approximately one third of the time the SPS is not used as a proton accelerator.
  
\vskip 2mm
  The Collaboration considers that the beamline could be available conservatively in 2025 (or even a few years earlier depending on
  CERN priorities) and that this would accommodate comfortably the time needed for the final design and
  construction of the detector. Hence, data taking could start in 2025 (or earlier), and be completed
  within a few years, as little as 1-2 years for the most optimistic luminosity scenarios.
  In addition to the LDMX experiment itself, the main construction needs are the electron linac as
  injector to the SPS, a 50~m tunnel for last path of the extracted beam, and a small experimental
  hall. The potential of such a primary electron beam facility goes beyond LDMX: (i) It also opens for
  a beam-dump search for visibly decaying dark photons, (ii) gives a Jefferson laboratory type facility
  with extended energy range for Nuclear Physics, and (iii) would be a significant Accelerator
  Physics R\&D-asset at CERN.

\vskip 0.5cm
\noindent
{\it Open questions, planned feasibility studies}\\
  The design studies up to now~\cite{Akesson:2018vlm} 
  have been based on the assumption of a 4~GeV beam with on
  average one electron at a rate of 46~MHz. They have demonstrated the experiment’s ability to
  reach close to 0 background for $4 \times 10^{14}$ eot. Within this scenario, in-depth studies of the
  simulation of photo-nuclear backgrounds are progressing, in order to refine the hadron calorimeter
  design. This will be followed by detector prototyping in 2019/20.

\vskip 1mm
  The sensitivities for the other BSM phenomena outlined in Ref.~\cite{Akesson:2018vlm} 
  will be studied in the near future.
  Other plans for the near future include further studies of multi-electron events (starting now with 2e/bunch) as well as
  16 GeV beam energy; how many electrons/bunch can be tolerated in terms of triggering, reconstruction and identification,
  how high a granularity is needed and feasible, and how short a bunch spacing can be handled.
  This will feed into the determination of the exact run conditions in terms of the beam parameters described above,
  in order to achieve a luminosity of $10^{16}$ EOT, which will allow to probe all thermal targets below a few hundred MeV.
  A further handle on the effective luminosity especially for the study of high-mass signals (where degradation in
  momentum resolution is tolerable) is the target material and thickness, that can be modified from
  the default 10\% $X_0$ W. The exploration of these parameters has only just begun.

\vskip 0.4cm
\noindent
{\it Status of the Collaboration}\\
  LDMX@eSPS is currently being proposed by 78 physicists from 23 Institutions
  as listed in the Letter of Intent submitted to the SPSC\footnote{CERN-SPSC-2018-023/SPSC-EOI-018.} in September 2018.
  A condensed version of the LOI will be submitted for the next update of ESPP.

\subsubsection{AWAKE}
\label{ssssec:awake}
\noindent
    {\it Brief presentation, unique features}
    
\vskip 1mm
    The AWAKE experiment is placed underground at point 4 of the SPS, at the former site of the
    CNGS target complex. The AWAKE phase-I consisted of a 10~m long plasma cell impinged by 400 GeV proton bunches
    extracted from the SPS. A laser pulse, co-propagating with a proton bunch, creates a plasma in a column of rubidium vapour and seeds
    the modulation of the bunch into microbunches. Recently electrons have been accelerated in the wakefield of the proton microbunches.
    Based on the success of AWAKE phase-I, the collaboration is currently investigating the possibility of accelerating an electron beam
    to 5-10~GeV in a 10-20~m plasma cell. A possible implementation of this phase
    is an electron beam dump experiment where electrons are accelerated to ${\mathcal{O}}(50)$ GeV using SPS bunches
    with $3.5 \times 10^{11}$ ppp every 5 sec.

\vskip 2mm
    Electron bunches of $5 \times 10^9$ electrons/bunch can be impinged upon a tungsten target
    where a Dark Photon could be produced and detected
    by an NA64-like experiment downstream. The experiment aims to detect visible dark photon decays to $e^+ e^-$ initially,
    with the possibility of extending to $\mu^+ \mu^-$ and $\pi^+ \pi^-$ final states.

\vskip 0.5cm
\noindent
    {\it Key requirements for detector, beam, beam time, timeline}
    
\begin{figure}[h]
\centerline{\includegraphics[width=\linewidth]{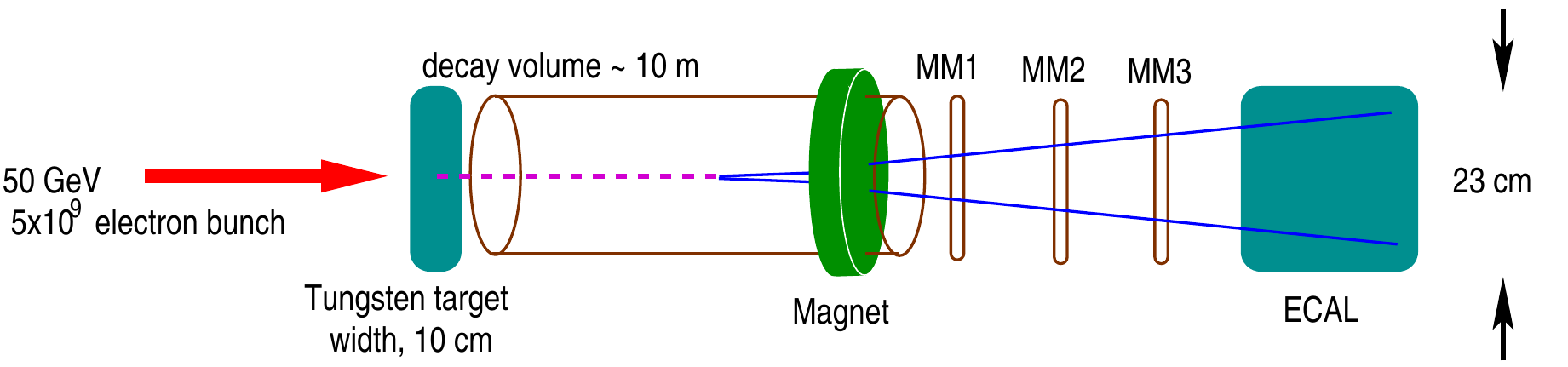}}
\caption{The AWAKE/NA64 conceptual layout.}
\label{fig:awake}
\end{figure}

\vskip 1mm
The dark photons decay in a decay volume of order 10 m long, and
the decay products are detected in three micromegas trackers MM1, MM2, MM3 as
well as a tungsten plastik shashlik calorimeter, ECAL and the further possible addition
of a HCAL. A downstream magnet separates decay products and allows the momentum
reconstruction. A schematic layout of the experiment is shown in Figure~\ref{fig:awake}.

The advantage of this experimental setup is the luminosity gain provided by deploying
bunches of electrons. This enables a larger eot in a shorter time frame and
results in an extended coverage of the sensitivity parameter space. Taking into account
the LIU-SPS with upgraded extraction kickers and a 12 week experimental run with a
70\% SPS duty cycle, AWAKE/NA64 expects to integrate $10^{16}$ eot in one year of operation.
This is more than three orders of magnitude larger than the expected integrated eot by NA64 in Run 3.

The proposed experiment requires a location accessible to SPS protons that drive
the AWAKE accelerator and tunnel length long enough to accommodate a 50-100 meters
long plasma cell as well as 20 metres of dump, drift volume and detectors.

A possible location is in the former CNGS target hall and decay tunnel.
This project relies on the successful implementation of the AWAKE acceleration concept and could be installed at earliest during LS3.

\vskip 0.5cm
\noindent
{\it Open questions, planned feasibility studies}\\
Ongoing feasibility studies will include full reconstruction of the dark photon mass,
as well as GEANT studies which incorporate realistic AWAKE electron bunches at different
average beam energies.

\vskip 2mm
The simulation of a NA64-like experiment on a possible AWAKE-based beam line is still in a very early stage:
the evaluation of the background rates and experimental efficiencies is still to be done and therefore is not contained in the sensitivity
curves shown in Section~\ref{sec:phys-reach-MeV-GeV}. 

\vskip 0.5cm
\noindent
{ \it Status of the Collaboration}\\
The AWAKE/NA64 team consists of the following:
E.~Gschwendtner, A.~Caldwell, M.~Wing, A.~Hartin, J.~Chappell,
A.~Petrenko, P.~Mugli and A.~Pardons for AWAKE. S.~Gninenko, P.~Crivelli and E.~Depero for NA64.

The AWAKE collaboration has also submitted a separate document for the next update of the ESPP.

\subsubsection{KLEVER}
\label{ssssec:klever_mev_gev}
The main goal of the KLEVER experiment is look for New Physics in the multi-TeV mass range via a measurement
of the rare decay $K_L \to \pi^0 \nu \overline{\nu}$ and is discussed in Section~\ref{ssec:exps_multi_TeV}.
However, the experiment may also be sensitive to specific signatures of hidden sector physics at the
MeV-GeV scale, as discussed in Section~\ref{sec:phys-reach-MeV-GeV}.

\subsubsection{SHiP @ BDF}
\label{ssssec:ship}
\vskip 2mm
\noindent
    {\it Brief presentation, unique features}\\
    The Search for Hidden Particles (SHiP) experiment
    has been proposed to study a wide variety of models containing
    light long-lived particles with masses below ${\mathcal{O}}(10)$ GeV with unprecedented sensitivity.
   
    This will be achieved through two main points. Firstly, using the copious
    amounts of charm, beauty, $\tau$ leptons and photons produced in an interaction of the intense
    beam designed to be operated at the {\it Beam Dump Facility}
    (BDF)~\cite{Pbc:001} 
    at the SPS, which in turn can produce hidden sector
    particles such as a Heavy Neutral Leptons, Dark Scalars, Dark Photons, Axion Like Particles, Light Dark Matter,
    R-parity violating neutralinos etc.. The BDF will be able to provide $4\times 10^{13}$ 400 GeV protons per 1-sec long spill,
    corresponding to an integrated yield of $2\times 10^{20}$ pot in 5 years of operation.
    Secondly, by reducing the background to zero over the experiment
    lifetime through the combination of a magnetic {\it muon shield} to sweep away muons from reaching
    the detector acceptance, decay volume under vacuum, veto systems surrounding the detector, timing coincidence through a
    dedicated fast timing detector, and a magnetic spectrometer within the decay volume.

\vskip 0.5cm
\noindent
{\it Detector description, key requirements for detector}\\
The main experimental challenge concerns the requirement of highly efficient reduction of beam-induced backgrounds
to below 0.1 events in the projected sample of $2 \times 10^{20}$ protons on target.
To this end, the experimental configuration includes a long target made of heavy material
to stop pions and kaons before their decay, a decay volume in vacuum, a muon shield based
on magnetic deflection able to reduce the 
flux of muons emerging from the target by six orders of magnitude in the detector acceptance, and
a hermetic veto system surrounding the whole decay volume.

\vskip 2mm
The SHiP experiment incorporates two complementary apparatuses. The first detector
immediately downstream of the muon shield consists of an emulsion based spectrometer optimised
for recoil signatures of hidden sector particles and $\tau$ neutrino physics.
The second detector system aims at measuring the decays of Hidden Sector particles to fully
reconstructible final states as well as partially reconstructible final states that involve neutrinos.
The spectrometer is designed to accurately reconstruct the decay vertex, the mass, and the
impact parameter of the hidden particle trajectory at the proton target. A set of calorimeters
and muon stations provide particle identification. A dedicated timing detector with $\sim$100 ps
resolution provides a measure of coincidence in order to reject combinatorial backgrounds.
The decay volume is surrounded by background taggers to tag neutrino and muon interactions in the
vacuum vessel walls and in the surrounding infrastructure.

\begin{figure}[h]
  \centering
\includegraphics[width=0.9\textwidth]{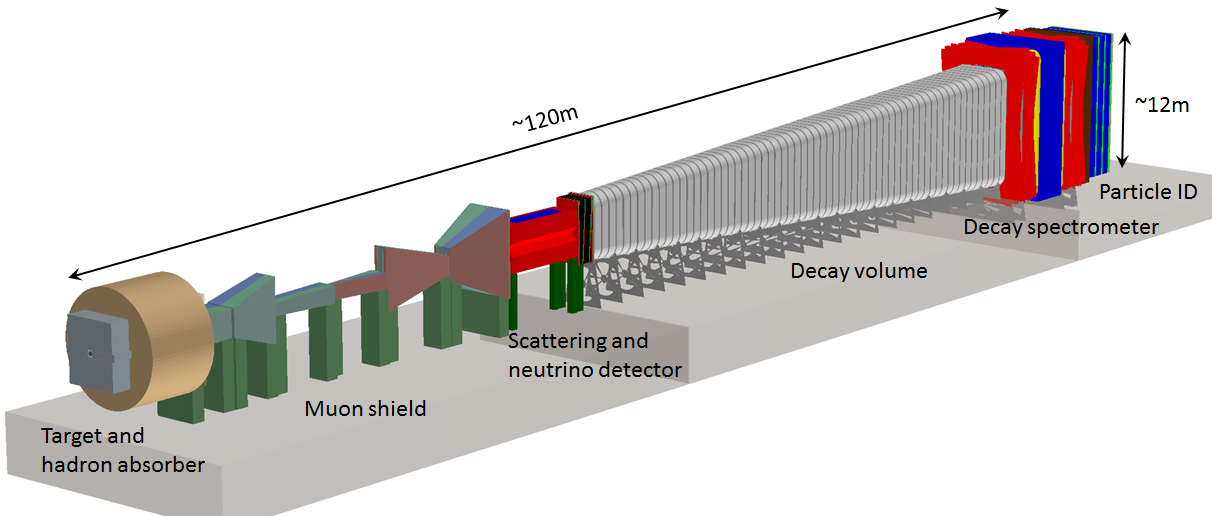} 
      \caption{Layout of the SHiP detector.}
      \label{fig:ship}
\end{figure}

A schematic of the detector layout is shown in Figure~\ref{fig:ship}.

\vskip 0.5cm
\noindent
    {\it Beam requirements, beam time, timeline}\\
    The BDF facility is described in a separate report~\footnote{Beam Dump Facility Report in preparation for the ESPP}.
    It consists of a 
    400 GeV momentum primary proton beam line slowly extracted from the SPS in spills 1 sec long.
    It is able to provide up to $4.0 \times 10^{13}$ protons per cycle, 7.2~sec long.
    The SHiP operational scenario is based on a similar fraction of beam time as the past CERN
    Neutrinos to Gran Sasso (CNGS) program. In the baseline scenario, the beam sharing delivers an annual yield of
    $4 \times 10^{19}$ protons to the SHiP experimental facility and a total of $10^{19}$ to the other physics programs at the CERN
    North Area, while respecting the beam delivery required by the LHC and HL-LHC. The physics
    sensitivities are based on acquiring a total of $2 \times 10^{20}$ protons on target, which may thus be
    achieved in five years of nominal operation.

    \vskip 2mm
    CERN's North Area has a large space next to the SPS
    beam transfer lines which is largely free of structures and underground galleries, and is entirely
    located on the current CERN territory. The proposed implementation is based on minimal
    modification to the SPS complex and maximum use of the existing beam lines. The design
    foresees space for future extensions.
    SHiP profits from the unique feature in the SPS of slow extraction of a de-bunched beam over
    a timescale of around a second. It allows tight control of combinatorial background, and allows
    diluting the large beam power deposited on the proton target both spatially and temporally.
    Should an observation require consolidation, a second mode of operation with slow extraction of
    bunched beam is also foreseen in order to further increase the discrimination between the signature
    of a Light Dark Matter object, by measuring their different times of 
    flight, and background
    induced by neutrino interactions.

    \vskip 2mm
    The schedule for the SHiP experiment and the experimental facility is largely driven by the
    CERN long-term accelerator schedule. Accordingly, the schedule aims at profiting as much as
    possible from data taking during Run 4 (currently 2027-2029).
    Most of the experimental facility can be constructed in parallel to operating the North Area
    beam facilities. The connection to the SPS has been linked to Long Shutdown 3 (i.e. for LHC
    2024-2026) but requires that the stop of the North Area is extended by one year (2025-2026).
    The schedule requires preparation of final prototypes and the TDRs for both the detector and
    the facility by beginning 2022, and construction and installation between 2023 and beginning
    2027.

\vskip 0.5cm
\noindent
{\it Background and feasibility studies}\\
An extensive simulation campaign was performed to optimise the design of the muon shield, detector setup as
well as develop a selection that reduces all possible sources of background to $< 0.1$ events over
the experiment lifetime. The backgrounds considered were: neutrinos produced through the
initial collision that undergo deep inelastic scattering anywhere in the SHiP facility producing
$V^0$s; muons deflected by the shield that undergo deep inelastic scattering in the experimental
hall or anywhere within the decay volume producing $V^0$s; muons in coincidence from the same
spill (combinatorial muons) escaping the shield; cosmic muons interacting anywhere in the decay
volume or with experimental hall.

\vskip 2mm
The rate and momentum spectrum of the muon halo
obtained with the full simulation is being calibrated using data from  a dedicated 1-month long run
performed in July 2018 where a smaller replica of the SHiP target was exposed to ${\mathcal{O}}(5 \times 10^{11})$
400 GeV protons. Results are expected by the Comprehensive Design Report, due by the end of 2019.

\vskip 2mm
All samples rely on GEANT4 to simulate the entire SHiP target, muon shield, detector, and
experimental hall (walls, ceiling, floor). In addition, neutrino interactions were simulated through
GENIE.

\vskip 2mm
A highly efficient selection is devised to reject all types of backgrounds and is detailed in
the SHiP Technical Proposal. This selection requires two good quality tracks reconstructed in
the SHiP spectrometer. Additional criteria are placed on the vertex quality, distance of closest
approach, and impact parameter of the two-track system. In addition, candidates are
rejected if the veto systems either at the front or around the decay vessel are compatible with an
interaction within them. Tracks are also required to be in coincidence within a 300 ps timing
window ($\sim 3 \sigma$).

\vskip 2mm
A neutrino sample equivalent to ten years of SHiP operation resulted in exactly zero
events surviving a basic selection. In order to ensure this background source is negligible, a sample
corresponding to fifty years of operation is being simulated. The combinatorial, deep inelastic and
cosmic muon backgrounds are expected to produce $\leq 10^{-3}$ events over the experiment lifetime.
Further studies will be conducted with even larger samples to further optimise the selection. In
addition, backgrounds to Light Dark Matter signatures are currently under evaluation.

\vskip 2mm
In addition to simulation studies, a thorough R\&D campaign on all sub-detectors has been carried out in the
last three years with the aim to have realistic estimate of detector performance  obtained with suitable
technological choices.

\vskip 0.5cm
\noindent
    {\it Open questions:}\\
    The main challenges concern the beam losses and activation during the slow extraction process,
    the design of the large muon shield, and the exact knowledge of the spectrum of the muon
    halo.

    \begin{itemize}
\item[-]
Significant progress has been made in the studies of techniques to reduce the beam losses and activation.
Studies in 2017 confirmed the intensity reach to within a factor of two. Deployment of crystal channeling in conjunction with
modified optics to reduce the  beam density at the end of 2018, both in MD and in operation, now
shows that the baseline proton yield is realistically within reach.

\item[-] The design and performance of the muon shield poses certain technological challenges.
These include how to best assemble sheets of Grain Oriented (GO) steel without disrupting the
magnetic circuit, how to cut the GO sheets into desired configurations, and how to best
connect the GO sheets to achieve the desired stacking factor. In order to address these
questions a prototyping campaign is underway.

\item[-] 
The design of the muon shield and the residual rate of muons depends on the momentum
distribution of the muons produced in the initial proton collision. The latest shield
optimisation and rate estimates were performed using PYTHIA simulations. In order to
validate these simulations a test beam campaign was performed in July to measure the muon
flux using a replica of SHiP's target. Further details can be found in Ref. [SHiP-EOI-016].
The data are currently being analysed.
Depending on the outcome of this test beam campaign, a further optimisation of the shield
configuration will be performed.

\end{itemize}

\vskip 0.5cm
\noindent
    {\it Status of the collaboration}\\
The SHiP Expression of Interest was submitted to SPSC in October 2013. This was followed by
the Technical Proposal submitted to the SPSC in April 2015. The SHiP Technical Proposal was
successfully reviewed by the SPSC and the CERN RB up to March 2016, with a recommendation
to prepare a Comprehensive Design Study report by 2019.

\vskip 2mm
SHiP is currently a collaboration of 295 members from 54 institutes (out of which 4 are associate Institutes)
representing 18 countries, CERN and JINR.
The status of the collaboration is kept up-to-date in the CERN greybook\footnote{
See {\rm https://greybook.cern.ch/greybook/experiment/detail?id=SHiP}.}.
In addition to the experimental groups, about 40 people from the CERN Accelerator Division are currently working
on the design and R\&D of the Beam Dump Facility.

\vskip 2mm
The formal organisation of SHiP consists of a Country Representative Board
(CRB), Interim Spokesperson, Technical Coordinator and Physics Coordinator, and the group of
project conveners as elected and ratified by the CRB. The organisation has been adopted for the
Comprehensive Design Study phase. A report which summarizes the simulation studies and
R\&D activities is in preparation for the SPSC\footnote{SHiP Collaboration, {\em Status of the SHiP experiment}, CERN CDS,
will be submitted to the SPSC in January 2019}.

A contribution which summarizes the status of the experiment will be
submitted to the ESPP update~\footnote{SHiP Collaboration, {\em The Search for Hidden Particles experiment at the
CERN SPS accelerator}, has been submitted to ESPP update.}.

\clearpage
\subsection{Proposals at the LHC interaction points}
\label{sssec:SPS}

\subsubsection{FASER}
\label{ssssec:faser}
\noindent
{\it Brief presentation, unique features}\\
FASER (ForwArd SEarch expeRiment at the LHC) is a proposed small and inexpensive experiment designed to
search for light, weakly-interacting particles at the LHC. Such particles are dominantly
produced along the beam collision axis and are typically long-lived particles (LLPs), traveling
hundreds of meters before decaying. To exploit both of these properties, FASER is to be
located along the beam collision axis, 480~m downstream from the ATLAS interaction point (IP).
At this location, FASER and a larger successor, FASER 2, will enhance the LHC's
discovery potential by providing sensitivity to dark photons, dark Higgs bosons,
heavy neutral leptons, axion-like particles, and many other proposed new particles.

\vskip 2mm
The FASER signal is LLPs that are produced at or close to the IP, travel along the beam collision axis, and decay visibly in FASER:
\begin{equation}
  p p  \to \text{LLP} +X, \quad  \text{LLP travels} \ \sim 480~\text{m}, \quad \text{LLP} \to \text{charged tracks} + X \text{ (or $\gamma \gamma + X$)} \ .
\end{equation}
These signals are striking: two oppositely charged tracks (or two photons) with very high energy ($\sim \text{TeV}$)
that emanate from a common vertex inside the detector and which have a combined momentum that points back through 100~m
of rock and concrete to the IP.  

\vskip 2mm
The sensitivity reach of FASER has been investigated for a large number of new physics
scenarios~\cite{Feng:2017uoz, Feng:2017vli, Batell:2017kty, Kling:2018wct, Helo:2018qej, Bauer:2018onh, Cheng:2018vaj, Feng:2018noy, Hochberg:2018rjs, Berlin:2018jbm, Dercks:2018eua,Ariga:2018uku}.
FASER will have the potential to discover a broad array of new particles, including dark photons,
other light gauge bosons, { heavy neutral leptons with dominantly $\tau$ couplings}, and axion-like particles.
FASER 2 will extend FASER's physics reach in these models to larger masses and also probe currently uncharted
territory for dark Higgs bosons, {other types of heavy neutral leptons}, and many other possibilities.

\begin{figure}[h]
      \centering
      \includegraphics[width=0.5\textwidth]{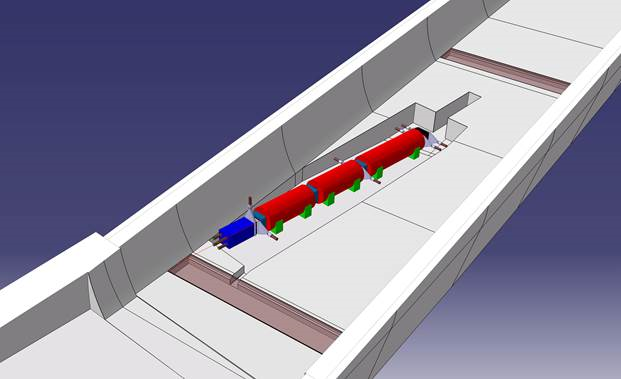}
\caption{
View of FASER in tunnel TI12.  The trench lowers the floor by 45~cm at the front of FASER to allow FASER
to be centered on the beam collision axis. Credit: CERN Site Management and Buildings Department. } 
      \label{fig:Infrastructure}
    \end{figure}

\vskip 0.5cm
\noindent
    {\it Location, beam requirements, beam time, timeline}\\
    FASER will be located 480 m downstream from the ATLAS IP in service tunnel {TI12}
    as shown in Figure~\ref{fig:Infrastructure}.
    { TI18 is also a possibility.}
    Both TI12 and TI18 were formerly used to connect the SPS to the LEP tunnel, but {they are} currently empty and unused.

\vskip 2mm
    The proposed timeline is for FASER to be installed in TI12 during Long Shutdown 2 (LS2), in time to collect
    data during Run 3 of the 14 TeV LHC from 2021-23.  FASER's cylindrical active decay volume has a radius
    $R = 10~\text{cm}$ and length $L = 1.5~\text{m}$, and the detector's total length is under 5~m.
    To allow FASER to maximally intersect the beam collision axis, the floor of {TI12} should be lowered by {45~cm}.
    This will not disrupt essential services, and no other excavation is required.
    FASER will run concurrently with the LHC and require no beam modifications.
    Its interactions with existing experiments are limited only to requiring bunch crossing
    timing and luminosity information from ATLAS.

\vskip 2mm
    If FASER is successful, a larger version, FASER 2, with an active decay volume with
    $R = 1~\text{m}$ and $L = 5~\text{m}$, could be installed during LS3 and take data in the 14 TeV HL-LHC era.
    FASER 2 would require extending {TI 12 or TI18} or widening the {staging area UJ18 adjacent to TI18}. 

\begin{figure}[h]
      \centering
      \includegraphics[width=0.89\textwidth]{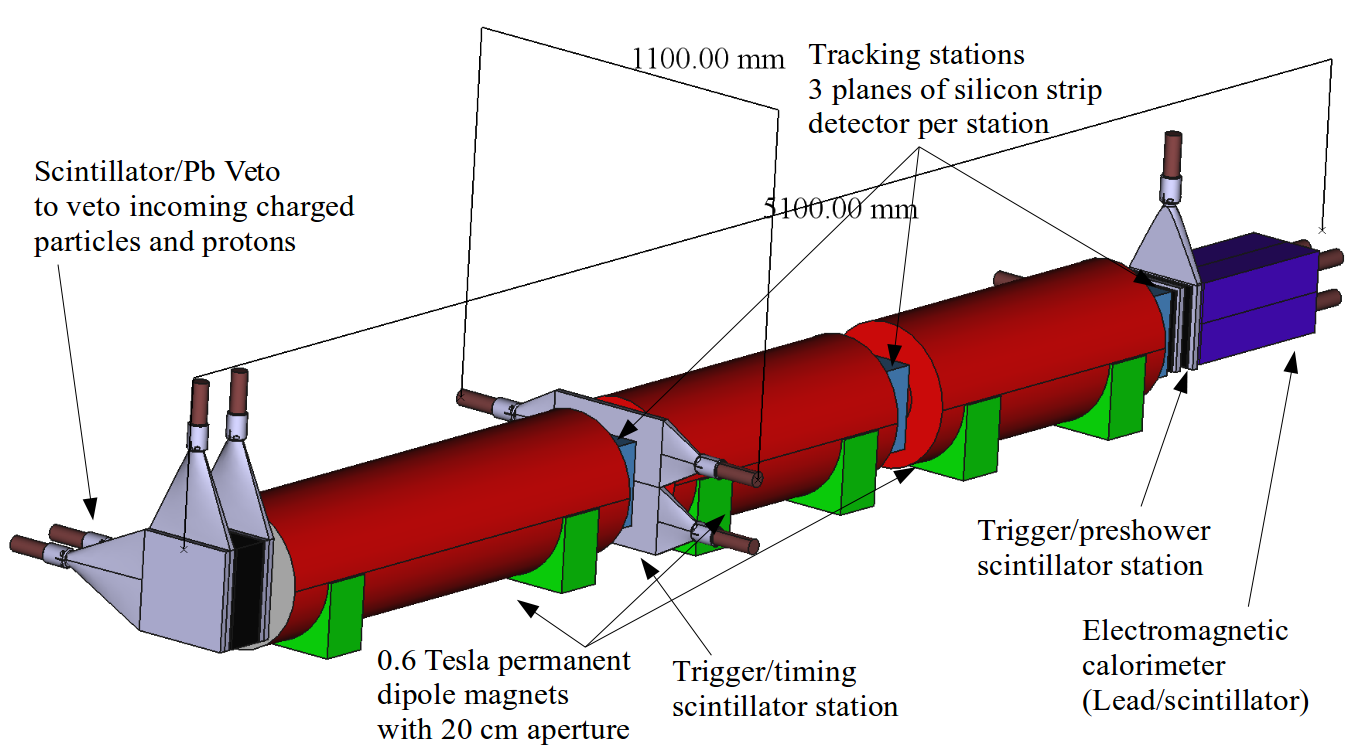} 
      \caption{Layout of the FASER detector. See text for description of the detector components.}
      \label{fig:DetectorLayout}
    \end{figure}

\vskip 0.5cm
\noindent
    {\it Detector description, key requirements for detector}\\
    The layout of the FASER detector is illustrated in Figure~\ref{fig:DetectorLayout}.
    At the entrance to the detector on the left is a double layer of scintillators (gray) to veto charged particles
    coming through the cavern wall from the IP, primarily high-energy muons.
    The veto layer is followed by a $1.5~\text{m}$ long, {0.6}~T permanent dipole magnet (red) with a $20~\text{cm}$
    aperture. This serves as the decay volume for LLPs decaying into a pair of charged particles,
    with the magnet separating these to a detectable distance.
    Next is a spectrometer consisting of two $1~\text{m}$ long, {0.6}~T dipole magnets with three tracking stations (blue),
    each composed of layers of precision silicon strip detectors located at either end and in between the magnets.
    Scintillator planes (gray) for triggering and precision time measurements are located at the entrance
    and exit of the spectrometer.
    The final component is an electromagnetic calorimeter (purple) to identify high energy electrons and
    photons and measure the total electromagnetic energy. 

\vskip 0.5cm
\noindent
    {\it Open questions, feasibility studies}\\
    The FASER signals are two extremely energetic ($\sim \text{TeV}$) coincident tracks or photons that start at a
    common vertex and point back to the ATLAS IP.  Muons and neutrinos are the only known particles that can transport
    such energies through {100~m of rock and concrete} between the IP and FASER. Preliminary estimates show that muon-associated
    radiative processes and neutrino-induced backgrounds may be reduced to negligible levels.

\vskip 2mm
    Recently a FLUKA study~\cite{Ferrari:2005zk, Bohlen:2014buj, FLUKAstudy} from the CERN Sources,
    Targets and Interactions group has been carried out to assess possible backgrounds and the radiation level in the
    FASER location. The study shows that no high energy ($>100$~GeV) particles are expected to enter FASER from proton
    showers in the dispersion suppressor or from beam-gas interactions.
    In addition, the radiation level expected at the FASER location is very low due to the dispersion function in the LHC
    cell closest to FASER.

\vskip 2mm
    {Emulsion detectors and battery-operated radiation monitors were installed in TI12 and TI18 during Technical Stops in 2018.} 
    The results from these {\it in situ} measurements {have validated the FLUKA estimates, confirming that the
    high-energy particle background is highly suppressed and radiation levels are also very low and not expected
    to be problematic for detector electronics.  Additional work is ongoing to refine background estimates,
    evaluate signal efficiencies, and optimize the detector.}

\vskip 0.5cm
\noindent
    {\it Status of the collaboration}\\
   {FASER submitted a Letter of Intent (CERN-LHCC-2018-030/LHCC-1-032~\cite{Ariga:2018zuc}) to the LHCC in July 2018.
   At its September meeting, the LHCC reviewed the LoI favorably and encouraged the FASER Collaboration to submit a
   Technical Proposal.  This was submitted to the LHCC in November 2018, and based on a positive review,
   the LHCC has recommended approving the FASER proposal.} 
    A working group has also been created within the PBC activities to study the interplay between the detector,
    the civil engineering, the backgrounds and radiation levels at the FASER installation point.
    {Two private foundations have expressed their intent to support FASER's construction and operation costs.}

\vskip 2mm
    { The FASER group currently (December 2018) consists of 27 collaborators (22 experimentalists and 5 theorists)
    from 16 institutions in China, Germany, Israel, Japan, Poland,
    Switzerland, the United Kingdom, and the United States. 
    The list of collaborators is the following:
    
\vskip 2mm
     \noindent {\it -- Tsinghua University, China}: G. Zhang; \\
    \noindent{\it -- University of Mainz, Germany}: M. Schott;\\
    \noindent{\it -- Technion, Israel}: E. Kajomovitz;\\
    \noindent{\it -- Weizmann Institute, Israel}: L. Levinson;\\
     \noindent {\it -- KEK, Japan}: Y. Takubo; \\
     \noindent{\it -- Kyushu University, Japan}: T. Ariga (Kyushu/Bern), H. Otono; \\
     \noindent {\it -- Nagoya University, Japan}: O. Sato; \\
     \noindent {\it -- National Centre for Nuclear Research, Warsaw, Poland}: S. Trojanowski (Sheffield/Warsaw); \\
    \noindent {\it -- University of Bern, Switzerland}:  A. Ariga and T. Ariga (Kyushu/Bern);\\
    \noindent {\it -- CERN, Switzerland}: J.~Boyd (contact with PBC accelerator group), S.~Kuehn, and B.~Petersen;\\
    \noindent {\it -- University of Geneva, Switzerland}: F.~Cadoux, Y.~Favre, D.~Ferrere, S.~Gonzalez-Sevilla,
    P.~Iacobucci, and A.~Sfyrla;\\
     \noindent {\it -- University of Sheffield, United Kingdom}: S. Trojanowski (Sheffield/Warsaw); \\
    \noindent{\it -- Rutgers University, United States}: I. Galon; \\
     \noindent{\it -- University of California, Irvine, United States}: D. Casper, J.~L.~Feng (contact with the PBC BSM group),
     F. Kling, J. Smolinsky, A. Soffa;\\
    \noindent{\it -- University of Oregon, United States}: E. Torrence; \\
    \noindent{\it -- University of Washington, United States}: Shih-Chieh Hsu.\\ }

\noindent
The updated status of the collaboration and experiment are available at:\\
{https://twiki.cern.ch/twiki/bin/viewauth/FASER/WebHome.}
FASER has submited also a separate document for the next ESPP update.

\subsubsection{MATHUSLA}
\label{ssssec:mathusla}
\noindent
{\it Brief presentation, unique features}\\
The basic motivation for the MATHUSLA detector (MAssive Timing Hodoscope for Ultra-Stable neutraL pArticles)~\cite{Chou:2016lxi}
is the search for LLPs produced in $\sqrt{s} = 14$ TeV HL-LHC collisions, with lifetimes much greater than the size of the main detectors
and up to the BBN constraint of $\sim 0.1$~sec, with the peak sensitivity near $\beta c \tau \sim $ 100~m.
MATHUSLA also has a secondary physics
case as a cosmic ray telescope.

This proposal has been the subject of several
studies~\cite{Co:2016fln, Dev:2016vle, Dev:2017dui, Caputo:2017pit, Curtin:2017izq, Evans:2017kti, Helo:2018qej, DAgnolo:2018wcn, Berlin:2018pwi, Deppisch:2018eth, Jana:2018rdf,Dev:2018sel}, and the physics motivation from both a
bottom-up and top-down point of view, including connections to naturalness, dark matter, baryogenesis and neutrino masses,
has been explored in a comprehensive white paper~\cite{Curtin:2018mvb}.
The MATHUSLA collaboration has also recently presented its Letter of Intent~\cite{Alpigiani:2018fgd} to the LHCC.
Given that some overlap exists between the MATHUSLA physics case and the PBC framework, the LHC Committee recommended
MATHUSLA to be discussed within the PBC framework as well.

\vskip 2mm
\noindent
    {\it Location, beam requirements, beam time, timeline}\\
    The size of the detector and the corresponding location is not yet finalized.
    All sensitivity estimates in this document assume the  MATHUSLA200 benchmark geometry
    from the Letter of Intent \cite{Alpigiani:2018fgd}, which was also the original layout proposed in~\cite{Chou:2016lxi,Curtin:2018mvb}.
    This geometry assumes a very large detector, $200 \times 200$~m$^2$ area detector built on the surface,
    situated 100~m horizontally and vertically
    away from a LHC interaction point IP (either ATLAS or CMS IP), and a decay volume height of 20~m above the ground.

    \vskip 2mm
    It is very unlikely that a detector with these large dimensions can be implemented at CERN,
    hence the sensitivity plots shown in this document
    should be properly rescaled once the final geometry will be finalized and the exact distance from the ATLAS/CMS IP points determined.
    An integrated luminosity of 3 ab$^{-1}$ corresponding to the full HL-LHC period is assumed, with a hypothetical start of the data taking
    during Run 4. The timeline for the construction of the detector is under study.
    
\vskip 0.5cm
\noindent
    {\it Detector description, key requirements for detector}\\
    The MATHUSLA detector is essentially a large tracker, situated  above an air-filled decay fiducial volume on the surface
    above ATLAS or CMS, that is able to robustly reconstruct displaced vertices (DVs) from the decay of neutral LLPs into two ore
    more charged particles. 
    The tracker should have on the order of 5 planes to provide robust tracking with $\sim$~ns timing and cm spatial resolution. 
    This is vital for rejecting cosmic ray (CR) and other backgrounds, and allows for the reconstruction
    of multi-pronged DVs for LLPs with boost up to $\sim 10^3$, corresponding to minimum LLP masses of $\mathcal{O}(10$ Mev)
    if  the LLP is produced in exotic $B$-meson decays and $\mathcal{O}(0.1 - 1 {\rm GeV})$
    for weak or TeV scale production~\cite{Curtin:2018mvb}.
    Analyzing the geometry and multiplicity of the DV final states also allows the LLP decay mode and mass to be determined
    in many scenarios~\cite{Curtin:2017izq}.
    A layer of detectors in the floor is also considered, since this will improve LLP reconstruction and provide additional
    veto capabilities that may be necessary to reject  upwards-going backgrounds like high-energy muons from the HL-LHC.

    \vskip 2mm
    For the current MATHUSLA design, the focus is on proven and relatively cheap technologies
    to allow for MATHUSLA's construction in time for the HL-LHC upgrade.
    Therefore, the trackers are envisioned to be implemented with Resistive Plate Chambers (RPCs),
    which have been used for very large area experiments in the past~\cite{Aielli:2006cj, Iuppa:2015hna}, or extruded scintillators
    which have also been used extensively~\cite{Aliaga:2013uqz,Aushev:2014spa}.

    Assuming the baseline dimensions of 200$\times$200~m$^2$, with five active layers,
    this would correspond to 200,000 m$^2$ of active detectors
    that have to provide time and space coordinates with $\sim$~ns time resolution and $\sim$~cm space resolution.
    
 \vskip 0.5cm
\noindent
    {\it Open questions, feasibility studies}\\
    The main open questions for MATHUSLA are related to its large dimensions and to its capacity of controlling
    the backgrounds mostly coming from the tens of MHz of cosmic rays crossing the detector in all directions, with a total integrated
    rate of $\sim 10^{15}$ charged particle trajectories over the whole HL-LHC run.

\begin{itemize}
      \item[-] {\it Cost:}
    The Collaboration has not provided an official estimate of the cost of the detector because of ongoing design optimizations.
    MATHUSLA requirements on resolution and rate are significantly lower compared to past experiments using
    similar detector technologies.
    The scale of the detector area is a further opportunity for cost optimization by employing mass production techniques.
    The detector size and location are currently being optimized to take
    into account land constraints and opportunities, with the hope to be able to reduce the size while keeping
    similar sensitivity. The detector design is modular for a staged implementation. The total cost will be driven by
    civil engineering and the large area tracking detectors. The Collaboration is investigating low-cost
    solutions with the challenging goal to keep the overall cost of the full
    size detector below 100~MCHF.

  \item[-] {\it Background:}
    As was argued in detail in \cite{Chou:2016lxi}, it is crucial for the projected sensitivities that MATHUSLA can search for LLP decays
    without backgrounds. 
    The surface location shields MATHUSLA from ubiquitous QCD backgrounds from the LHC collision.
    It was quantitatively demonstrated that muon and neutrino backgrounds from the IP can be sufficiently rejected.
    Extremely stringent signal requirements and 4-dimensional DV reconstruction would limit the probability of cosmic rays
    to fake the hadronic or even leptonic LLP decays. 
    Background estimates using a combination of detailed Monte Carlo studies with full detector simulation,
    the known cosmic ray spectrum, and empirical measurements at the LHC using a test stand detector,
    are currently in progress.
    The outcome of these studies will quantitatively determine whether the proposed background
    rejection strategies are sufficiently effective to reach the zero-background regime.
    However, to date, no quantitative analysis based on the full GEANT4 simulation of the detector with large Monte Carlo samples
    has been shown, and, as such, the assumption that MATHUSLA200 is a zero-background experiment is still to be demonstrated.

    \end{itemize}

    The Collaboration is currently studying a modular detector design, evaluating possible experimental sites at CERN and
    developing simulations of background and signal acceptance.
    Crucial to this endeavor is the data from the MATHUSLA test stand, a $\sim (3 \times 3 \times 5)$~m$^3$ MATHUSLA-type detector that
    is currently taking data on local cosmic rays and LHC muon backgrounds at CERN Point 1,
    allowing simulation frameworks to be calibrated and
    reconstruction strategies to be verified.

    \vskip 2mm
    Precise timelines for the full detector proposal are still being established, but the aim is to have the full detector operating
    roughly by the time the HL-LHC goes online, around 2025 or shortly thereafter. 
    The MATHUSLA collaboration has also prepared a separate stand-alone submission to the ESPP update.

\vskip 0.5cm
\noindent
    {\it Status of the Collaboration}\\
    A snapshot of the MATHUSLA collaboration is provided by the author list of the Letter of Intent~\cite{Alpigiani:2018fgd}.
    It includes 64 authors, of which 48 are experimentalists and 16 are theorists. Of the 48 experimentalists, the breakdown
    of the Institutions is the following: \\

    \noindent  {\it - Bolivia:} Universidad Mayor de San Andr\'es (2); \\
    \noindent {\it - Brazil:} University of Campinas (1); \\
    \noindent {\it - China:} Institute of High Energy Physics, Beijing (1); Shanghai Jiao Tong University (1); \\
    \noindent {\it - Israel:} Tel Aviv University (7);\\
    \noindent {\it - Italy:} INFN (1), Politecnico di Bari (1); Sezione di Roma Tor Vergata (5); Universit\`{a} degli Studi di Roma La Sapienza (2); \\
    \noindent {\it - Mexico:} Benem\'erita Universidad Aut\'onoma de Puebla (4), Universidad Aut\'onoma de Chiapas (1), Universidad Michoacana de San Nicol\'as de Hidalgo (1); \\
    \noindent {\it - Switzerland:} CERN (2); \\
    \noindent {\it - USA:} Boston University (1), NYU (1), Ohio State (1), Rutgers (4), SLAC (2), University of Arizona (1), University of Maryland (2), University of Washington (7) .

\subsubsection{CODEX-b}
\label{ssssec:codexbx}

\noindent
{\it Brief presentation, unique features}\\
The CODEX-b detector \cite{Gligorov:2017nwh} is proposed as a new, shielded subdetector for LHCb to be
placed in what is currently the LHCb data acquisition room.
The purpose of the detector is to search for new, neutral long-lived particles (LLPs) which would
penetrate the shield and decay in the detector volume. The largest gain in reach is for relatively light LLPs
-- i.e. $m \leq 10$\,GeV --  for which the backgrounds in ATLAS and CMS are prohibitive.
The LLPs can be produced from hadron or Higgs decays, or as decay products from other, beyond the standard model states.
Due to its proximity to the IP, \mbox{CODEX-b} is competitive with MATHUSLA200 in the low lifetime regime,
despite its smaller acceptance and luminosity. The close distance to LHCb also means that CODEX-b can be interfaced
with the LHCb trigger and reconstruction streams, as a true subdetector of the experiment.

\begin{figure}[h]
      \centering
	\includegraphics[width = 0.75\linewidth]{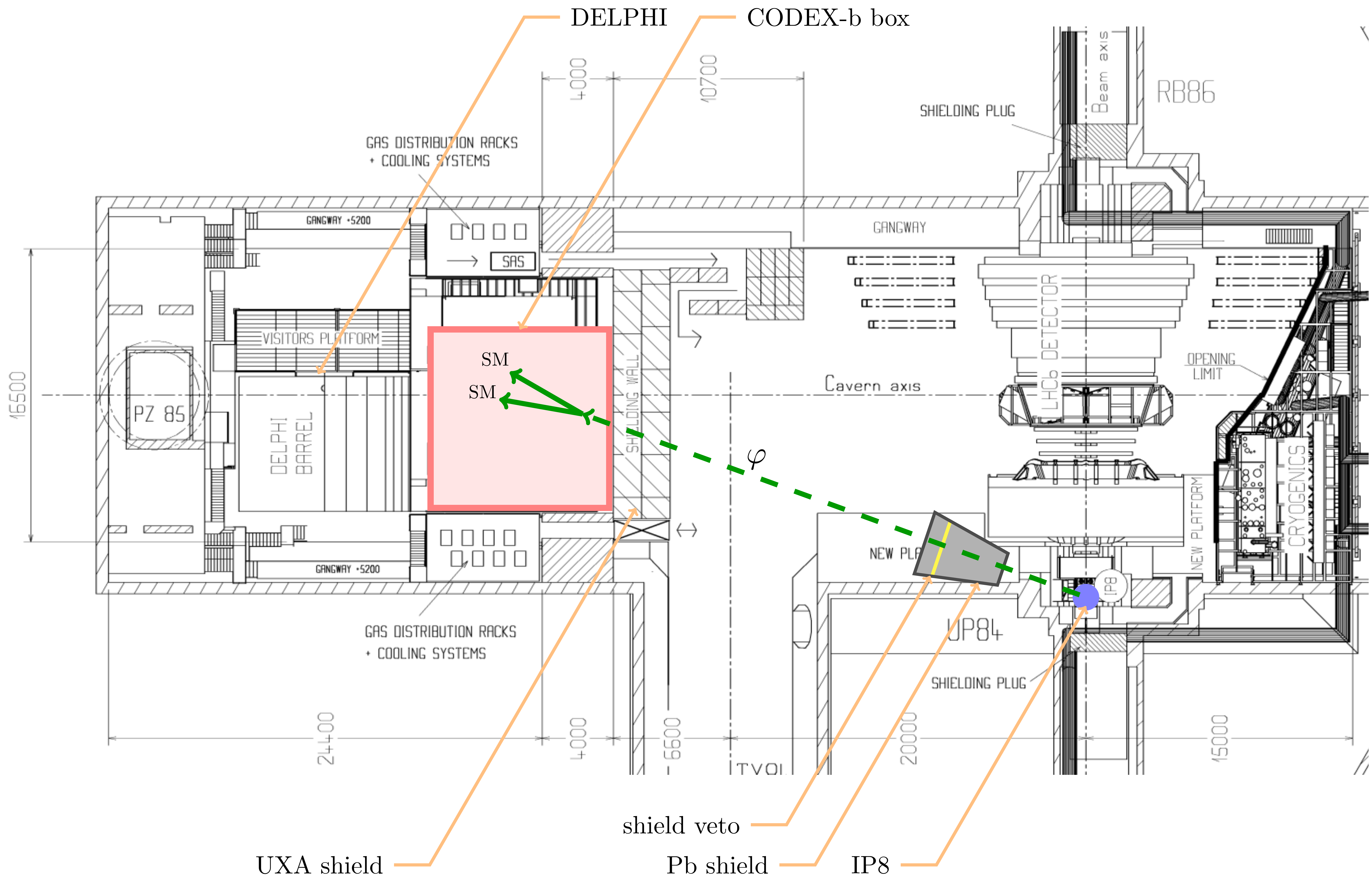}
	\caption{Layout of the LHCb experimental cavern UX85 at point 8 of the LHC~\cite{cavern}, overlaid with the proposed CODEX-b location.} 
	\label{fig:LHCbCav}
    \end{figure}

\vskip 0.5cm
\noindent
    {\it Location, beam requirements, beam time, timeline}\\
    In more detail, the proposal is to house a tracking detector in the UXA hall roughly $25$\,m from the interaction point (IP8),
    behind the 3~m thick concrete UXA shield wall.
    The UXA shield would supplemented with an additional lead or steel shield near the IP.
    The layout of the cavern and the proposed location of CODEX-b is shown in Figure~\ref{fig:LHCbCav}.
    The proposed location for the detector is currently occupied by the LHCb data acquisition system, but will be available from
    the beginning of Run 3. The size of the fiducial volume, and therefore the sensitivity, could be doubled if the DELPHI exhibit
    can be removed, but this is not essential. The necessary power supplies and services are already present in the cavern,
    and no further modifications to the cavern and/or beamline would be needed. 

    To reach the required sensitivity, CODEX-b has to integrate 300 fb$^{-1}$. This is the dataset proposed for
    the LHCb phase-II upgrade to start in Run 5,  which is still under discussion in the LHCC.
    
\vskip 0.5cm
\noindent
    {\it Detector description, key requirements for detector}\\
    The detector itself would consist of a $10$$\times10$$\times10$\,m$^3$
    volume instrumented with RPC tracking layers or alternative
    off-the-shelf tracking technology.
    The shield consists of $25$ nuclear interaction lengths of shielding near IP8 -- e.g. $4.5$\,m of Pb or steel.
    Combined with an additional $7$ interaction lengths of shielding from the UXA wall, this should suffice to suppress
    primary and secondary $K_L$, neutron and other hadronic backgrounds \cite{Gligorov:2017nwh}, as verified through a preliminary
    GEANT4 simulation of the shielding response.
    An active muon veto with an efficiency of $\mathcal{O}(10^{-5})$ is embedded in the shield, in order to reject backgrounds
    from muon-induced secondaries in the downstream parts of the shield.
    The veto is located several metres within the shield to avoid a prohibitively large veto rate from charged primaries.

\vskip 0.5cm
\noindent
    {\it Open questions, feasibility studies}\\
    For CODEX-b to have the desired sensitivity, the LHCb high luminosity upgrade should be approved and an additional
    passive shield must be installed, as discussed above.
    One of the concerns related to this project is related to the approval of the LHCb high-luminosity upgrade
    which is still pending.
    The group behind the CODEX-b proposal will require additional funds and person-power, in order
    to further develop and eventually integrate this additional large sub-detector into the LHCb framework.

    \vskip 2mm
    The CODEX-b detector geometry has been integrated into the LHCb simulation,
    with the help of the LHCb simulation team. This allows for a full simulation of collisions in IP8,
    including both the particles passing through the LHCb and CODEX-b detector volumes, and allows
    both realistic tracking studies to be performed and for studies
    of correlations between signals in CODEX-b and activity in LHCb.
    A baseline reconstruction algorithm is being worked on, and a detailed report on the baseline geometry performance is
    foreseen for end of 2018.

    \vskip 2mm
In parallel, a two-scintillator setup has been used 
    to perform a measurement of backgrounds in the DELPHI cavern during nominal LHC collisions at IP8.
    Measurements were taken at various points along the nominal CODEX-b geometry, and work is ongoing to relate these to the GEANT
    background estimates in the CODEX-b paper.
    
\vskip 2mm
This data-driven background estimate is expected to be ready on a similar timescale
    to the nominal geometry performance report. As a consequence, the assumption of zero-background based on preliminary GEANT4 simulations and
    assumed in the compilation of the sensitivity curves in the following Sections, is still to be demonstrated.

\vskip 0.5cm
\noindent
    {\it Status of the Collaboration}\\
    The CODEX-b Collaboration consists currently of 12 experimentalists and five theorists.
    A preliminary list of Institutions and collaborators is the following:

\vskip 2mm
    \noindent
    \emph{Experimentalists:}\\
        \noindent {\it - China:} Institute of Particle Physics, Central China Normal University, Wuhan, Hubei: Biplab Dey;\\
         \noindent {\it - Poland:} Henryk Niewodniczanski Institute of Nuclear Physics Polish Academy of Sciences, Krakow: A. Dziurda and M. Witek; \\
    \noindent {\it - France: } LPNHE, Sorbonne Universit\'e, Paris Diderot Sorbonne Paris Cit\'e, CNRS/IN2P3, Paris:
      V. Gligorov and E. Ben Haim; \\
    \noindent {\it - France: } Clermont Universit\'e, Universit\'e Blaise Pascal, CNRS/IN2P3, LPC: V. Tisserand. \\
    \noindent {\it - UK:} University of Birmingham, Birmingham: P. Ilten; \\
    \noindent {\it - UK:} School of Physics and Astronomy, University of Manchester, Manchester: M. Williams. \\
    \noindent {\it - US:} University of Cincinnati, Cincinnati, OH, United States: M. Sokoloff.  \\
    \noindent {\it - US:} Syracuse University, Syracuse, NY: S. Stone.\\
    \noindent {\it - US:} Massachusetts Institute of Technology, Cambridge, MA: M. Williams; \\
    \noindent {\it - Italy:} INFN, sezione di Bologna, V. Vagnoni. \\

    \noindent  \emph{Theory collaborators:}\\
    \noindent {\it US:} University of Cincinnati, Cincinnati:  J. Evans.\\
    \noindent {\it US:} Institute for Advanced Study, Princeton: S. Knapen.\\
    \noindent {\it US:} Berkeley and Lawrence Berkeley National Lab, Berkeley: M. Papucci and H. Ramani. \\
    \noindent {\it US:} UC Santa Cruz, Santa Cruz, United States and Lawrence Berkeley National Lab, Berkeley: D. Robinson.

\clearpage
\section{Proposals sensitive to New Physics in the multi-TeV mass range}
\label{ssec:exps_multi_TeV}

The lack of an unambiguous evidence of NP  so far
could indicate that NP physics can be at a very high mass scales, and therefore well beyond the reach of direct detection at
the LHC or any other envisageable future high-energy collider but, perhaps, accessible via indirect effects.
These can arise as modification in 
branching fractions, angular distributions, CP asymmetries in decays of strange, charm, beauty hadrons, or
as a presence of measurable LFV decays in charged leptons
or as presence of permanent EDMs in elementary particles containing quarks of the first (proton and deuteron) or
second (charmed and strange baryons) generation.

\begin{itemize}
\item[-] {\bf Ultra rare meson decays} \\
Weak flavour-changing neutral current (FCNC) decays are very sensitive
to contributions from heavy physics beyond the SM as they are both Cabibbo-Kobayashi-Maskawa
(CKM) and loop-suppressed.
In particular, the branching fractions (BRs)  for the decays $K \to \pi \nu \overline{\nu}$ are among
the observables in the quark-flavor sector most sensitive to NP.
Because they are strongly suppressed and calculated very
precisely in the Standard Model, these BRs are potentially sensitive to mass scales of hundreds of
TeV, surpassing the sensitivity of $B$ decays in most Standard Model extensions~\cite{Buras:2014zga}.
Observations of lepton-flavor-universality-violating phenomena are mounting in the $B-$ sector.
Measurements of the $K \to \pi \nu \overline{\nu}$ BRs are critical to interpreting the data
from rare $B$ decays, and may demonstrate that these effects are a manifestation of new degrees of
freedom such as leptoquarks~\cite{Buttazzo:2017ixm,Bordone:2017lsy,Fajfer:2018bfj}.

\vskip 2mm
The {\it KLEVER} project aims at measuring the BR of the very rare decays $K_L \to \pi^0 \nu \overline{\nu}$ with 20\%
accuracy, assuming the SM branching fraction. It will complement the result that will be obtained in the
next few years by the NA62 Collaboration on the charged mode,
with an upgraded beam line and detector.

\item[-] {\bf LFV decays of charged leptons} \\
Lepton-flavor-violating (LFV) charged lepton decays are also an excellent probe of physics BSM:
in fact within the SM with zero neutrino masses they are stricly forbidden, but many theories beyond the
SM~\cite{Babu:2002et,Brignole:2003iv,Paradisi:2005fk,Hays:2017ekz}
predict a non zero branching fraction, depending on the mechanism of neutrino masses generation.

Although strong constraints exist in the muon sector, those involving the third generation are
less stringent and need to be improved. Added impetus comes from the recent hints for the
violation of lepton universality in B-meson decays, as this phenomenon, in general, implies
LFV, with many theorists predicting effects just below the current experimental
bounds~\cite{Feruglio:2016gvd,Crivellin:2015era,Greljo:2015mma,Feruglio:2017rjo}.

\vskip 2mm
The {\it TauFV} proposal 
wants to search for LFV processes in $\tau$ and $D$-meson decays, exploiting the huge production of $\tau$ leptons
and $D$ meson occuring in the interactions of a high intensity 400 GeV proton beam with a target. TauFV aims at using
$\sim 2 \%$ of the total proton yield of the proposed Beam Dump Facility in the North area.

\item[-] {\bf Searching for permanent Electric Dipole Moments (EDMs)}\\
Permanent electric dipole moments (EDMs) are forbidden by parity and time reversal symmetries and
with the assumption of CPT invariance, they also violate CP invariance.
For fundamental reasons of quantum mechanics an EDM ($\vec{d_X}$) needs to be proportional to
the spin ($\vec{s}$) of a quantum mechanical particle $X$,
$\vec{d_X} =  \eta \cdot \mu_X \cdot \vec{s} $ , where $\mu_X = \frac{e \hbar}{m_X}$ 
is the magneton associated with particle $X$ of mass $m_X$ and charge $e$.
The constant $\eta$ contains all relevant (new) physics. The dependence  of  $\vec{d_X}$ on the inverse of the
particle mass causes that sensitivities to New Physics of EDM search experiments are different for the same numerical
values of established or future limits and it roughly scales with the mass of the tested particle.
Typical mass limits corresponding to, e.g., electron EDMs are  $\approx 5$ TeV for two loop processes
such as in multi Higgs scenarios, $\approx 60$ TeV  for one loop processes such as in supersymmetry
and $\approx 1000$ TeV in loop-free particle exchange such as for leptoquarks.

\vskip 2mm
Two PBC proposals aim at studying permanent EDMs in proton/deuteron and in charmed and strange baryons: these are the
{\it CPEDM} and the {\it LHC-FT} proposals, respectively.

\end{itemize}

In the following paragraphs a brief description of the {\it KLEVER}, {\it TauFV}, {\it CPEDM} and {\it LHC-FT}
proposals is reported. Their physics reach, also in connection to a multi-TeV new physics scale, is
discussed in Section~\ref{sec:phys-reach-multi-TeV}.

\subsection{KLEVER}
\label{ssssec:klever}
%
\noindent
{\it Brief presentation, unique features}\\
The NA62 experiment at the CERN SPS is expected to measure $BR(K^+ \to \pi^+ \nu \overline{\nu} )$
to within 10\% by the end of LHC Run 3. In order to fully constrain the CKM matrix, or possibly,
distinguish between different NP scenarios,
it is necessary to measure $BR(K_L \to \pi^0 \nu \overline{\nu}$) as well.
The KOTO experiment at J-PARC, should have enough data for the
first observation of the $K_L \to \pi^0 \nu \overline{\nu}$
decay by late 2020s\footnote{T. Yamanaka, presentation at the 26th J-PARC Program Advisory Committee, 18 July 2018,
  https://kds.kek.jp/indico/event/28286/contribution/11/material/slides/1.pdf},
but a next-generation experiment is needed in order to measure the BR.

\vskip 2mm
As far as KOTO is concerned, a new detector and an upgraded beam line would be required to go to ${\mathcal{O}}(100)$
events sensitivity:
an extension of the J-PARC hadron hall is currently being considered by the Japan Science Council with KOTO$^{++}$ as a priority.

\vskip 2mm
The KLEVER experiment aims to measure $BR(K_L \to \pi^0 \nu \overline{\nu})$ to $\sim$ 20\% accuracy assuming the SM
branching fraction, corresponding to the collection of 60~SM events with an S/B ratio of $\sim$ 1
using a high-energy neutral beam at the CERN SPS starting in Run 4.

\vskip 2mm
Relative to KOTO, which uses a neutral beam with a mean momentum of about 2 GeV, the boost from the high-energy beam
in KLEVER facilitates the rejection of background channels such as $K_L \to \pi^0\pi^0$
by detection of the additional photons in the final state.
On the other hand, the layout poses particular challenges for the design of the small angle
vetoes, which must reject photons from $K_L$ decays escaping through the beam pipe amidst an
intense background from soft photons and neutrons in the beam. Background from $\Lambda \to n \pi^0$ decays
in the beam must also be kept under control.

\vskip 0.5cm
\noindent
{\it Beam, beam time, timeline}\\
KLEVER would make use of the 400-GeV SPS proton beam to produce a neutral secondary beam with a mean $K_L$ momentum of 40~GeV,
leading to a fiducial volume acceptance of 4\%, and a $K_L$ yield of $2 \times 10^{-5}$ $K_L$/pot.
With a selection efficiency of 5\%, collection of 60 SM events would require a total primary flux
of $5\times 10^{19}$ pot, corresponding to
an intensity of $2\times 10^{13}$ ppp under NA62-like slow-extraction conditions. This is a six-fold increase
in the primary intensity relative to NA62. The feasibility of an upgrade to provide this intensity on
the T10 target is under study in the Conventional Beams working
group~\cite{Pbc:002}.
Preliminary indications are positive: there is general progress on issues related to the slow extraction
of the needed intensity to the North Area (including duty cycle optimization); a workable solution for T4-to-T10 bypass has
been identified. The ventilation in the TCC8 cavern appears to be reasonably hermetic, obviating
the need for potentially expensive upgrades. A four-collimator neutral beamline layout for ECN3
has been developed and simulation studies with FLUKA and Geant4 are in progress to quantify the
extent and composition of beam halo, muon backgrounds, and sweeping requirements.

\vskip 2mm
KLEVER would aim to start data taking in LHC Run 4 (2026). Assuming a delivered proton
intensity of $10^{19}$ pot/yr, collection of 60 SM events would require five years of data taking.
To be ready for the 2026 start date, detector construction would have to begin by 2021 and be substantially
concluded by 2025, leaving three years from the present for design consolidation and R\&D.

\vskip 0.5cm
\noindent
{\it Key requirements for detector}\\
\begin{figure}[h]
\centerline{\includegraphics[width=\linewidth]{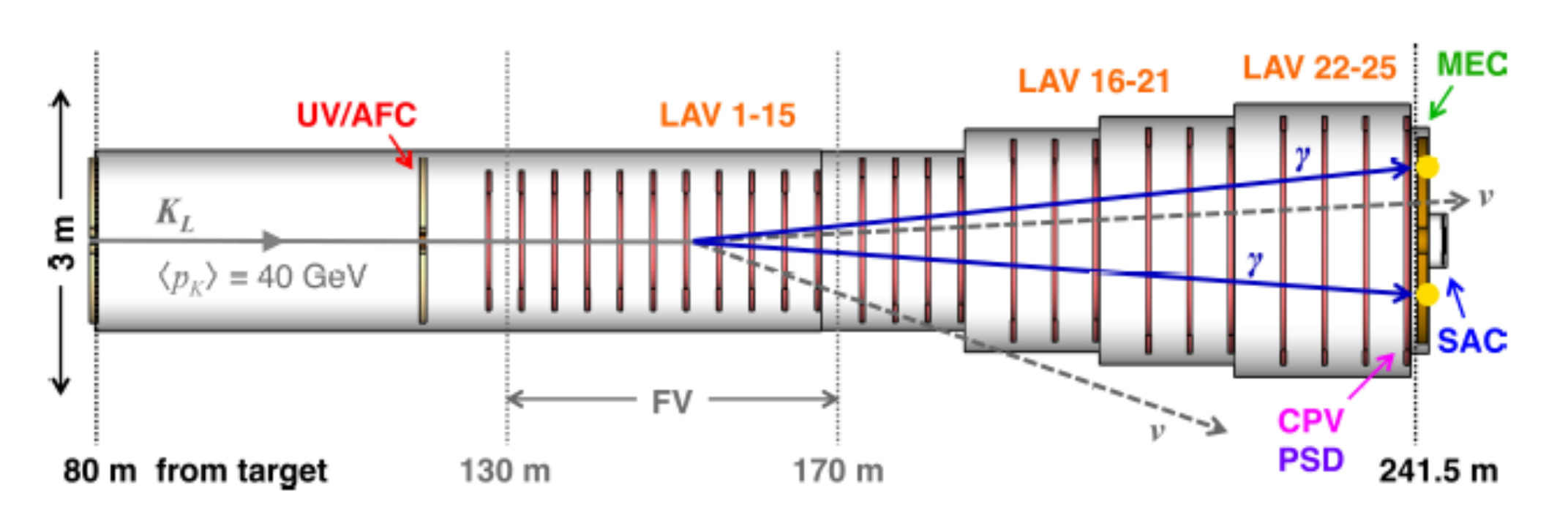}}
\caption{KLEVER experimental apparatus: upstream veto (UV), active final collimator
(AFC), large-angle photon vetoes (LAV), main electromagnetic calorimeter (MEC), smallangle
calorimeter (SAC), charged particle veto (CPV), pre-shower detector (PSD).}
\label{fig:klever}
\end{figure}
A schematic layout of the experiment is shown in Figure~\ref{fig:klever}.
Most of the subdetector systems for KLEVER will have to be newly constructed. Early studies
indicated that the NA48 liquid-krypton calorimeter (LKr) could be reused as the Main Electromagnetic Calorimeter (MEC),
and indeed, the efficiency and energy resolution of the LKr appear to be satisfactory for KLEVER.

However, the LKr timing resolution would be a major liability. The LKr would measure the event time in KLEVER
with 500 ps resolution, while the total rate of accidental vetoes from the SAC could be 100 MHz.
The LKr time resolution might be improved via a comprehensive readout upgrade, but concerns
about the service life of the LKr would remain, and the the size of the inner bore would limit the
beam solid angle (and hence kaon flux). The collaboration is investigating the possibility of replacing the LKr
with a shashlyk-based MEC patterned on the PANDA FS calorimeter (in turn, based on the KOPIO
calorimeter~\cite{Atoian:2007up}). This is a shashlyk design incorporating “spy tiles” for longitudinal sampling
of the shower development, resulting in additional information for 
$\gamma/n$ separation. A first test of this
concept was carried out with a prototype detector at Protvino in April 2018.

\vskip 2mm
The upstream veto (UV), which rejects $K_L \to \pi^0 \pi^0$ decays upstream of the fiducial volume,
would use the same shashlyk technology as the MEC. The active final collimator (AFC), inserted
into the hole in the UV for passage of the beam, is a LYSO collar counter with angled inner surfaces.
This provides the last stage of beam collimation while vetoing photons from $K_L$ that decay in transit
through the collimator itself.
Because of the boost from the high-energy beam, it is sufficient for the large-angle photon vetoes
(LAVs) to cover polar angles out to 100 mrad. The LAVs are lead/scintillating-tile detectors based
on the CKM VVS~\cite{Ramberg:2004en}.
Extensive experience with this type of detector (including in prototype tests for NA62)
demonstrates that the low-energy photon detection efficiency will be sufficient for KLEVER~\cite{Atiya:1992vh,Ambrosino:2007ss}.

\vskip 2mm
As far as the rejection of charged particles is concerned, simulations indicate that the needed
rejection can be achieved with two staggered planes of charged-particle veto (CPV) each providing
99.5\% detection efficiency, supplemented by the $\mu$ and $\pi$ recognition capabilities of the MEC
(assumed in this case to be equal to those of the LKr) and the current NA62 hadronic calorimeters
and muon vetoes.

\vskip 2mm
Finally, a pre-shower detector (PSD) featuring $0.5~X_0$ converter and two planes of tracking with
$\sigma_{x,y} \sim 100~\mu$m (assumed to be large-area MPGDs) would allow angular reconstruction of at least
one $\gamma$ from $K_L \to \pi^0 \pi^0$ events with two lost 
$\gamma$'s to be reconstructed in 50\% of cases.

\vskip 0.5cm
\noindent
{\it Open questions, planned feasibility studies}\\
Simulations of the experiment carried out with fast-simulation techniques (idealized geometry,
parameterized detector response, etc.) show that the target sensitivity is achievable (60 SM events
with S/B = 1). Background channels considered at high simulation statistics include $K_L\to \pi^0 \pi^0$
(including events with reconstructed photons from different $\pi^0$s and events with overlapping photons
on the MEC), $K_L \to 3 \pi^0$ and $K_L \to \gamma \gamma$. 

\vskip 2mm
Background from $\Lambda \to n \pi^0$ and from decays with charged
particles is assumed to be eliminated on the basis of studies with more limited statistics. An effort is
underway to develop a comprehensive simulation and use it to validate the results obtained so far.
Of particular note, backgrounds from radiative $K_L$ decays, cascading hyperon decays, and beam-gas
interactions remain to be studied, and the neutral-beam halo from more detailed FLUKA
simulations needs to be incorporated into the simulation of the experiment. Preliminary studies
indicate that the hit and event rates are similar to those in NA62, with the notable exception of the
SAC, which will require an innovative readout solution. Offline computing resources required are
similarly expected to be on the scale of NA62.

\vskip 2mm
A PBC concern is related to the overall cost of the project 
if compared to the current strength of the Collaboration. The Collaboration is well aware that
success in carrying out the KLEVER experimental program will
require the involvement of new institutions and groups, with resources to contribute to the project,
and initiatives to seek new collaborators are a major focus at present.

The proton sharing with existing or potential users in the North Area
(as, eg., SHiP), is also a concern: this  will require a proper schedule and prioritization among the proposals.

\vskip 0.5cm
\noindent
{\it Status of the Collaboration}\\
About 13 Institutions currently participating in NA62 have
expressed support for and interest in the KLEVER project. These are:\\
{\it - Bulgaria: } University of Sofia (3 people);\\
{\it - Czech Republic:} Charles University (1+ person);\\
{\it - Germany:} Mainz (1+ person);\\
{\it - Italy:} University and INFN Ferrara (4 people), University and INFN Firenze (1 person),
INFN Frascati (3 people), University Guglielmo Marconi and INFN-Frascati (2 people);
University and INFN Naples (5 people);
University and INFN Pisa (3+ people);
University and INFN Tor Vergata (9 people);
University and INFN Torino (5+);\\
{\it - Russia:} INR Moscow (10 people), IHEP Protvino (5 people); \\
{\it - USA:} George Mason University (1 person).

\noindent
Individuals from UK institutions participating in NA62 have indicated an interest in the KLEVER
project and are exploring the possibility of joining.
In addition 5 people from CERN EN-EA are currently dedicated to the study of the KLEVER beam line.

\vskip 2mm
In addition to direct KLEVER input for the European Strategy update, an Expression of Interest to the SPSC is in preparation and
will serve as an opportunity to consolidate project membership.

\subsection{TauFV}
\label{ssssec:taufv}
\noindent
{\it Brief presentation, unique features}\\
The TauFV Collaboration aims at exploiting the high intensity of the Beam Dump Facility (BDF) at CERN
and install a detector, upstream of the proposed SHiP beam-dump target, which
will have world-leading sensitivity to many LFV decay modes, for example probing for $\tau \to \mu \mu \mu$
decays down into the $10^{-10}$ regime.
For the $\tau \to \mu \mu \mu$ mode, a limit of $2.1 \times 10^{-8}$ at the 90\%
confidence level has been set by the Belle collaboration~\cite{Hayasaka:2010np}.
Results of similar sensitivity have been obtained by BaBar~\cite{Lees:2010ez} and LHCb~\cite{Aaij:2014azz}.
The Belle-II experiment expects to reach a sensitivity of $1\times 10^{-9}$~\cite{Abe:2010gxa},
but may be able to go lower if all background is suppressed.

\vskip 2mm
TauFV will be well suited to other LFV studies in tau decays, for example $\tau^- \to e^- e^+ e^-$,
$\tau^- \to \mu^- e^+e^-$, $\tau^- \to e^- \mu^+ \mu^-$, $\tau^- \to \mu^+ e^- e^-$ and $\tau^- \to \mu^+ e^- e^-$.
Particularly high sensitivity is expected for the latter two modes, where the initial level of contamination will
be lower. Lepton number violation (LNV) searches will be performed with decays such as
$\tau^- \to h^- h^- \ell^+$ ($h$ = any hadron, $\ell = e$ or $\mu$ ). The experiment will also have access to an
enormous number of charm decays (e.g. $5 \times 10^{15}$ $D^0$ mesons), which will allow a parallel
programme of LFV and LNV study with modes such as $D \to h \mu^- e^+$ and $D \to  h \ell^- \ell^-$.
World-leading measurements will be possible in the field of charm physics, many enabled
by the excellent calorimetry of the experiment, including CP-violation measurements and
searches for suppressed decays such as $D^0 \to \mu^+ \mu^-$ and $ D^0 \to \gamma\gamma$. 

\vskip 0.5cm
\noindent
    {\it Location, beam, beam time, timeline}\\
    The baseline scenario is to use 2\% of the protons currently intended for the SHiP experiment,
    which could be achieved with an integrated target thickness of 2~mm of tungsten. A five
    year period of operation would produce $4 \times 10^{18}$ pot, which would result in
    $8\times 10^{13}$ $D^-_s  \to \tau^- \overline{\nu}$ decays. This enormous yield is two orders of magnitude larger than
    the number of $\tau$ leptons so far produced at the LHCb interaction point, and five orders of
    magnitude larger than that produced at Belle.
    
\vskip 2mm
    The timescale for installing and operating TauFV is dictated both by the construction of
    the BDF, and by the development of the challenging sub-detector technology, in particular the front-end ASICs.
    The TauFV experimental hall could be prepared in 2026-27, in parallel with the installation of SHiP.
    If the project proceeds rapidly it would be possible to deploy the full detector at this time.
    Alternatively, a first-stage experiment, capable of demonstrating the possibility of performing high-precision 
    flavour physics in this new environment, could be installed instead.
    The full scale experiment would then be assembled in LS4, currently foreseen for 2030.
    An attractive feature of TauFV is that the physics reach is not limited by the intensity
    of the available beam. Therefore, it is conceivable that future upgrades could be planned,
    integrating significantly more pot, depending on how both the detector technology and the
    physics landscape evolve.

 \vskip 2mm   
 From the beam optics point of view, several locations can provide the required beam conditions and
 the beam drift space to accommodate the detector along the new 200~m transfer
 line between the TDC2 switch-yard cavern and the BDF target station, without either
 affecting the location of the BDF experimental area or requiring significant changes to the
 beam-line configuration. The choice is instead driven by considerations related to the civil
 engineering in the vicinity of the existing installations, radiological protection, and access
 and transport requirements, both above ground and underground. Lateral space is required
 on both sides for shielding in order to limit the radiation exposure of the surrounding underground area to
 levels typical for the rest of the beam line.
 The currently preferred location is situated 100~m upstream of the BDF target bunker.
 An access and service complex for the transfer line is already foreseen at this location. This
 complex will be extended and reconfigured to include a bypass tunnel, the detector bunker,
 service cavern and the required surface infrastructure.

\begin{figure}[h]
  \centerline{
    \includegraphics[width=0.3\linewidth]{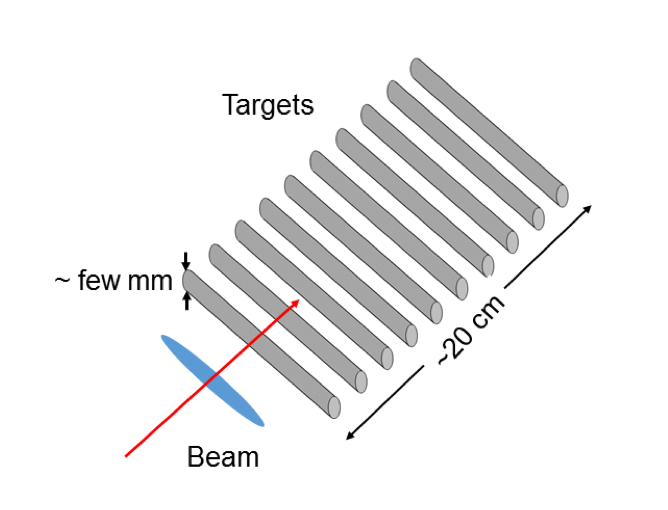}
    \includegraphics[width=0.65\linewidth]{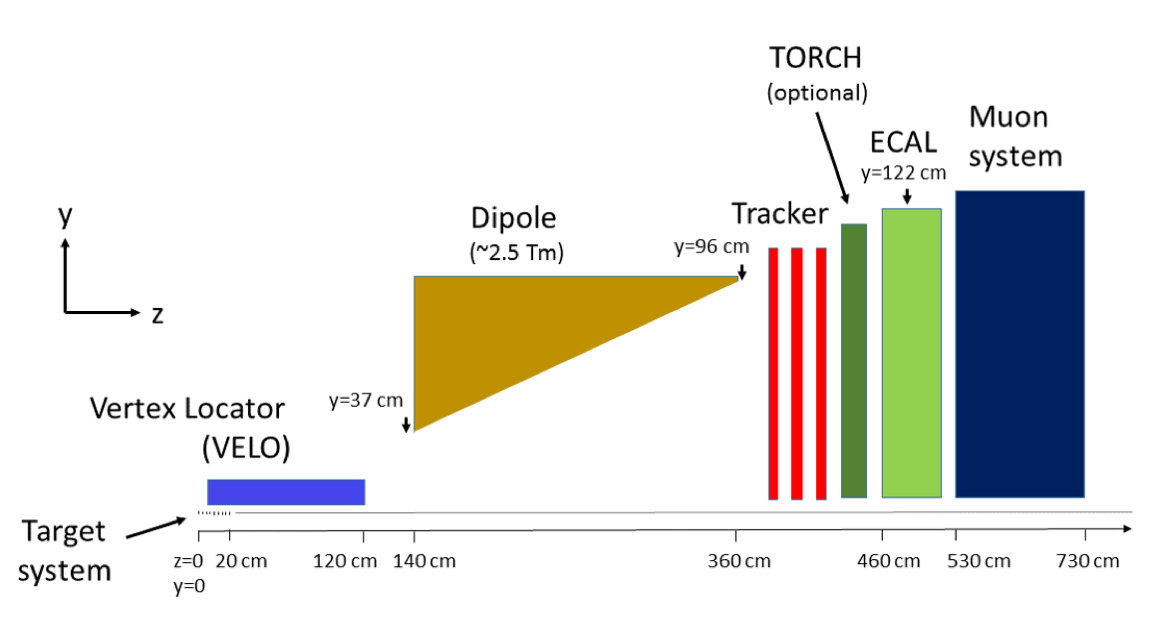}
  }
\caption{Schematic layout of the target system (left) and the spectrometer (right) of the TauFV experiment.}
\label{fig:taufv}
\end{figure}

\vskip 0.5cm
\noindent
{\it Key requirements for detector}\\
The target system of TauFV will consist of a set of thin tungsten blades, matched to an
elliptical beam profile of vertical size $\sim$1~mm, each separated by $\sim$2~cm and distributed over
a length of 10-20~cm (Figure~\ref{fig:taufv}, left). This layout will ensure that interactions will be well
spread both longitudinally and transversally, which is desirable for background rejection.
Furthermore, the majority of the $\tau$ leptons will decay in free space, and there will be a low
probability of a decay track passing through a downstream target.

\vskip 2mm
The spectrometer design (Figure~\ref{fig:taufv}, right) has an acceptance in polar angle
between 20 and 260~mrad, and length of around 7~m. A Vertex Locator (VELO), comprising planes of
silicon-pixel detectors broadly similar to the LHCb VELO, interleaves the target system, and
continues downstream of it. Bending of charged tracks is provided by a dipole of integrated
field of $\sim 2.5$~Tm, which is followed by a tracker, a TORCH detector, a high performance ECAL and a muon system.
All detector components will have fast-timing capabilities, good radiation hardness and
high granularity.
The TORCH detector provides time resolution for charged tracks of $<20\,$ps, which is a key weapon in the
suppression of combinatoric background, and also brings hadron identification capabilities,
which will enhance the charm-physics programme of the experiment.

\vskip 2mm
R\&D is beginning on the most critical elements of the detector, in particular the VELO
and the ECAL. Here there is very close synergy with the requirements of Upgrade II of
LHCb~\cite{Bediaga:2018lhg}. The VELO stations will be built from hybrid pixel sensors, and discussions
have begun with the MediPix collaboration concerning the requirements of the ASIC. A
promising solution for the ECAL would be to employ crystal modules, based on YAG or
GAGG as a scintillator, and using the leading edge of the light pulse, or alternatively a
silicon preshower, to provide the fast-timing information. Crystal samples have already been
acquired, with the aim of constructing and evaluating a prototype module later this year.

\vskip 0.5cm
\noindent
{\it Physics reach and background considerations}\\
Evaluation of the physics reach of the TauFV experiment has so far focused on the benchmark
mode $\tau^- \to \mu^-\mu^+\mu^-$.  Studies are still ongoing, but preliminary results suggest that excellent
control of combinatoric background will be achievable, mainly due to the distributed target system,
which suppresses the likelihood of fake combinations, and the fast-timing provided by the TORCH and
other sub-detectors.  Hence, combinatorics will certainly be the sub-dominant source of background,
and will not significantly impact upon the measurement down to signal branching ratios of $1 \times 10^{-10}$, and maybe lower.
Background from same topology decays of $D^+$ and $D^s$ mesons involving three leptons will be a greater concern,
but will be combatted through good mass resolution, kinematic requirements involving the direction between
the interaction and decay vertices, and the possibility to tag the soft photon from $D_s^{*} \to D_s \gamma$ decays,
thereby rejecting backgrounds from $D^+$ mesons.
Restrictions on the invariant mass of each dimuon pair can isolate ultra-pure regions of phase space,
but at the expense of introducing model-dependence into the interpretation of the results.
All these methods will be combined in a multivariate analysis to obtain maximum discrimination.
Although final results are not yet available, it seems probable that sensitivity to branching ratios of a few $10^{-10}$
will be attainable.  The physics reach in modes of the sort $\tau^- \to e^+ \mu^-\mu^-$, which are afflicted by
combinatoric background alone, will be even better by an order of magnitude.

\vskip 2mm
The potential of TauFV in charm physics can be assessed by a direct comparison with LHCb,
as the sources of background are the same, and are generally dominated by combinatorics, in the suppression of which
TauFV has clear advantages.  The ECAL, optimised for soft photons, will give TauFV exciting possibilities in radiative decays.
TauFV will therefore provide charm measurements of, similar or higher precision to those of the proposed
LHCb Upgrade II across a wide range of decay topologies, including modes that are complementary to the collider experiment.

\vskip 0.5cm
\noindent
{\it Open questions, planned feasibility studies}\\
The project is currently at a very early stage: no results from simulation are still available, and the evaluation
of the background is currently ongoing. This will be a mandatory step to address a definitive estimate of the physics reach.
In order to pursue the proposed physics programme, a strengthening of the Collaboration is also necessary.

\vskip 0.5cm
\noindent
{\it Status of the Collaboration}\\
The current TauFV collaboration consists of nine physicists from the University of Bristol (1)
CERN (2), Imperial College London (1) , the University of Oxford (1), the University of Zurich (3) and
ETH Zurich (1). 
Before the end of this year these groups will complete the initial
optimisation of the layout and determine the physics reach for benchmark channels. In parallel,
discussions will take place with additional potential collaborators.

\subsection{CPEDM and LHC-FT}
\label{ssssec:cpedm}

\noindent
{\bf Experimental landscape} \\
There has been a substantial number of dedicated search experiments for permanent electric dipole moments
in a variety of systems over the past 60 years. All of them were well motivated and with a clear potential
to discover new physics.
They can be distinguished in four types depending on the particles studied.
I.e. there are experiments on: \\
\noindent
- free elementary particles, such as electron, muon, tau, but also approximately free elementary particles as proton, neutron; \\
- atoms, such as Hg, Xe, Tl, Cs; \\
- molecules such as YbF, ThO, BaF, HfF$^+$; \\
- condensed matter samples, such as ferroelectric materials, liquid Xe. \\

\noindent
Each of these lines of research has its own merits. 
Since a finite value of the not yet fully understood $\Theta_{QCD}$ parameter could cause EDM in hadrons,
only an EDM found in a lepton would  immediately indicate non-SM physics.

\vskip 2mm
Any first discovered EDM would call for further experiments to unravel the potentially different sources of the
underlying new process of CP violation. Several hadronic EDMs could be used to 
demonstrate or disprove a $\Theta_{QCD}$ explanation, the combination with leptons will be indispensable
to disentangling new physics.
Because of the known CP-violation in the Standard Model,
permanent EDMs of fundamental particles are predicted which arise, e.g., for neutrons from three loop processes
and for leptons from at least four loop processes.
The Standard Model EDM values are of order $10^{-32}$ ecm for neutrons and $10^{-40}$ ecm for electrons~\cite{Khriplovich:1997ga}.
Such small values are by orders of magnitude below present experimental possibilities and they therefore open
large windows of opportunity for observing New Physics.
For almost all particles speculative models exist which can provide for EDMs almost as large as the present experimental
limits~\cite{Chupp:2014gka}.

\vskip 2mm
Motivation to carry out experiments to search for EDMs in one or another system therefore require judgment
calls on the viability of such speculative models. Independent of this, the non-observation of any EDM has
ruled out more speculative theories than any other known experimental approach\footnote{N. Ramsey, at ”Breit Symposium”, Yale (1999).}.
As one example of future power of EDM experiments,  searches for EDMs on baryons and light nuclei,
i.e. neutron, proton, deuteron and $^3$He, have particular potential to unravel different models of
CP violation~\cite{Dekens:2014jka}.
Below we present in a synthetic way the main techniques currently used to search (directly or indirectly)
for EDM in elementary particles as neutrons, protons, electrons and muons.

\begin{itemize}
  
\item[-] {\it Neutron EDM using Ultra-Cold Neutrons} \\
Experiments to search for the EDM of the free neutron ($d_n$) have been conducted since the
1950ies~\cite{Smith:1957ht}.
A long chain of experiments with ever increasing sensitivity, first with neutron beams and later with stored
ultracold neutrons (UCN), has yielded the present best limit of  $|d_n| < 3 \cdot 10^{-26}$ecm~\cite{Afach:2015sja}.
Presently there are at least five different sizeable efforts\footnote{
See e.g. nedm2017 Workshop, Harrison Hot Springs, Oct 15-20, 2017, organized by R. Picker et al.}
aiming at improving the sensitivity to the neutron EDM in steps to $10^{-27}$ecm and then to $10^{-28}$ecm
over the next 5-10 years. Several efforts (at PSI, ILL, LANL, TRIUMF) will use improved intensities
of UCN stored in vacuum and at room temperature. One effort (at SNS) aims at conducting the
measurement with UCN inside cold superfluid He.
The SFHe environment offers advantages~\cite{Golub:1994cg}
of potentially larger numbers of UCN exposed to larger electric fields, however, at the cost of
considerable complication of setup and handling.
A first measurement in the cryogenic environment has still to be demonstrated. Beyond 10 years, some proposed or
ongoing R\&D efforts might succeed with cryogenic setups.
Alternatively, also an experiment at a pulsed cold neutron beam of ESS has been proposed~\cite{Piegsa:2013vda}.

\item[-] {\it Neutral atoms as probe of neutron and proton EDMs} \\
EDMs have also been searched for in neutral atoms. From EDM searches in diamagnetic atoms numerous limits on
parameters describing physics within the Standard Model or beyond it could be extracted.
The most recent table top experiment on $^{199}$Hg has established $|d_{Hg}| < 7.4 \cdot 10^{-30}$ ecm (95\% C.L.)~\cite{Graner:2016ses}.
From this value various other limits have been derived when assuming that there was for each case only one
process that can cause an EDM in $^{199}$Hg. Amongst others a best neutron EDM limit
of $|d_n| < 1.6 \cdot 10^{-26}$ ecm~\cite{Dmitriev:2003sc}
proton EDM limit of $|d_p| < 2.0 \cdot 10^{-25}$ ecm~\cite{Dmitriev:2003sc}
have been established as well as a limit on the QCD $\Theta$ parameter at
$|\Theta_{QCD}| < 1.1 \cdot 10^{-10}$~\cite{Sahoo:2016zvr}.

\item[-] {\it Paramagnetic atoms and molecules as probe of electron EDM  } \\
EDM searches in paramagnetic atoms have yielded limits primarily on the electron EDM.
Those early limits have been superseded since by searches in molecules and in molecular ions,
where internal electric fields in these molecules give rise to some $10^5 \to 10^9$ fold enhancement for an electron EDM
for example by using excited states of ThO in a molecular beam~\cite{Andreev:2018ayy}
or the ground state HfF$^+$ in an rf-particle trap~\cite{Cairncross:2017fip}
Bounds could be established at $|d_e| < 1.1 \cdot 10^{-29}$ ecm and $|d_e|< 1.3 \cdot 10^{-28}$ ecm (90\% C.L.),
respectively, with these experiments which are exploiting significantly different techniques.
Further improvements are expected soon from projects using these and also further molecules,
such as YbF~\cite{Hudson:2011zz}
and BaF~\cite{Aggarwal:2018pru}
It is highly realistic to expect that within the coming decade sensitivities better than $10^{-30}$ ecm will
be achieved for the EDM on the electron.

\item[-] {\it Muon EDM} \\
The most sensitive EDM search experiments so far have been conducted on systems involving particles from the first
particle generation exclusively. Yet limits on higher generation particles could be established as well.
Along with measurements of the muon magnetic anomaly (muon g-2) always EDM values could be obtained,
the best limit being now $|d_{\mu}| < 1.8 \cdot 10^{-19}$ ecm (95\% C.L.)~\cite{Bennett:2008dy}.
The series of muon g-2 experiments since the 1960ies has exploited the strong motional magnetic fields
muons experience when moving at high velocities (close to the speed of light) through static magnetic fields.
This basic concept underlies a muon EDM experiment proposed for  the Paul Scherrer Institute (PSI)~\cite{Adelmann:2010zz}.
As a major improvement in the experimental concept a radial electric field is installed in the storage volume
to compensate the particle’s g-2 value related spin precession. An EDM on the muon manifests itself as an
out of orbit plane precession of the muon spin, which can be detected via the time evolution of spatial distribution of
decay electrons.

\vskip 2mm
At existing muon facilities a statistics limited sensitivity of $|d_{\mu}| \approx  7 \cdot 10^{-23}$ ecm can be
achieved within 1 year of data taking. At this precision the viability of the technique to directly search for an
EDM on even short-lived charged particles can be demonstrated. Further, already at this sensitivity a number
of Standard Model extensions can be tested~\cite{Adelmann:2010zz}
which in particular account for the fact that the muon is a second generation particle.

\end{itemize}

\begin{figure}[h]
\centerline{\includegraphics[width=\linewidth]{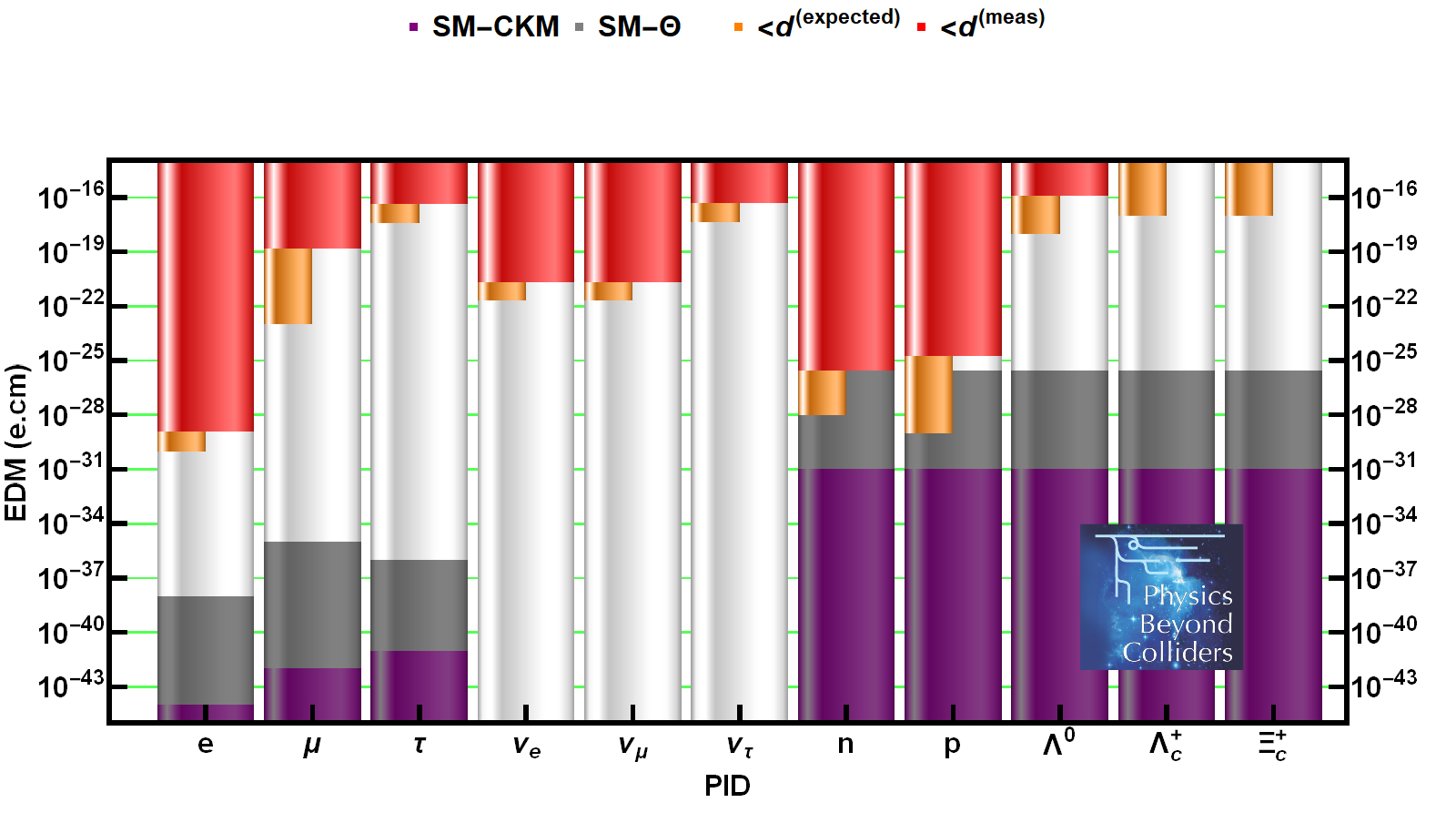}}
\caption{Overall status of EDM measurements: current limits on EDMs of fundamental particles
  are displayed as red bars from the top. From below come the SM estimates from CKM CP violation and $\Theta_{QCD}$.
  White regions indicate safe BSM discovery territory for the experiments.}
\label{fig:EDM-limits}
\end{figure}

Further limits on higher generation particles have been established.
Figure~\ref{fig:EDM-limits}
displays limits on the electric dipole moments of fundamental particles.
Muon and neutron limits have been deduced from measuring directly on these free particles,
while e.g. the limit on the electron EDM results from the ACME experiment on ThO~\cite{Andreev:2018ayy}
assuming the electron EDM as the sole CP violating source.

The limits on the neutrino EDMs are together with limits on magnetic moments deduced from cross sections
which would be affected by electro-magnetic couplings. The experimental limits are displayed as red bars
from the top. From below come the SM estimates from CKM CP violation and $\Theta_{QCD}$.
White regions indicate safe BSM discovery territory for the experiments.
The range of ongoing or proposed experimental projects is indicated in orange.

\vskip 2mm
\noindent
    {\bf PBC proposals: CPEDM and LHC-FT} \\
Improved sensitivities can in several cases be obtained with the projects proposed for CERN within this PBC study:
the proton EDM is the topic of the CPEDM Collaboration,
and the strange and charm baryons might be improved or measured for the first time at all~\cite{Botella:2016ksl,Bagli:2017foe} with the experiment
proposed by the LHC-FT group:

\begin{itemize}
\item[-] {\it LHC-FT: measurement of EDMs in charmed and strange baryons} \\
Interest in hadronic EDM of second and even third generation quarks comes, e.g., from the fact that the indirect constrains on the
charm EDM are rather weak, of order $4 \times 10^{-17}$ecm~\cite{Sala:2013osa}
only. As no finite EDM has been observed so far and no source of BSM CP violation is 
known yet, experimental efforts covering uncharted territory are necessary.
The charm quark as well as the muon might via unexpectedly large EDM give clues on specific flavor structure of new physics.  

\vskip 2mm
The experiment concept relies on a bent crystal to extract protons from the LHC beam halo. These protons will then hit 
a dense target and produce charged heavy and strange baryons that will then be channelled in bent crystals positioned in front of the
detector. The intense electric field between the crystal atomic planes is able to induce a sizeable
spin precession during the lifetime of the particle. The EDM, along with the magnetic dipole
moment\footnote{see Report of the Physics Beyond Colliders QCD WG, to appear.},
can be determined by analysing the
angular distribution of the decay particles.
Recently, the possibility to use the same technology for measuring the EDM (and MDM) of the tau lepton
has been discussed in Refs.~\cite{Fu:2019utm,Fomin:2018ybj}.

\vskip 2mm
The LHC interaction point IP8, where the LHCb detector~\cite{Alves:2008zz,LHCb-DP-2014-002} sits, has been identified as a suitable
location of the experiment. A main challenge is represented by the limited coverage of the detector
in the very forward region, requiring a secondary crystal with a large bending exceeding 15 mrad.
A W target of $\approx 2$ cm thickness hit by a proton flux of $\approx 10^7$ protons/s is the upper limit for a
parallel detector operation.
R\&D is ongoing to assess the feasibility of the secondary crystal along other challenges of the proposal, which include
the compatibility with the machine, its operation mode, maximum reachable proton flux and the design of the absorber downstream the
detector.

\vskip 2mm
About $2.4\times10^{14}$~proton on target could be reached with three years of data taking
after the installation during an LHC technical stop during Run 3, either with two weeks per year of dedicated detector running 
at $10^8$~proton/sec or with parallel detector operation at $10^7$~proton/sec.
This would lead to EDM sensitivities of about $10^{-17}\,ecm$ for charm baryons.
Extending the detector coverage down to $10$~mrad along with an increase of the proton flux during LHC Run~4 and Run~5,
either at LHCb or at a dedicated experiment would improve sensitivity by about one order of magnitude.

Figure~\ref{fig:edm_lhcft} shows the EDM sensitivity for different baryons in two different scenarios,
scenario 1 (S1) corresponds to data collected at the LHCb interaction point in a first phase at low luminosity (about $2\times 10^{14}$ pot);
scenario 2 (S2) corresponds to data collected at a possible next-generation experiment at higher luminosity ($\sim 10^{17}$)
and enhanced coverage.

\begin{figure}[h]
\centerline{\includegraphics[width=0.7\linewidth]{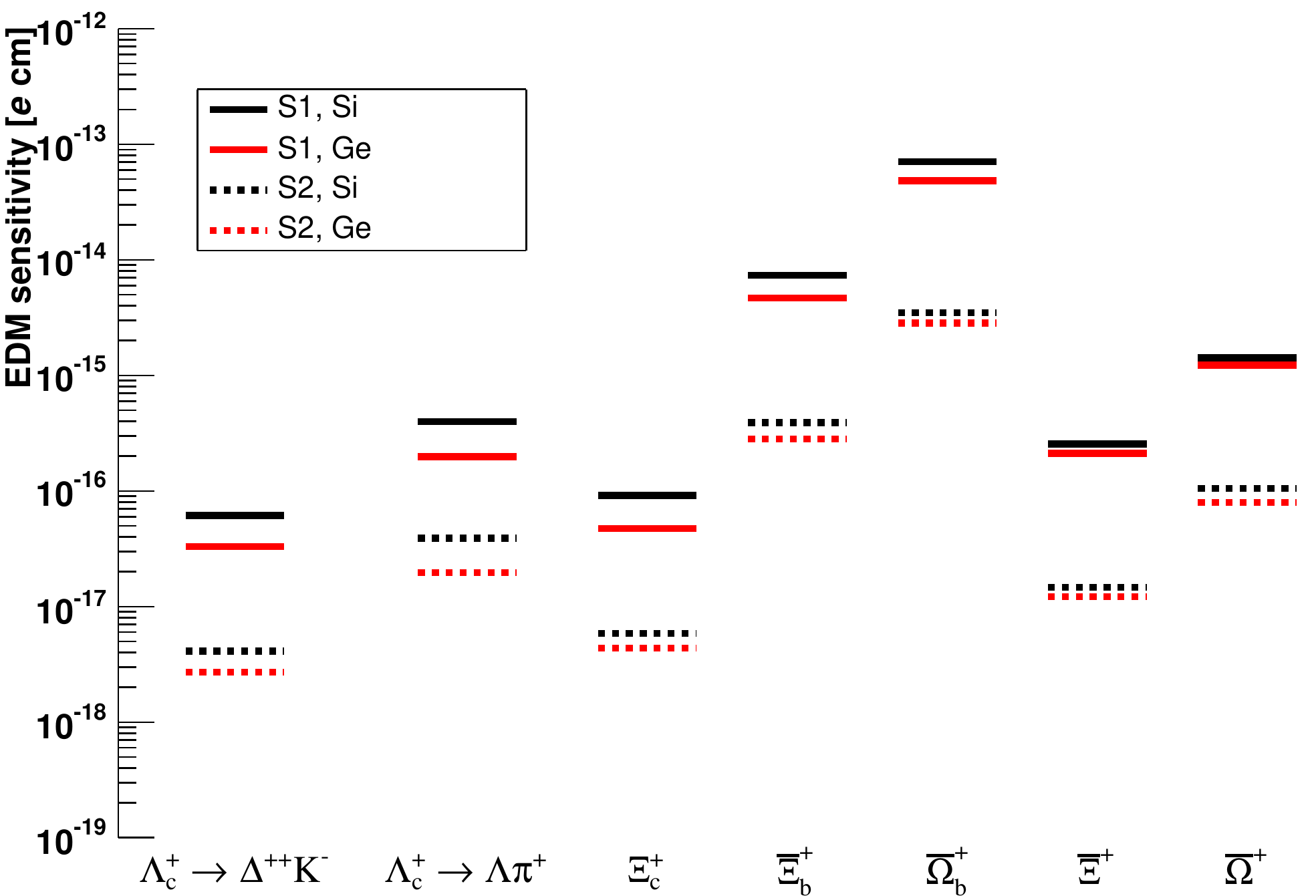}}
\caption{EDM sensitivities for different baryons in two different scenarios,
scenario 1 (S1) corresponds to data collected at the LHCb interaction point in a first phase at low luminosity (about $2\times 10^{14}$ pot);
scenario 2 (S2) corresponds to data collected at a possible next-generation experiment at higher luminosity ($\sim 10^{17}$)
and enhanced coverage. Figure revisited from Ref.~\cite{Bagli:2017foe}.}
\label{fig:edm_lhcft}
\end{figure}

\item[-] {\it CPEDM: measurement of proton and deuteron EDMs} \\
The same experimental concept as for muons, i.e., exploiting a magnetic storage ring and motional electromagnetic fields,
underlies the proposed deuteron EDM experiment for CERN.
The spin analysis in this case is achieved by a newly developed deuteron polarimeter.
Numerous preparations including polarimetry and spin manipulation are already being studied by the JEDI collaboration.
The COSY experiment is an indispensable proof of principle at $\sim 10^{-24}$~ecm
sensitivity for a  ring experiment using a dedicated magnetic storage ring for deuterons (or protons) at CERN,
which is needed to observe or establish a limit on the deuteron EDM at the level of $10^{-29}$ ecm.

\vskip 2mm
For a deuteron experiment at CERN a new magnetic storage ring is required with some 80~m circumference
to store polarized deuterons and observe the time evolution of their polarization.
The precursor experiment at COSY is expected to develop all required detectors with sufficient
sensitivity and to test the viability of the approach for hadrons.
Both the muon and deuteron EDM experiment concepts take advantage of the fact that the magnetic anomaly
is rather small and therefore magnetic spin precession in a magnetic storage ring can be compensated effectively by
radial electrostatic fields.

\vskip 2mm
The proton EDM experiment proposed by the CPEDM collaboration for CERN uses as a contrast a purely
electrostatic dedicated storage ring of some 400-500~m circumference and with alternating field gradients,
since the magnetic anomaly is much larger than for deuteron amd muon.
For sensitivity $10^{-29}$ ecm it exploits a proton beam of 233~MeV energy.
The device needs provision for  clockwise and counter-clockwise particle injection to minimize systematics.
External magnetic fields at the experimental site need to be compensated to some 10~nT all over the
particle storage volume and through the experimental running time.
The substantial necessary expenses require a full structured programme of stepwise testing of all
essential concepts and necessary devices.
The proton EDM project at CERN is a joint effort of the proton and deuteron EDM communities.
It appears that a small size proof of principle experiment would be indispensable.

\vskip 2mm
An experiment on the proton EDM tests to a large part the same speculative models as the neutron EDM,
except for such that are constructed with isospin dependence.
Therefore a proton EDM experiment will need to exceed the prospected future sensitivity values expected
for neutron experiments in order to justify the expenditures.
Here one expects  some $10^{-27}$ ecm by 2025 and  $10^{-28}$ ecm by 2030.
Note, for the deuteron an EDM can arise from either a proton or a neutron EDM  (or both) and in addition an EDM may
be due to CP violating parts in the proton-neutron interaction of the deuteron binding.
Both experiments, once they have proven sufficient sensitivity, are therefore strongly motivated and they have
robust discovery potential.
Yet, the speed of progress in the area of molecular EDM searches and the significantly
lower costs of table top experiments need to weighted against those of the storage ring approaches.
The CeNTREX experiment at
Yale\footnote{D. DeMille et al., https://www.physics.umass.edu/sites/default/files/ attachments/page/20470/fie-kawall-centrex.pdf}
aims for a 30 times improvement for the proton EDM (and 100-fold improvement on $\Theta_{QCD}$) as compared
to limits established for the proton to
date\footnote{https://demillegroup.yale.edu/research/centrex-search-electric-dipole-moment-edm-proton}.

\end{itemize}

\clearpage
\section{Physics reach of PBC projects}
\label{sec:phys-reach}
In the following Sections we review the physics reach of the experiments proposed in the PBC-BSM study group
and the impact that CERN could have in the search for New Physics at mass scales different from the TeV scale
in the next 10-20 years. Their physics reach is compared to the existing results and to the projections of experiments
either operating at existing facilities or proposed to future facilities beyond those considered in this study.
The results are presented following the scheme outlined in Section~\ref{sec:exps} where the experiments were classified
in terms of their sensitivity to New Physics in the sub-eV (Section~\ref{ssec:exps_sub_eV}),
MeV-GeV (Section~\ref{ssec:exps_MeV_GeV}), and multi-TeV (Section~\ref{ssec:exps_multi_TeV}) mass scales.

\section{Physics reach of PBC projects in the sub-eV mass range}
\label{sec:phys-reach-sub-eV}
Experiments searching for axions/ALPS in the sub-eV presented in the PBC-BSM study group exploit their possible
coupling to photons, and, as such, are sensitive to the benchmark case  BC9 discussed in Section~\ref{sec:portals}.

The photon regeneration experiment can be sensitive
to milli-charged particles (benchmark case BC3) and hidden photons (benchmark case BC2), however no sensitivity
estimate has been given for the first case BC3. For the hidden photons, their production in a LSW apparatus
is not related to the presence of the static magnetic field; since one of the major improvements of the proposed experiment
is related to the increase of the magnetic field amplitude, a smaller advancement over the present sensitivity
to hidden photons is expected.

\subsection{Axion portal with photon dominance  (BC9)}
\label{ssec:subev-axions-photon}

\noindent
    {\bf Current bounds} 
    
\begin{figure}[h]
\begin{center}
\includegraphics[width=0.8\linewidth]{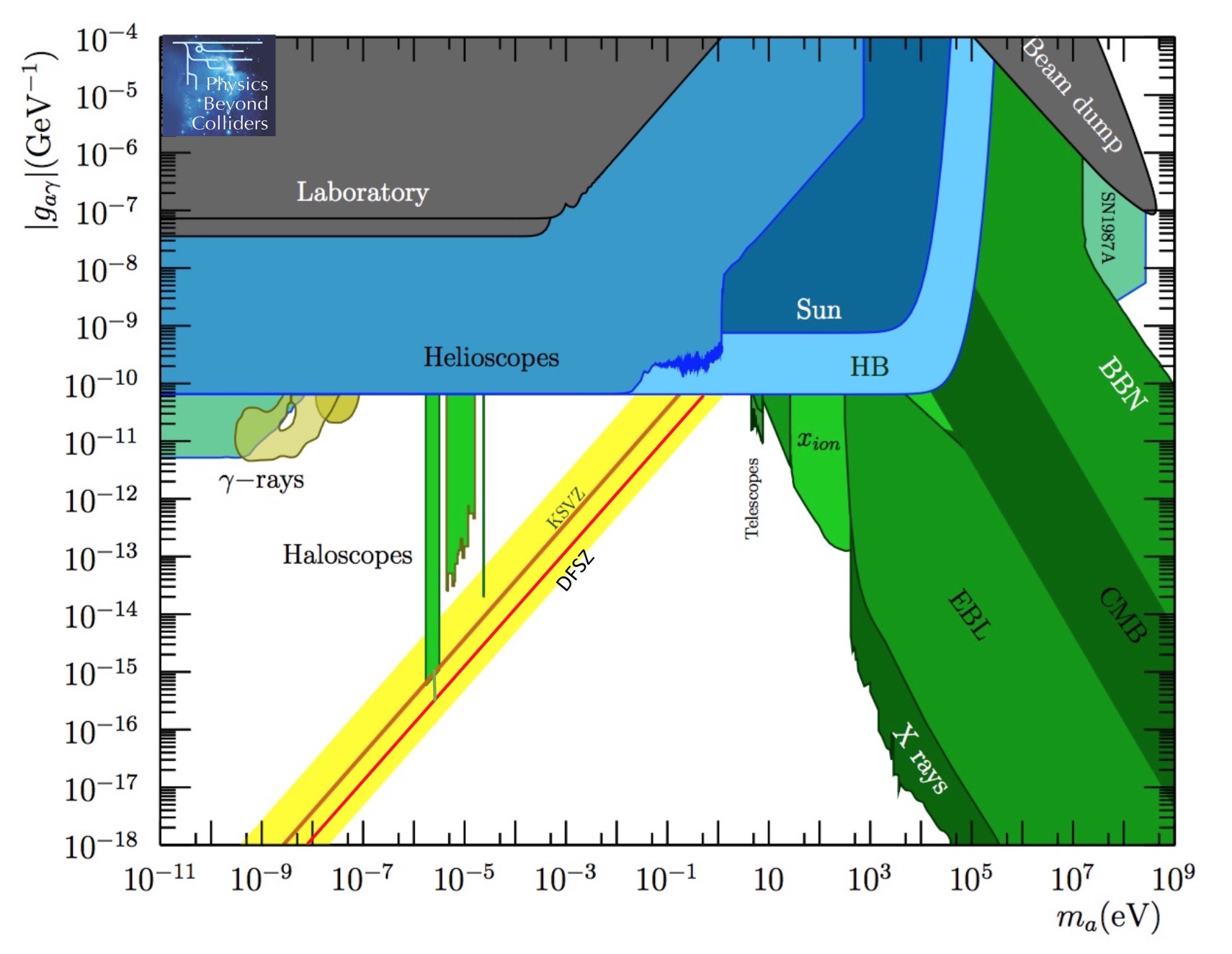}
\caption{Current constraints for the axion photon coupling $g_{a\gamma}$ versus axion mass $m_a$.
Revised from~\cite{Irastorza:2018dyq}.
See text for details.}
\label{fig:lim}
\end{center}
\end{figure}

  The most updated review on the laboratory searches for axions and ALPs has been given by the recent paper
  by Irastorza and Redondo \cite{Irastorza:2018dyq}. Figure~\ref{fig:lim} shows the current constraints
  for the axion-photon coupling $g_{a\gamma}$ versus axion mass $m_a$ in the sub-eV mass range.
  The Figure has been updated with the recent result of ADMX~\cite{Du:2018uak}.

\vskip 2mm
  The Figure follows a colour scheme to present results obtained with different methods:
  black/grey for laboratory results, bluish colours for helioscope searches and bounds related to stellar physics,
  greenish for haloscopes or cosmology dependent arguments.
  Hinted regions, like the QCD axion, are in yellow/orange.

 Laboratory limits (dark grey area in Figure~\ref{fig:lim}) are essentially due to the results of OSQAR  (region below 1~meV),
 and PVLAS  (region above 1~meV). OSQAR~\cite{Ballou:2015cka} is a CERN based light shining through a wall experiment
 based on a protoype LHC magnet. PVLAS~\cite{DellaValle:2015xxa} is a sensitive polarimeter employing two rotating 2.5~T permanent
 magnets and an ultra high finesse Fabry Perot cavity to search for the magnetic birefringence of the vacuum~\cite{DellaValle:2013xs}.
 A possible next generation magnetic polarimeter to study this effect is under discussion within the
 PBC Technology Group~\cite{Pbc:006} under the name VMB@CERN.

The bounds from helioscopes and haloscopes experiments are mostly driven by CAST~\cite{Anastassopoulos:2017ftl}
and ADMX~\cite{SLOAN201695,Du:2018uak} results, respectively.

\begin{itemize}
\item[-] {\bf CAST}\\ 
CAST is an helioscope, searching from axions/ALPs with photon-coupling produced in the sun
through Primakoff conversion of plasma photons in the electrostatic field
of a charged particles.

\vskip 2mm
The most efficient way to detect solar axion is through their reconversion into photons in the presence of a static
electromagnetic field (normally magnetic dipole field)~\cite{Du:2018uak}.
Reconverted photons are then detected by using low background X-ray devices.
The achievable sensitivity in terms of the axion photon coupling constant is proportional to
\begin{equation} 
{\rm sens} (g_{a\gamma \gamma}) \propto \frac{b^{1/8}}{B^{1/2} L^{1/2} A^{1/4} t^{1/8}}
\label{eq:sens}
\end{equation}
where $L$ is the length of the magnetic field of amplitude $B$, $A$ is the area of the useful bore,
$b$ the background rate and $t$ the integration time. Large volume magnets are then a primary ingredient for such a research.

\vskip 2mm
By using a prototype LHC dipole magnet with 9~T magnetic field ove a 9.3~m length,
CAST for the first time was able to explore solar axion in the QCD model range, at least in the mass region 0.1 - 1 eV.
To maximize observing efficiency, the magnet was supported by a  structure capable to track the sun for a fraction of the day.
The last CAST result~\cite{Anastassopoulos:2017ftl} set the current best limit on the axion-photon
coupling strength ($0.66 \times 10^{- 10}$ GeV$^{- 1}$ at 95\% confidence level),
thus competing with the most stringent limits from astrophysics on this coupling.

CAST has also searched for other axion production channels in the Sun, enabled by the axion-electron or the axion-nucleon couplings.
The project is now over, and the magnet may be utilized for building a haloscope (project RADES~\cite{Melcon:2018dba}).
Most of the CAST collaboration will be entering IAXO. Among the key element of the CAST apparatus are the use of X-ray
focusing optics and very low background micromegas detector.

\item[-] {\bf ADMX} \\
Axion or ALPs can be the main component of the dark matter halo of our galaxy and produce measurable signals in a
suitable terrestrial detector. Such a detector normally exploits the long coherence length of these low mass particles,
which are thermalized inside the galactic halo, in such a way to obtain detectability in spite of their very weak
interactions with ordinary matter.
Under the assumption that the searched for particle is the only constituent of the DM halo, limits on the coupling can
be obtained in the absence of a detected signal.
Strictly speaking, the limit is on the product between the coupling and the fraction of the local DM density in the case
of a subdominant component. The oldest strategy to search for axions is the Sikivie or Primakoff
haloscope~\cite{PhysRevLett.51.1415},
which has given almost all current limits for direct detection of dark matter in the sub eV range.

\vskip 2mm
In a Sikivie type detector, a high Q tunable microwave resonator is immersed in a strong static magnetic field.
DM axions can be converted into real photons via a Primakoff process and deposit energy into the resonant mode of the cavity.
In the last two decades the Axion Dark Matter Experiment - {\it ADMX} -
has implemented this method for cavities in the GHz range. Under the assumption of dominant DM component for the axion,
ADMX has excluded the KSVZ axion in the 1.91 - 3.69 $\mu$eV mass range~\cite{SLOAN201695},
and very recently the DFSZ one in the 2.66 - 2.81 $\mu$eV
range~\cite{Du:2018uak}. The apparatus is based on a large volume high Q tunable copper cavity, operated in the sub K temperature
range and read by a SQUID based detection chain. 
Coverage of masses up to 40 $\mu$eV  (10 GHz) is envisioned for the near  future by combining the outputs
of multiple co-tuned cavity resonators in the current 8 T superconducting magnet.

\end{itemize}

\noindent
For the stellar and cosmology dependent limits shown in Figure~\ref{fig:lim} the acronyms
are as follows (see Refs.~\cite{1475-7516-2011-02-003,1475-7516-2012-02-032}):
  
\begin{itemize}
\item[-] {\it HB, Sun, SN1987a:} 
  limits from stellar evolution obtained by studying
  the ratio of horizontal branch (HB) to red giants in globular clusters
  (GCs)~\cite{PhysRevLett.113.191302}, by a combined fit of solar data
  (Sun)~\cite{1475-7516-2015-10-015}, and by the study of the SN1987A neutrino pulse duration~\cite{PhysRevD.52.1755};
  
\item[-]  {\it Telescopes, X-rays, $\gamma$-rays:}
  photons produced in axions decays inside galaxies show up as a peak
  in galactic spectra that must not exceed the known background;
\item[-]  {\it $x_{\rm ion}$: } 
  the ionization of primordial hydrogen caused by the decay photons of axions must not contribute significantly
  to the optical depth after recombination;

\item[-] {\it EBL:} 
  photons produced in ALP decays when the universe is transparent must not exceed the
  extragalactic background light (EBL);
  \item[-] {\it CMB:} 
  axions decay photons must not cause spectral distortions in the CMB spectrum;
  \item[-] {\it  BBN:} 
  the decay of high mass ALPs produces electromagnetic and hadronic showers that must not spoil
  the agreement of big bang nucleosynthesis with observations of primordial
  nuclei.
\end{itemize}

\noindent
    {\bf Experimental landscape and physics reach of PBC projects in the next 10 years } 

\begin{figure}[h]
\begin{center}
\includegraphics[width=0.8\linewidth]{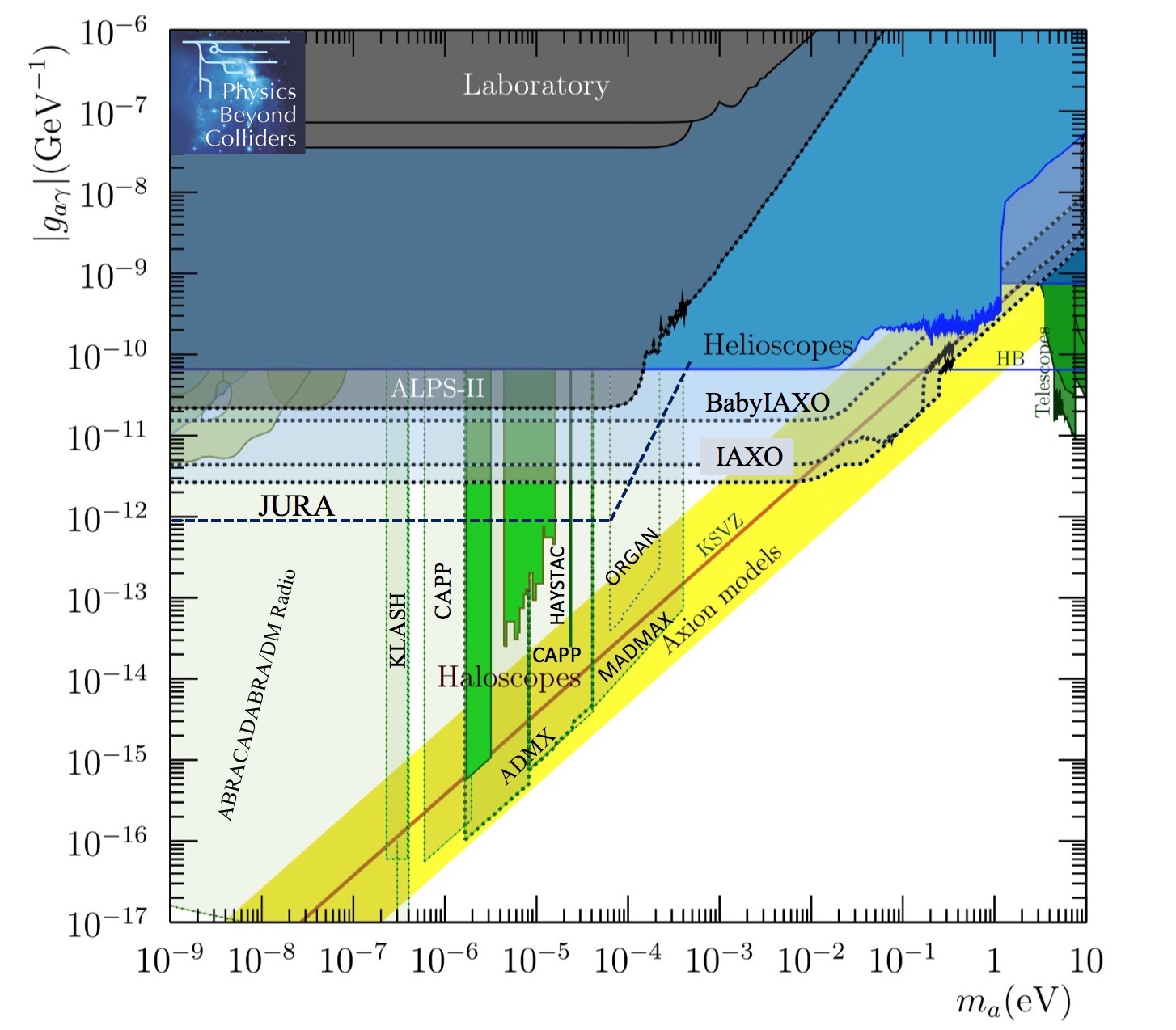}
\caption{Physics reach of Baby-IAXO, IAXO and JURA
compared with other experiments currently proposed and/or planned in the world. Revised from \cite{Irastorza:2018dyq}.
See text for details.}
\label{fig:lim2}
\end{center}
\end{figure}

\vskip 2mm
Figure~\ref{fig:lim2} shows the physics reach of the proposed PBC experiments as Baby-IAXO, IAXO and JURA
compared with other experiments currently proposed and/or planned in the world.
Both IAXO and JURA projects could be operated on a ${\mathcal{O}}(10)$ year timescale.    
Table~\ref{tab:helio} shows the list of the relevant parameters of the IAXO project, together with the Baby-IAXO setup
and other past or competing experiments. Table~\ref{tab:jura} shows the key parameters for the JURA proposal.

\begin{table}[htb]
\caption{List of the relevant parameters of the IAXO project, together with the Baby-IAXO setup
    and other past or competing experiments. For the meaning of the parameters see Equation~\ref{eq:sens}.}
    \label{tab:helio}
\begin{center}
\begin{tabular}{|c|c|c|c|c|c|c|}
\hline \hline
Experiment & Status & $B$(T) & $L$(m) & $A$(cm$^2$) & $N$(counts keV$^{-1}$ cm$^{-2}$ s$^{-1}$) & $t$(years) \\
\hline
BNL E840~\cite{PhysRevLett.69.2333}  & End & 2.2 & 1.8 & 130 & &\\
SUMICO~\cite{Moriyama:1998kd} & End & 4 & 2.5 & 18 & & \\
CAST~\cite{Anastassopoulos:2017ftl} & Running & 9 & 9.3 & 30 & $10^{-6}$ &  1.1 \\
TASTE~\cite{1748-0221-12-11-P11019} & Proposal & 3.5 & 12 & $2.8 \times 10^3$ & $5\times10^{-7}$ & 3 \\
BabyIAXO & Design & 2.5 & 10 & $2.8 \times 10^3$ & $1\times10^{-7}$ & 1 \\
IAXO & Design & 2.5 & 22 & $2.3 \times 10^4$ & $1\times10^{-8}$ & 3 + 3(gas) \\
\hline
\end{tabular}
\end{center}
\label{Tab:helio}
\end{table}

\begin{table}[htb]
\caption{Key parameters of the JURA proposal.}
\begin{center}
\begin{tabular}{|c|c|}
  \hline \hline
  parameter  & value \\ \hline
  Magnetic field & 13T $\times$ 426 m \\
  Laser wavelength & 1064 nm \\
  Production cavity circulating power & 2.5 MW \\
  Amplification in regeneration cavity & $10^5$ \\
  Detector noise & $10^{-4}$ s$^{-1}$ \\
  Measuring time  & 4 weeks \\ 
\hline
\end{tabular}
\end{center}
\label{tab:jura}
\end{table}

The experiments planned or proposed in the world that could be able to produce results earlier or on the same timescale
of the PBC projects are listed below.

\vskip 2mm
\noindent {\bf LSW experiments} 

\begin{itemize}
\item[-] {\it ALPS~II} \\
In a photon regeneration experiment axions are produced by an electromagnetic beam
(laser or microwave) traversing  an external dipolar magnetic field.
These axions are then reconverted into photons after a wall and can be detected with very
sensitive detector fighting only technical and thermal noise.

\vskip 2mm
The pioneer experiment was conducted in Brookhaven by the BFRT Collaboration~\cite{Cameron:1993mr}, and the two most recent results
are those of the experiments ALPS~\cite{Ehret:2010mh} and OSQAR~\cite{PhysRevD.92.092002}.
ALPS is DESY based and used a decommissioned HERA magnet.
It is currently performing a major improvement to phase II, where a set of 10 + 10 HERA magnets
will be coupled to two 100 long Fabry Perot cavities.
ALPS~II\cite{1748-0221-8-09-T09001} will in fact take advantage of a resonant regeneration apparatus,
thus expecting a major improvement of the current limit on LSW experiment given by OSQAR.
ALPS~II will represent the current state of the art LSW experiment, and for this reason its activities
are monitored with interest by the PBC since they will give key elements to judge the proposal JURA.

\end{itemize}

\vskip 4mm
\noindent {\bf Haloscopes} 
\begin{itemize}
\item[-] {\it HAYSTAC} \\
HAYSTAC is a high frequency version of the Sikivie detector, born on a group that was collaborating with ADMX.
Its most notable feature is the use of a Josephson parametric amplifier with very low noise temperature,
allowing the experiment to reach cosmological sensitivity in the mass region around 20 $\mu$eV. Ref.~\cite{PhysRevLett.118.061302}. 

\item[-] {\it KLASH} \\
{WISPDMX} and {KLASH} proposals aim at studying the low mass region (0.1 - 1 $\mu$eV), by employing large resonator
and refurbished magnets from high energy physics experiments. Ref.~\cite{Alesini:2017ifp}.

\item[-] {\it CAPP} \\
Activities on the axion searches are also pushed by the South Korean Center for Axion and Precision Physics - { CAPP}.
The initiative {\it CULTASK}~\cite{Chung:2018wms} is a CAPP based standard haloscope  whose strength is the development of very
high field large bore magnets, with fields up to 35~T and above.
A CAPP-CAST collaboration~\cite{Miceli:2015xas} is also ongoing to use rectangular cavities embedded inside
the CAST magnet,
while the CAPP initiative ACTION~\cite{PhysRevD.96.061102} study the use of toroidal geometry.

\item[-] {\it ORGAN} \\
{ ORGAN} plans to study the higher mass region in the 50 - 200 $\mu$eV range, with specially designed resonant
systems. Ref.~\cite{MCALLISTER201767}.
\end{itemize}

\noindent {\bf Other techniques with photon-coupling} \\
The search for axions with masses above tens of $\mu$eV is very challenging 
when using resonant cavity detectors. Typically the useable cavity volume
is reduced but also other factors like a decrease in the technically
achievable resonant enhancement are a challenge.
In view of this, new initiative are being developed where
the detectors are broadband and instrumenting large volumes.
The explored coupling is still the one with the photon, and again there is the need for large volume of high static magnetic field. 

\begin{itemize}

\item[-] {\it BRASS and MADMAX} \\
By exploiting the axion induced electric field on a boundary immersed in a static magnetic field,
the BRASS experiment will use a magnetized 8 m radius disk immersed 
in a 1 T static magnetic field to study the mass region 10~$\mu$eV to 10~meV simultaneously.
At the moment it is at very preliminary stage.
The same concept of radiating disk is at the base of MADMAX experiment~\cite{PhysRevLett.118.091801}, where however a multiple disk
configuration is used to obtain again some sort of broad resonance enhancement of the signal.
This collaboration is already being developed and it is in the R\&D phase.

\item[-] {\it DM Radio and ABRACADABRA} \\
Another method for ultra low mass dark matter axion detection is the use of a lumped element LC resonator inside a strong magnet,
where an alternating current is induced by the axion field.
Studies are underway to implement such idea within the ADMX magnet for a detector with sensitivity in the mass region
below 1 $\mu$eV. The {\it DM radio} experiment~\cite{Silva-Feaver:2016qhh} is based on the same idea but uses a tunable LC resonator
shielded by a superconducting structure and read by a SQUID. {\it ABRACADABRA}~\cite{PhysRevLett.117.141801} is a 1~m scale broadband detector
based on a toroidal magnet with a superconducting pick up loop inside and read by  SQUID.
Again, the best sensitivity is obtained for masses below 1 $\mu$eV.
All these efforts are just finalizing their R\&D phase and should come out with first data in a few years. 

\end{itemize}

\clearpage
\section{Physics reach of PBC projects in the MeV-GeV mass range}
\label{sec:phys-reach-MeV-GeV}

\vskip 2mm
As detailed in Section~\ref{ssec:exps_MeV_GeV}, the PBC examines the comprehensive physics case for 6 different proposals
that aim to study the hidden sector in the MeV-GeV mass range exploiting the PS and SPS accelerator complex.
In addition, this is compared to the physics reach in the same mass range of several proposed experiments at the LHC interaction points.
In this Section their physics reach is presented, compared against each other and put in the worldwide context.
The presentation of the results follows the scheme outlined in Section~\ref{sec:portals} where 11 benchmark cases
were identified as theoretically well motivated  target areas to explore.
The 11 benchmark cases do not pretend to be exhaustive, but only to provide a common ground to compare
different sensitivities from different experiments. These benchmark cases should be considered
as the starting point towards a comprehensive
investigation of hidden sector models in the MeV-GeV mass range that could be performed in the future.

\vskip 2mm
The results are shown in the next Sections as 90~\% CL exclusion limits and compared to the existing bounds and the physics reach
of other similar initiatives proposed worldwide in the same timescale. 

\vskip 2mm
It is important to remark that the level of maturity in compiling these curves is highly non homogeneous among the
PBC proposals. As a matter of fact, the physics reach of upgrades of existing experiments (as NA62$^{++}$ or NA64$^{++}$)
can already rely on a deep understanding of the experimental effects and a realistic analysis of the levels of the backgrounds
based on collected data. New, but already consolidated projects (as, e.g., LDMX and SHiP) can profit of
detailed Monte Carlo simulations and a thorough level of understanding of possible background sources.
More recent proposals, instead, are in the process of implementing a full simulation and for this study
they have evaluated their physics reach
mostly based on toy Monte Carlo or fast simulation. As a consequence, they should be taken with many caveats.

\vskip 2mm
The 90\% CL exclusion curves can be interpreted as $3 \sigma$ discovery in case the backgrounds are mantained below a fraction of event.
In case of discovery in the visible channels, only experiments equipped with spectrometers providing mass measurements and
particle identification will be able to understand the physics behind the signature.  

\vskip 2mm
The experiments are described in detail in Sections 5 and 6.
The relevant facts pertaining the current situation on the level of maturity of each project
are collected below and summarized in Table~\ref{tab:pbc_exps_studies}.
These considerations should be taken into account when comparing sensitivity curves across the proposals.

\vskip 5mm
\noindent
{\bf PBC proposals on a 5 year timescale}
\begin{itemize}

\item[-] {\it NA64$^{++} (e)$}\\
The NA64$^{++}(e)$ sensitivity curves assume to collect
$5\times 10^{12}$ eot at the current H4 line where the existing NA64 experiment has already
collected ${\mathcal{O}}(10^{12})$ eot.
The projection is based on the knowledge of the experimental efficiencies and background levels
measured in the current
run and assumes an upgrade of the detector that must be able to cope with the increased $\times (5-10)$
beam intensity.

\item[-] {\it NA64$^{++} (\mu)$}\\
The NA64$^{++}(\mu)$ sensitivity curves assume an integrated yield of $5\times 10^{13}$ muons-on-target (mot)
that can be collected in $\sim 1.5$ years. This data taking  is supposed to start during Run 3 (Phase I) and finalized in Run 4
(Phase II).
The status of the proposal, along with a thorough evaluation of the beam purity and the main background sources,
is summarized in a recent Addendum\footnote{CERN-SPSC-2018-024/SPSC-P-348-ADD-3.} sent to the SPSC.

\item[-] {\it NA62$^{++}$}\\
The NA62$^{++}$ sensitivity curves assume to collect ${\mathcal{O}}(10^{18})$ pot in dump mode by 2023.
The backgrounds and experimental efficiencies have been partially included in the curves: their 
estimate is based on $\sim 3\times 10^{16}$ pot dataset already collected in
a few days of operation in dump mode during the current 2016-2018 run.

\item[-] {\it FASER 150 fb$^{-1}$}\\
FASER in its initial phase will be a small detector of 10 cm radius and 1.5~m length.
It is planned to be installed during LS2 in TI18
480 m downstream of the ATLAS IP and shielded by 90 m of rock.
The sensitivity curves assume 100\% detection and reconstruction efficiency and zero background.
While a full simulation of the detector is still do be done, a preliminary study with FLUKA
has shown that possible backgrounds of high-energy ($>$ 100 GeV) particles  and radiation levels
at the FASER location are very low.
Moreover an emulsion detector and a battery-operated radiation monitor installed at the FASER site
in June 2018 is helping to validate and complement the current background estimates.

\end{itemize}

\vskip 2mm
\noindent
{\bf PBC proposals on a $\sim 10-15$ timescale}

\begin{itemize}

\item[-] {\it REDTOP}\\
The REDTOP sensitivity curves assume a dataset of $2 \times 10^{17}$ pot
that can be collected in two years of run at the PS,
one year at the energy corresponding to the $\eta$ threshold of 1.7-1.9 GeV and
one year at the $\eta'$ threshold, 3.5 GeV. Detector efficiency and backgrounds have been evaluated with the full
Monte Carlo and included in the results.
The fact that the detector, including the optical TPC, could be ready
in order to take data during Run 3, as claimed by the Collaboration, is instead an open question.
REDTOP main physics goal is to search for BSM physics in ultra-rare $\eta$
and $\eta'$ decays~\footnote{See http://redtop.fnal.gov/the-physics/.}.
As part of that physics program, REDTOP can also explore hidden sector physics in a
similar parameter space as NA62$^{++}$ and SeaQuest experiments,
but using a very different experimental technique (the $\eta/\eta'$ decays)
with respect to  beam dump methods and thus with different systematic
uncertainties and background sources.

\item[-] {\it SHiP}\\
An extensive simulation campaign was performed to optimise the design of the muon shield as
well as develop a selection that reduces all possible sources of background to $< 0.1$ events over
the experiment lifetime. The backgrounds considered were: neutrinos produced through the
initial collision that undergo deep inelastic scattering anywhere in the SHiP facility producing
$V^0$s; muons deflected by the shield that undergo deep inelastic scattering in the experimental
hall or anywhere within the decay volume producing $V^0$s; muons in coincidence from the same
spill (combinatorial muons) escaping the shield; cosmic muons interacting anywhere in the decay
volume or with experimental hall. The rate and momentum spectrum of the muon halo
obtained with the full simulation is being calibrated using data from  a dedicated 1-month long run
performed in July 2018 where a smaller replica of the SHiP target was exposed to $\sim 5 \cdot 10^{11}$
400~GeV protons.

\vskip 2mm
All samples relied on GEANT4 to simulate the entire SHiP target, muon shield, detector, and
experimental hall (walls, ceiling, floor). In addition, neutrino interactions were simulated through
GENIE. A comprehensive study  of background sources and other experimental effects is reported in the SHiP
document in preparation for the SPSC.

\item[-] {\it KLEVER} 
The results obtained in this study are based on the fast simulation described in Section~\ref{ssssec:klever}.
Particle production in the target and propagation of the neutral beam through the beamline
elements has been studied with a detailed FLUKA simulation and parameterized for the fast simulation.
An effort is underway to develop a comprehensive simulation based on the NA62 Monte Carlo and reconstruction
framework with the new detectors added and input from the FLUKA simulation of the neutral beam.
A preliminary version of this simulation was used to validate the acceptance calculation for signal events.

\item[-] {\it LDMX} \\
A thorough investigation of all the possible background sources and experimental effects
has been performed by the LDMX collaboration~\cite{Akesson:2018vlm} for a 4~GeV electron beam, with on
average 1 electron per bunch and 46~MHz repetition rate.
These are the baseline conditions for LDMX at the DASEL facility on LCLS-II at SLAC.
LDMX @ eSPS should be operated with a 16 GeV electron beam energy, a
higher repetition rate, and higher $e^-$ multiplicity per bunch.
The evaluation of the background in this operation mode is still to be done
but no major showstopper is expected.

\item[-] {\it CODEX-b}\\
    The CODEX-b detector geometry has been integrated into the LHCb simulation,
    with the help of the LHCb simulation team. This allows for a full simulation of collisions in IP8,
    including both the particles passing through the LHCb and CODEX-b detector volumes, and allows
    both realistic tracking studies  and studies
    of correlations between signals in CODEX-b and activity in LHCb to be performed.
    In parallel a measurement of the backgrounds in  the DELPHI cavern during nominal LHC operation at IP8
    has been carried out in summer 2018 in order to calibrate the GEANT4 simulation.
   A lot of work is ongoing but, as to date,  the assumption of zero-background assumed
    in the compilation of the sensitivity curves in the following Sections is still to be proven.

\item[-] {\it MATHUSLA200}\\
The assumption of zero backgroung for a large (200$\times$ 200)~m$^2$ surface detector that is
crossed by tens of MHz of cosmic rays in all  directions is a strong assumption that has to be proven.
The surface location shields MATHUSLA from ubiquitous QCD backgrounds from the LHC collision.
and it was quantitatively demonstrated that muon and neutrino backgrounds from the LHC IP can be sufficiently rejected.
Background estimates using a combination of detailed Monte Carlo studies with full detector simulation,
the known cosmic ray spectrum, and empirical measurements at the LHC using a test stand detector,
are currently in progress.
However no quantitative analysis based on the full GEANT4 simulation of the detector geometry has been shown.

\item[-] {\it FASER2} \\
If FASER is successful, a larger version of the detector, with an active volume
of 1 m radius and 5 m length could be installed during LS3 and integrate 3 ab$^{-1}$ during the HL-LHC era.
However this installation would require not negligible engineering, as the extension of the TI18 or the widening
the adjacent staging area UJ18. This makes FASER2 at the moment very uncertain.
All the considerations related to background estimates done for FASER apply to FASER2, with the additional caveat that
an increase of background is expected during the HL-LHC operation.

\end{itemize}

\begin{sidewaystable}
  \centering
  \caption{Current status of the understanding of the main backgrounds and experimental efficiencies
  considered by the PBC proposals in the evaluation of their physics reach.}
  \label{tab:pbc_exps_studies}
\begin{tabular}{cccc} 
  {\rm Proposal}  & Background  & Efficiency & Based on   \\
        \hline \hline
        {\bf at the PS: }  &  &  & \\ \hline
        {\small RedTop  }   & {\small included}         &  {\small included}   & {\small full simulation} \\

        {\bf at the SPS: }  &  &  & \\ \hline
        {\small KLEVER}     & {\small $K_{\rm L} \to \pi^0 \nu \overline{\nu}, K_{\rm L} \to \pi^0 \pi^0$  bkgs included } & {\small included} &
         {\small Main backgrounds and efficiencies }\\
                            &                           &             & {\small evaluated with fast simulation and } \\
                            &                           &             & {\small partly validated with the full (NA62-based) Monte Carlo }   \\ \hline      
         {\small LDMX }      & {\small background included} &   {\small included}            & {\small full Geant4 simulation for 4 GeV beam}\\

        {\small NA62$^{++}$} & {\small zero background} &  {\small partially included}   & {\small analysis of $\sim 3\cdot 10^{16}$ pot in dump mode } \\
                            & {\small proven for fully reconstructed final states} &   &   \\
        {\small NA64$^{++}(e)$} & {\small included} & {\small included} & {\small background, efficiencies evaluated from data } \\
        {\small NA64$^{++}(\mu)$}  & {\small in progress}  & {\small in progress}  & {\small test of the purity of the M2 line with COMPASS setup} \\
        {\small NA64$^{++}(K_{S,L},\eta,\eta')$}  & {\small to be done}  & {\small to be done}  & {\small -- } \\
        {\small AWAKE/NA64 }   &   {\small to be done}      & {\small to be done } & { \small --} \\

        {\small SHiP}       & {\small zero background }    & {\small included} &  {\small Full Geant4 simulation, digitization and reconstruction}\\
                            &                              &          &  {\small $\nu-$ interactions based on $2\times 10^{20}$ pot} \\
                            &                              &          &  {\small $\mu-$ combinatorial and $\mu-$ interactions based on $\sim 10^{12}$ pot} \\
                            &                              &          &  {\small measurement of the muon flux at H4 performed in July 2018} \\  \hline          

        {\bf at the LHC: }  &  &  & \\ \hline
        {\small CODEX-b }   & {\small zero background assumed} & {\small not included} & {\small Evaluation of background in progress with full MC}  \\
                            & {\small (preliminary GEANT simulation)} & &  \\
        {\small FASER   }   & {\small zero background assumed} & {\small not included}  & {\small Fluka simulation and in-situ measurements}  \\
        {\small MATHUSLA200 } & {\small zero background assumed}  & {\small not included}  &
                             {\footnotesize FLUKA, Pythia and MadGraph simulation for }\\
       &     &  & {\footnotesize $\nu-$, $\mu-$ fluxes from the LHC IP and cosmic rays background.} \\
        {\small MilliQan  } &  {\small included}                 & {\small included}  & {\small full Geant4 simulation of the detector }  \\
        
     \hline \hline
\end{tabular}
\end{sidewaystable}

\clearpage
\subsection{Vector Portal}
\label{ssec:vector_portal}

In the case of a vector mediator or {\it dark photon}, several contraints have been set
depending on the assumption that the mediator can decay directly to dark matter (DM) particles ($\chi$)
({\it invisible decays}) or has a mass
below the $2\cdot m_{\chi}$ threshold and therefore can decay only to SM particles ({\it visible decays}).

\subsubsection{Minimal Dark Photon model (BC1)}
\label{sssec:bc1}

{In the Minimal Dark Photon model, the SM is augmented by a single new state $A'$. DM is assumed
to be either heavy or contained in a different sector.
In that case, once produced, the dark photon decays back to the SM states.
The parameter space of this model is then ($m_{A'}, \epsilon$) where $m_{A'}$ is the mass of the dark photon and
$\epsilon$ the coupling parameter of the Dark Photon with the standard photon.}

\vskip 5mm
\noindent {\bf Current bounds}
\vskip 2mm
Visible decays of vector mediators 
are mostly constrained from searches for di-electron or di-muon
resonances~\cite{Lees:2014xha,Batley:2015lha, Merkel:2014avp} and from the
re-interpretation of data from fixed target or neutrino experiments
in the low ($<$ 1 GeV) mass region~\cite{Riordan:1987aw, Bjorken:1988as,Bross:1989mp}.
NA48/2~\cite{Batley:2015lha}, A1~\cite{Merkel:2014avp} and BaBar~\cite{Lees:2014xha} experiments
put the strongest bounds for $\epsilon > 10^{-3}$ in the $0.01 - 10 $ GeV mass range.
They search for a bump in the $e^+ e^-$ or $\mu^+ \mu^-$ invariant mass distribution over a smooth background.
These experiments consider a variety of dark photon production mechanisms, such as meson decays (NA48/2~\cite{Batley:2015lha}),
bremsstrahlung (A1~\cite{Merkel:2014avp}),
and annihilation (BaBar~\cite{Lees:2014xha},
KLOE~\cite{Archilli:2011zc, Babusci:2012cr, Babusci:2014sta, Anastasi:2016ktq}, LHCb~\cite{Aaij:2017rft}).
These results are complemented by those from beam dump
experiments, such as E141~\cite{Riordan:1987aw}
and E137~\cite{Batell:2014mga, Bjorken:1988as} at SLAC,
E774 at Fermilab~\cite{Bross:1989mp},
CHARM~\cite{Bergsma:1985qz,Gninenko:2012eq}
and NuCal~\cite{Blumlein:1990ay}.
The KOTO experiment has also set recently a limit on the $BR(K_L \to \pi^0 X) < 2.4 \cdot 10^{-9}$ (90\% CL)
\cite{Ahn:2018mvc} that could fill a little bit of the hole of the E949 coverage at $m = m_{\pi^0}$.

\begin{figure}[htb]
\centerline{\includegraphics[width=0.8\linewidth]{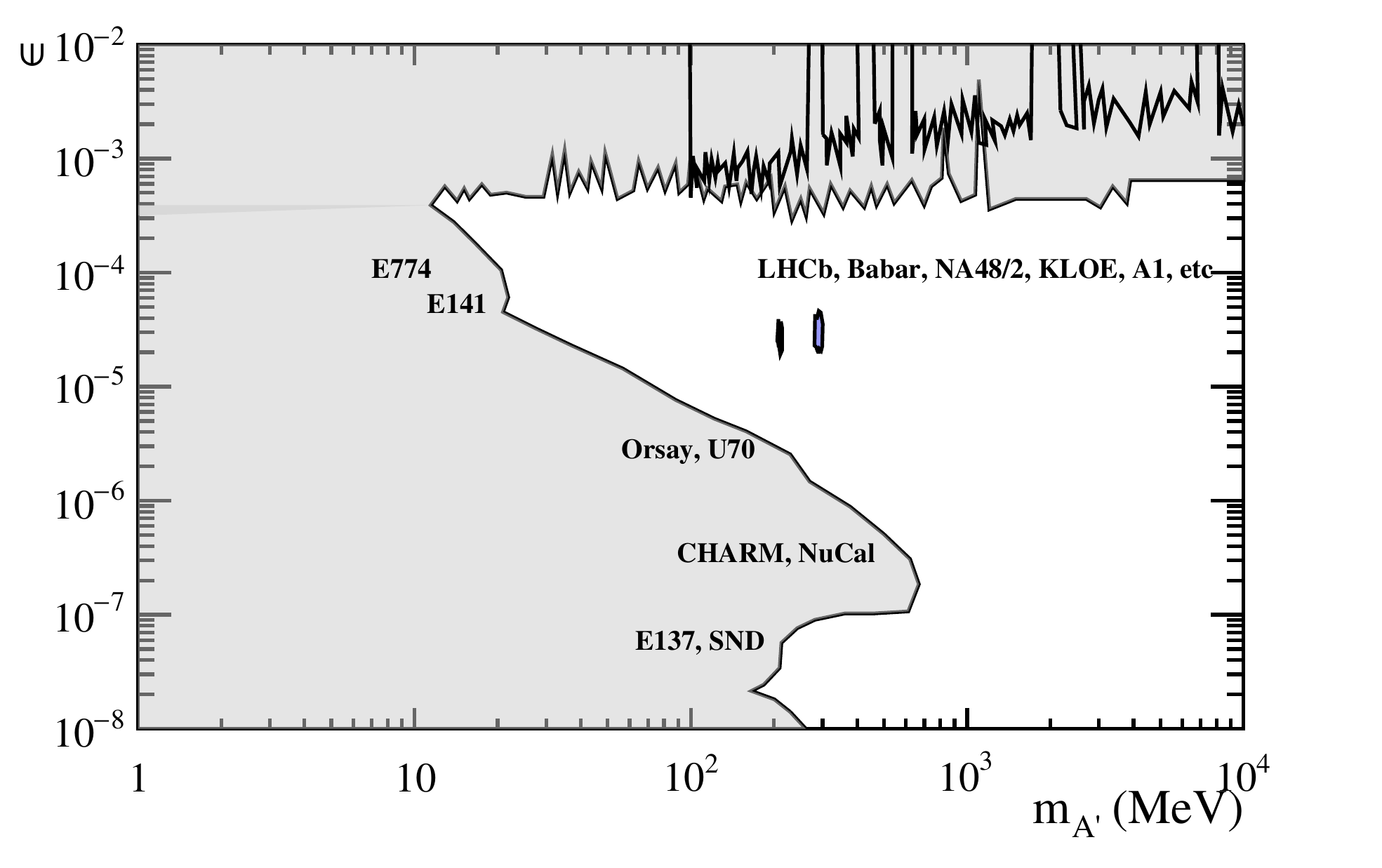}}
\caption{Current limits for Dark Photon in visible decays in the plane mixing strength
  $\epsilon$ versus mass of the Dark Photon $m_{A'}$. }
\label{fig:DP_bc1_past}
\end{figure}

\vskip 2mm
Existing limits in the plane mixing strength versus mass of the dark photon are
shown in Figure~\ref{fig:DP_bc1_past}.
In the following we briefly detail the contributions by classifying them
as a function of the experimental technique used.

\vskip 2mm
\noindent
{\it 1. Searches for dilepton resonances:}

\begin{itemize}
\item[-] {\it NA48/2 @ CERN:}
searches for dark photons decays to $e^+ e^-$ final state in the decay chain $\pi^0 \to \gamma A'$ using
$\sim 2\times 10^7$  fully reconstructed $\pi^0 \to \gamma e^+ e^-$ decays collected in 2003-2004.
Ref.~\cite{Batley:2015lha}.

\item[-] {\it BaBar @ KEK:}
searches for a dark photon in the reaction
$e^+ e^- \to \gamma A'$, $A' \to e^+e^-, \mu^+ \mu^-$ using 514 fb$^{-1}$ of data.
Ref.~\cite{Lees:2014xha}.

\item[-] {\it KLOE @ DAFNE:}
searches for dark photon in visible final states using a large variety of production modes, such as
meson decay ($\phi \to \eta A'$), annihilation
($e^+e^- \to \gamma A'$), and dark-higgsstrahlung ($e^+ e^- \to A' h'$).
Refs.~\cite{Archilli:2011zc, Babusci:2012cr,Babusci:2014sta, Anastasi:2016ktq}.

\item[-] {\it LHCb: } 
inclusive di-muon search in $pp$ collisions at $\sqrt{s}=13$ TeV performed with the current Run 2 LHCb data
above the dimuon threshold.
Ref.~\cite{Aaij:2017rft}.

\end{itemize}

\vskip 0.5cm
\noindent
{\it 2. Reinterpretation of data of fixed target experiments:}

\begin{itemize}

\item[-] {\it E137 @ SLAC (electron beam dump):}
E137 was an experiment conducted at SLAC in 1980–1982 where a 20 GeV electron beam
was dumped on a target.
Dark matter interacting with electrons
(e.g., via a dark photon) could have been produced in the electron-target collisions and scattered off
electrons in the E137 detector, producing the striking, zero-background signature of a high-energy
electromagnetic shower that points back to the beam dump.
Refs.~\cite{Batell:2014mga,Bjorken:1988as}.

\item[-] {\it CHARM @ CERN (proton beam dump):}
the CHARM Collaboration performed a search for axion-like particles decaying to photon, electrons or muons pairs
using the 400 GeV, $2.4 \times 10^{18}$ protons-on-target (pot)
dumped on a thick copper target distant 480 m from the 35 m long decay volume.
Ref.~\cite{Bergsma:1985qz}.

\item[-] {\it E141 @ SLAC (electron beam dump):}
the E141 Collaboration searched for high-energy positron signals from a hypothetical $X^0 \to e^+ e^-$ decay,
produced in the interactions of 2$\times 10^{15}$ 9 GeV electrons dumped on a 10- and 12-cm long $W$-targets.
Ref.~\cite{Riordan:1987aw}.

\item[-] {\it E774 @ Fermilab (electron beam dump)}:
the E774 Collaboration used $5.2 \times 10^9$ eot from an electron beam of 275~GeV at FNAL.
A hypothetical $X^0$ particle could have been produced by bremsstrahalung in the dump and then decays to $e^+ e^-$ pairs in the
$\sim 2$~m long decay volume.
Ref.~\cite{Bross:1989mp}.

\end{itemize}

\vskip 5mm
\noindent {\bf Future experimental landscape}
\vskip 2mm

\begin{figure}[h]
\centerline{\includegraphics[width=0.8\linewidth]{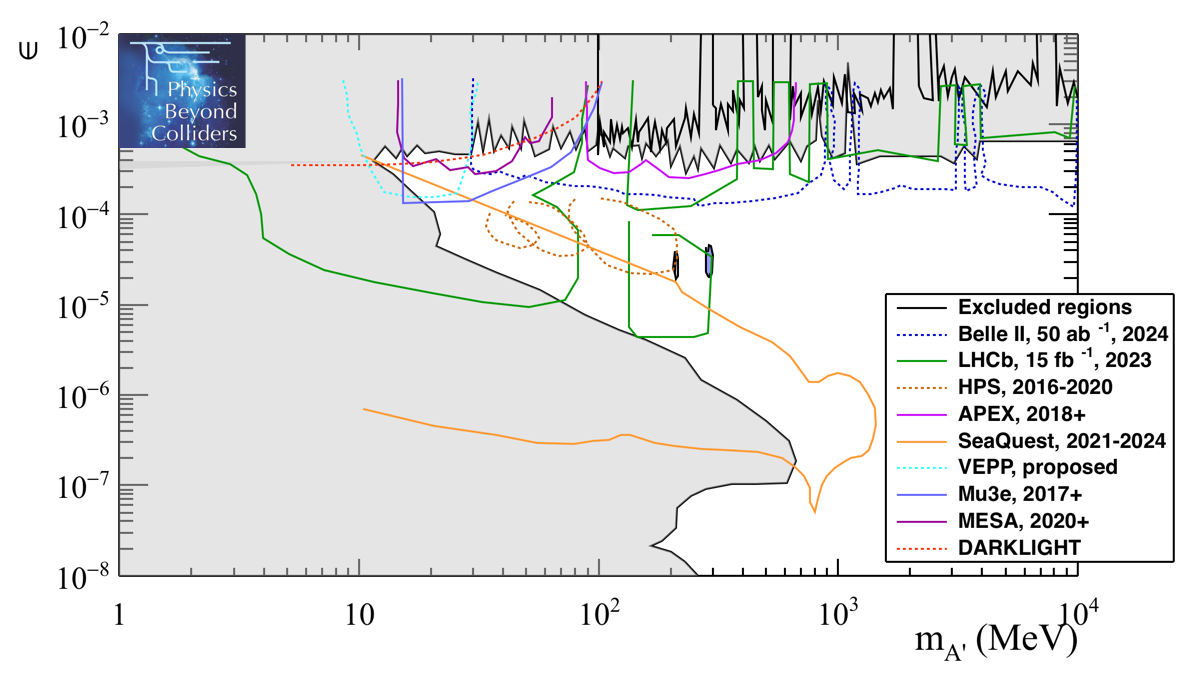}}
\caption{Future upper limits at 90 \% CL for dark photond in visible decays in the plane mixing strength
  $\epsilon$ versus mass $m_{A'}$ from experiments and proposals not related to the PBC activity.}
\label{fig:DP_bc1_context}
\end{figure}

Several experiments and proposals not considered in the PBC activity
will search for dark photons when decays to visible final states using different types
of beams and experimental techniques. The projections of their sensitivity
in the near future are shown in Figure~\ref{fig:DP_bc1_context}. The
status of these projects is briefly reported in the following.

\begin{itemize}
\item[-] {\it Belle-II @ KEK:}
will search for visible Dark Photon decays $A' \to e^+ e^-, \mu^+ \mu^-$ where $A'$ is produced
in the process $e^+ e^- \to A’ \gamma$. Projections are based on 50~ab$^{-1}$.
Timeline: data taking started in 2018, expected about 50~ab$^{-1}$ by 2025.
Ref.~\cite{Kou:2018nap}.

\item[-] {\it LHCb @ CERN:}
LHCb will search for dark photon in visible final states both using
 the inclusive di-muon production~\cite{Ilten:2016tkc}
and the $D^{*0} \to D^0 e^+ e^-$ decays~\cite{Ilten:2015hya}.
The $D^{0*}$ search will cover
dark photon masses from the di-electron mass threshold up to 1.9 GeV. The $D^{*0}$ search
requires the upgrade of the current LHCb trigger system, currently scheduled during 2019-2020.
The projections are based on 15 fb$^{-1}$, 3 years data taking with 5 fb$^{-1}$/year
with an upgraded detector after Long Shutdown 2. 

\item[-] {\it HPS @ JLAB:} electron beam-dump at CEBAF electron beam (2.2-6.6 GeV, up to 500 nA),
search for visible ($A' \to e^+ e^-$) dark photon (prompt and displaced) decays  produced via bremsstrahlung production  in a thin $W$ target.

The experiment makes use of the 200~nA electron beam available in Hall-B at Jefferson Lab. Over the 180
PAC-days~\footnote{ 1 PAC-day = 2 calendar days.} granted, HPS
collected data for 1.7 PAC-days in 2015 (engineering run) at 1.06 GeV
beam energy and 5.4 PAC-days for a physics run in 2016 at 2.3 GeV.
Results of the 2015 analysis are reported in Ref.~\cite{Adrian:2018scb}. A 28 PAC-days data
taking at 4.55 GeV beam energy  is expected in Summer 2019. Ref.~\cite{Battaglieri:2014hga}.

\item[-] {\it APEX @ JLAB:} electron beam dump at CEBAF electron beam, search for visible dark
photon decays. Status: planned one-month physics run in 2018-2019. Refs.~\cite{Abrahamyan:2011gv,Essig:2010xa}.

\item[-] {\it SeaQuest @ FNAL} : will search for visible dark photon decays $A' \to e^+ e^-$ 
  at the 120 GeV Main Injector beamline at FNAL.
  SeaQuest plans to install a refurbished electromagnetic calorimeter (ECAL)
  from the PHENIX detector at Brookhaven National Laboratory within the next few years.
  The collaboration plans to submit an official proposal to the Fermilab physics advisory committee
  in 2019 to install the ECAL at the end of the next polarization target run in 2021 and
  acquire $\sim 10^{18}$ ($10^{20}$) pot by 2024 (2030s). The $10^{20}$ pot yield could be collected as a result of the
  Fermilab Proton Improvement Plan~\cite{Shiltsev:2017mle}.
  Ref.~\cite{Berlin:2018pwi}.

\item[-] {\it VEPP3 at BINP:}
    missing mass method and visible decay searches at BINP at Novosibirsk.
    Dark photons are produced by colliding a 500 MeV positron beam on
    an internal gaseous hydrogen target, and both visible and invisible (via the missing mass mode)
    final state are identified. 
    Timeline: First run is anticipated for 2019-2020. Ref.~\cite{Wojtsekhowski:2012zq}.    

\item[-] {\it Mu3e @ PSI:}
Search for $\mu \to e e e $ decay at PSI. Phase I: sensitivity $2 \times 10^{-15}$ with the existing muon line,
from proton cyclotron of 2.4~mA protons at 590~MeV.
Phase II: sensitivity of $10^{-16}$ with upgraded muon line.

\item[-] {\it MAGIX at MESA (Mainz, Germany):}
is a step beyond  the traditional visible dark photon decay searches with a dipole spectrometer
 at the 105~MeV polarized electron beam at A1/MAMI.
 The MESA accelerator has $E_{\max} = $ 155 MeV energy, and up to 1~mA current.
 The MAGIX detector is a twin arm dipole spectrometer placed around a gas target.
 Production mechanism similar to HPS and identification through a di-electron resonance.
 The possibility of a beam dump setup similar to BDX is under study.
 Timeline: Proposal in 2017 with targeted operations in 2021-2022 and 2 years of data taking.
 Ref.~\cite{Doria:2018sfx}.

\end{itemize}

\noindent {\bf Physics reach of PBC projects on 5 and 10-15 year timescales}

\begin{figure}[h]
\centerline{\includegraphics[width=0.8\linewidth]{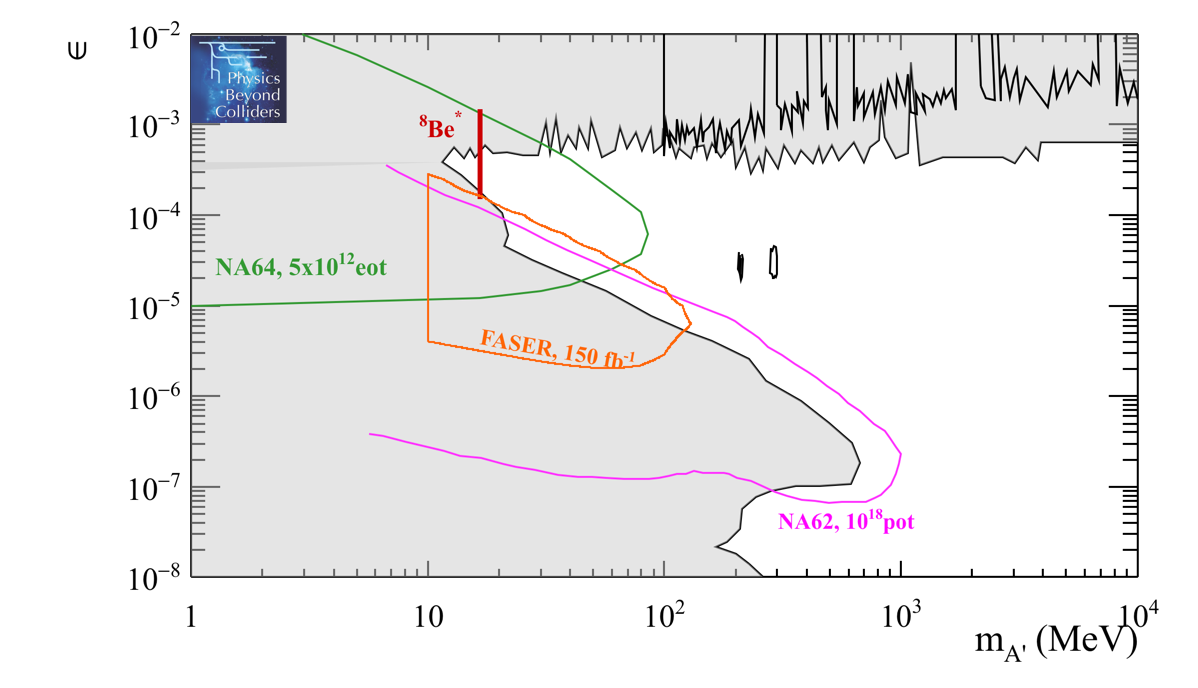}}
\caption{PBC projects on $\sim 5$ year timescale:
upper limits at 90 \% CL for Dark Photon in visible decays in the plane mixing strength
  $\epsilon$ versus mass $m_{A'}$. The vertical red line shows the allowed range of $e-X$ couplings of a
  new gauge boson $X$ coupled to electrons that could explain the $^8$Be anomaly~\cite{Feng:2016jff,Feng:2016ysn}. }
\label{fig:DP_bc1_pbc_1}
\end{figure}

\vskip 2mm
Figure~\ref{fig:DP_bc1_pbc_1} shows the 90\% CL exclusion limits for searches for dark photons
decaying to visible final states performed by PBC proposals that might produce results on $\sim 5 $ year
timescale: NA64$^{++}(e)$, NA62$^{++}$ and FASER with 150 fb$^{-1}$.
These projects will be competing with other initiatives in the same timescale, as for example SeaQuest, HPS and LHCb,
as discussed in the previous paragraph and shown in Figure~\ref{fig:DP_bc1_context}.

\begin{figure}[h]
\centerline{\includegraphics[width=0.8\linewidth]{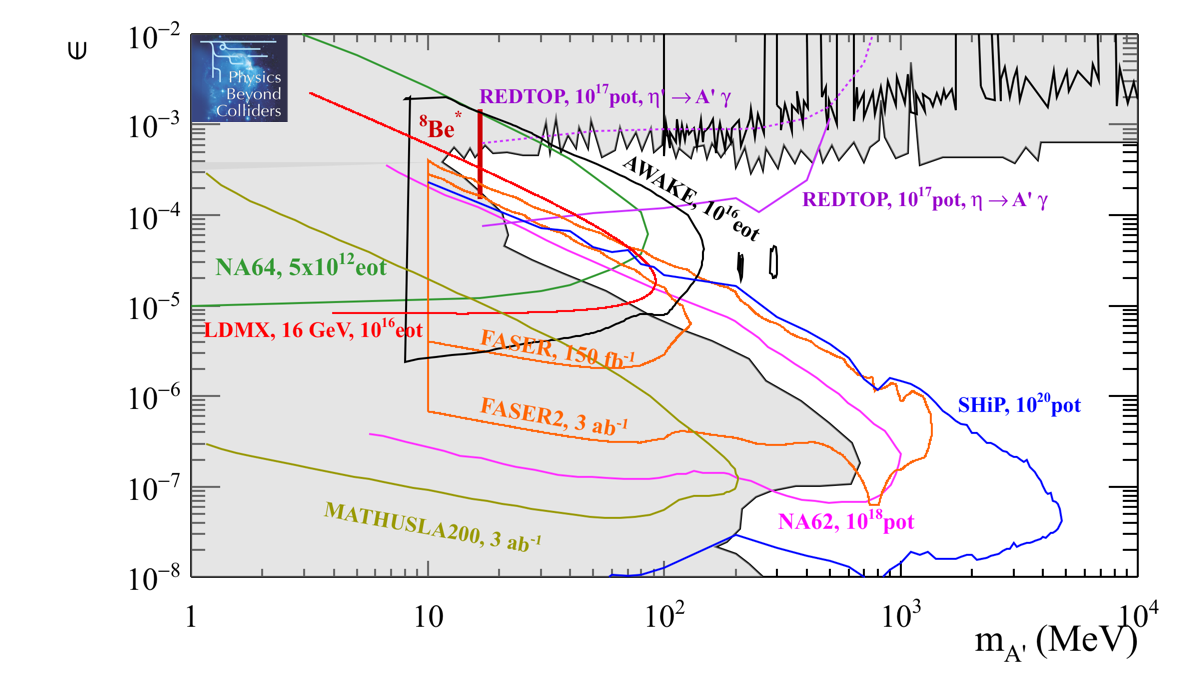}}
\caption{Future upper limits at 90 \% CL for Dark Photon in visible decays in the plane mixing strength
  $\epsilon$ versus mass $m_{A'}$ for PBC projects on a $\sim$ 10-15 year timescale.
  The vertical red line shows the allowed range of $e-X$ couplings of a
  new gauge boson $X$ coupled to electrons that could explain the $^8$Be anomaly~\cite{Feng:2016jff,Feng:2016ysn}.}
\label{fig:DP_bc1_pbc_2}
\end{figure}

\vskip 2mm
The physics reach of PBC projects on a 10-15 year timescale is shown in Figure~\ref{fig:DP_bc1_pbc_2}.
In this timescale several projects could be ready and operated, as REDTOP, SHiP, FASER2, MATHUSLA200, AWAKE, and LDMX.
The sensitivity for dark photons decaying in visible final states will be dominated by SHiP,
while FASER2, LDMX and AWAKE will be directly competing with SeaQuest,
LHCb, HPS, and others as shown in Figure~\ref{fig:DP_bc1_context}.
MATHUSLA200 in this scenario is instead not competitive, mostly due to the fact that the Dark Photon is produced forward.

\clearpage
\subsubsection{Dark Photon decaying to invisible final states (BC2)}
\label{sssec:bc2}

\vskip 2mm
{
This is the model where minimally coupled viable WIMP dark matter model can be constructed with a Dark Photon as light mediator.
Preferred values of the dark coupling $\alpha_D = g^2D/4\pi$
is such that the decay of $A’$ occurs predominantly into DM $\chi \bar\chi$ states.
These states can further rescatter on electrons and nuclei due to $\epsilon$-proportional
interaction between SM and DM states mediated by the mixed $AA'$ propagator.
The parameter space for this model is $(m_{A’}, \epsilon; m_{\chi}; \alpha_D)$
with further model-dependence associated with the properties of the Dark Matter candidate $\chi$  (boson or fermion). }

\vskip 2mm
\noindent
The sensitivity plots for this benchmark case can be shown in two ways:
\begin{itemize}
\item[(a)] the plane $\epsilon$ versus $m_{A'}$ where $\alpha_D^2 >> \epsilon \alpha_D$
  and $m_{A'} > 2 m_{\chi}$;
\item[(b)] the plane $y$  versus $m_{\chi}$ plot where the ``yield'' variable $y$,
$y  = \alpha_D \epsilon^2  (m_{\chi}/m_{A'})^4$ , is argued to contain a combination
of parameters relevant  for the freeze-out and DM-SM particles scattering cross section.
Here $\alpha_D$ is the dark fine structure constant that describes the interactions between Dark Photon and Dark Matter.
The coupling of the dark photon to SM particles happens via the {\it millicharge} $\epsilon$e.
The choice adopted by the PBC is $\alpha_D  = 0.1$ and $m_{A'}/m_{\chi} = 3$.
\end{itemize}

\noindent
In case (b), the yield variable $y$ can be put in direct connection to the DM thermal relic abundance.
In fact, the direct DM annihilation responsible of the thermal relic abundance, is driven by the same couplings
that define the direct DM scattering, leading to rather well defined predictions:
\[
\langle \sigma \cdot v \rangle \sim  { y \over m_{\chi} }.
\]

The measured Dark Matter abundance imposes a minimum bound on this cross-section,
$\langle \sigma \cdot v \rangle > \langle \sigma \cdot v \rangle_{\rm relic}$.
This lower bound can be translated in turn into a lower bound on the strength of the SM-mediator and DM-mediator
couplings, and, as a consequence, opens up the possibility to link results obtained at accelerator-based experiments
to those coming from DM direct detection experiments, depending on the nature of the DM candidate.
Two cases considered in this study  are Elastic Scalar and  Pseudo-Dirac fermion Dark Matter.

\vskip 5mm
\noindent {\bf \large Current bounds and future experimental landscape}

\vskip 1mm
In case of dark photon with invisible decays, the stronger limits on the coupling strength of DM with light vector
mediator for DM and mediator masses in the MeV-GeV range are provided by the
NA64 experiment~\cite{Banerjee:2016tad},
and from a recent result on mono-photon search from BaBar~\cite{Lees:2017lec}.
Limits in the low mass range come from a re-interpretation by theorists of old results
from the LSND~\cite{deNiverville:2011it}
and E137~\cite{Batell:2014mga} experiments, and as such, should be taken with many caveats.
A re-analysis of electron-scattering
data from direct detection experiments 
has led to constraints in the sub-GeV DM region~\cite{Essig:2012yx,Essig:2017kqs}

\begin{figure}[h]
\centerline{\includegraphics[width=0.8\linewidth]{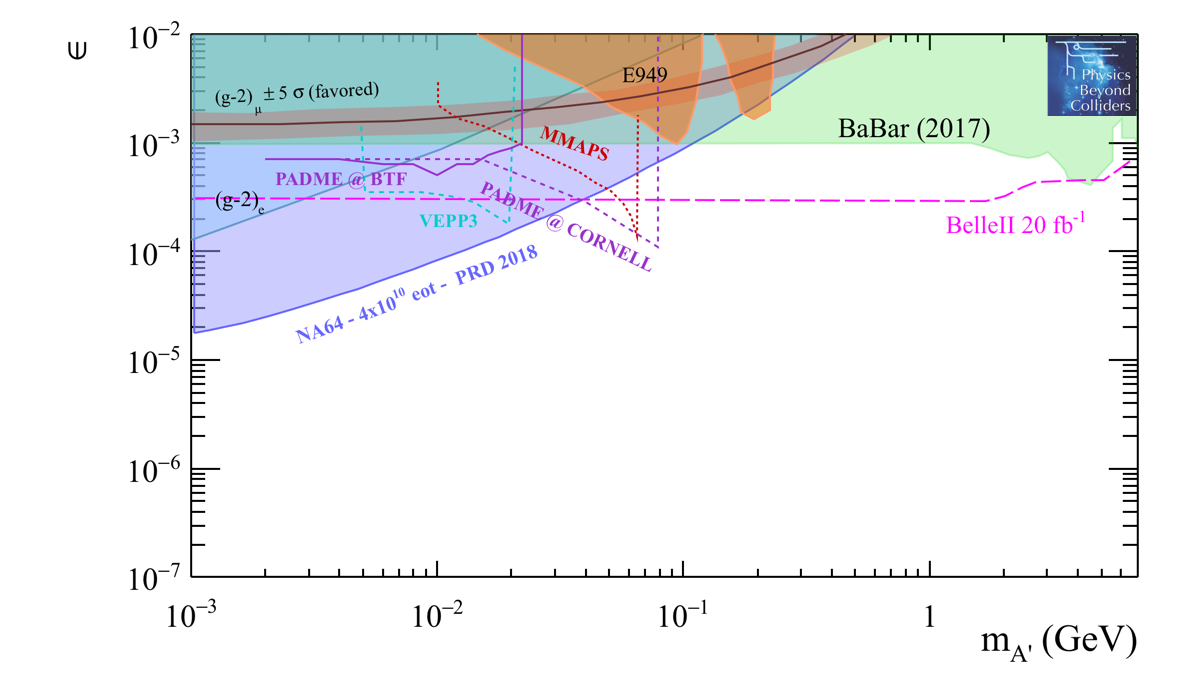}}
\caption{Current limits (filled areas) and experimental landscape for projects not PBC related (solid or dashed lines)
for Dark Photon in invisible decays  in the plane mixing strength $\epsilon$ versus dark photon mass $m_{A'}$.} 
\label{fig:DP_bc2_epsilon_context}
\end{figure}

\vskip 2mm
\noindent
{\it (a) Plane $\epsilon$ versus $m_{A'}$} \\
Figure~\ref{fig:DP_bc2_epsilon_context} 
shows the current 90\% CL upper limits in the plane $\epsilon$ versus $m_{A’}$
from BaBar~\cite{Lees:2017lec}, E787/E949~\cite{Adler:2001xv,Artamonov:2009sz} and NA64~\cite{Banerjee:2016tad} as filled areas and future perspectives
from projects not PBC related as solid or dashed curves. The region preferred by the $(g-2)_{\mu}$ puzzle~\cite{Bennett:2006fi} is also shown in  the plot.
Most of the future projections come from experiments using the missing momentum/missing mass techniques, as explained below.

\begin{itemize}
 \item[-] {\it Belle II @ KEK:} search for dark photons in the process $e^+ e^- \to  A' \gamma, A' \to $ invisible
 relies on a L1 trigger sensitive to mono-energetic ISR photon  with energy $E = (E^2_{CM} - M^2_{A'})/2 E_{cm}$.
 A trigger threshold as low as 1.2~GeV is anticipated to be applied for the  2018-2019 dataset, corresponding to
 $\sim 20 $ fb$^{-1}$ of data. Ref.~\cite{Kou:2018nap}.
 
 \item[-] {\it MMAPS @ Cornell:}
 MMAPS aims at searching for dark photons in the process $e^+ e^- \to A'\gamma$
 using the interactions of a 5.3~GeV positron beam extracted from the Cornell synchrotron
 with a fixed Be target.
 The measure of the outgoing photon kinematics with a CsI calorimeter allows to infer
 the $A'$ mass.
This method provides sensitivity to all possible decay modes. The main limitations arise from the
detector resolution and QED backgrounds, such as $e^+ e^- \to \gamma \gamma$ or $e^+ e^- \to e^+ e^- \gamma$
 where charged final particle(s) sometimes escape undetected.\\
Timeline: proposal stage, no starting date (>2020).

 \item[-] {\it PADME @ LNF:}
missing momentum searches at the Beam Test Facility (BTF) in LNF. The principle is similar
to the MMAPS experiment, using a 550 MeV positron beam on a diamond target. In
addition to invisible $A'$ decays, PADME is studying its sensitivity to diphoton decays
of axion-like particles and dark Higgs decays.
Timeline: Expected to collect $10^{13}$ positron on target by end of 2019. Proposal to
move PADME at Cornell if new positron beamline is approved.\\

\end{itemize}

We do expect NA62 will be able to produce results in the next few years as a by-product of the $K^+ \to \pi^+ \nu \overline{\nu}$
analysis, but no sensitivity curves have been provided by the Collaboration so far.

\begin{figure}[h]
\centerline{\includegraphics[width=0.8\linewidth]{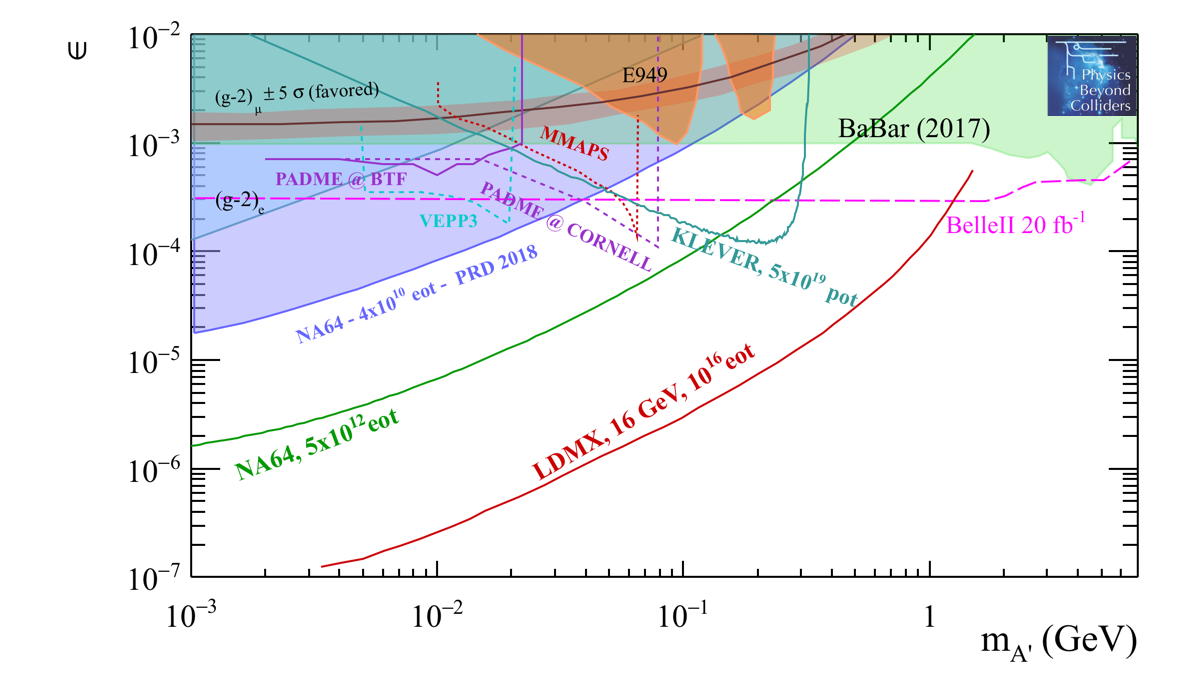}}
\caption{Dark Photon decaying to invisible final states.
Prospects for PBC projects on a timescale of 5 years (NA64$^{++}(e)$, green line) and 10-15 years (LDMX, red line and  KLEVER, cyan line)
compared to the current bounds (solid areas) and future experimental
landscape (other solid and dashed lines) as explained in Figure~\ref{fig:DP_bc2_epsilon_context}.}
\label{fig:DP_bc2_epsilon_pbc_2}
\end{figure}

\vskip 2mm
Figure~\ref{fig:DP_bc2_epsilon_pbc_2} show the projections from the PBC experiments, NA64$^{++}(e)$ with $5 \times 10^{12}$ eot will be able to
explore a large part of the parameter space on a 5-year timescale; KLEVER could further push the investigation of dark photons
in invisible final states in the mass region between 100-200 MeV as a by-product of the analysis of the $K_L \to \pi^0 \nu \overline{\nu}$
rare decay. The ultimate sensitivity can be reached by LDMX, either at the DASEL facility with 8 GeV electron beam and even further at the
eSPS facility with 16 GeV electron beam at CERN.

\begin{figure}[h]
\centerline{
\includegraphics[width=0.7\linewidth]{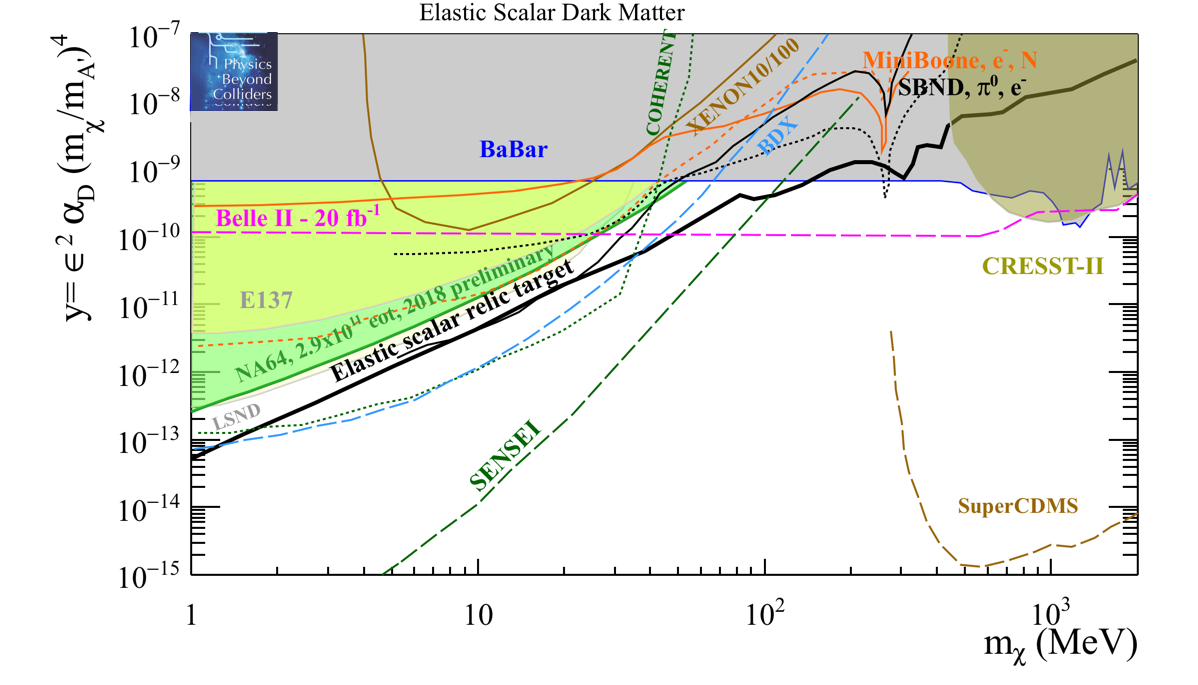}}
\centerline{
\includegraphics[width=0.7\linewidth]{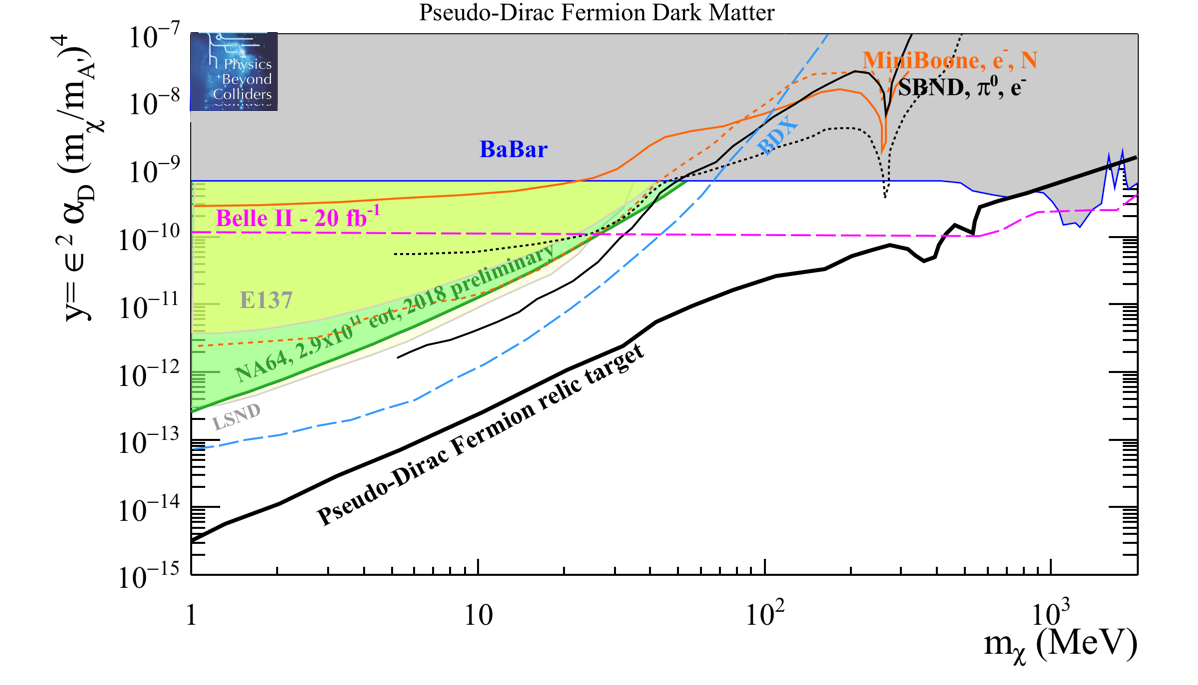}}
\caption{Current limits (filled areas) and experimental landscape for projects not PBC related (solid, dashed, and dotted lines)
for Dark Photon decaying to light Dark Matter  in the plane of the ``yield'' variable (see text) versus
Dark Matter mass $m_{\chi}$, assuming DM as an Elastic Scalar particle (top) or Pseudo-Dirac Fermion (bottom).
In the limit computation we assume a Dark coupling constant value $\alpha_D = 0.1$ and a ratio between the Dark Photon $A'$
and LDM $\chi$ masses $m_{A'} / m_{\chi} = 3$.} 
\label{fig:DP_bc2_y_scalar_pseudo_context}
\end{figure}

\vskip 2mm
\noindent
{\it (b) Plane $y$ versus $m_{\chi}$} \\
The current bounds and future perspectives in the plane $y$ versus DM mass are shown
in Figure~\ref{fig:DP_bc2_y_scalar_pseudo_context} for two different
hypotheses on the Dark Matter nature, Elastic Scalar and Pseudo-Dirac fermion.

In this plot, the lower limits for the thermal relic targets are also shown, under that hypothesis that a single DM candidate
is responsible for the whole DM abundance. Under the hypothesis of an elastic scalar DM candidate,
limits from direct detection DM experiments,
as CRESST-II~\cite{Angloher:2015ewa}, XENON 10/100~\cite{Angle:2011th,Aprile:2016wwo} and Super-CDMS~\cite{Agnese:2016cpb}
enter in the game, which is not the case in  the hypothesis of a DM as Pseudo-Dirac fermion.

\vskip 2mm
On the contrary, results from accelerator-based experiments, are largely independent of the assumptions on a specific DM nature
as DM in this case is produced in relativistic regime and the strength of the interactions with light mediators and SM particles
is only fixed by thermal freeze-out.

\vskip 2mm
Future initiatives that could explore a still uncovered parameter space in the plane
$y$ versus DM mass for DM masses below 1~GeV are all those who have sensitivity in the plane $\epsilon$ versus Dark Photon mass
and, in addition, experiments exploiting DM scattering with nucleons and/or electrons,
both accelerator-based and from direct detection searches.
Among the accelerator-based experiments, there are BDX at JLab~\cite{Battaglieri:2016ggd},
MiniBooNE at FNAL~\cite{Aguilar-Arevalo:2018wea} and COHERENT at ORNL~\cite{deNiverville:2015mwa},
as explained below.

\begin{itemize}
\item[-] {\it BDX at JLab (electron beam-dump):}
the Beam Dump eXperiment (BDX) is aiming to detect Light Dark Matter
produced in the interaction of an intense electron beam with the dump.
The experiment is sensitive to elastic
DM scattering $e^- \chi \to e^- \chi$ in the detector after production in
$e^- Z  \to  e^- Z (A’ \to \chi \chi )$.
A detector placed $\sim$20~m downstream of the Hall-A beam-dump at Jefferson
lab is expected to identify a Dark Matter scattering  by measuring a
~GeV electromagnetic shower produced by  the DM interaction with
atomic electrons of the detector. The BDX detector is composed by an
electromagnetic calorimeter surrounded by two layers of active plastic
scintillator vetos. The calorimeter re-use $\sim$800 CsI(Tl) crystals
formerly used in the BaBar EM Cal, upgraded with sipm-readout and
triggerless data acquisition. The experiment makes use of the high
energy ($\sim$ 10 GeV) and high intensity ($\sim$100 $\mu$A) electron beam available
in Hall-A running in parallel (parasitically) with the scheduled
hadron physics program. BDX has been approved with the maximum
scientific rating (A) by the JLab PAC and granted with 285 PAC-days of
data taking, corresponding to an integrated yield of $10^{22}$ eot.
The BDX collaboration is currently seeking for funds to
build the new experimental hall that will host the BDX detector. The
experiment is expected to be deploied and take data  in 2-3 years from
now. Ref.~\cite{Battaglieri:2016ggd}.

\item[-] {\it MiniBooNE @ FNAL (proton beam dump):}
neutrino detector at the 8 GeV Booster Neutrino Beamline at FNAL.
MiniBooNE is a 800 ton mineral oil Cherenkov detector situated 490 m downstream of the beam dump.
The DM is searched for via the chain $p + p \to X \pi^0/\eta$, $\pi^0/\eta \to \gamma A'$ and $A' \to \chi \chi$.
The results are based on $1.8 \times 10^{20}$ pot and 
have been published for DM-nucleon and electron-elastic scattering.
Ref.~\cite{Aguilar-Arevalo:2018wea}.

\item[-] {\it COHERENT (proton beam dump):}
the COHERENT Collaboration aims to measure Coherent Elastic Neutrino-Nucleus Scattering using the high-quality
pion-decay-at-rest neutrino source at the Spallation Neutron Source in Oak Ridge, Tennessee.
The SNS provides an intense flux of neutrinos in the energy range of few tens-of-MeV.
The beam has a sharply-pulsed timing structure that is beneficial for background rejection.
The current experimental appartus includes ${\mathcal{O}}$(10~kg) NaI(Tl)
and CsI(Tl) detectors, and a 35~kg single-phase LAr scintillation detector.
The same apparatus can be used to search for dark matter mainly
produced via the process $\pi^0/\eta \to  \gamma A'$ with $A' \to \chi \chi^*$, where the $\pi^0/\eta$ are produced
out of collisions from the primary proton beam.
The DM candidates are identified through coherent scattering leading to a detectable nuclear recoil.
Timeline: currently taking data, upgrade after 2019.  Refs.~\cite{deNiverville:2015mwa}.
See also https://sites.duke.edu/coherent/.

\item[-] {\it SBN @ FNAL (proton beam-dump)}:
The SBN program consists of
three LAr-based detectors of 112~ton (SBND), 89~ton (microBooNE), and 476~ton (ICARUS-T600) situated at 110~m, 470~m
and 600~m downstream the beam dump, respectively, of the 8 GeV primary proton beam
of the  Booster Neutrino Beamline at FNAL.
Dark Matter could be primarily produced via pion decays created in the collisions of the protons with the dump and scatter
in LAr TPC detectors. 
SBND is expected to yield the most sensitive results and could
improve upon MiniBooNE by more than an order of magnitude with $6 \times 10^{20}$ pot.
Projections shown in Figure~\ref{fig:DP_bc2_y_scalar_pseudo_context} are based on $2\times 10^{20}$ pot.
\end{itemize}

Several experiments designed to perform direct detection DM searches will be able to put bounds.

\begin{itemize}
\item[-] {\it SENSEI:}
is a direct detection experiment that will be able to explore DM candidated with masses
in the 1 eV and few GeV range, by detecting the signal released in DM-electron scattering interactions in a fully depleted silicon CCD.
For the first time, it has been demonstrated that the charge in each
pixel of a CCD -  in a detector consisting of millions of pixels  -  can be measured with
sub-electron noise. A 1-gram detector is already operating in the NUMI access
tunnel. A larger project (100 grams) can be deployed at a deeper site on a 1-2 year timescale
if funding is obtained. Ref.~\cite{Tiffenberg:2017aac}.

\item[-] {\it CRESST-II:}
uses cryogenic detectors to search for nuclear recoil events induced by elastic scattering of DM particles in CaWO$_4$ crystals.
Because of its low-energy threshold, the sensitivity to DM was extended in the sub-GeV region.
Current bounds are derived from a dataset corresponding to 52~kg live days. Ref.~\cite{Angloher:2015ewa}.

\item[-] {\it XENON 10/100:}
DM-electron scattering searches have already illustrated their potential, probing down to
$m_{\chi} >$ 5 MeV~\cite{Essig:2012yx,Essig:2017kqs} using XENON10 data~\cite{Angle:2011th} sensitive to single electrons and down
to $m_{\chi} >$ 35 MeV~\cite{Essig:2017kqs} using XENON100 data~\cite{Aprile:2016wwo}.

\end{itemize}

\clearpage
\noindent {\bf Physics reach of PBC projects on 5 and 10-15 years timescales}

\begin{figure}[h]
\centerline{
\includegraphics[width=0.7\linewidth]{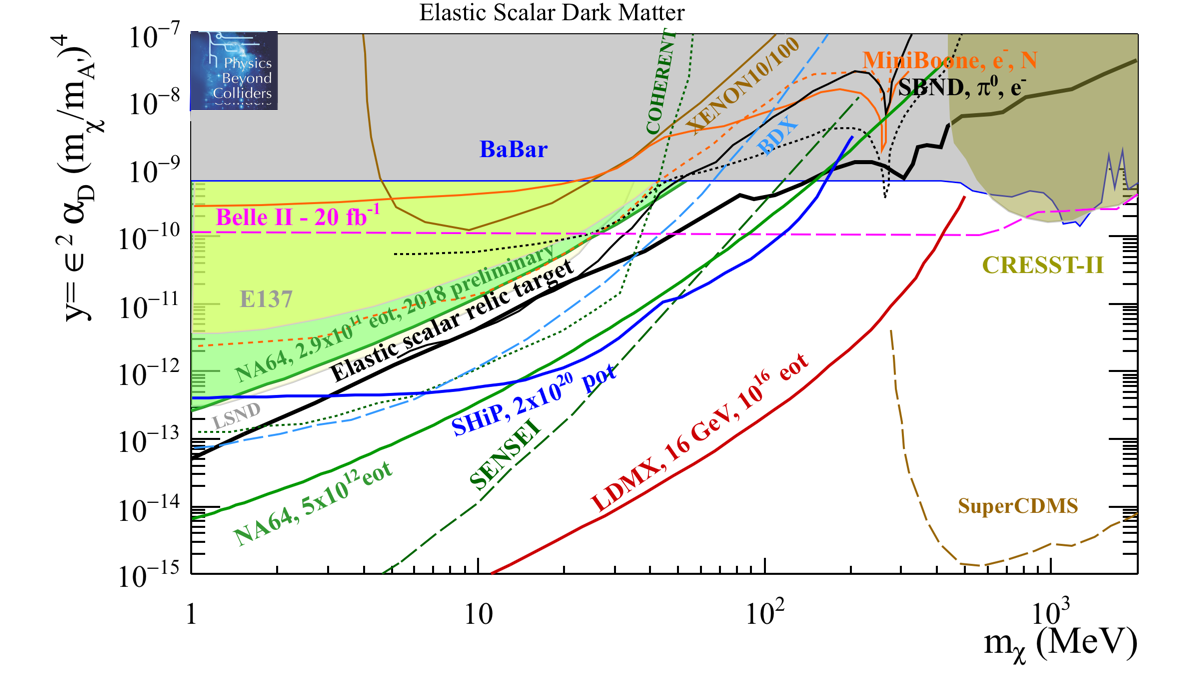}}
\centerline{
\includegraphics[width=0.7\linewidth]{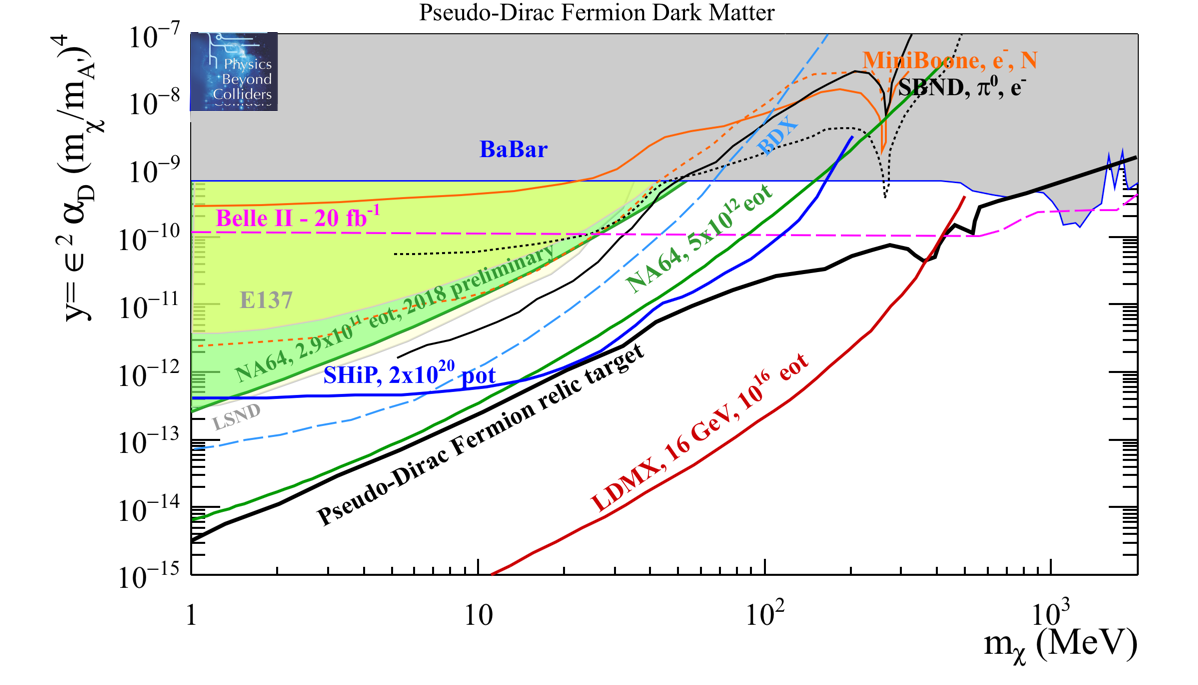}}
\caption{Dark Photon decaying to DM Elastic Scalar (top) or Pseudo-Dirac fermion (bottom) particle.
Prospects for PBC projects on a timescale of 5 years (NA64$^{++}$, green line) and 10-15 years (LDMX, red line and  SHiP, blue line)
are compared to the current bounds (solid areas) and future experimental
landscape (other solid and dashed lines). In the limit computation we assume a dark coupling constant value $\alpha_D = 0.1$
and a ratio between the dark photon $A'$ and LDM $\chi$ masses $m_{A'} / m_{\chi} = 3$.} 
\label{fig:DP_bc2_y_scalar_pseudo_pbc_2}
\end{figure}

\vskip 1mm
PBC projects able to put bounds on the $y$ versus $m_{\chi}$ plane are NA64$^{++}(e)$ on a 5-year timescale
and LDMX and SHiP on a 10-15 year timescale, as shown in
Figure~\ref{fig:DP_bc2_y_scalar_pseudo_pbc_2}.
NA64$^{++}(e)$ and LDMX will use the missing energy/missing momentum techniques, respectively. SHiP, instead, will exploit the
elastic scattering of DM candidates with the electrons in the medium of the emulsion-based neutrino detector.
As such, SHiP is fully complementary to the other two.

\clearpage

\subsubsection{Milli-charged particles (BC3)}
\label{sssec:bc3}

\vskip 2mm
Milli-charged particles (mCP) can be seen as a specific limit of the vector portal when $m_{A'}$ goes to zero and the
parameter space simplifies to the mass ($m_{\chi}$) and effective charge ($|Q| = |\epsilon g_De|$) of milli-charged particles.
The suggested choice of parameter space is $(m_{\chi},Q_{\chi}/e)$
and $\chi$ can be taken to be a fermion.
The searches for millicharged particles can be performed  either through missing energy techniques or through minimum
ionizing (milli-charged) signals.

\vskip 2mm
A recent review~\cite{Magill:2018tbb} shows the potential of the existing data
from MiniBooNE~\cite{Aguilar-Arevalo:2018gpe} and the Liquid Scintillator Near
Detector (LSND)~\cite{Athanassopoulos:1996ds}, and the soon to be released data
from MicroBooNE, the ongoing SBN program~\cite{Antonello:2015lea},
the Deep Underground Neutrino Experiment (DUNE)~\cite{Acciarri:2015uup}, beyond the standard electron beam-dump experiments
\cite{Prinz:1998ua,Davidson:2000hf}
to probe the milli-charged  particles model. In the following sections we stick on experimental published bounds
and official projections of current experiments and future proposals.

\vskip 5mm
\noindent
{\bf Current bounds and future experimental landscape}
\vskip 2mm

The most stringent current experimental bounds on millicharged particles arise from the
SLAC milliQ experiment~\cite{Prinz:1998ua}, EDGES experiment~\cite{Monsalve:2018fno} and colliders~\cite{Davidson:2000hf}.

\begin{itemize}
\item[-] {\it SLAC milliQ experiment}\\
A dedicated search for mQ’s has been carried out at SLAC in the late 90's.
This search was sensitive to particles with electric charge in the range $(10^{-1} - 10^{-5})e$
and masses between 0.1 and 1000 MeV. The experiment, located near the positron-production target of the SLC beam,
looked for extremely feebly ionization and/or excitation signals in scintillators counters (down to a single scintillation photon)
that might arise from millicharged particles surviving the 110~m sandstone filling the distance between the detector and the
positron source. Ref.~\cite{Prinz:1998ua}.

\item[-] {\it Colliders}\\
In the mass region above 100 MeV the strongest direct bounds arise from colliders, 
mainly from the  constraint  from  the  invisible width of the $Z$, as well as direct searches for fractionally
charged particles at LEP. Ref.~\cite{Davidson:2000hf}.

\item[-] {\it EDGES experiment} \\
The unexpected strength of 21~cm absorption signal measured by the Experiment to Detect the Global Epoch of Reionization
(EDGES) could be naturally explained~\cite{Munoz:2018pzp, Barkana:2018lgd} if even only a fraction (less than 0.4\%) of the Dark Matter is in
form of milli-charged particles, due to CMB constraints~\cite{Kovetz:2018zan}\footnote{Even with subcomponent DM, it is likely
excluded by direct detection~\cite{Barkana:2018qrx,Crisler:2018gci}. However this constraint is somewhat uncertain,
as it is possible that supernovae evacuate the DM from the disc, and this would nullify the direct detection
constraint~\cite{Chuzhoy:2008zy}.}.
Data from Ref.~\cite{Monsalve:2018fno}, interpretation from Ref.~\cite{Kovetz:2018zan}.

\item[-] {\it SuperNovae 1987A bounds}\\
The number of neutrinos detected at Earth during the explosion of the  SN 1987A
agree roughly with theoretical expectations. This allows us to use the generic stellar energy-loss argument
that if other particles were contributing to the cooling of the proto neutron star
these would have reduced the neutrino fluxes and duration of the neutrino signal.

\item[-] {\it milliQan}\\
Future experimental bounds outside the PBC projects will be set by the milliQan experiment.
milliQan has been proposed to be sited in the PX56 Observation and Drainage gallery above the CMS underground
experimental cavern (UXC). It consists of one or more scintillator detector layers
of roughly 1~m$^3$ each. The experimental signature would consist of a few photo-electrons arising from
the small ionization produced by the mCPs that travel without interacting through material after
escaping the CMS detector. milliQan plans to integrate $\sim 300$ fb$^{-1}$ during Run 3 and 3 ab$^{-1}$ in the HL-LHC era.
Ref.~\cite{Ball:2016zrp}.

\item[-] {\it FerMINI}\\
  The FerMINI project~\cite{Kelly:2018brz} is a milliQan-type detector proposed to be installed downstream of the proton
  target of a neutrino experiment, either in MINOS near detector hall and the proposed DUNE near detector hall, both at Fermilab.
  FerMINI can achieve unprecedented sensitivity for milli-charged particles with mass in the MeV-GeV range
  and fractional charge $Q_{\chi}/e$ in the $10^{-4} -10^{-1}$ range.

\end{itemize}

\vskip 5mm
\noindent
{\bf PBC projects on 5 and 10-15 years timescale}

\begin{figure}[h]
\centerline{\includegraphics[width=0.8\linewidth]{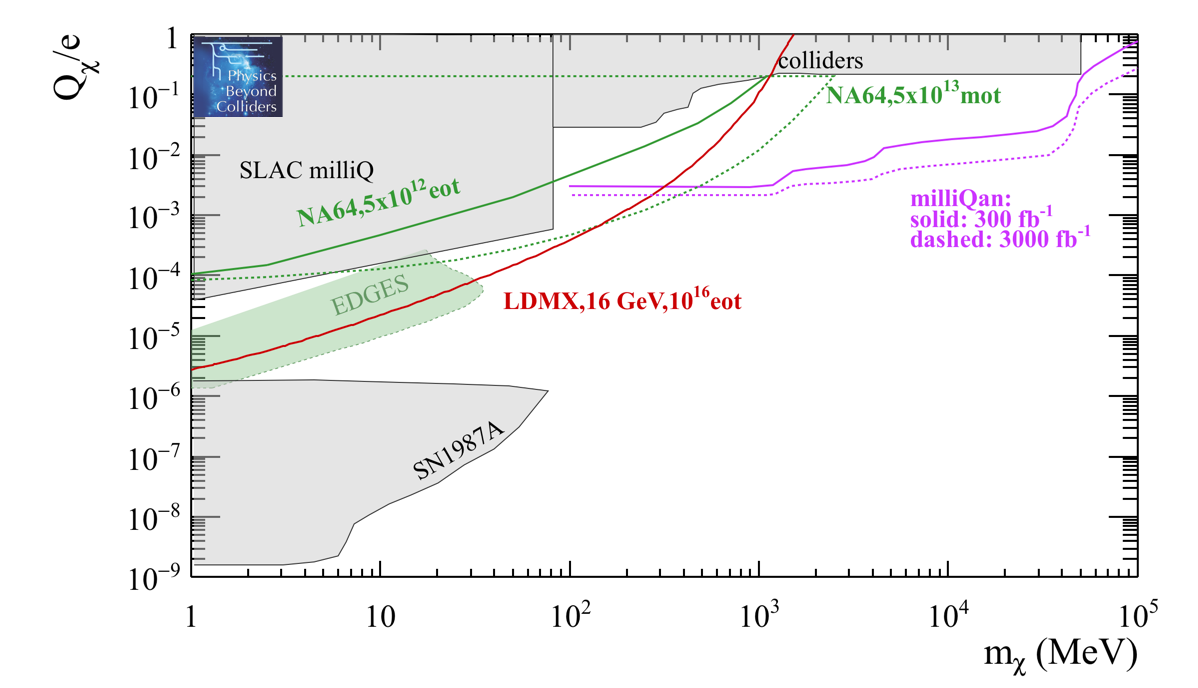}}
\caption{BC3: milli-charged particles. Current bounds (filled areas), experimental landscape and physics reach of PBC projects
in 5 years time scale (NA64$^{++}(e)$ and NA64$^{++}(\mu)$), and on 10-15 years time scale (LDMX @ eSPS).} 
\label{fig:milliQ_bc3_pbc_2}
\end{figure}

\vskip 2mm
Three PBC projects have sensitivity to search for milli-charged particles, as shown in Figure~\ref{fig:milliQ_bc3_pbc_2}:
NA64$^{++}(e)$ and NA64$^{++}(\mu)$ on 5-10 year timescale and LDMX on a $\sim 10-15$ year timescale.
NA64$^{++} (e)$ with $5 \times 10^{12}$ eot collected during Run 3 can explore the region with masses between 100-1000 MeV
and fractional charge $Q/e = 10^{-3} -10^{-2}$; NA64$^{++}(\mu)$ with $5 \times 10^{13}$ mot can improve over NA64$^{++}(e)$
by pushing down the limit of the fractional charge by almost an order of magnitude.
LDMX, with an electron beam of 16 GeV momentum and a collected yield of $10^{16}$ eot will further improve the search in particular in the
intermediate (100-1000 MeV)  mass region.

\clearpage
\subsection{Scalar Portal}
\label{ssec:scalar_portal}

\subsubsection{Dark scalar mixing with the Higgs (BC4 and BC5)}
  \label{sssec:bc4_bc5}

A light scalar particle mixing with the Higgs with the angle $\theta$ can be a mediator between
DM and SM particles.  The Langrangian to be added to the SM one is in the form:
\begin{equation}
\label{scalar}
{\cal L}_{\rm scalar} = {\cal L}_{\rm SM} + {\cal L}_{\rm DS} - (\mu S+ \lambda S^2)H^\dagger H.
\end{equation}
The minimal scenario (BC4) assumes for simplicity that $\lambda = 0$
and all production and decay processes of the dark scalars are controlled by the
same parameter $\mu = \sin \theta$.
Therefore, the parameter space for this model is $(\theta, m_S)$.
A more general approach (BC5) consists in having both $\lambda$ and $\mu$ being different from zero:
in this case, the parameter space is $\{\lambda, \theta, m_S\}$, and $\lambda$ is assumed to dominate
the production via {\em e.g.} $h\to SS$, $B \to K^{(*)}SS$, $B^0 \to SS$ etc.
In the following we will assume the branching fraction $BR(h \to SS) \sim 10^{-2}$
in order to be complementary to the LHC searches for the Higgs to invisible channels.

\vskip 2mm
A key feature of the scalar portal is that its production is often proportional to one of the larger
Yukawa couplings, $y_t$, in the case of the electro-weak penguin, while its decay is controlled by one
of the smaller Yukawa's or the induced gluon coupling. This means that it is natural for dark scalars
to be both long-lived and be produced at a relatively large rate, which makes them an excellent target
for the proposals discussed in this study.

\vskip 5mm
\noindent
 {\bf Current bounds and future experimental landscape}

\begin{figure}[h]
\centerline{\includegraphics[width=0.8\linewidth]{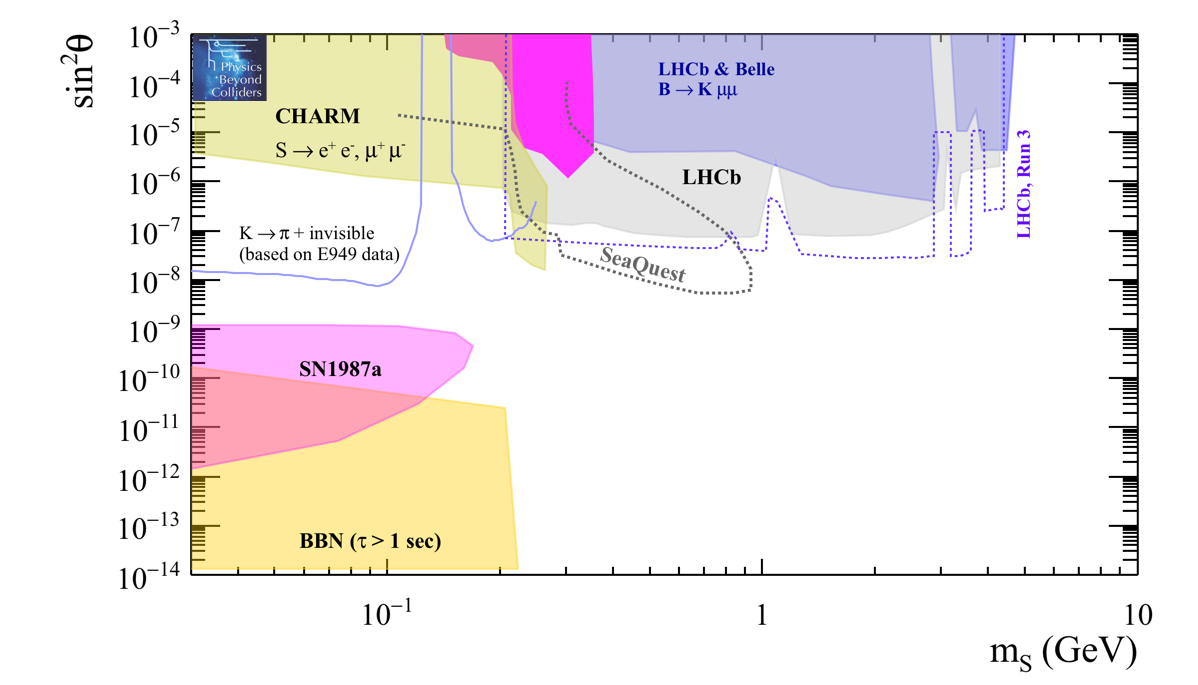}}
\caption{BC4: Dark scalar mixing with the Higgs. Current limits (filled areas)
and experimental landscape (solid and dotted curves) for searches for dark scalar in the plane
  coupling strength ($\sin^2\theta$) versus dark scalar mass $m_{S}$.}
\label{fig:BC4_context}
\end{figure}

\vskip 2mm
Figure~\ref{fig:BC4_context} shows the current bounds on the mixing parameter $\sin^2 \theta$
versus mass of the dark scalar $m_S$. Bound on this scenario come from re-interpratation of data from old beam dump
experiments~\cite{Clarke:2013aya, Winkler:2018qyg},
bump hunt in visible $B$ meson decays~\cite{Aaij:2016qsm,Aaij:2015tna, Wei:2009zv}
and cosmological and astrophysical arguments, as explained below.

\begin{itemize}
\item[-] {\it CHARM}\\
The CHARM Collaboration has put bounds
on light axion-like particles using a 400 GeV proton beam impinging on a copper
target~\cite{Bergsma:1985qz}.
Figure~\ref{fig:BC4_context} shows the reinterpretation of the CHARM data from
Ref.~\cite{Winkler:2018qyg} as  yellow shaded region. 

\item[-] {\it Visible Meson Decays}\\
A visibly decaying scalar mediator can
contribute to the processes $B^+ \to K^+ \mu^+ \mu^-$ and $B^0 \to K^{*0} \mu^+ \mu^-$, which are
tightly constrained by LHCb~\cite{Aaij:2016qsm,Aaij:2015tna} and Belle~\cite{Wei:2009zv} measurements.
In the same parameter space,
we also show bounds computed by us based on the measurement of
the $K^+ \to \pi^+ \nu \overline{\nu}$ branching fraction from E949 experiment~\cite{Artamonov:2008qb}.

\item[-] {\it  BBN}\\
A sufficiently light (m $<$ 10 MeV), weakly coupled scalar particle with a thermal number density
can decay appreciable during Big Bang Nucleosynthesis (BBN) and spoil the successful
predictions of light element yields accumulated in the early universe.

\item[-] {\it Supernovae}\\
A light, weakly coupled scalar mediator can be produced on shell
during a supernova (SN) explosion and significantly contribute to its energy loss,
thereby shortening the duration of the observable neutrino pulse emitted during core collapse.
The most significant constraint arises from SN1987a which has been used to constrain
the parameter space for axions and axion-like
particles~\cite{Turner:1987by,Frieman:1987ui,Burrows:1988ah,Essig:2010gu}.

\end{itemize}

Searches in the near ($\sim$ 5 years) future will be performed by:
 {SeaQuest at FNAL}~\cite{Berlin:2018pwi},
using the same dataset for the search for a Dark Photon into $e^+ e^-$ final state
as explained in Section~\ref{ssec:vector_portal};
{ LHCb}, that will update the bump hunt in $B \to K \ell^+ \ell^-$ decays
with an integrated luminosity of $\sim$ 15 fb$^{-1}$ which is expected to be collected during Run 3.
{NA62 in kaon-mode} will be able to explore the mass range below the kaon mass,
as a side product of the measurement at ${\mathcal{O}}(10\%)$ accuracy of the rare decay $K^+ \to \pi^+ \nu \overline{\nu}$,
by interpreting it as $K^+ \to \pi^+ S$. The NA62 search should be able to push down the limit
currently set by the E949 experiment by, at least, an order of magnitude, even if official projections have not yet been made by the collaboration.

\vskip 5mm
\noindent {\bf Physics reach of PBC projects on 5 and 10-15 year timescale} 
\vskip 2mm

\begin{figure}[h]
\centerline{\includegraphics[width=0.8\linewidth]{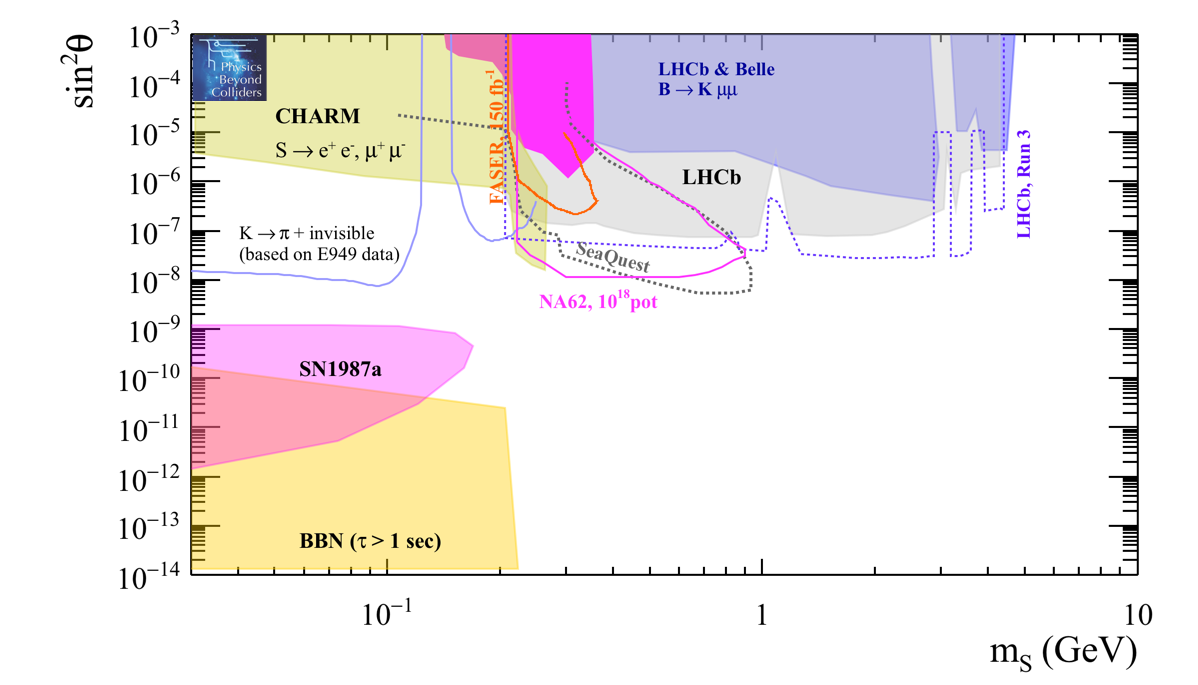}}
\caption{BC4: prospects on 5 year timescale for PBC projects for the dark scalar mixing with the Higgs
  in the plane mixing angle $\sin^2\theta$ versus dark scalar mass $m_{\rm S}$. The two PBC projects that can provide
  results on 5-year timescale are NA62$^{++}$ and FASER.}
\label{fig:bc4_pbc_1}
\end{figure}

Figure~\ref{fig:bc4_pbc_1} shows the sensitivity of FASER during its first phase of data taking during Run 3,
 and NA62$^{++}$ in dump mode with $10^{18}$pot collected
in about 100 days of data taking during Run 3. These results could be obtained on a $\sim 5$ year timescale. 
NA62$^{++}$ in dump mode should be able to improve
the limit between the di-muon mass and $\sim 1 $ GeV range and will compete with SeaQuest in the same timescale;
FASER, in its first phase, is instead not competitive.

\begin{figure}[h]
\centerline{\includegraphics[width=0.8\linewidth]{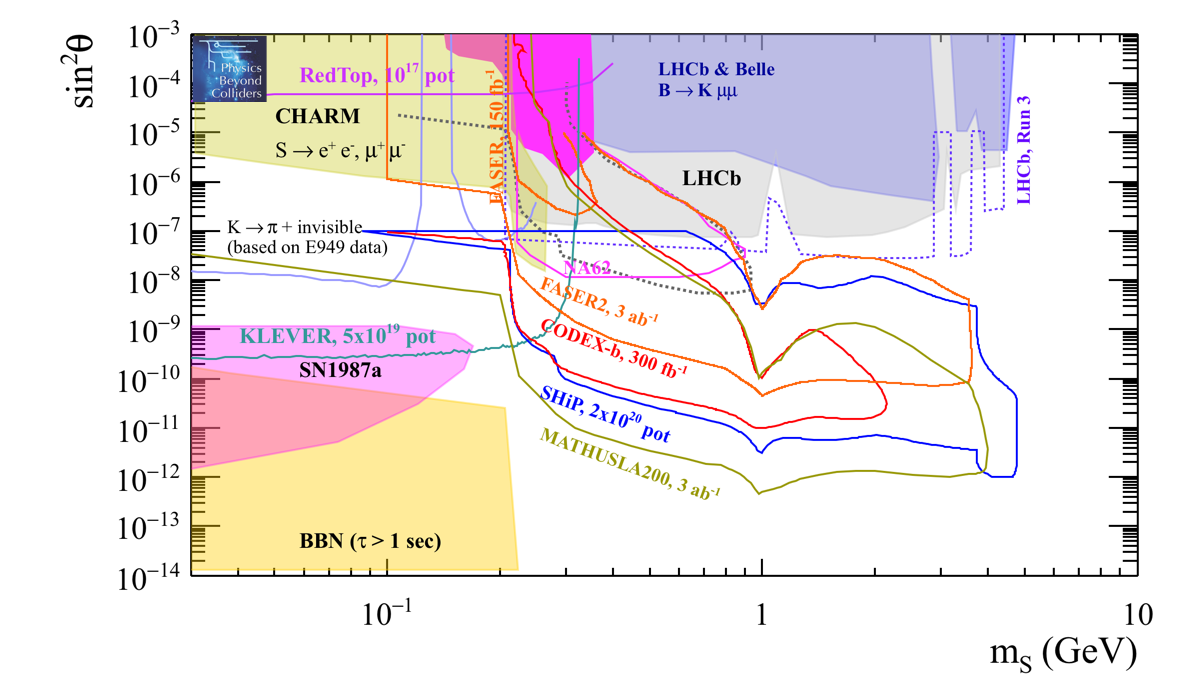}}
\caption{BC4: prospects on 10-15 year timescale for PBC projects for the Dark Scalar mixing with the Higgs in the plane mixing
  angle $\sin^2\theta$ versus dark scalar mass $m_{\rm S}$.}
\label{fig:bc4_pbc_2}
\end{figure}

\vskip 2mm
On a longer timescale (10-15 years) the explored parameter space will be significantly extended
by bigger PBC projects, as SHiP with $2 \times 10^{20}$ pot, KLEVER with $5 \times 10^{19}$ pot delivered, REDTOP with $10^{17}$ pot collected at the
$\eta$ threshold,
MATHUSLA200 and FASER2 with 3 ab$^{-1}$, running parasitically at the ATLAS or CMS interaction points,
and CODEX-b with 300 fb$^{-1}$, if the LHCb phase-II upgrade
will be approved during the HL-LHC era.
Above the di-muon threshold SHiP, FASER2, MATHUSLA200 and CODEX-b have comparable sensitivity.
Below the kaon mass, KLEVER will be able to close
the gap between the recasted limit from the data of the E949 experiment (and possible future result from NA62) and the Super Novae bound.
REDTOP with $10^{17}$  pot at the $\eta$ threshold instead is not competitive with respect to the others.
This is shown in Figure~\ref{fig:bc4_pbc_2}.

\begin{figure}[h]
\centerline{\includegraphics[width=0.8\linewidth]{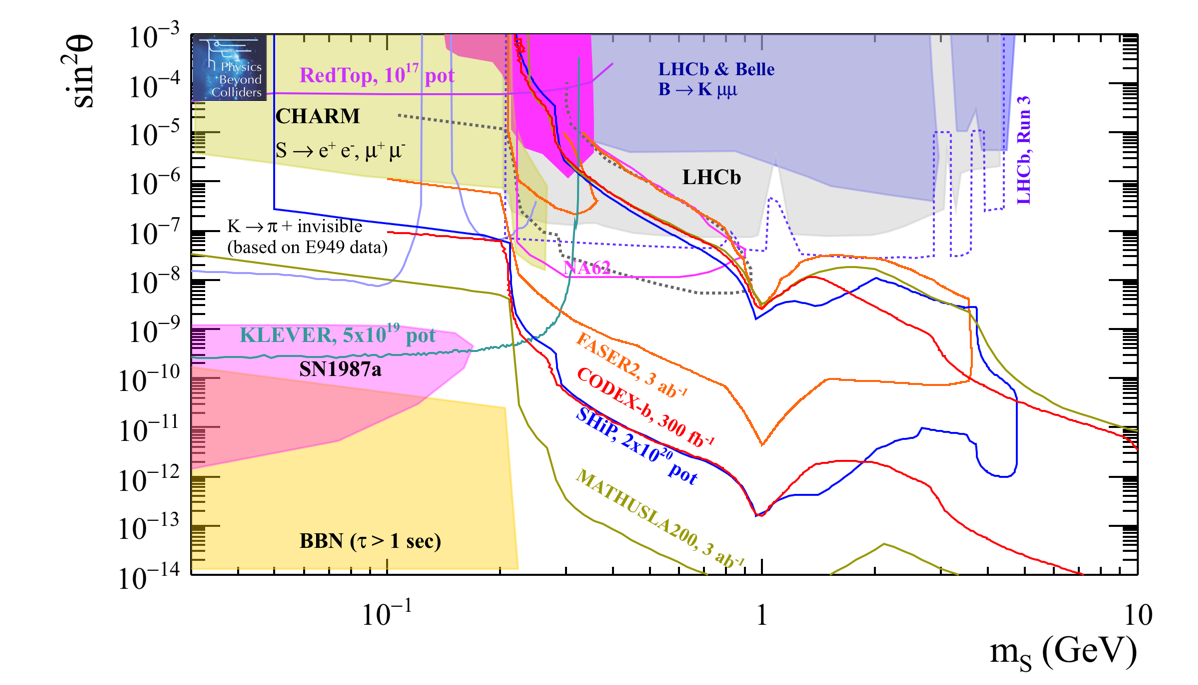}}
\caption{BC5: prospects on 10-15 year timescale for PBC projects for the dark scalar mixing with the Higgs in the plane mixing
  angle $\sin^2\theta$ versus dark scalar mass $m_{\rm S}$  under the hypothesis that both parameters $\lambda$
  and $\mu$ are different from zero. The sensitivity curves have been obtained assuming $BR(h \to SS) = 10^{-2}$.
  The NA62$^{++}$ and KLEVER curves correspond still to the case of $\lambda=0$,
  and, hence, should be considered conservative.}
\label{fig:BC5_pbc_2}
\end{figure}

\vskip 2mm
The extended version of the minimal Higgs-Dark Scalar model, with both couplings $\mu$ and $\lambda$
different from zero, allows to cover a larger fraction of the parameters space, as shown in Figure~\ref{fig:BC5_pbc_2},
due to the new contributions arising from a virtual (as in the $B \to K S S$ mode) or real (as in the case $h \to S S$) Higgs
in the chain. Also in this case the larger impact is provided by the bigger experiments, MATHUSLA200, SHiP, FASER2 and CODEX-b
which  will be able to explore the region well above the GeV mass scale in a fully uncharted range of couplings.
The experiments at central location near the LHC interaction points, MATHUSLA and CODEX-b, will have sensitivity all the way up
to $\sim 60$ GeV, if the assumption of zero background is valid.

\clearpage
\subsection{Neutrino Portal}
\label{ssec:neutrino_portal}

All fermions in the Standard Model with the exception of neutrinos
are known to exist with both, left handed and right handed chirality. A particularly strong
motivation for the existence of right handed neutrinos $\nu_R$ or {\it Heavy Neutral Leptons} (HNLs)
comes from the fact that they
can explain the light neutrino flavour oscillations via the type I seesaw
mechanism~\cite{Minkowski:1977sc,GellMann:1980vs, Mohapatra:1979ia, Yanagida:1980xy,Schechter:1980gr, Schechter:1981cv}.

\vskip 2mm
Another motivation for the existence of the $\nu_R$  comes from cosmology.
Couplings between $\nu_R$ generally violate $CP$, and the interactions of the $\nu_R$ in
the early universe can potentially generate a matter-antimatter asymmetry in the primordial plasma. At temperatures
above $T_{\rm sphaleron}$ = 130 GeV~\cite{DOnofrio:2014rug} this asymmetry can be converted into a net baryon number by weak
sphalerons~\cite{Kuzmin:1985mm}. This process called {\it leptogenesis} can either occur during the ``freeze-out'' and decay
of the $\nu_R$ ~\cite{Fukugita:1986hr} (“freeze-out scenario”) or during their production
\cite{Akhmedov:1998qx,Asaka:2005pn,Hambye:2016sby} (“freeze-in scenario”).
It is one of the most promising explanations for the baryon asymmetry of the universe (BAU),
which is believed to be the origin of baryonic matter in the present day universe, see ~\cite{Canetti:2012zc} for a
discussion.

\vskip 2mm
Heavy neutral leptons have been studied in connection to
Large Scale Structure formation~\cite{Viel:2005qj},
Big Bang Nucleosynthesis~\cite{Ruchayskiy:2012si},
cosmic microwave background, diffuse extragalactic background radiation,
supernovae~\cite{Fuller:2009zz}. 
In general, the scale of the HNL masses  is entirely unknown; different choices can have a wide range of
implications for particle physics, astrophysics and cosmology, see e.g. ~\cite{Drewes:2013gca} for an overview.
In the Neutrino Minimal Standard Model ($\nu$MSM)~\cite{Asaka:2005pn}
two HNLs are expected to be in the range MeV-GeV while a third HNL,  is a DM candidate and has masses as low as a few keV.
This model is particularly interesting from a phenomenological viewpoint because it
is feasible for masses of as low as 10~MeV~\cite{Canetti:2012kh}, which are well within reach of accelerator-based
experiments.

\vskip 2mm
Moreover, the decay width of the HNLs is suppressed by both the small mixing angle and $G_F^2$, while the latter factor
drops out in their production. As for the dark scalar, this means that the HNL's are naturallly long lived, but have a relatively
unsuppressed production rate, which makes them ideal targets for the PBC proposals.

\vskip 5mm
\noindent {\bf Current bounds on HNLs: general considerations}
\vskip 2mm

Mixing of heavy neutrinos with both $\nu_e$  and $\nu_{\mu}$  can be probed searching for bumps in the missing-mass distribution
of pions and kaons leptonic decays, eg. $K^+ \to \ell^+ \nu_{\ell}$, $(\ell = e, \mu)$.
These bounds are very robust because they assume only that a heavy neutrino exists and
mixes with $\nu_e$ and/or $\nu_{\mu}$. 
Another strategy to search for heavy neutrinos mixed with $\nu_e$, $\nu_{\mu}$ and $\nu_{\tau}$,
is via searches of their decay products. If the HNLs exist, they would be produced in every process
containing active neutrinos with a branching fraction proportional to the mixing parameters $|U_{e,\mu,\tau}|^2$.
Then the HNLs would decay via Charged Current (CC) and Neutral Current (NC) interactions into active neutrinos and other visible
final states, as pions, muons and electrons.
In beam-dump experiments, the HNLs would be produced in meson decays.
Other ways to constrain the couplings of HNLs in a relatively high mass regime
is using possible $Z^0$ decays into heavy neutrinos from LEP data~\cite{Abreu:1996pa}.
In this case, only large values of the mixing angle can be explored.

The bounds obtained from searches for HNLs with visible decays are in general less robust than the ones from
searches that use the missing mass technique.
In fact, the bounds obtained with HNLs in visible decays would be largely weakened
if the HNLs have other dominant decay modes into invisible particles.

\vskip 2mm
In the following we will consider only benchmark scenarios
in which a HNL couples to one SM generation at the time. This choice is driven by simplicity and allows us to ease
the comparison with bounds provided by past experiments that in most cases were
sensitive to one flavor coupling only. Other combinations of ratios of couplings are certainly possible
but they are beyond the present study.

Strong constraints on couplings for HNLs with masses below the kaon mass are set by past experiments,
in particular PS191~\cite{Bernardi:1987ek}, CHARM~\cite{Bergsma:1985is},
NuTeV~\cite{Vaitaitis:1999wq},
E949~\cite{Artamonov:2014urb},
PIENU~\cite{PIENU:2011aa},
TRIUMF-248~\cite{Britton:1992xv} and NA3~\cite{Badier:1985wg}. An interesting search has been also performed recently by the
NA62 collaboration~\cite{CortinaGil:2017mqf}.

\vskip 2mm
A significant improvement in the entire mass range below the
$B$-meson mass could be achieved by SHiP, as will be shown in the following.
The same mass range could also be probed by a FASER2~\cite{Kling:2018wct}, CODEX-b~\cite{Gligorov:2017nwh}
and MATHUSLA200~\cite{Chou:2016lxi}.
Above the $B-$meson mass, displaced vertex searches at high energy
hadron~\cite{Helo:2018qej,Izaguirre:2015pga,Antusch:2016vyf,Gago:2015vma,Antusch:2017hhu,Antusch:2016ejd}
or lepton~\cite{Antusch:2017pkq,Antusch:2016vyf} colliders would be more sensitive, see e.g., Ref.~\cite{Antusch:2016ejd}
for a summary.
Neutrinos that are heavy enough to decay promptly can leave
distinct lepton number and flavour violating signatures in high energy collisions,
see~\cite{Deppisch:2015qwa,Cai:2017mow} for a recent review.

\subsubsection{Neutrino portal with electron-flavor dominance (BC6)}
\label{sssec:bc6}

In this Section we consider  the case in which HNLs couple only to
first SM generation and the sensitivity plots are shown in the plane
\{$|U_e|^2, m_{N}$\}.

\vskip 5mm
\noindent
    {\bf Current bounds, experimental landscape and PBC projects on 5 year timescale} 

\begin{figure}[h]
      \centerline{\includegraphics[width=0.8\linewidth]{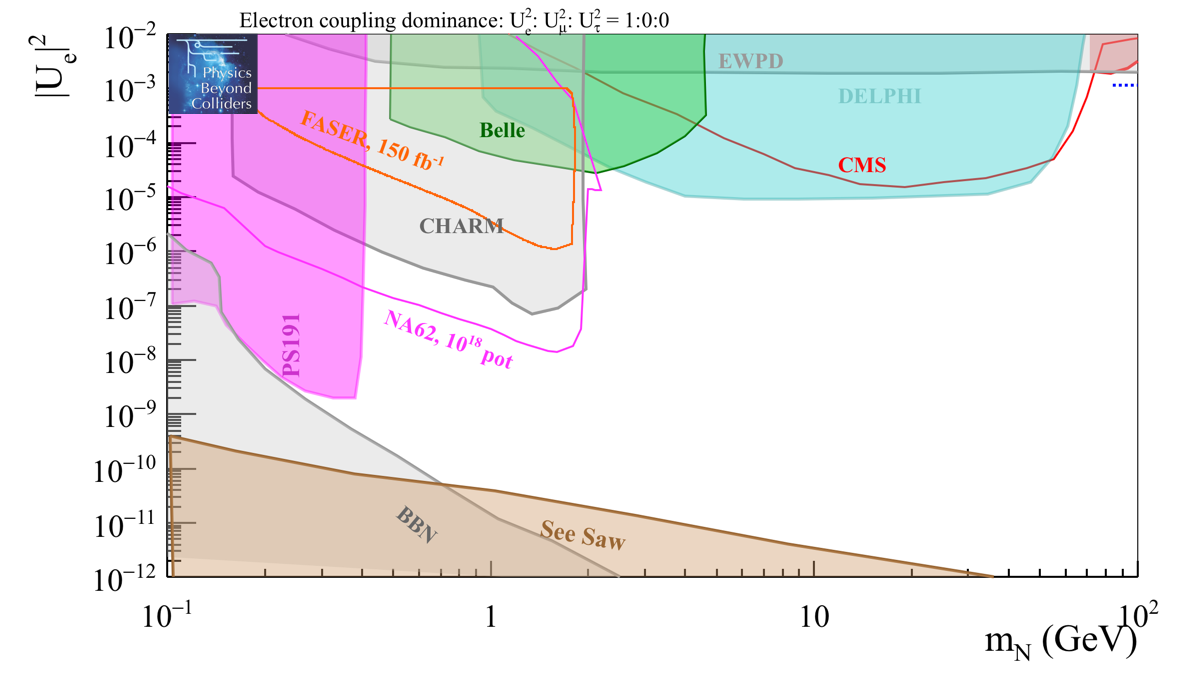}}
      \caption{BC6: Sensitivity to Heavy Neutral Leptons with coupling to the first lepton generation only.
        Current bounds (filled areas) and near ($\sim$ 5 years) future physics reach of two PBC projects, FASER and NA62$^{++}$ (solid lines).
      See text for details.}
      \label{fig:bc6_pbc_1}
    \end{figure}

    \vskip 2mm
    Current bounds and future experimental landscape in the next $\sim$ 5 years,
    including some PBC projects, is shown in Figure~\ref{fig:bc6_pbc_1}
    for the case of HNLs with couplings only to the first lepton generation and
    masses in the MeV-GeV range. 

    \vskip 2mm
    Existing bounds, shown as filled coloured areas, for masses below the charm mass
    arise mostly from beam dump experiments (PS191~\cite{Bernardi:1987ek} and CHARM~\cite{Bergsma:1985is})
    while those above the charm mass
    from LEP data, dominated by the DELPHI result~\cite{Abreu:1996pa}, from Belle~\cite{Liventsev:2013zz} and more recently
    from CMS~\cite{Sirunyan:2018mtv}.
    The allowed range of couplings is bounded from below by the BBN constraint~\cite{Ruchayskiy:2012si},
    and the see-saw limit~\cite{Canetti:2010aw}.  

    \begin{itemize}
    \item[-] {\it PS 191 @ CERN: } 
      the PS191 CERN experiment was specifically designed to search for neutrino decays
      in a low-energy neutrino beam. The apparatus consisted of 10~m long nearly empty decay volume instrumented
      by flash chambers, calorimeter and scintillator hodoscope. Ref.~\cite{Bernardi:1987ek}.
    \item[-] {\it CHARM @ CERN:}
    a search for heavy neutrinos was performed by the CHARM collaboration by
      dumping ${\mathcal{O}}(2 \cdot 10^{18})$
      400 GeV protons on a thick Copper beam dump and looking for visible decays with
      electrons in the final state in the 35 m long
      decay volume with a spectrometer of $3 \times 3$ m$^2$ cross section. Ref.~\cite{Bergsma:1985is}.
    \item[-] {\it Belle @ KEK:} 
      Belle performed a search for heavy neutrinos with 772 M of $B \overline{B}$ pairs using
      leptonic and semileptonic $B$ mesons decays, $B \to X l \nu_R$, where $\ell = e,\mu$ and $X$ was a
      charmed meson $D^{(*)}$, a light meson $(\pi, \rho, \eta, etc.)$ or nothing (purely leptonic decays),
      in a range of masses between the kaon and the $B$ mass.
      Ref.\cite{Liventsev:2013zz}.
    \item[-] {\it LEP data:}
     the most stringent limits above the $B$ meson mass have been put by DELPHI~\cite{Abreu:1996pa}.
      HNLs have been searched for using data collected by the DELPHI detector corresponding to $3.3 \times 10^6$
      hadronic $Z^0$ decays at LEP1.
    \item[-] {\it CMS @ LHC:}
    CMS searched for HNLs in three prompt charged leptons sample in any combination of electrons
      and muons collected at a center-of-mass energy of 13~TeV and corresponding to an integrated luminosity
      of 35.9~fb$^{-1}$.  The search is performed in the HNL mass range between 1~GeV and 1.2~TeV.
      Ref.~\cite{Sirunyan:2018mtv}.
      \item[-] {\it BBN constraint:} \\ a HNL with parameters
        to the left of the Big Bang Nucleosynthesis (BBN) line would live sufficiently long in the early Universe to
        result in an overproduction of primordial Helium-4~\cite{Ruchayskiy:2012si}.
     \item[-] {\it See-saw limit:} below the see-saw limit, the mixing of the HNL with active neutrinos becomes too weak
        to produce the observed pattern of neutrino flavour oscillations~\cite{Canetti:2010aw}.   
    \end{itemize}

    Only two PBC projects with a $\sim 5$ year timescale can contribute to this benchmark case:
    {FASER} with 150 fb$^{-1}$, which unfortunately is not competitive with the current bounds 
    set by CHARM, and more interestingly, {NA62$^{++}$ } that can push down the CHARM limits
    by about one order of magnitude in the same mass range by collecting $\sim 10^{18}$ pot in dump mode.
    The NA62$^{++}$ projections correspond to the 90 \% CL
    exclusion limits in case all the final states with at least two charged tracks are considered~\cite{Drewes:2018gkc}. In the projections
    the zero background hypothesis is assumed. Studies performed with the already acquired $3 \times 10^{16}$ pot
    dataset in dump mode show that the background can be reduced to zero with the current setup
    for fully reconstructed final states,
    while for open final states the addition of an Upstream Veto in front of the decay volume is required.
    The addition of this detector is currently under study by the Collaboration. 
    
\clearpage
\noindent {\bf Physics reach of PBC projects on 10-15 year timescale} 
\vskip 2mm

\begin{figure}[h]
  \centerline{\includegraphics[width=0.8\linewidth]{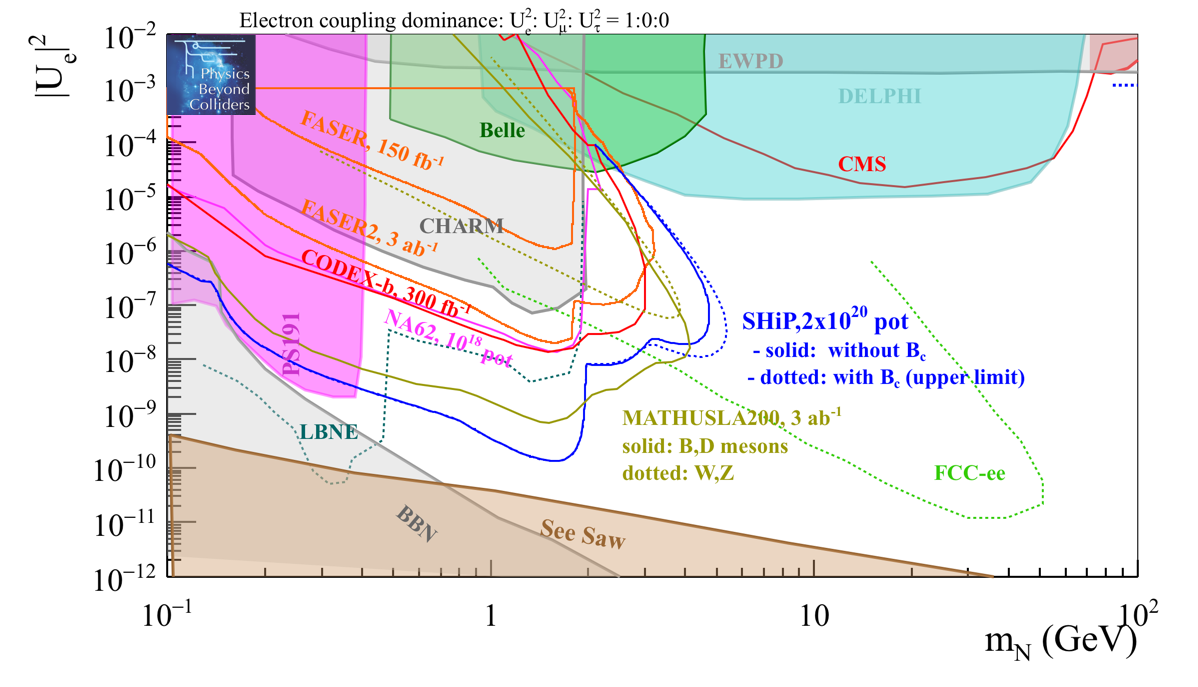}}
  \caption{BC6: Sensitivity to Heavy Neutral Leptons with coupling to the first lepton generation only. Current bounds (filled areas)
  and 10-15 years prospects
  for PBC projects (SHiP, MATHUSLA200, CODEX-b and FASER2) (solid lines).
  Projections for a LBNE near detector with $5 \times 10^{21}$ pot and from
  FCC-ee with $10^{12}$ $Z^0$ decays are also shown.}
  \label{fig:bc6_pbc_2}
\end{figure}

On 10-15 year timescale many PBC projects can contribute to this benchmark case,
as shown in Figure~\ref{fig:bc6_pbc_2}: MATHUSLA200, FASER2, CODEX-b and SHiP.
For MATHUSLA200 we show separately the contributions from heavy mesons and gauge bosons decays.
The SHiP sensitivity curve is obtained without (solid curve) and with (dashed curve) a particular assumption for the contribution from $B_c$.
This is because the $\sigma(B_c)/\sigma(B)$ fraction at the SPS energies is not reliably known. We therefore show an upper sensitivity
limit provided by assuming the fraction measured at the LHC energy.

Also in this case the plot shows the 90\% CL exclusion limits under the hypothesis of zero background.
This hypothesis is a strong assumption that has been properly validated only by SHiP, so far, using a full GEANT4
simulation of the detector, including digitization and reconstruction,  and large samples of Monte Carlo data.
The background evaluation for  MATHUSLA200, CODEX-b and FASER2 is still work in progress and will be carried
on in the coming years. Figure~\ref{fig:bc6_pbc_2} shows also projections from the {\it LBNE near detector} as
5-year sensitivity corresponding to an exposure of $5 \times 10^{21}$ protons on
target for a detector length of 30~m and assuming a normal hierarchy of neutrinos masses~\cite{Adams:2013qkq}
and from {\it FCC-ee} with $10^{12}$ $Z^0$ decays and HNLs decaying 
between 10-100~cm from the interaction vertex~\cite{Blondel:2014bra}.

\subsubsection{Neutrino portal with muon-flavor dominance (BC7)}
\label{sssec:bc7}

In this Section we consider the case in which HNLs couple only to
second SM generation and the sensitivity plots are shown in the plane
\{$|U_{\mu}|^2, m_{N}$\}.

\vskip 1cm
\noindent {\bf Current bounds, experimental landscape and PBC projects on 5 year timescale} 

\begin{figure}[h]
  \centerline{\includegraphics[width=0.8\linewidth]{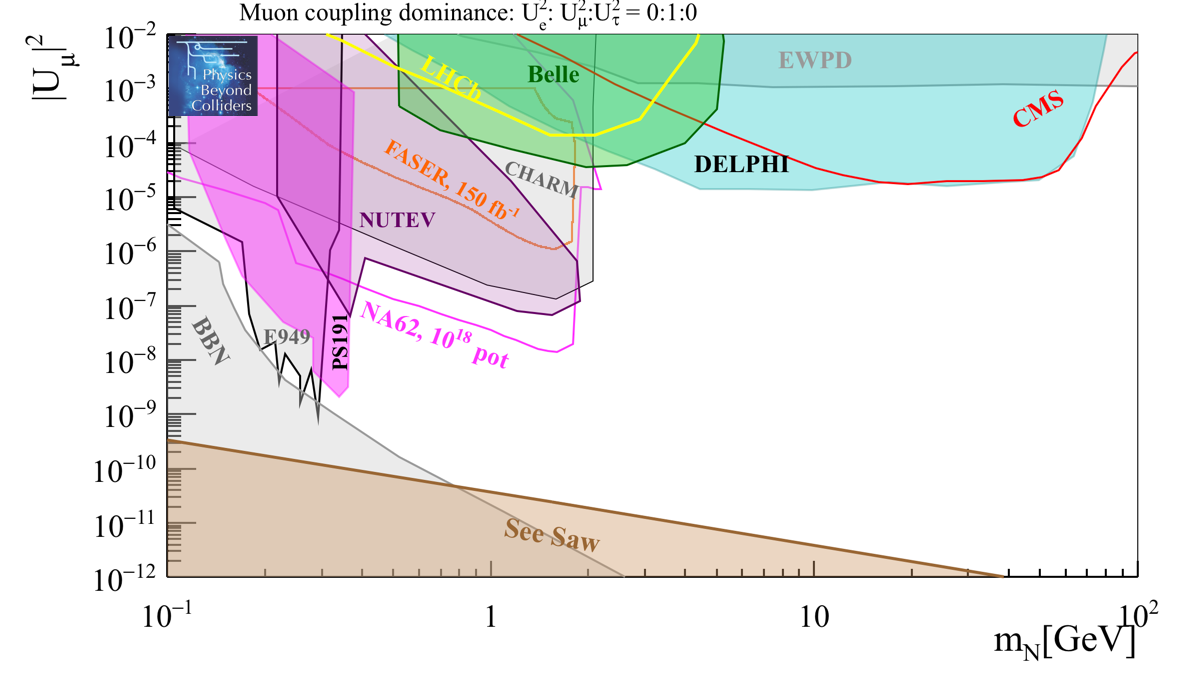}}
  \caption{BC7: Sensitivity to Heavy Neutral Leptons with coupling to the second lepton generation only.
    Current bounds (filled areas) and near ($\sim$ 5 years) physics reach of two PBC projects, FASER and NA62$^{++}$ (solid lines).
    See text for details.}
  \label{fig:bc7_pbc_1}
\end{figure}

\vskip 1mm
Current bounds and the future experimental landscape in the next $\sim$ 5 years,
including some PBC projects, are shown in Figure~\ref{fig:bc7_pbc_1}
for the case of HNL with couplings only to the second lepton generation and
masses in the MeV-GeV range. 

\vskip 2mm
Also in this case the allowed range of couplings is bounded from below by the BBN constraint~\cite{Ruchayskiy:2012si},
and the see-saw limit~\cite{Canetti:2010aw}.  
Existing experimental limits are shown as filled coloured areas:  for masses below the charm mass
they arise mostly from the same beam dump experiments contributing to the sensitivity
for electron-flavor dominance (PS191~\cite{Bernardi:1987ek} and CHARM~\cite{Bergsma:1985is}, as explained in the previous
paragraph) with the addition of NuTeV~\cite{Vaitaitis:1999wq}, and E949~\cite{Artamonov:2014urb}:

\begin{itemize}
\item[-] {\it NuTeV @ Fermilab:} a search for HNLs decaying in muonic final states has been performed at the neutrino
  detector NuTeV at Fermilab in 1996-1997, using $2 \times 10^{18}$ 800 GeV protons
  interacting with a Berillium-oxide target and a proton dump. Ref.~\cite{Vaitaitis:1999wq}.
\item[-] {\it E949 @ BNL:} Evidence of a heavy neutrino, in the  process $K^+ \to \mu^+ \nu_R$ was sought
  by the E949 collaboration using $1.7 \times 10^{12}$ stopped kaons.
\end{itemize}

Above the charm mass, current bounds are set by DELPHI~\cite{Abreu:1996pa}, Belle~\cite{Liventsev:2013zz},
CMS~\cite{Sirunyan:2018mtv} with the same analysis used to set bounds for electron-dominance, and by LHCb with a dedicated
analysis to search for prompt and diplaced $\pi^- \mu^+$ vertices in $B^+ \to \pi^- \mu^+ \mu^+$ LNV
decays~\cite{Aaij:2014aba}.

\vskip 2mm
As in the benchmark case of HNLs with couplings only to the first generation, NA62$^{++}$ in dump mode and
FASER will be able to perform this search with competitive physics reach in a time scale of $\sim$ 5 years.

\vskip 5mm
\noindent {\bf Physics reach of PBC projects on 10-15 year timescale} 

\begin{figure}[h]
  \centerline{\includegraphics[width=0.8\linewidth]{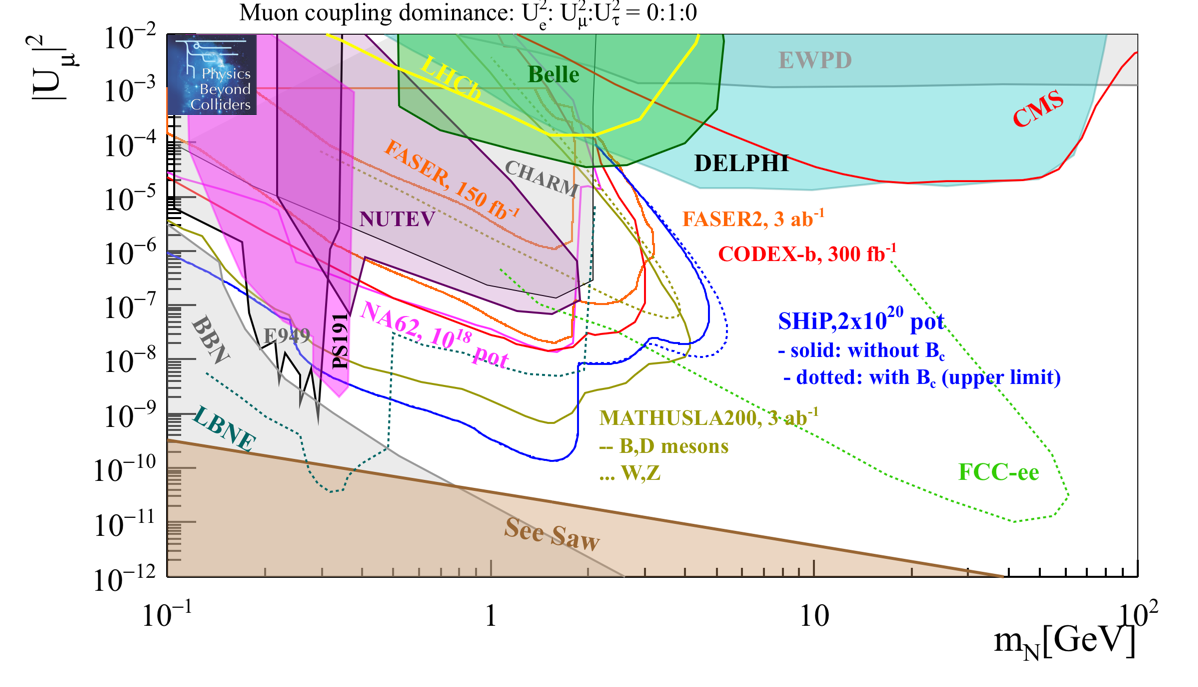}}
  \caption{BC7: Sensitivity to Heavy Neutral Leptons with coupling to the second lepton generation only.
    Current bounds (filled areas) and 10-15 years prospects for PBC projects (SHiP, MATHUSLA200, CODEX-b and FASER2) (dotted and solid lines).
    Projections for the LBNE near detector with $5 \times 10^{21}$ pot and  FCC-ee with $10^{12}$ $Z^0$
    decays are also shown.}
  \label{fig:bc7_pbc_2}
\end{figure}

\vskip 2mm
Figure~\ref{fig:bc7_pbc_2} shows the 90 \% CL exclusion limits from MATHUSLA200, FASER2, CODEX-b and SHiP
in a 10-15 years time scale. Also in this case the curves are obtained under the assumption of zero background,
for which the same considerations drawn in the previous paragraph hold.

\subsubsection{Neutrino portal with tau-flavor dominance (BC8)}
\label{sssec:bc8}

In this Section we consider the case in which HNLs couple only to
third SM generation and the sensitivity plots are shown in the plane
\{$|U_{\tau}|^2, m_{N}$\}.

\vskip 5mm
\noindent {\bf Current bounds and experimental landscape}
\vskip 2mm

\begin{figure}[h]
  \centerline{\includegraphics[width=0.8\linewidth]{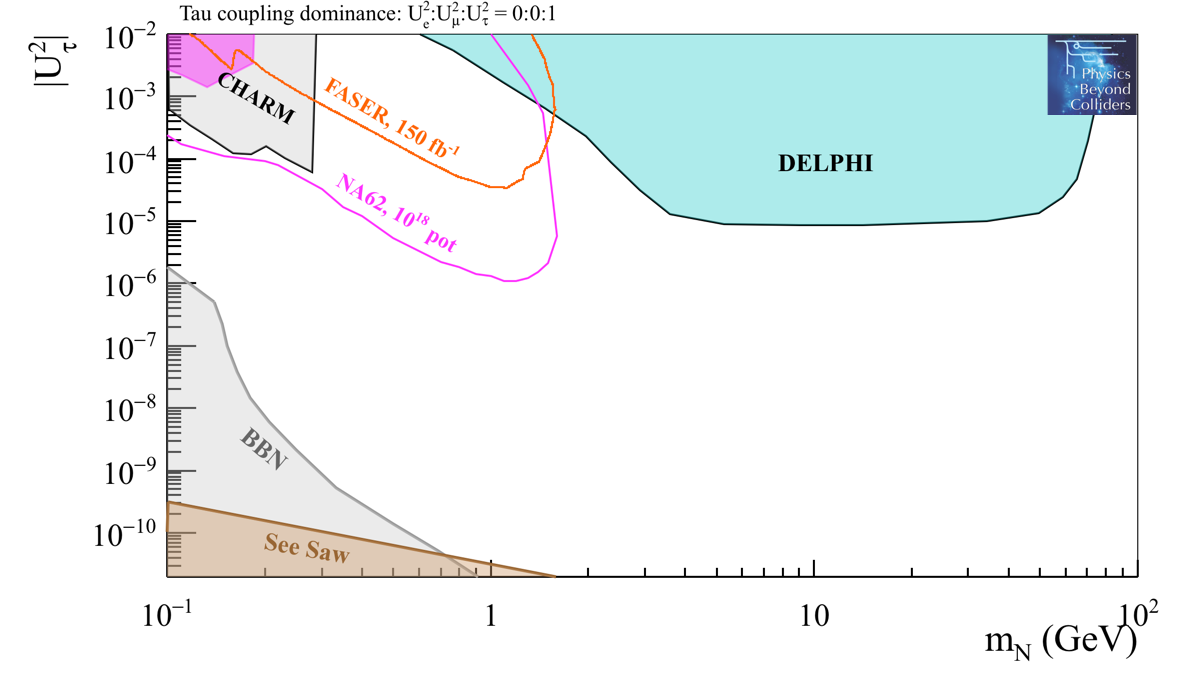}}
    \caption{BC8: Sensitivity to Heavy Neutral Leptons with coupling to the third lepton generation only.
    Current bounds (filled areas) and near ($\sim$ 5 years) future physics reach of two PBC projects, FASER and NA62$^{++}$ (solid curves).
    See text for details.}
  \label{fig:bc8_pbc_1}
\end{figure}

Current bounds and future experimental landscape in the next $\sim$ 5 years,
including some PBC projects, is shown in Figure~\ref{fig:bc8_pbc_1}
for the case of HNL coupling only to the third lepton generation and masses in the MeV-GeV range. 
Also in this case the allowed range of couplings is bounded from below by the BBN constraints~\cite{Ruchayskiy:2012si},
and the see-saw limit~\cite{Canetti:2010aw}.

Main bounds in this benchmark case arise from CHARM~\cite{Orloff:2002de},
NOMAD~\cite{Astier:2001ck}, and again the same data from DELPHI~\cite{Abreu:1996pa} used for the
other two benchmark cases (BC6 and BC7).

\begin{itemize}
\item[-] {\it CHARM:} limits on the square mixing strength $|U_{\tau}|^2$ in a mass range 10-290 MeV were set by
  re-interpreting the null result of a search for events produced by the decay of neutral particles into
  two electrons performed by the CHARM experiment using the neutrino flux produced by ${\mathcal{O}}(2\times 10^{18})$ 400~GeV protons
  on a solid copper target. Ref.~\cite{Orloff:2002de}.
  
\item[-] {\it NOMAD:} a search for heavy neutrinos was performed using $4.1 \times 10^{19}$ 450 GeV protons on target
  at the WANF facility at CERN in 1996-1998. The HNLs were searched in the 
 process $D_s \to \tau \nu_R$ followed by the decay $\nu_R \to \nu_{\tau} e^+ e^-$ in the NOMAD detector.
 This allowed to derive an upper limit on the mixing strength between
 the heavy neutrino and the tau neutrino in the $\nu_R$ mass range from 10 to 190 MeV. Ref.~\cite{Astier:2001ck}.
\end{itemize}

\noindent {\bf Physics reach of PBC projects on 5 and 10-15 year timescale} 
\vskip 2mm

\begin{figure}[h]
  \centerline{\includegraphics[width=0.8\linewidth]{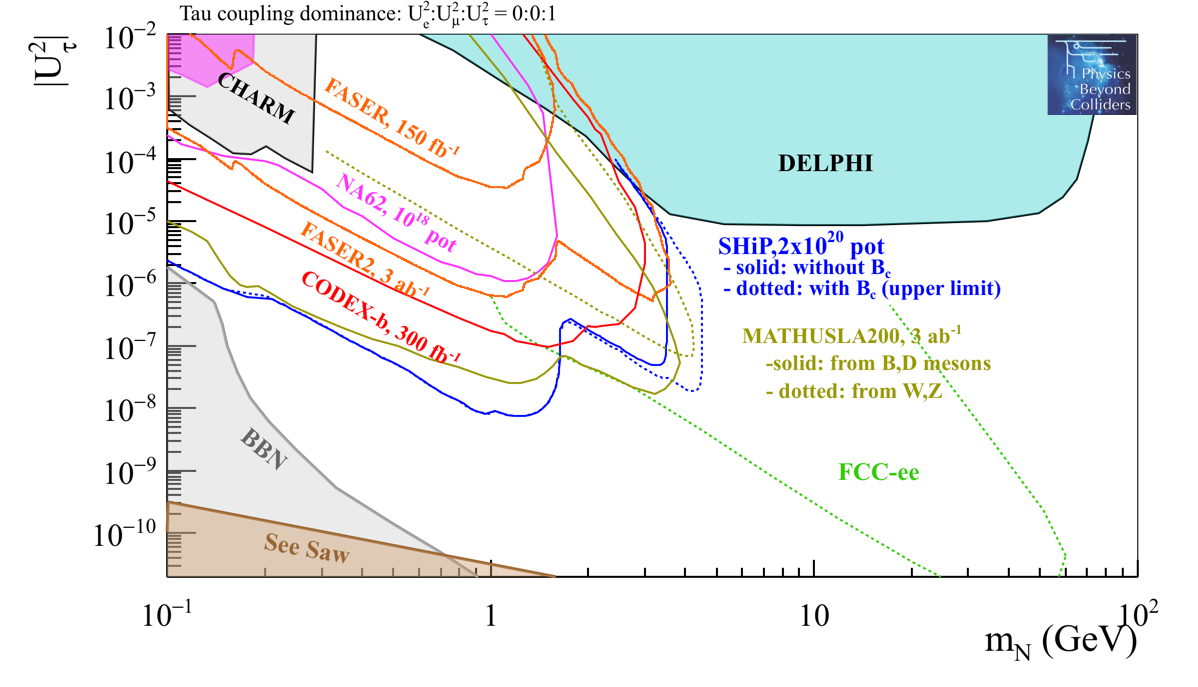}}
    \caption{BC8: Sensitivity to Heavy Neutral Leptons with coupling to the third lepton generation only.
    Current bounds (filled areas) and 10-15 years prospects for PBC projects (SHiP, MATHUSLA200, CODEX-b and FASER2) (solid and dotted curves).
    Projections from FCC-ee with $10^{12}$ $Z^0$ decays are also shown.}
  \label{fig:bc8_pbc_2}
\end{figure}

Among the PBC projects the only two contributing on a 5-year timescale are, again, FASER with 150 fb$^{-1}$
and NA62$^{++}$, as shown in Figure~\ref{fig:bc8_pbc_1}.
Figure~\ref{fig:bc8_pbc_2} shows the 90 \% CL exclusion limits from MATHUSLA200, FASER2, CODEX-b and SHiP
in a 10-15 years time scale. The physics reach from FCC(ee) with $10^{12}$ $Z^0$ is also shown.

\clearpage
\subsection{Axion Portal}
\label{ssec:axion_portal}

The discovery of the Higgs boson shows clearly that elementary scalar bosons exist in
nature. Therefore it is timely and well-motivated to search for further light scalar or pseudoscalar
particles.
 Pseudo-scalar particles can arise as pseudo-Nambu-Goldstone bosons
 (PNGB) of a spontaneously broken {\emph{U(1)} symmetry at a scale $f_A$.
The principal example of very light pseudo-Goldstone bosons is the
 axion~\cite{Peccei:1977hh,Peccei:1977ur,Wilczek:1977pj,Weinberg:1977ma}
 introduced to solve the strong {\emph CP} problem in QCD. Natural extensions of the axion paradigm
bring to a wide range of interesting pseudoscalar particles which typically
 have very similar interactions as the axion, but without a strict relation between the mass and the coupling,
 the Axion-Like Particles or ALPs.

\vskip 2mm
 ALPs also provide an interesting connection to the puzzle of dark matter, because they
 can mediate the interactions between the DM particle and SM states and allow for additional
 annihilation channels relevant for the thermal freeze-out of DM.
 In fact in presence of an additional pseudoscalar particle
 that mediates the interactions of DM with the SM sector~\cite{Freytsis:2010ne,Dienes:2013xya},
 constraints from direct detection experiments~\cite{Akerib:2013tjd,Agnese:2013jaa,Agnese:2014aze,Angloher:2014myn}
 and invisible Higgs width~\cite{Aad:2014iia,Chatrchyan:2014tja}
 on the scalar portal  can be easily evaded~\cite{Freytsis:2010ne,Dienes:2013xya}.

Moreover, if the pseudoscalar mass is in the sub-GeV range
it can furthermore evade detection at the LHC, as, e.g., in monojet searches~\cite{Chatrchyan:2011nd}.
Another advantage of a very light pseudoscalar $a$ is that it allows for the
possibility that DM can obtain the correct relic density from thermal freeze-out even if it is very weakly
coupled to SM particles.
This is due to the fact that, provided the pseudo-scalar mass is less
than twice the mass of the DM particle $\chi$, the annihilation process
$\chi \overline{\chi} \to a a$, followed by decays of
the pseudoscalars into SM particles, allows for a highly efficient annihilation of DM
particles. The only important constraint is that such pseudoscalar particles 
must decay before BBN.
%
As explained in Section~\ref{sec:portals}, ALPs can mediate interactions between DM and the SM sector
via three different couplings, photon-, gluon-, and fermion-coupling.

\subsubsection{Axion portal with photon-coupling (BC9)}
\label{sssec:bc9}

Assuming a single ALP state $a$, and the predominant coupling to photons,
all phenomenology (production, decay, oscillation in the magnetic field) can be determined
as functions on ($m_a; g_{a \gamma \gamma} = f_{\gamma}^{-1}$) parameter space.

\vskip 2mm
\noindent {\bf Current bounds and near future experimental landscape}
\vskip 1mm

The current bounds for ALPs with photon coupling are shown in Figure~\ref{fig:ALPS_gg_context}, left.
A zoom on the region of interest for experiments at accelerators is shown in the right panel and covers a range of masses
between MeV and GeV.
We note that this is also the mass region of interest in models where ALPs serve as
mediators to a DM sector.

\vskip 2mm
Searches for ALPs with photon coupling have been performed using monophoton searches at LEP,
$e^+ e^- \to \gamma^* \to a \gamma$, mono-photon searches at BaBar ($e^+ e^- \to \gamma a, a \to$ invisible),
radiative $\Upsilon$ decays ($\Upsilon (nS) \to \gamma^* \to \gamma a$), 
radiative $Z-$boson decays, and electron- and proton beam dump experiments, where the ALPs are produced
mainly via the Primakoff effect, i.e. the conversion of a photon into an ALP in the vicinity of
a nucleus~\cite{Halprin:1966zz}.

\begin{figure}[h]
  \centerline{
  \includegraphics[width=0.47\linewidth]{AtPPlot.jpg}     
  \includegraphics[width=0.47\linewidth, height=5cm]{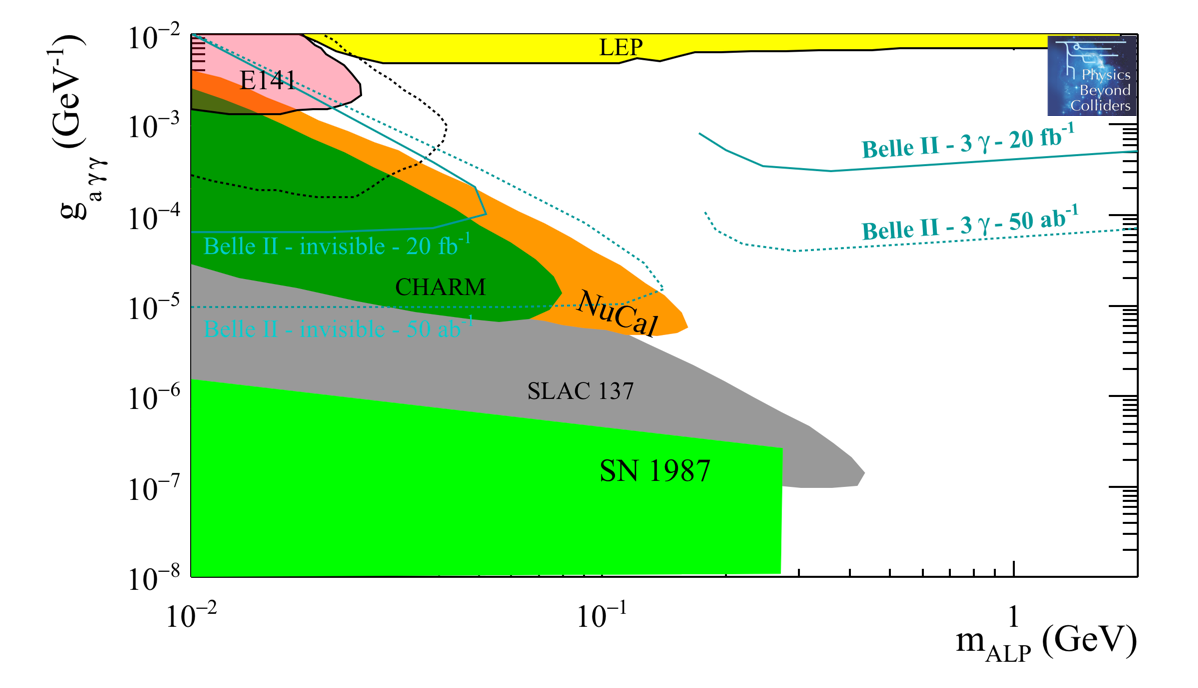} 
  }
  \caption{Left panel: current limits on axion with photon coupling in the plane coupling ($g_{a\gamma\gamma}$) versus mass ($m_a$).
  The right panel shows the zoom in the range of masses interesting for accelerator-based experiments (note the different units
  for the mass ranges in the two panels).
  In the left pane, current bounds are shown as filled coloured areas. A possible sensitivity from Belle-II
  from a phenomenological study~\cite{Dolan:2017osp} is also shown.}
\label{fig:ALPS_gg_context}
\end{figure}

\begin{itemize}
\item[-] {\it  E141 @ SLAC (electron beam dump):} primarily searched for long-lived particles decaying
into the $e^+ e^-$ final state~\cite{Riordan:1987aw}
  but the addition of a photon converter in front of the detector for a 
  a certain period of data taking
  opened the possibility to be sensitive also to photons from an ALP decay.
  Reinterpretation of the E141 results was performed in Refs.~\cite{Krasny:1987eb,Dobrich:2017gcm},
  leading to a somehow more conservative bounds
  with respect previous interpretations~\cite{Hewett:2012ns}.

\item[-]
{\it E137 @ SLAC (electron beam dump):} dedicated search for ALPs coupling only to
photons~\cite{Bjorken:1988as}. 
However the exclusion limits presented in the paper did not contain the turnover towards large couplings due to
the exponential suppression of the number of ALPs that reach the detector, this has been added in Ref.~\cite{Dolan:2017osp}.

\item[-]
  {\it CHARM (proton beam dump):}
  a search for a ALP in 400 GeV proton interactions with a thick copper target was performed with the CHARM detector~\cite{Bergsma:1985qz}.
  The target was placed 480 m apart from the 35 m long decay volume and the hypothetical decay
  $a\to \gamma \gamma$ has been sought using a fine-grained calorimeter of 9 m$^2$ active area.

\item[-]
  {\it NuCal (proton beam dump):}
  The production and decay of a light scalar and pseudoscalar particles has been investigated in a proton-iron beam
  dump experiment at the 70 GeV Serpukhov accelerator.
  Ref.~\cite{Blumlein:1990ay}.

\item[-]
  {\it LEP:}
  Limits from LEP data in the ALPs mass range  MeV - GeV arise from a reinterpretation~\cite{Jaeckel:2015jla}
  of the LEP $Z^0 \to \gamma \gamma$ data~\cite{Acciarri:1994gb,Abreu:1991rm,Abreu:1994du,Acciarri:1995gy}
  where one of the two neutral clusters was considered the result of a merging of two highly collimated photons
  from the ALP decay in the process  $Z^0 \to a \gamma, a \to \gamma \gamma$.
  Mono-photon searches, i.e. searches for highly-energetic photons in
  association with missing energy resulting from the process $e^+ e^- \to \gamma + a (a \to$ invisible)
  have been performed as well~\cite{Abdallah:2008aa}
  but they are not sensitive to ALPs with mass in the sub-GeV range~\cite{Dolan:2017osp}.

\item[-]
{\it Bound from Astrophysics:}
Supernova 1987A. Weakly coupled particles such as axions or ALPs with masses up to
about 100~MeV can be copiously produced in the hot core of a supernova.
Because of their weak couplings these particles stream out of the core and thereby constitute
a new energy loss mechanism. In the absence of such new particles the main cooling mechanism
is due to neutrino emission. The corresponding neutrino signal has been observed in the case of
SN 1987A, placing a bound on possible exotic energy loss mechanisms, which should not
exceed the energy loss via neutrino emission~\cite{Dolan:2017osp}. 

\end{itemize}

We note that mono-photon searches have also been carried out at the LHC (for the
most recent analyses see e.g. ref.~\cite{Aaboud:2017dor},
but their sensitivity does not significantly improve
on the bound from LEP.

\vskip 2mm
Near future bounds will come from Belle-II where the ALP can be searched in the invisible and 3 $\gamma$ decay modes.
A phenomenological study has been performed in Ref.~\cite{Dolan:2017osp}
where the authors consider  ALP decays into dark matter (invisible) and  two (resolved) photons.
The bounds have been shown in Figure~\ref{fig:ALPS_gg_context}. However the $a \to$ invisible
sensitivity  heavily relies on the possibility to use the single-photon trigger with a low threshold
(1.8 GeV), that in not guaranteed at the nominal Belle-II luminosity regime.

\vskip 4mm
\noindent {\bf PBC projects on 5 and 10-15 years timescale} 
\vskip 1mm

Three PBC experiments can perform searches of ALPs with photon coupling on a 5 year timescale:
NA62$^{++}$ in dump mode and FASER, will look for visible ALP decays, $a \to \gamma \gamma$,
while NA64$^{++} (e)$ will be able to perform a search into visible and invisible decays.
The contour plots are shown in Figure~\ref{fig:ALPS_gg_pbc_1}.
On a longer timescale the big PBC projects can enter in the game  further extending the physics reach
in a still uncharted parameter space. On this respect SHiP and LDMX will be fully complementary,
the first covering larger masses and smaller coupling, the latter filling the uncovered phase space between the old
beam-dump experiments and the colliders, in the 10-300 MeV mass range.

\begin{figure}[h]
  \centerline{
  \includegraphics[width=0.8\linewidth]{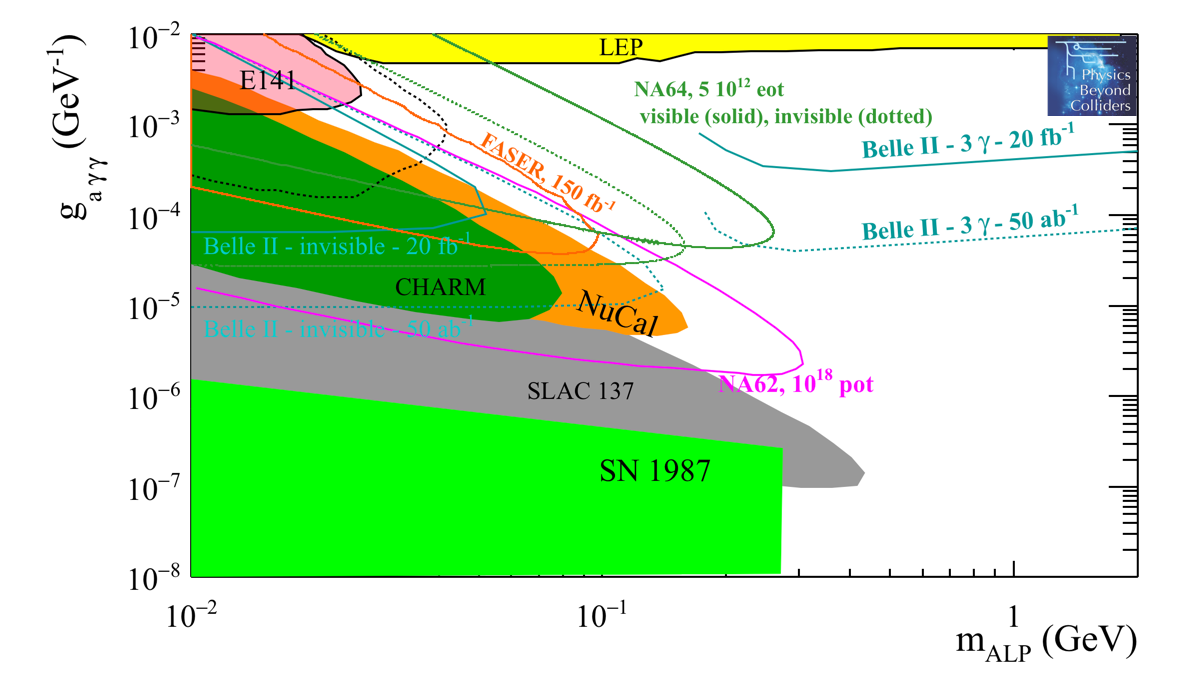} }
  \caption{BC9: ALPs with photon coupling. Current bounds (filled areas) and prospects for PBC projects on 5 years timescale (solid lines)
  in the plane coupling $g_{a\gamma\gamma}$ versus mass $m_{\rm ALP}$. The results from a phenomenological study for Belle-II~\cite{Dolan:2017osp}
  is also shown.}
\label{fig:ALPS_gg_pbc_1}
\end{figure}

\begin{figure}[h]
  \centerline{
  \includegraphics[width=0.8\linewidth]{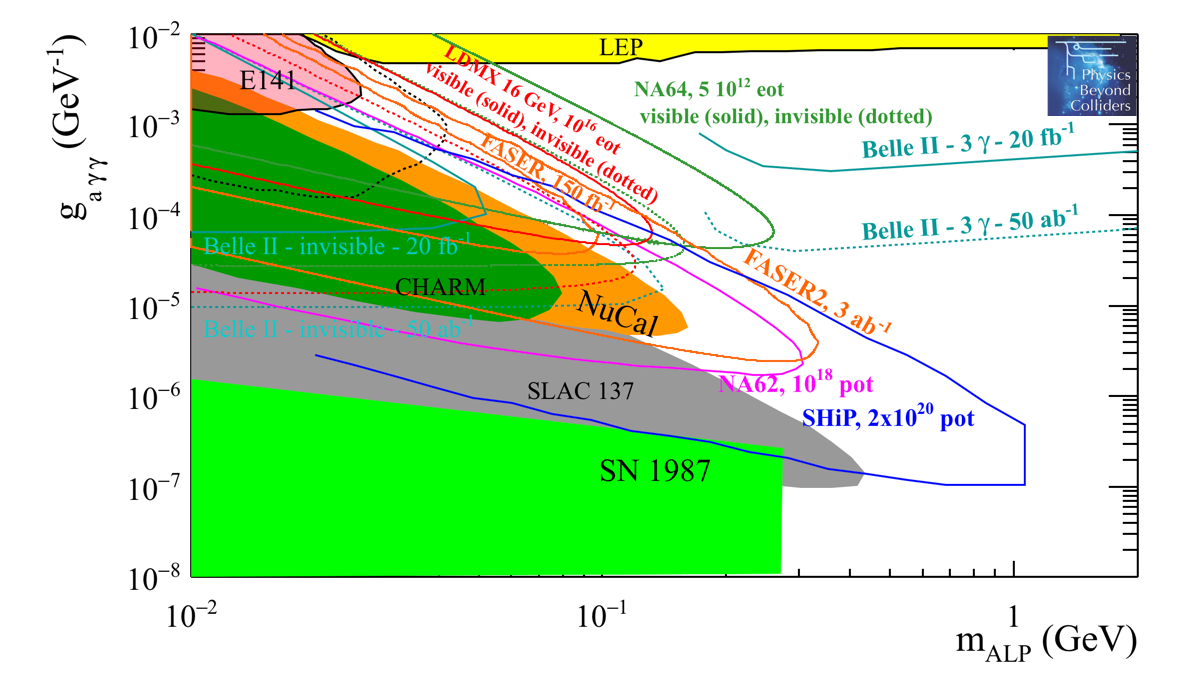} }
    \caption{BC9: ALPs with photon coupling. Current bounds (filled areas) and prospects for PBC projects on 10-15 years timescale (solid lines)
  in the plane coupling $g_{a\gamma\gamma}$ versus mass $m_{\rm ALP}$. The results from a phenomenological study for Belle-II~\cite{Dolan:2017osp}
  is also shown.}
\label{fig:ALPS_gg_pbc_2}
\end{figure}

\clearpage
\subsubsection{Axion portal with fermion-coupling (BC10)}
\label{sssec:bc10}

Assuming a single ALP state $a$, and the predominant coupling to fermions, all
phenomenology (production and decay) can be determined as functions on ($m_a; g_Y = 2 v f^{-1}_{\ell}, 2 v f^{-1}_{q}$),
with $v$ the vev of the Higgs.
Furthermore, for the sake of simplicity, we take $f_q = f_{\ell}$.
Details about approximations and assumptions
in computing sensitivities for this benchmark case are reported in Appendices A and B.

The effective Yukawa coupling ALP-SM fermions is proportional to the mass
of the SM fermions. Hence: a possible ALP with fermion coupling is mostly originated from meson decays
and only very rarely from electrons.
Heavy mesons can be produced in $e^+ e^-$ colliders, $pp$ colliders and
in the interactions of a proton beam with a target. 

\vskip 2mm
Searches for ALPs with fermion coupling are being pursued at the LHC,
namely in the analysis of rare $B$ decays
as for example $B^+ \to K^{*0} \mu^+ \mu^-$~\cite{Aaij:2015tna}.
The geometry of the LHC experiments, on the other hand, is such that
a search can be performed only if the ALP decays more or less instantaneously, hence has large couplings.
For ALPs with smaller couplings and longer lifetimes these searches are much less effective
even though ALPs may still be produced in abundance.

Beam-dump experiments in contrast are particularly sensitive to long-lived and very weakly coupled light new states,
which can travel through the hadron absorber before decaying. Several constraints already exist,
mostly coming from old beam dump experiments as CHARM~\cite{Bergsma:1985qz},
NuCal~\cite{Blumlein:1990ay}, and E613~\cite{Duffy:1988rw}.
Other constraints are derived from $K$ and $B$ mesons experiments, as explained below.

\vskip 5mm
\noindent
{\bf Current bounds and near future prospects, including PBC projects}
\vskip 1mm    

The current status of the exclusion limits for ALPS with fermion coupling in the MeV-GeV range is shown
in Figure~\ref{fig:ALPS_bc10_pbc_1}, as filled coloured areas.

Most of the current bounds arise from a re-interpretation of experimental results from CHARM~\cite{Bergsma:1985qz},
E949~\cite{Anisimovsky:2004hr, Artamonov:2009sz}, KTeV~\cite{AlaviHarati:2000hs} 
performed by theorists~\cite{Dolan:2014ska,Dobrich:2018jyi}.
As such, these bounds should be taken with many caveats.
A few searches are instead coming directly from experiments, as for example BaBar~\cite{Lees:2012iw} and
LHCb~\cite{Aaij:2015tna,Aaij:2016qsm}.

\begin{itemize}
\item[-] {\it $K^+ \to \pi^+ + X:$} the $K_{\mu}2$ experiment has studied the momentum distribution
  of charge pions produced in the decay $K^+ \to \pi^+$~\cite{Yamazaki:1984vg}.
  In presence of a light pseudoscalar, the decay channel $K^+ \to \pi^+ a$ would lead to a bump in the spectrum.
  
\item[-] {\it $K^+ \to \pi^+ +$ invisible:} reinterpretation of the E949 results~\cite{Anisimovsky:2004hr, Artamonov:2009sz}
 as an upper bound on the process $K^+ \to \pi^+ a$ performed in Ref.~\cite{Dolan:2014ska},
 and  cross-checked by the KLEVER collaboration. The curve assumes that the ALP escapes the decay volume.
 
\item[-] {\it $B^0 \to K_S$ + invisible:}
  This search is the analogous as the $K^+ \to \pi^+$ + invisible search. The strongest bounds come from Cleo~\cite{Ammar:2001gi}.
  
\item[-] {\it $K_{L} \to \pi^0  \ell^+ \ell^-$:} in the mass range $210 < m < 420$ MeV the pseudoscalar will decay predominantly
  to muon pairs. The KTeV/E749 collaboration has set an upper limit on the $K_L \to \pi^0 \ell^+ \ell^-$ decay~\cite{AlaviHarati:2000hs}
  and this result has been converted into an upper limit for the branching fraction of the decay
  $K_L \to \pi^0 a$ under the hypothesis that the ALP decays instantaneously~\cite{Dolan:2014ska}.

\item[-] {\it Radiative $\Upsilon$ decays:} pseudo-scalar particles have been sought by the BaBar collaboration
   in radiative $\Upsilon$ decays $\Upsilon \to a \gamma$, with $a \to \mu^+ \mu^-$ for $a < 2 m_{\tau}$~\cite{Lees:2012iw}
   and $a \to \tau^+ \tau^-$ for $m_a$ above threshold~\cite{Lees:2012te}.
   
 \item[-] {\it $B \to K \mu^+ \mu^-$:} the measurement of the branching fraction of the $B^+ \to K^+ \mu^+ \mu^-$ decay
   as a function of the dimuon mass performed by LHCb~\cite{Aaij:2012vr}
   has been interpreted in Ref.~\cite{Dolan:2014ska} as
   an upper bound for the process $B^+ \to K^+ a, a \to \mu^+ \mu^-$  in each $\mu^+ \mu^-$ mass bin, under the hypothesis
   that $a$ decays instantaneously.
   Dedicated searches have been instead performed by the Collaboration, allowing also for displaced dimuon vertices in the
   $B^0 \to K^{0*} \mu^+ \mu^-$ and $B^+ \to K^+ \mu^+ \mu^-$ processes, and the results reported in Refs.~\cite{Aaij:2015tna} and
   ~\cite{Aaij:2016qsm}, respectively. These data have been adapted to the model prescriptions used in this study
   by F.~Kahlhoefer.
   
\end{itemize}

\begin{figure}[h]
  \centerline{ \includegraphics[width=0.8\linewidth]{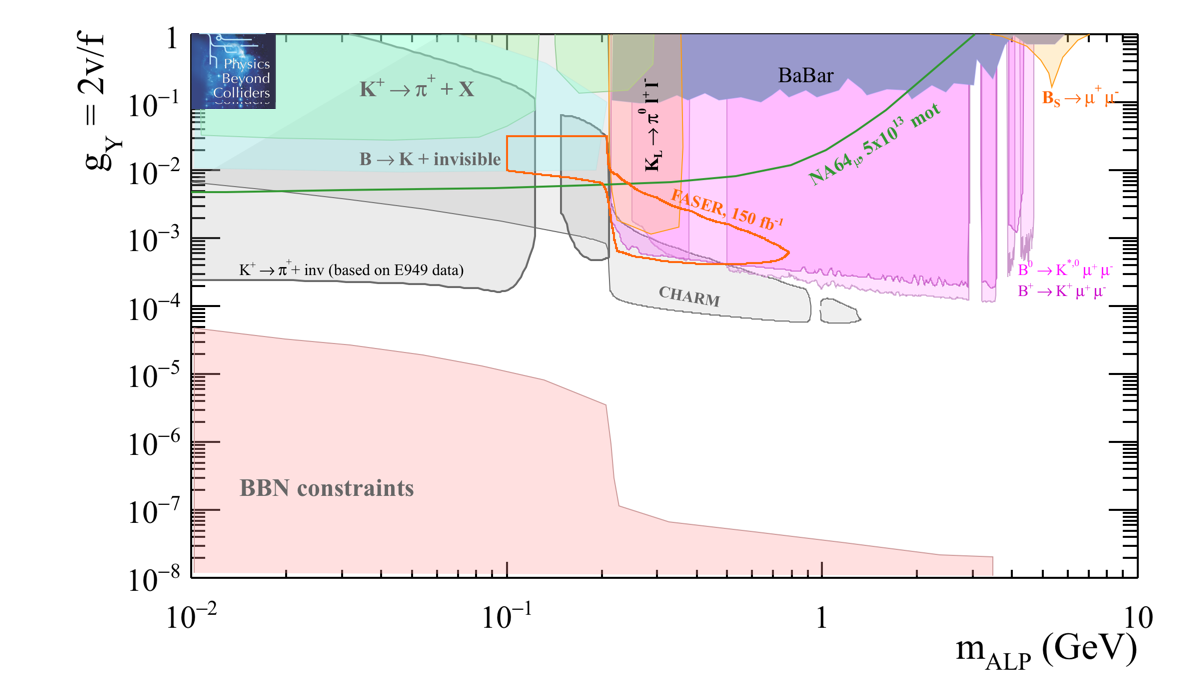} }
  \caption{BC10: ALPs with fermion coupling. Current bounds (filled areas)  and near ($\sim$ 5 years) prospects for  PBC projects (solid lines).
   CHARM and LHCb filled areas have been adapted
  to PBC prescriptions by F.~Kahlhoefer, following Ref.~\cite{Dobrich:2018jyi}. E949 area has been computed
  by the KLEVER collaboration and M.~Papucci based on E949 data. All other exclusion regions have been
  properly re-computed by M.~Papucci, following Ref.~\cite{Dolan:2014ska}.}
\label{fig:ALPS_bc10_pbc_1}
\end{figure}

\begin{figure}[h]
  \centerline{ \includegraphics[width=0.8\linewidth]{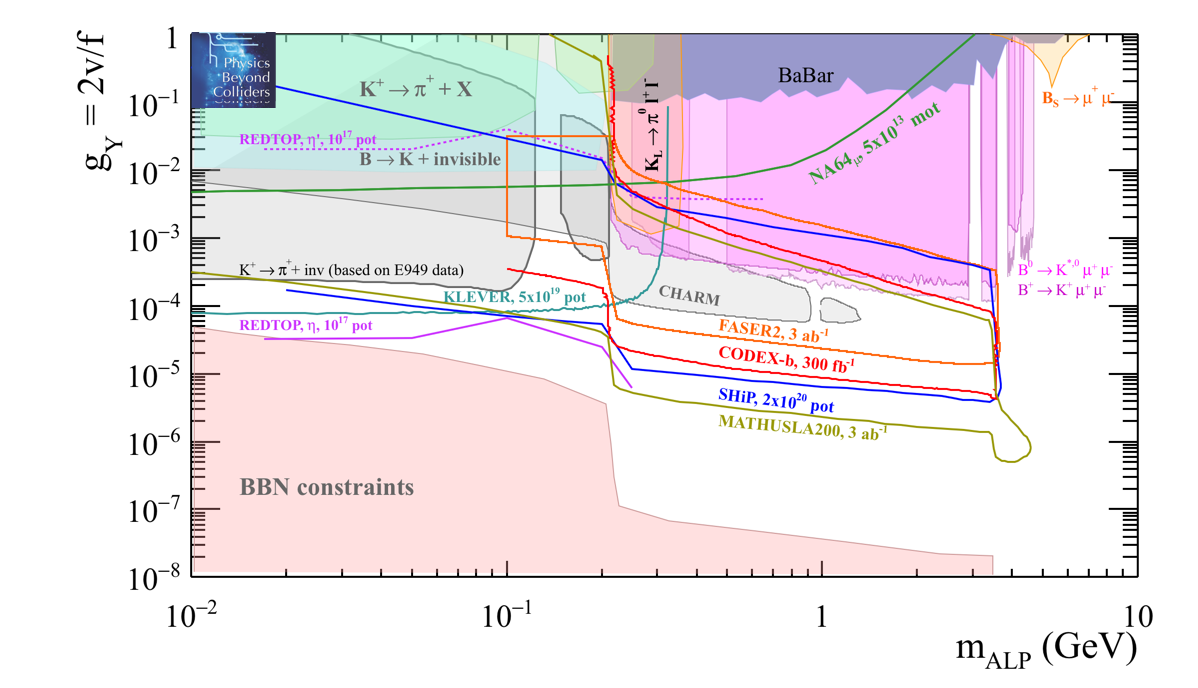} }
    \caption{BC10: ALPs with fermion coupling. Current bounds (filled areas) and medium-far  ($\sim$ 10-15 years) prospects for PBC projects
    (solid lines) for ALPs with fermion coupling. 
    CHARM and LHCb filled areas have been adapted to PBC prescriptions by F.~Kahlhoefer,
    following Ref.~\cite{Dobrich:2018jyi}. E949 area has been computed
  by the KLEVER collaboration and M.~Papucci based on E949 data. All other exclusion regions have been
  properly re-computed by M.~Papucci, following Ref.~\cite{Dolan:2014ska}.}

\label{fig:ALPS_bc10_pbc_2}
\end{figure}

\clearpage
\subsubsection{Axion portal with gluon-coupling (BC11)}
\label{sssec:bc11}

This benchmark case considers a scenario in which the ALP $a$ only couples to the gluon field
at a scale $\Lambda$ = 1 TeV. One can write down the corresponding low-energy Lagrangian at the tree level as
\begin{equation}
{\cal L} = {\cal L_{SM}} + {\cal L_{DS}} + a {g^2_s \over 8 f_G} G^b _{\mu \nu} \tilde{G^{b \; \mu \nu}} .
\end{equation}

Because the ALP mixes with the neutral pseudoscalar mesons, it is produced
in any process that produces such mesons. Moreover it can be produced also in $B$ mesons decays, as explained in
Section~\ref{ssec:portals_axion}. Details about approximations and assumptions
assumed in computing sensitivities for this benchmark case are reported in Appendices A and B.

\vskip 2mm
Figure~\ref{fig:ALPS_bc11_pbc_2} shows the current bounds (as coloured filled areas) and the prospects for
PBC projects (solid lines) both on 5- (FASER) and 10-15 years (CODEX-b, MATHUSLA200, FASER2) timescale.
Below the three pion threshold, the CODEX-b and MATHUSLA200 reach for this benchmark is conditional upon
the eventual detectors being sensitive to the di-photon final state.
Production from K and B decays depend on UV completion and the results shown
assume $\approx [{\rm log \Lambda^2_{UV}}/m^2_t \pm {\mathcal{O}}(1)] \Rightarrow 1$.
The CODEX-b curve has been obtained considering B-decays only, hence it is conservative.
NA62$^{++}$ and SHiP are also expected to be sensitive to this benchmark case but they did not provide
the sensitivity curves on the timescale of this paper.

\vskip 2mm
Current bounds arise from flavor physics, old beam-dump experiments and LEP data.
A comprehensive reinterpretation of these data has been performed in Ref.~\cite{Aloni:2018vki}
in the $m_{\pi} < m_a < 3$~GeV mass region, namely:
\begin{enumerate}
\item data from LEP~\cite{Knapen:2016moh,Abbiendi:2002je} and old beam dump experiments, E137\cite{Bjorken:1988as}
and NuCal~\cite{Blumlein:1990ay}, have been used to recast limit on the $a \gamma \gamma$ vertex and translated
into a limit in the $BR(a \to \gamma \gamma)$;

\item the limits on the branching fractions of the decays
$\phi \to \pi \pi \gamma \gamma$ and $\eta' \to \pi^+ \pi^- \pi^+ \pi^- \pi^0$~\cite{Tanabashi:2018oca} are
used to set a limit on the rate of the processes $\phi \to \gamma a(\pi \pi \gamma)$ and
$\eta' \to \pi^+ \pi^- a (\pi^+ \pi^- \pi^0)$, assuming that all the rate is due to ALPs;

\item  decays driven by the $b \to s a$ penguin diagram are considered and a recast of results
is performed while analyzing:
\begin{itemize}
\item[-] the $m_{\eta \pi \pi}$ spectrum of the decay $B^{\pm} \to K^{\pm} \eta \pi^+ \pi^-$, interpreted
as $B^{\pm} \to K^{\pm} a (\eta \pi^+ \pi^-)$, from Ref.~\cite{Aubert:2008bk};
\item[-] the $m_{K^* K}$ spectrum of the decay $B^{\pm} \to K^{\pm} K^{\pm} K_{\rm S} \pi^{\mp}$, interpreted as
$B^{\pm} = K^{\pm} a (K^{\pm} K_{\rm S} \pi^{\mp})$, from Ref.~\cite{Aubert:2008bk};
\item[-] the measurement of the two decay rates, $BR(B^0 \to K^0  \phi \phi)$~\cite{Lees:2011zh}
and $BR(B^{\pm} \to K^{\pm} \omega (3\pi))$~\cite{Chobanova:2013ddr},
to put a constraints on the processes $B^0 \to K^0 a (\phi \phi)$ and
$B^{\pm} \to K^{\pm} a (\pi^+ \pi^- \pi^0)$, respectively.
\end{itemize}

\item measurements on processes driven by the $s \to d$ penguin diagram,
as $K^{\pm} \to \pi^{\pm} \gamma \gamma$~\cite{Ceccucci:2014oza} and $K_{\rm L} \to \pi^0 \gamma \gamma$~\cite{Abouzaid:2008xm},
are used to recast limits on ALPs. 
\end{enumerate}

\vskip 2mm
For cases 2) and 3) listed above,
at one loop, the $a g g$ vertex generates an axial-vector $a tt$ coupling~\cite{Bauer:2017ris} which enhances
the rate for $B \to K^{(*)} a$ decays~\cite{Batell:2009jf,Hiller:2004ii,Bobeth:2001sq,Choi:2017gpf}.
Following Ref.~\cite{Aloni:2018vki} the UV-dependent factor contained in the loop,
$\approx [{\rm log \Lambda^2_{UV}}/m^2_t \pm {\mathcal{O}}(1)]$, is approximated to unity (which corresponds to a UV scale $\sim $ TeV).

\begin{figure}[h]
  \centerline{ \includegraphics[width=0.8\linewidth]{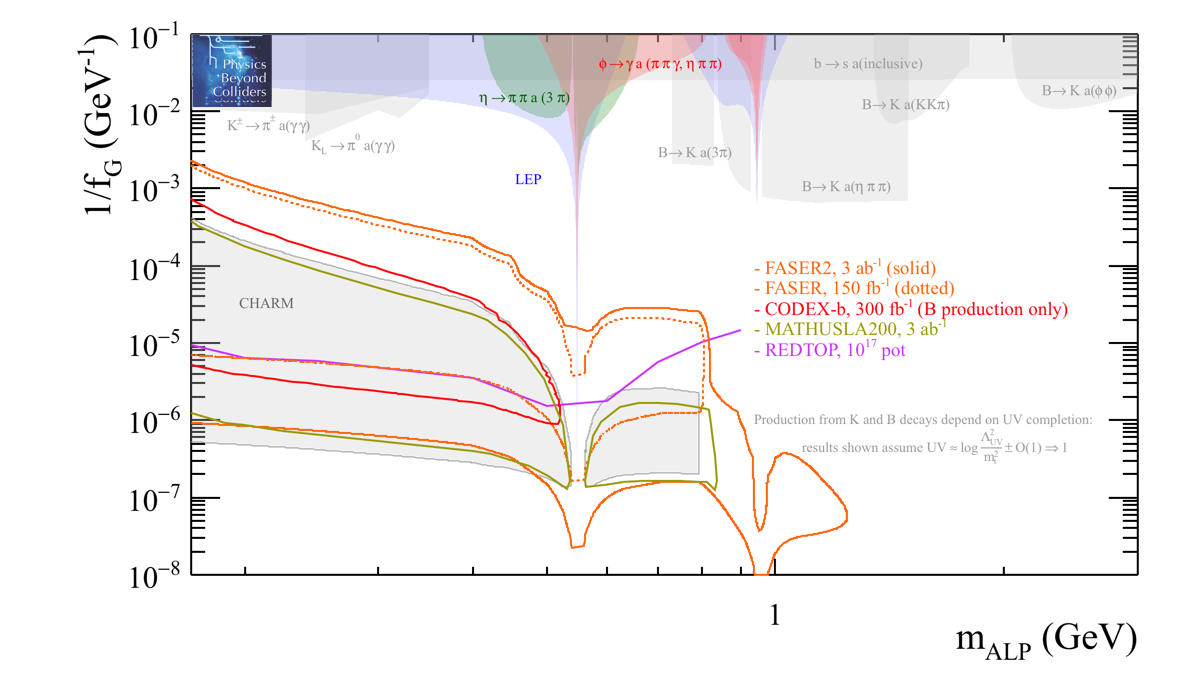} }
  \caption{Current bounds (as coloured filled areas) and the prospects for
PBC projects (solid lines) both on 5- (FASER) and 10-15 year (CODEX-b, MATHUSLA200, FASER2) timescale.
The CHARM gray filled area has been computed by F.~Kling, recasting the search for long-lived particles
decaying to two photons performed at CHARM~\cite{Bergsma:1985qz}.
Other coloured filled areas are kindly provided by Mike Williams and revisited from Ref.~\cite{Aloni:2018vki}.
The gray areas depend on UV completion and the results shown
assume $\approx [{\rm log \Lambda^2_{UV}}/m^2_t \pm {\mathcal{O}}(1)] \Rightarrow 1$.
The CODEX-b curve has been obtained considering B-decays only, hence it is conservative.
Both NA62$^{++}$ and SHiP are sensitive to this benchmark case too, the curves are currently being compiled.}
\label{fig:ALPS_bc11_pbc_2}
\end{figure}

\clearpage
\section{Physics reach of PBC projects in the multi-TeV mass range}
\label{sec:phys-reach-multi-TeV}

The PBC projects have sensitivity to physics beyond the Standard Model at and above  the TeV mass scale. 
Since the center of mass energy in the collisions for the PBC experiments is small 
compared to the LHC experiments, this sensitivity comes through modifications of 
known particle properties through virtual exchanges of New Physics particles. 
In some cases, when new physics violates exact or approximate symmetries of the 
SM (such as $CP$ symmetry, and/or lepton flavour), the standard model backgrounds are very low.
As a result precision measurements can be sensitive to NP in the multi-TeV range. 

\subsection{Measurement of EDMs as probe of NP in the multi TeV scale}
Measurements of, and constraints on, Electric Dipole Moments (EDMs) of elementary particles and atoms are a very powerful way
of probing theories of New Physics. One of the key puzzles of the Standard Model,
the smallness of $CP$-violation in the QCD sector, originates from the tight bounds on neutron 
and atomic EDMs. The axion solution to this strong $CP$ problem implies the existence of heavy Peccei-Quinn sectors at high-energy scales 
that cannot be directly accessed in high-energy experiments, and so an alternative approach is needed to understand the solution to this problem.
New physics at the weak scale (or more generically, TeV scale and beyond) physics can also induce EDMs. Famously, the Kobayashi-Maskawa  
$CP$ violation mechanism does not induce neutron or proton EDMs above $10^{-32} e{\rm cm}$, which is firmly
outside the reach of current and next generation EDM experiments. This opens up the  possibility of exploring the TeV frontier with EDMs 
by increasing the experimental sensitivity. One of the long-term proposals to measure EDMs is the proton (and other charged nuclei)
storage ring where EDMs can be probed to unprecedented precision. 

\vskip 2mm
If new $CP$-violating physics is heavy, for the purpose of the EDM description, one can encapsulate its effect
in form of the SM effective operators. For example, the following operators can be interpreted as up- and down- quark electric dipole 
moments $d_{u(d)}$:
\begin{equation}
\label{CP}
{\cal L} = \frac{ v \times \sin(\phi^{(u)})}{\Lambda_u^2} \times \frac{ie}{2}\bar u F_{\mu\nu} \sigma_{\mu\nu} \gamma_5 u + 
\frac{ v \times \sin(\phi^{(d)})}{\Lambda_d^2} \times \frac{ie}{2} \bar d F_{\mu\nu} \sigma_{\mu\nu} \gamma_5 d +...
\end{equation}
The insertion of the SM vacuum expectation $v=246$\,GeV is necessitated by the $SU(2)\times U(1)$ gauge invariance. 
Possible small Yukawa couplings, loop factors etc have been subsumed into energy scale coefficients $\Lambda_{u}$ 
and $\Lambda_{d}$. $\phi^{(d)}$ and $\phi^{(u)}$ indicate CP-violating phases. 

\vskip 2mm
There is, of course, a wide variety of possible dimension six operators, and more independent EDM measurements and constraints 
are required to limit them all. Non-perturbative methods can be used to relate proton/neutron EDMs to the quark EDM 
and other $CP$-odd operator coefficients. Suppressing $u,d$ flavour dependence, and taking for simplicity $d_p\sim {\mathcal{O}}(d_q)$, 
one arrives at the maximum expected sensitivity of protons EDM being reinterpreted as the sensitivity to $\Lambda$, 
\begin{equation}
\frac{ |\sin(\phi^{(q)}) | }{\Lambda_q^2} \sim \frac{1}{(7\times 10^5 \, {\rm TeV})^2} \times \left(    
\frac{d_p  }{  10^{-29} e {\rm cm}}    \right)^{1/2}\;.
\end{equation}
It is likely that operators in (\ref{CP}) are proportional to small Yukawa couplings and a loop factor, so that the sensitivity to the 
actual energy scales of new physics are several orders of magnitude  lower than this estimate indicates. Even then, the suggested target of 
$10^{-29} e {\rm cm}$ for $d_p$ will cover models with CP-violation in the multi-100~TeV range, thereby exploring, for example,  
the scalar quark mass range which would be expected if the measured value of the Higgs mass, $m_h \simeq 125$\,GeV, is interpreted within SUSY models.

\subsection {Experiments sensitive to Flavour Violation}
Of particular interest for the PBC program is the search for  flavour violation which is almost entirely  absent in the SM,
but introduced in many beyond the SM scenarios, including theories with supersymmetry. 
Two PBC experiments aim to explore the sensitivity of flavour-violating processes to TeV scale physics, TauFV and KLEVER.
 
\vskip 2mm
\noindent
{\bf \large TauFV}\\
In case of the TauFV experiment, the physics goal is to observe and measure, or alternatively set the upper bound on, 
the several lepton-flavour-violating (LFV) $\tau$ or $D$-meson decays.
To estimate the sensitivity to a multi-TeV New Physics scale, the LFV process $\tau^\pm \to \mu^+\mu^\pm\mu^-$ is considered.
Such process is almost entirely forbidden
in the Standard Model, and any attempt to measure this  small branching will automatically probe the New Physics that violates 
approximate $\tau$ and $\mu$ flavour conservation. 

To quantify the New Physics reach, one can introduce a series of effective operators that mediate such transitions.
For this particular decay process, at lowest order,  one can have
\begin{equation}
{\cal L} = \frac{e^{i\phi}}{\Lambda_{\mu \tau}^2} \times (\bar \mu \gamma_\alpha \mu) (\bar \mu \gamma_\alpha \tau)  ~+ (h.c.)
+ ~{\rm other~ Lorentz~ structures}.
\end{equation}
In this expression, all coupling constants have been subsumed into the definition of the energy scale $\Lambda_{\mu \tau}$, 
apart from a possible phase $\phi$. 

The resulting branching ratio for the $\tau \to 3 \mu $ decay in the $m_\mu \ll m_\tau$ limit is given by
\begin{equation}
\Gamma_{\tau\to 3 \mu} = \frac{m_\tau^5}{256\pi^3\Lambda_{\mu \tau}^4}.
\end{equation}
Given the stated goal of the TauFV experiment (in the absence of positive signal) is to reach 
the exclusion level of ${\rm BR}_{\tau\to 3 \mu}\sim 10^{-10}$, one can translate the above 
formula to $\Lambda_{\mu\tau}$ sensitivity,
\begin{equation}
\Lambda_{\mu\tau} > 55\,{\rm TeV} \times \left( \frac{10^{-10}}{{\rm Br}_{\tau\to 3 \mu}} \right)^{1/4}.
\end{equation}

\clearpage
\noindent
{\bf \large KLEVER}\\
 The KLEVER proposal seeks to complement the NA62 experiment by measuring $K_L \to \pi^0 +\,\mbox{missing\,energy}$. In the Standard Model, 
 the missing energy is carried by neutrinos, $K_L \to \pi^0\nu \bar\nu$, and the corresponding branching 
 ratio is predicted to be ${\rm BR_{SM}} = (3.4 \pm 0.6) \times 10^{-11}$ ~\cite{Buras:2015qea}.
 
\vskip 2mm
The branching ratios for the decays $K\to\pi\nu\bar{\nu}$
are among the observables in the quark-flavor sector most sensitive to NP.
Because the SM decay amplitudes are strongly suppressed by the GIM
mechanism and the CKM hierarchy and dominated by short-distance physics,
the SM rates are small and predicted very precisely, making the
$K\to\pi\nu\bar{\nu}$ BRs potentially sensitive to NP at
mass scales of hundreds of TeV, in general surpassing the sensitivity
of $B$ decays in SM extensions~\cite{Buras:2014zga}.
Observations of lepton-flavor-universality-violating phenomena are
mounting in the $B$ sector. Most explanations for such phenomena
predict strong third-generation couplings and thus significant changes
to the $K\to\pi\nu\bar{\nu}$ BRs through couplings to final states with
tau neutrinos~\cite{Bordone:2017lsy}.

\vskip 2mm
The BR for the decay $K_L\to\pi^0\nu\bar{\nu}$ has never been measured.
The current experimental result,
 ${\rm BR}(K^+\to\pi^+\nu\bar{\nu}) = 1.73^{+1.15}_{-1.05}\times10^{-10}$,
obtained at Brookhaven from $K^+$ decays at rest with seven candidate events,
\cite{Artamonov:2009sz},
together with considerations of isospin symmetry
\cite{Grossman:1997sk},
leads to the model-independent bound
${\rm BR}(K_L\to\pi^0\nu\bar{\nu}) < 1.4\times10^{-9}$.
This limit has to be compared to the direct limit set by the KOTO experiment,
${\rm BR}(K_L\to\pi^0\nu\bar{\nu}) < 1.4\times10^{-9}$ at 90 \% CL~\cite{Ahn:2018mvc}.
Because the amplitude for $K^+\to\pi^+\nu\bar{\nu}$
has both real and imaginary parts while the amplitude for
$K_L\to\pi^0\nu\bar{\nu}$ is purely imaginary, the decays have different sensitivity to new
sources of $CP$ violation.

\vskip 2mm 
In general, the measurement of the $BR(K_L \to \pi^0 \nu \overline{\nu}$) is
sensitive to additional sources of flavour violation coming from NP at, or above, the TeV scale.
Parametrizing the effective Lagrangian for new physics in terms of effective operators as before,
and taking one flavour of neutrinos for simplicity, 
 \begin{equation}
{\cal L} = \frac{e^{i\phi}}{\Lambda_{ds}^2} \times (\bar \nu \gamma_\alpha (1-\gamma_5)\nu) (\bar d \gamma_\alpha (1-\gamma_5) s)  ~+ (h.c.)
+ ~{\rm other~ Lorentz~ structures},
\end{equation}
one can quantify the sensitivty of KLEVER to NP. 
The decay $K_L \to \pi^0\nu \bar\nu$ is $CP-$violating, and therefore the amplitude is proportional
to $\sin(\phi)$, being $\phi$ the phase of NP contributions.
In contrast, the $K^+\to\pi^+\nu\bar{\nu}$ branching fraction is phase-independent,
so it can also be used as a probe of TeV physics independently from $K_L \to \pi^0\nu\bar{\nu}$.
Should NA62 discover deviations from the Standard Model, a KLEVER-type of measurement would be required to
further investigate their origins.

If the  sensitivity of KLEVER can reach the SM branching ratio level, it would also entail sensitivity to New Physics
at approximately 
\begin{equation}
\frac{|\sin\phi|}{\Lambda_{ds}^2} \sim \frac{1}{(500\,{\rm TeV})^2}.
\end{equation}

\vskip 2mm
Figure~\ref{fig:KpnnBSM}, reproduced from \cite{Buras:2015yca},
illustrates a general scheme for the expected
correlation between the charged and neutral decays under various scenarios.

\begin{figure}[ht]
\centering
\includegraphics[width=0.5\textwidth]{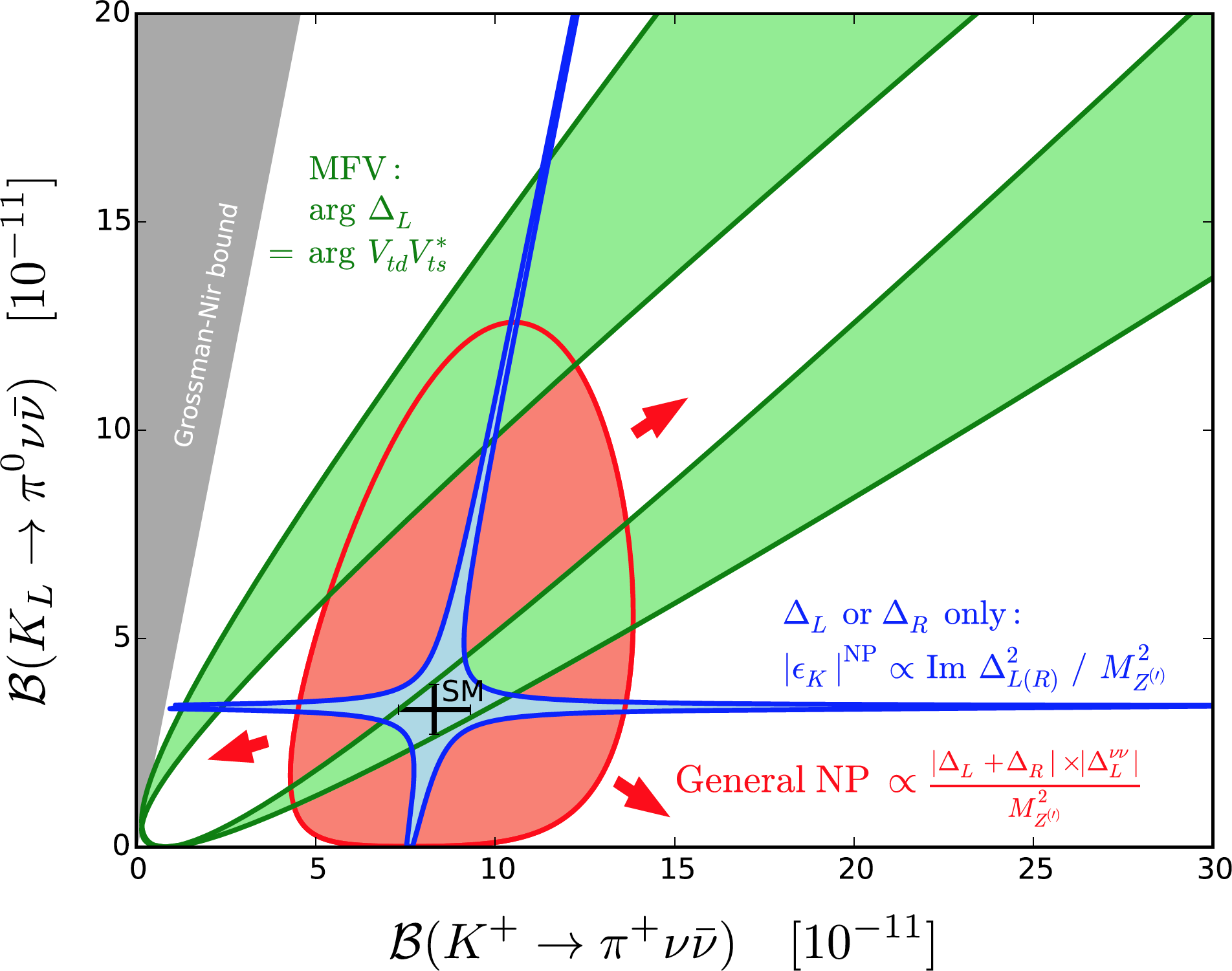}
\caption{Scheme for BSM modifications of $K\to\pi\nu\bar{\nu}$ BRs.
  Reproduced from Ref.~\cite{Buras:2015yca}.} 
\label{fig:KpnnBSM}
\end{figure}

If the NP has a CKM-like structure of flavor interactions, the $K_L$
and $K^+$ BRs will lie along the band of correlation shown in green.
In models with only left-handed or only right-handed couplings to the
quark currents (e.g., models with modified $Z$ couplings or littlest-Higgs models with
$T$ parity), because of constraints from $\epsilon'_K$, the BRs must lie along one
of the branches shown in blue. If the NP has an arbitrary flavor
structure and both left-handed and right-handed couplings (e.g., in Randall-Sundrum models), there is little
correlation, as illustrated in red.

\vskip 2mm
In a recent breakthrough, the RBC-UKQCD Collaboration obtained the
first result for ${\rm Re}\,\epsilon'_K/\epsilon_K$  from a lattice calculation thought to have
reliable systematics: ${\rm Re}\,\epsilon'_K/\epsilon_K = (1.38\pm5.15\pm4.59)\times10^{-4}$,
2.1~$\sigma$ less than the experimental value \cite{Bai:2015nea}.
Estimates from large-Nc dual QCD support the lattice result~\cite{Buras:2015xba}.
With this result for ${\rm Re}\,\epsilon'_K/\epsilon_K$, the correlation between
$\epsilon_K$ and ${\rm BR}(K_L\to\pi^0\nu\bar{\nu})$ has been examined in
various SM extensions at energy scales $\Lambda$ in the neighborhood of
1--10~TeV by several authors, in many cases, with
constraints from $\epsilon'_K$, $\Delta m_K$, and ${\rm BR}(K_L\to\mu\mu)$
considered as well.
The results of these studies are summarized in Table~\ref{tab:bsm}.
In general, an observed value of $\epsilon_K$ that is larger than
expected in the SM implies a suppression of
${\rm BR}(K_L\to\pi^0\nu\bar{\nu})$ to below the SM value.
However, it is possible to construct models in which $\epsilon_K$ and
${\rm BR}(K_L\to\pi^0\nu\bar{\nu})$ are simultaneously enhanced.
With moderate parameter tuning (e.g., cancellation among
SM and NP interference terms to the 10--20\% level),
${\rm BR}(K_L\to\pi^0\nu\bar{\nu})$ may be enhanced by up to an order of
magnitude.

\begin{sidewaystable}
  \centering
  {\small
  \begin{tabular}{lcccc}\hline\hline
    Model & $\Lambda$ [TeV] & Effect on ${\rm BR}(K^+\to\pi^+\nu\bar{\nu})$ & Effect on ${\rm BR}(K_L\to\pi^0\nu\bar{\nu})$ & Refs. \\ \hline
    Leptoquarks, most models & 1--20 & \multicolumn{2}{c}{Very large enhancements; mainly ruled out} & \cite{Bobeth:2017ecx} \\
    Leptoquarks, $U_1$ & 1--20 & +10\% to +60\% & +100\% to +800\% & \cite{Bobeth:2017ecx} \\
    Vector-like quarks & 1--10 & $-90$\% to +60\% & $-100$\% to +30\% & \cite{Bobeth:2016llm} \\
    Vector-like quarks + $Z'$ & 10 & $-80$\% to +400\% & $-100$\% to 0\% & \cite{Bobeth:2016llm} \\
    Simplified modified $Z$, no tuning & 1 & $-100\%$ to +80\% & $-100$\% to $-50$\% & \cite{Endo:2016tnu} \\
    General modified $Z$, cancellation to 20\% & 1 & $-100$\% to +400\% & $-100$\% to +500\% & \cite{Endo:2016tnu} \\
    SUSY, chargino $Z$ penguin & 4--6 TeV & & $-100$\% to $-40$\% & \cite{Endo:2016aws} \\
    SUSY, gluino $Z$ penguin & 3--5.5 TeV & 0\% to +60\% & $-20$\% to +60\% & \cite{Endo:2017ums} \\
    SUSY, gluino $Z$ penguin & 10 & Small effect & 0\% to +300\% & \cite{Tanimoto:2016yfy} \\
    SUSY, gluino box, tuning to 10\% & 1.5--3 & $\pm10$\% & $\pm20$\% & \cite{Crivellin:2017gks} \\
    LHT & 1 & $\pm20$\% & $-10$\% to $-100$\% &\cite{Blanke:2015wba} \\
    \hline\hline
  \end{tabular}
  }
  \caption{Effects on BRs for $K\to\pi\nu\bar{\nu}$ decays in various SM
    extensions, with constraints from other kaon observables, including
    in particular ${\rm Re}\,\epsilon'_K/\epsilon_K$.}
  \label{tab:bsm}
\end{sidewaystable}

\vskip 2mm
The KLEVER experiment aims to use a high-energy neutral beam
at the CERN SPS to achieve 60-event sensitivity for the decay
$K_L\to\pi^0\nu\bar{\nu}$ at the SM BR with an $S/B$ ratio of 1.
At the SM BR, this would correspond to a relative uncertainty of about 20\%,
demonstrating a discrepancy with $5\sigma$ significance if the observed
rate is a bit more than twice or less than one-quarter of the SM rate,
or with $3\sigma$ significance if the observed rate is less than half of the
SM rate. These scenarios are consistent with the rates predicted for
many different SM extensions, as seen in Table~\ref{tab:bsm}.

\clearpage
\subsection{$B$ physics anomalies and $BR(K \to \pi \nu \bar \nu)$}
A number of anomalies have been observed, some of which are $3\sigma$ deviations from Standard Model predictions, in semi-leptonic
$B$ decays~\cite{Aaij:2014ora, Aaij:2017vbb}.
The upcoming analysis of the full LHC Run II data set (as well as future Belle II experiments) will go a long way towards clarifying
the  status of these anomalies: Are they evidence for new physics, or just statistic fluctuations?
Taken together the anomalies hint at a violation of Lepton Flavour Universality.
PBC experiments such as NA62 and KLEVER can therefore shed complementary light on explanations for these anomalies. 
Explanations for the B anomalies include  models with flavour violation only in the third generation~\cite{Bordone:2017lsy}, 
theories with an additional $Z^{\prime}$~\cite{Kamenik:2017tnu}, 
and theories with leptoquarks~\cite{Barbieri:2015yvd}.

In most such models the decay $K_L \rightarrow \pi^0 \nu \bar{\nu}$, as probed by KLEVER, could be as sensitive to the physics
responsible to the anomalies as $K^+ \rightarrow \pi^+ \nu \bar{\nu}$. The key question then  is to which level of
precision can one measure these branching ratios  relative to the SM expectation and the timescale within which such
sensitivity can be reached.
With the sensitivities discussed for the PBC program, both NA62 and KLEVER can shed light on many of the possible explanations
for the anomalies.

\clearpage
\section{Conclusions and Outlook}
\label{sec:conclusion}

In the past decade, one of the major accomplishments of particle physics has been the discovery of the Higgs boson
that has succesfully completed the experimental validation of the SM.
Beyond this outstanding achievement, a wealth of experimental results have been produced by the ATLAS, CMS and LHCb collaborations
during Run 1 and Run 2 of the LHC: these collaborations  have explored in depth the paradigm of NP at the TeV scale, required to
solve the hierarchy problem in case of the presence of an intermediate scale between the EW and the Planck scales.
The search for NP has been performed so far both via direct searches and through precision measurements in flavor: tremendous
progress has been achieved in understanding the SM structure in the last decades.

\vskip 2mm
This progress is expected to continue for the next two decades:
the upgrade of the LHCb experiment in 2019-2020 will allow a dataset of
50 fb$^{-1}$ to be collected in about five years of operation.
Major upgrades of the ATLAS and CMS detectors are also scheduled in 2023-2026 with the ultimate aim to reach an integrated
luminosity of about 3000 fb$^{-1}$ by around 2035.

\vskip 2mm
Away from the LHC, Belle II is expected  to collect an integrated luminosity of 50 ab$^{-1}$ by 2024.
This will provide a dataset that is about a factor of 50 times larger than that collected by BaBar and Belle in the recent past.
The Mu2E experiment at FNAL, and the Mu3e and the upgrade of MEG experiments at the PSI in the next decade
will advance tremendously in the investigation of NP in charged lepton-flavor-violating processes, nicely
completing and complementing the quest of NP perfomed at the LHC experiments and at the B-factory.

\vskip 2mm
However, the absence, so far, of unambiguous signal of NP from direct searches at the LHC,
indirect searches in flavour physics and direct detection Dark Matter experiments, along with
the absence of a clear guidance from the theory about the NP scale, imposes today, more than
ever, to broadening the experimental effort in the quest for NP and exploring different ranges
of interaction strengths and masses with respect to what is already covered by existing or
planned in itiatives.

\vskip 2mm
The CERN laboratory could offer an unprecendented variety of high-intensity, high-energy
beams and scientific infrastructures that could be exploited to this endevour. This effort would
nicely complement and further broaden the already rich physics programme ongoing at the
LHC and HL-LHC.

\vskip 2mm
The proposals presented in the PBC-BSM context 
can search for NP in a fully complementary range of masses and couplings with respect to those investigated at the LHC:
new particles with masses in the sub-eV range and very weakly coupled to the SM particles, can be explored by the IAXO and JURA proposals
or through the investigation of oscillating EDMs in protons or deuterons in a electrostatic ring (CP-EDM);
MeV-GeV hidden-sector physics can be explored by a multitude of experiments at the PS beam lines (REDTOP proposal), SPS beam lines
(NA62$^{++}$, NA64$^{++}$, SHiP, KLEVER, LDMX, and NA64/AWAKE proposals) and at the LHC interaction points
(FASER, CODEX-b, MATHUSLA200, and milliQan proposals).
The multi-TeV mass range ($\sim 100$ TeV) can be indirectly explored  both via ultra-rare or forbidden decays (KLEVER and TauFV) and
through the search for a permanent EDM in protons/deuterons (CP-EDM) or in strange/charmed baryons (LHC-FT).

\vskip 2mm
The Collaborations behind these proposals are backed up by a lively phenomenological and theoretical community,
and represent a fertile ground where New Physics models can be developped, discussed, and further improved.
These proposals will possibly compete with similar proposals 
planned in the world  (as, for example, at Jefferson Lab, FNAL, JPARC, KEK, Mainz, PSI, etc)
and complement the current effort in the search for NP in other domains
(as, for example, DM direct searches at Gran Sasso Laboratory, SNOLAB or elsewehere).
They will further enrich the ongoing effort at the LHC to discover NP at the TeV scale, increasing
the impact that CERN could have in the next 10-20 years on the international landscape.

\section*{Acknowledgements}
We are grateful to the colleagues of the Beam Dump Facility, Conventional Beams, AWAKE and eSPS Physics Beyond Colliders
working groups for useful discussions and synergies. We warmly thank F.~Kahlhoefer and M.~Williams for help with recasting some
old experimental results related to ALPs with fermion and gluon couplings.
We are grateful to Dean Robinson and Harikrishnan Ramani for help with computing the CODEX-b reach,
and to Marat Freytsis for helpful discussions related to the penguin calculations for the benchmark cases BC10 and BC11.
Individual authors of this report have received financial support from ERC Ideas Consolidator
Grant No. 771642 SELDOM (European Union); MINECO and GVA (Spain).

\clearpage
\appendix
\section{ALPS: prescription for treating the FCNC processes}
\label{sec:A}

The prescription for treating the FCNC processes for ALPS production and consequent decay
described below assume a certain number of approximations.
For an in-depth study of ALP production and decay,
see the recent work \cite{Aloni:2018vki} where the non-perturbative aspects of the problem
are treated using the data-driven approach, derived from meson production, interaction and decay.

\vskip 2mm
\noindent
{\bf \large ALPs with fermion coupling (BC10)}\\
There is a certain degree of UV dependence associated with the production through $B$-meson decays, and the PBC recommends
following the prescription in \cite{Batell:2009jf}. Concretely, the effective $b-s-a$ vertex upon integrating out the $W$ and $t$ is taken to be
\begin{equation}
\mathcal{L}\supset \frac{a}{f_q} \, \bar s_L b_R \times \frac{\sqrt{2} G_F m_t^2m_b V^\ast_{ts} V_{tb}}{8\pi^2}\times c^{(BC10)}_{fcnc} +(h.c.).
\end{equation}
and coefficient $c^{(BC10)}_{fcnc} $ is chosen to be
\begin{equation}
c^{(BC10)}_{fcnc} = \log\left(\frac{\Lambda_{UV}^2}{m_t^2}\right).
\end{equation}
 where the threshold (model dependent finite pieces) cannot be determined without UV completion and are dropped. 
 The generalization to $\bar d_Ls_Ra $ interactions is done by taking $V^\ast_{ts} V_{tb} \to V^\ast_{td} V_{ts}$.

 \vskip 2mm
 Taking $\Lambda_{UV}=1$ TeV and again following  \cite{Batell:2009jf}, the branching ratios are
 \begin{align}
 \mathrm{Br}(B\to K a)& \approx 1.5\times 10^{-5} \times \left(\frac{100\, \mathrm{TeV}}{f_q}\right)^2\times (\mathcal{F}_{K}(m_a))^2\,\lambda_{Ka}^{1/2}\\
  \mathrm{Br}(B\to K^\ast a) &\approx 1.8\times 10^{-5} \times \left(\frac{100\, \mathrm{TeV}}{f_q}\right)^2\times (\mathcal{F}_{K^\ast}(m_a))^2\,\lambda_{K^\ast a}^{3/2}
 \end{align}
 with form-factors extracted from $B$-physics literature:
 \begin{align}
 \mathcal{F}_{K}(m_a)&=\frac{1}{1-m_a^2/(38\, \mathrm{GeV}^2)}\\
  \mathcal{F}_{K^\ast}(m_a)&=\frac{3.65}{1-m_a^2/(28\, \mathrm{GeV}^2)}-\frac{2.65}{1-m_a^2/ (37\, \mathrm{GeV}^2)}\\
  \lambda_{ij}&=\left(1-\frac{(m_i+m_j)^2}{m_B^2}\right)\left(1-\frac{(m_i-m_j)^2}{m_B^2}\right).
 \end{align}
 
 For the inclusive rate, we assume
 \begin{equation}
 \mathrm{Br}(B\to X_s a) \approx 5 \times\Big( \mathrm{Br}(B\to K a)+ \mathrm{Br}(B\to K^\ast a)\Big),
 \end{equation}
 following arguments presented in Ref.~\cite{Aloni:2018vki}. 

\vskip 1cm
\noindent
{\bf \large ALPs with gluon coupling (BC11)}\\ 
In this benchmark case, the ALP can be produced directly in the hadronization process or through a $B$-meson decay.
Concretely, 1-loop operator mixing generates the effective coupling of already considered in BC10:
 \begin{equation}\label{eq:BC11radiative}
 \mathcal{L}\supset \delta c_{qq} \frac{\partial_\mu a} {f_G} \sum_\beta  \bar q_\beta \gamma_\mu\gamma_5 q_\beta 
 \end{equation}
 where $\delta c_{qq}$ is generated through a gluon loop. The corresponding log-enhanced 
 coefficient can be found in Ref.~\cite{Bauer:2017ris}.

 \vskip 2mm
 A full calculation would require specifying a UV model, especially since the log-enhanced coefficient is not parametrically large.
 For concreteness we follow the choice in \cite{Aloni:2018vki} by (1) dropping all logs and (2) setting $\delta c_{qq}$ equal
 to the coefficient of the leading log of the diagram which generates \eqref{eq:BC11radiative}. In formulas this corresponds to
 \begin{equation}
\mathcal{L}\supset  \frac{a}{f_q} \, \bar s_L b_R \times \frac{\sqrt{2} G_F m_t^2m_b V^\ast_{ts} V_{tb}}{8\pi^2}\times c^{(BC11)}_{fcnc} +(h.c.). 
\end{equation}
Out choice for $c^{(BC11)} _{fcnc}$ is
\begin{equation}
c^{(BC11)} _{fcnc} \simeq \alpha_s^2(m_t). 
\end{equation}
We note here that there is a significant UV-completion dependence,
and further work would be required to properly estimate the preferred range  for $c^{(BC11)} $.

\vskip 2mm
With this convention, again following  \cite{Batell:2009jf}, the branching fractions are
 \begin{align}
 \mathrm{Br}(B\to K a) &\approx 6.6\times 10^{-10}  \times \left(\frac{100\, \mathrm{TeV}}{f_G}\right)^2\times (\mathcal{F}_{K}(m_a))^2\,\lambda_{Ka}^{1/2}\\
  \mathrm{Br}(B\to K^\ast a)& \approx 7.9\times 10^{-10} \times \left(\frac{100\, \mathrm{TeV}}{f_G}\right)^2\times (\mathcal{F}_{K^\ast}(m_a))^2\,\lambda_{K^\ast a}^{3/2}.
 \end{align}
Again, the uncertainty in the amplitude 
could result in as much as  $\mathcal{O}(10)$ changes in the width, 
with more accurate calculation of RG effects and threshold corrections in UV complete models.

\vskip 0.5cm
\noindent
    {\bf \large Approximation for ALP lifetime} \\

For computing the ALP lifetime the PBC has taken the following approximations, depending on the mass range considered.    

\begin{itemize}

\item{\em Region 1, $m_a<3m_\pi$, photon decay $a\to \gamma\gamma$.} 
In this case the decay is dominated by the two-photon contribution.
\begin{eqnarray}
\label{region1}
\Gamma_{\rm tot}(m_a< 3m_\pi) \approx \Gamma_{\gamma \gamma} = \frac{m_a^3}{f_G^2} \times 
\frac{\pi\alpha^2}{4} \times \left( \frac{4 m_d+m_u}{3(m_u+m_d)}-\frac{m_a^2}{m_a^2-m_\pi^2}\frac{m_d-m_u}{2(m_u+m_d)} \right)^2\\
=\frac{m_a^3}{f_G^2} \times 
\frac{\pi\alpha^2}{4} \times \left( 1.0-0.18\times \frac{m_a^2}{m_a^2-m_\pi^2} \right)^2.
\end{eqnarray}
Notice that the $\frac{4 m_d+m_u}{3(m_u+m_d)}$ ratio  in (\ref{region1}), to good accuracy, is 1 for the physical point
of $m_u/m_d \simeq 0.48$. The second term in the bracket comes from the mixing with $\pi^0$ meson, see Ref.~\cite{Bauer:2017ris}. 
This formula can be further improved by including contributions 
from mixing with $\eta$ (and $\eta'$). 

\item{\em Region 2, $3m_\pi <m_a <2m_\pi + m_\eta$, $a\to 3\pi $ decay.} 

In this region, two new decay modes, $\pi^+\pi^-\pi^0$ and  $3\pi^0$, open up. 
Within 2-flavour chiral perturbation theory, these decays were treated in~\cite{Bauer:2017ris}.
The results are chirally suppressed, $\Gamma_{a\to 3 \pi}  \propto m_a m_\pi^2/(F_\pi f_G)^2$. 
Using formulae from \cite{Bauer:2017ris}, we adopt it to normalization used in these notes, 
\begin{eqnarray}
\label{region2}
\Gamma_{tot} = \Gamma_{\gamma\gamma} + \Gamma_{a\to 3 \pi};~~
\Gamma_{a\to 3 \pi} = \frac{\pi}{3\times 128} \frac{m_a m_\pi^4}{ F_\pi^2 f_G^2}\left( \frac{m_a^2}{m_a^2-m_\pi^2}\frac{ m_d-m_u}{m_u+m_d}\right)^2 \times I(m_\pi^2/m_a^2),\\
\nonumber ~~I(y) = \int_{4y}^{(1-\sqrt{y})^{2}}\sqrt{1-\frac{4y}{x}}\sqrt{(1-x-y)^2-4xy}\times [12(x-y)^2+2].
\end{eqnarray}
One can check that $\Gamma_{a\to 3\pi}$ is comparable to $\Gamma_{\gamma\gamma}$.
Asymptotically~\footnote{This formula is valid in the regime of $f_G \gg F_\pi$, $\theta_\pi \ll 1$.}, at large $m_a$, $I\to 2$.
Given the experience with $\eta$ decays \cite{Gasser:1984pr}, 
the validity of the leading chiral order answer is within a factor of $\sim 3$, and can be 
improved by including next orders in the chiral expansion. 
In this mass region,  $a\to \eta^* \to 3\pi$ mediated decay is also important, especially near $m_a=m_\eta$.

\item{\em Region 3, $2m_\pi + m_\eta<m_a <1.5\,{\rm GeV}$, multiple hadronic decays.}
  Above the $2m_\pi + m_\eta\sim 830$ MeV threshold many new hadronic contributions open up, ($\eta \pi\pi$, $\rho \pi$, also $f_0\pi$ etc)
  so that the result is much larger than chiral perturbaton theory answer for $a\to 3 \pi$.
One could use hadronic  resonance models to have a phenomenological description of $a$ decays, 
but for sake of simplicity we suggest an interpolating formula for the $a$ decay. The following formula, 
\begin{equation}
\label{region3}
\Gamma_{\rm tot}(m_1< m_a < m_2) \approx A(m_a-B)^3;~A=\frac{\Gamma_2(1-r)^3}{(m_2-m_1)^3};~
B = \frac{m_1-rm_2}{1-r}; ~ r = (\Gamma_1/\Gamma_2)^{1/3}
\end{equation}
interpolates in the region from $m_1=2m_\pi + m_\eta$ to $m_2=1.5$\,GeV where  
$\Gamma_1 = \Gamma_{\gamma\gamma}(m_1)+\Gamma_{3\pi}(m_1)$ and $\Gamma_2 = \Gamma_{gg}(m_a = 1.5\,{\rm GeV})$ is the inclusive decay to gluons (see below). This interpolation captures the rapidly growing decay rate by introducing $\propto m_a^3$ scaling.
$m_a=1.5$\,GeV is chosen to be the lower boundary of the perturbative description, and this choice bears significant uncertainty.

\item{\em Region 4, $m_a >1.5\,{\rm GeV}$, perturbative description.}
At $m_a \sim 1.1-1.5$\,GeV, many new additional hadronic decays of $a$ open up, 
$\pi_0 f_0(980)$, $\pi a_0$, $\eta f_0$, $\rho\rho$, $KK^*$ etc, 
quickly driving up the value for $\Gamma_{\rm tot}$. Asymptotically, the sum of all hadronic states
approaches the perturbative $a\to gluons$ answer. PBC recommends using the the  perturbative formula of $a$ decays to gluons
as a proxy for hadronic decays above $1.5$\,GeV:
\begin{equation}
\label{region4}
\Gamma_{\rm tot}(m_a>1.5\,{\rm GeV}) \approx \Gamma_{gg}  = \frac{m_a^3}{f_G^2} \times  \frac{\pi \alpha^2_s(m_a)}{2},
\end{equation}
This is an order-of-magnitude estimate, that cannot be improved using pertubation theory, and may only be improved with 
non-perturbative methods.

\item{\em Resonance regions, $m_a \sim m_\eta$, $m_a \sim m_{\eta'}$.}

In addition to the above expressions, one needs to add strong resonant contributions when $m_a$ becomes close to $m_\eta$ and $m_{\eta'}$. If the continuum - resonance interference is neglected, this is achieved via the following formulae, 
\begin{eqnarray}
\label{res}
\Gamma_{a-\eta,res} =  (2\pi^2\cos\theta_p)^2\times  \frac{F_\pi^2}{f_G^2}\times \frac{m_a^4\Gamma_\eta(m_a)}{(m_a^2-m_\eta^2)^2+m_\eta^2\Gamma_\eta^2(m_a)},~~ \\ \Gamma_{a-\eta',res} =  (2\pi^2\sin\theta_p)^2\times  \frac{F_\pi^2}{f_G^2}\times \frac{m_a^4\Gamma_{\eta'}(m_a)}{(m_a^2-m_{\eta'}^2)^2+m_{\eta'}^2\Gamma_{\eta'}^2(m_a)},
\end{eqnarray}
where $\Gamma_{\eta}(m)$ and $\Gamma_{\eta'}(m)$ are the energy-dependent widths $\Gamma(E)$ evaluated at $E=m_a$. 

Value for mixing angles, $\cos\theta_p \simeq 0.6$ and $\sin\theta_p \simeq 0.8$, are taken from  Ref. \cite{Fariborz:1999gr}. More details on mixing coefficients are given in the Appendix. 

Theoretical input is required in deriving $\Gamma_{\eta(\eta')}(E)$, where the main effect is due to the available phase space for $3\pi$ and 
$\eta 2\pi$ final states. PBC suggest using a simple approximate formula that reflects the growth of phases space. For $\eta$ meson we take
\begin{eqnarray}
\Gamma_\eta(E) = \Gamma_\eta\times \frac{f(E)}{f(m_\eta)};~~ f(E) = 
\frac{(E-3m_\pi)^{1.5}}{(E-3m_\pi)^{1.5}+(300\,{\rm MeV})^{1.5}}~{\rm for}~E>3m_\pi; \\ \nonumber
f(E) =0 ~{\rm for}~E< 3 m_\pi,
\end{eqnarray}
where $\Gamma_\eta$ is the total  width of [physical] $\eta$ meson.  
For $\eta'$, the same formula applies, with $\Gamma_\eta \to \Gamma_{\eta'}$ and $3m_\pi \to 2m_\pi +m_\eta$ substitutions. 
Better descrition can be achieved with the use of hadronic models. 

\end{itemize}

\vskip 0.5cm
\section{ALPs: production via $\pi^0, \eta, \eta'$ mixing}

If $m_a$ is below the hadronic scale of $\sim 4 \pi F_\pi$ ($F_\pi = 93$\,MeV), one can neglect heavy flavours and try
to use chiral perturbation theory by replacing $G\tilde G$ operator with using the anomaly equation. We use this equation for three light quarks 
($q_i = u,d,s$) in the following form, 
\begin{equation}
\frac{\alpha_s}{8\pi} G^b \tilde G^b = \sum_i \frac{m_*}{2m_i}\partial_\mu \bar q_i \gamma_\mu \gamma_5 q_i  -
m_*\sum_i \bar q_i i\gamma_5 q_i -\frac{\alpha}{4\pi} F\tilde F \sum_i\frac{N_cQ_i^2m_*}{m_i}
\end{equation}
where we suppress the Lorentz indices over the gluon and  photon fields strength, $G$ and $F$. Here $Q_i$ are the quark charges in units of $e$, $N_c=3$ and $m_* \equiv  (\sum_i m_i^{-1})^{-1}$. Dropping terms suppressed by $m_{u(d)}/m_s$, we have $m_* = m_um_d/(m_u+m_d)$ and
\begin{eqnarray}
\frac{\alpha_s}{8\pi} G^b \tilde G^b = \frac{m_d}{2(m_u+m_d)} \partial_\mu \bar u \gamma_\mu \gamma_5 u  
+ \frac{m_u}{2(m_u+m_d)} \partial_\mu \bar d \gamma_\mu \gamma_5 d  -
m_*\sum_i \bar q_i i\gamma_5 q_i -\frac{\alpha}{4\pi} F\tilde F \frac{4 m_d+m_u}{3(m_u+m_d)}
\end{eqnarray}
In the leading chiral order, the flavour-singlet $m_*\sum_i \bar q_i i\gamma_5 q_i $ combination can be neglected, 
and the total Lagrangian at low energy can be rewritten as 
\begin{equation}
\label{simplified}
{\cal L}_{\rm axion,l.e.} = {\cal L}_{\rm SM} + {\cal L}_{\rm DS} + 4\pi^2 \times \frac{a}{f_G} \frac{\alpha}{4\pi} \frac{4 m_d+m_u}{3(m_u+m_d)} F_{\mu\nu} \tilde F_{\mu\nu}  -
4\pi^2\times \frac{\partial_\mu a}{2f_G}\left(J_{A,\mu}^S +\frac{m_d-m_u}{m_u+m_d}J_{A,\mu}^T\right),
\end{equation}
where $J_{A,\mu}^{S}$ is the singlet, $\frac12 (\bar u \gamma_\mu \gamma_5 u  + \bar d \gamma_\mu \gamma_5 d) $,  
and $J_{A,\mu}^T$ is the triplet
axial-vector current, $\frac12 (\bar u \gamma_\mu \gamma_5 u  - \bar d \gamma_\mu \gamma_5 d) $. Interaction with 
$J_{A,\mu}^{S}$ leads to $a-\eta$ and $a-\eta'$ mixing, while interaction with $J_{A,\mu}^{T}$ results in 
$a-\pi^0$ mixing. 

Using the model that relates octet and singlet quark states to physical $\eta$ and $\eta'$ \cite{Fariborz:1999gr}, and usual rules 
for $\langle 0| J_{A,\mu} |pseudoscalar\rangle$ matrix elements, 
we transform the last term in (\ref{simplified}) to an on-shell 
mixing between $a$ and the pseudoscalars,
\begin{eqnarray}
{\cal L}_{\rm axion,l.e.} = ...
4\pi^2\times \frac{\partial_\mu a}{2f_G}\left(J_{A,\mu}^S +\frac{m_d-m_u}{m_u+m_d}J_{A,\mu}^T\right)\nonumber\\
\to 4\pi^2\times \frac{F_\pi}{2f_G}\left(
\partial_\mu a \partial_\mu \eta \times \cos\theta_p+ \partial_\mu a \partial_\mu \eta' \times \sin\theta_p
+ \partial_\mu a \partial_\mu \pi_0 \frac{m_d-m_u}{m_u+m_d} \right ),
\end{eqnarray}
where $\cos\theta_p \simeq 0.6$ and $\sin\theta_p \simeq 0.8$ related physical $\eta$ and $\eta'$ with octet and singlet
combinations \cite{Fariborz:1999gr}. 

\label{sec:B}

\clearpage
\bibliographystyle{JHEP}
\bibliography{biblio.bib}

\end{document}